\documentclass[preprint,12pt]{elsarticle}
\usepackage{makecell}
\usepackage{booktabs}
\usepackage{multirow}
\usepackage[T1]{fontenc}
\usepackage{amssymb}
\usepackage{appendix}
\usepackage{amsmath}
\usepackage{longtable}
\usepackage[utf8]{inputenc}
\usepackage[T1]{fontenc}
\usepackage{xcolor}
\usepackage{amsmath,amssymb,amsfonts}
\usepackage{textcomp}
\usepackage{float}
\usepackage{array}
\usepackage{booktabs} 
\usepackage{bm}
\usepackage{epsfig}
\usepackage{soul}
\usepackage{pgf}
\usepackage{multirow}
\usepackage{bigstrut}
\usepackage{comment}
\usepackage{rotating}
\usepackage{pdflscape}
\usepackage{amsmath} 
\usepackage{lscape}
\usepackage{subcaption}
\usepackage{setspace}
\usepackage{amssymb}
\usepackage{amsmath}
\usepackage{relsize}
\usepackage{xcolor}
\usepackage{url}
\usepackage{algorithm}
\usepackage{algpseudocode}
\usepackage{natbib}
\usepackage{graphicx}
\usepackage{multirow}
\usepackage{multicol}
\usepackage{comment}
\usepackage{booktabs} 
\usepackage{array} 
\usepackage{balance} 
\usepackage{soul}
\usepackage{siunitx}
\usepackage{graphicx}
\usepackage{subcaption} 
\usepackage{float}
\usepackage{xcolor}
\usepackage{comment}
\usepackage{graphicx}
\usepackage{tikz}
\usetikzlibrary{positioning, shapes, arrows.meta}
\usepackage{array}
\usepackage{xcolor}
\usepackage{pifont}

\newcommand{\cmark}{\textcolor{green!60!black}{\ding{51}}}
\newcommand{\xmark}{\textcolor{red}{\ding{55}}}

\newcolumntype{C}[1]{>{\small\centering\arraybackslash}m{#1}}
\newcolumntype{L}[1]{>{\small\raggedright\arraybackslash}m{#1}}

\usepackage[margin=2.5cm]{geometry}

\journal{a journal}

\begin{document}

\begin{frontmatter}

%\title{Quantum Optimization for the Steiner Traveling Salesman Problem with Time Windows and Pickup and Delivery} 
\title{Steiner Traveling Salesman Problem with Time Windows and Pickup–Delivery: integrating classical and quantum optimization}

%% Authors
\author[aff1]{Alessia Ciacco\corref{cor1}}
\ead{alessia.ciacco@unical.it}

\author[aff1]{Francesca Guerriero}
\ead{francesca.guerriero@unical.it}

\author[aff2]{Eneko Osaba}
\ead{eneko.osaba@tecnalia.com}

%% Corresponding author footnote
\cortext[cor1]{Corresponding author}

%% Affiliations
\affiliation[aff1]{organization={Department of Mechanical, Energy and Management Engineering, University of Calabria},
            addressline={Via Pietro Bucci},
           city={Rende},
            postcode={87036},
            state={CS},
            country={Italy}}

\affiliation[aff2]{organization={TECNALIA, Basque Research and Technology Alliance (BRTA)},
           addressline={Astondo Bidea, Edificio 700},
            city={Derio},
            postcode={48160},
            state={Bizkaia},
            country={Spain}}

%% Abstract
\begin{abstract}
%% Text of abstract
We propose the Steiner Traveling Salesman Problem with Time Windows and Pickup and Delivery, an advanced and practical extension of classical routing models. This variant integrates the characteristics of the Steiner Traveling Salesman Problem with time-window constraints, pickup and delivery operations and vehicle capacity limitations. These features closely mirror the complexities of contemporary logistics challenges, including last-mile distribution, reverse logistics and on-demand service scenarios.
To tackle the inherent computational difficulties of this NP-hard problem, we propose two specialized mathematical formulations: an arc-based model and a node-oriented model, each designed to capture distinct structural aspects of the problem. 
We further introduce a preprocessing reduction method that eliminates redundant arcs, significantly enhancing computational performance and scalability. 
We also introduce a benchmark instance generator designed to create diverse and scalable test scenarios, providing a flexible framework for assessing the performance and robustness of both classical and quantum optimization methods.
Both formulations are implemented using classical and quantum optimization approaches. In particular, the classical models are solved with Gurobi, whereas the quantum implementation is carried out on D-Wave’s \texttt{LeapCQMHybrid} platform, a hybrid quantum–classical environment that integrates quantum annealing with classical optimization techniques for constrained problem solving. Numerical experiments are conducted to validate the proposed formulations and the preprocessing reduction method. The analyses performed assess the structural properties of the two models, their computational behavior, and the impact of preprocessing on problem size and solution efficiency.

%\color{black}
%Experimental results demonstrate that hybrid quantum approaches are capable of solving problem instances of realistic size, underscoring their potential as a transformative tool for next-generation routing optimization.
\end{abstract}

%% Keywords

\end{frontmatter}

\section{Introduction}
Routing problems are among the most central and extensively studied challenges in combinatorial optimization, due to their wide range of real-world applications in domains such as logistics, transportation and urban distribution. In recent years, the relevance of these problems has grown significantly, driven by the digital transformation of supply chains, the exponential rise of e-commerce, and the increasing demand for reliable, customized solutions. Effective route optimization not only reduces operational costs but also contributes to lower emissions, improved fleet utilization and higher service quality.

The Vehicle Routing Problem (VRP) is a fundamental combinatorial optimization problem in logistics and transportation. It focuses on determining the most efficient routes for a fleet of vehicles to deliver goods to a set of customers, starting and ending at one or more depots \cite{ghiani2003real, braekers2016vehicle}.
Among the most studied extensions of the VRP is the variant involving time windows, where each customer must be served within a specified time interval. This reflects realistic delivery scenarios, where users often select preferred delivery slots when placing an order. These temporal constraints add realism to the model but also significantly increase its computational complexity, as feasible solutions must simultaneously satisfy both spatial and temporal requirements \cite{pugliese2017last,di2020trucks}.

Another practically relevant variant is the VRP with pickup and deliveries (VRP-PD), in which each request involves a pair of locations, one for pickup and one for delivery, subject to precedence and capacity constraints. This model reflects the growing complexity of modern logistics systems, where vehicles must not only deliver but also collect items, such as in reverse logistics, product returns or shared mobility services. VRP-PD based models have a wide range of applications, including ride-sharing, e-commerce, hospital logistics and urban freight distribution \cite{li2001metaheuristic,desaulniers2002vrp, ropke2006adaptive}.

The simultaneous integration of time windows and pickup and delivery constraints results in a highly challenging class of routing problems, due to the combined presence of spatial, temporal and logistical dimensions. These problems are known to be NP-hard and their complexity grows rapidly with problem size, making them particularly difficult for both exact and heuristic classical methods.

Furthermore, the Steiner Traveling Salesman Problem (STSP) is a combinatorial optimization problem that generalizes both the Traveling Salesman Problem (TSP) and the Steiner Tree Problem (STP). It aims to find a minimum-cost route that visits a subset of required nodes, referred to as required nodes, in a graph at least once. The route may include optional Steiner nodes, non-required vertices that help reduce the overall cost and allows edges to be traversed multiple times, offering flexibility not present in the classical TSP.

From the TSP, the STSP inherits the objective of constructing an efficient tour. However, unlike the TSP, which requires visiting all nodes exactly once, the STSP only requires covering the required nodes and permits the reuse of nodes and edges. From the STP, it adopts the idea of connecting key nodes while using intermediate ones to minimize cost, though the STSP seeks a tour rather than a tree structure.

This hybrid structure makes the STSP particularly suitable for applications in network design, logistics, and transportation, where only a subset of critical points must be connected efficiently within an existing infrastructure. As with the TSP and STP, the STSP is NP-hard, and solving large instances exactly remains computationally challenging.

%Furthermore, the STSP has been proposed as a generalization of the classical TSP, where only a subset of nodes, referred to as request nodes, must be visited. This model is especially suited to scenarios where certain locations must be visited while others serve purely as connection points. Typical applications of the STSP include infrastructure maintenance, waste collection, telecommunications networks and school transportation \cite{cornuejols1985traveling,fleischmann1985cutting,letchford2013compact,rodriguez2019steiner}.
Despite its practical potential, extending the STSP to account for real-world constraints such as time windows and pickup and delivery has received limited attention in the literature, possibly due to the high computational complexity these extensions introduce.  %In this work, we propose and study a novel and realistic variant, which we refer to as the Steiner Traveling Salesman Problem with Time Windows and Pickup and Delivery (STSPTWPD). This new formulation captures the partial coverage structure of the STSP while integrating the temporal constraints of customer availability and the precedence and capacity constraints typical of pickup and delivery operations. As such, the STSPTWPD can model a wide array of real-world problems in logistics and transportation that cannot be adequately represented by classical STSP or VRP formulations.\\

In this work, we address a significant gap in the current literature by introducing the Steiner Traveling Salesman Problem with Time Windows and Pickup–and–Delivery (STSP-TWPD), a novel and practically motivated routing problem that, to the best of our knowledge, has not been explored previously. This new variant extends the classical STSP by incorporating 
both temporal feasibility requirements and pickup–delivery relations, thereby capturing key features of modern urban logistics scenarios.

To model this problem, we propose two mathematical formulations. The first, an Arc-Based Formulation (ABF), builds upon traditional STSP models by assigning routing variables to arcs and incorporating constraints to enforce service times, time-window consistency, and pickup–delivery precedence. The second, a Node-Based Formulation (NBF), introduces a different modeling perspective in which routing decisions are expressed at the node level.

The literature currently lacks formal definitions or dedicated analyses of the Steiner TSP with either time windows (STSP-TW) or pickup–and–delivery constraints (STSP-PD). We therefore provide the first explicit definitions of these variants and evaluate them within a unified modeling and experimental framework. This allows us to disentangle and quantify the individual effects of temporal and precedence constraints, while establishing new benchmark problems and a methodological baseline for future research on Steiner-based routing.

To support the solution of these problems, we introduce a classic preprocessing procedure that removes redundant arcs and reduces overall model complexity. This reduction step is designed to handle the additional structural intricacies introduced by time-window and pickup–delivery constraints in the STSP-TWPD, thereby improving tractability and computational efficiency.

%In addition, we introduce a preprocessing-based aimed at eliminating unnecessary arcs and reducing model complexity. This step %significantly improves the efficiency and scalability of the formulation, especially when solved with hybrid quantum-classical solvers.
%is designed as a preprocessing procedure applicable to both classical and hybrid quantum–classical optimization frameworks. The proposed reduction method has been specifically designed to handle the additional structural and operational complexities introduced by time windows and pickup-and-delivery constraints in the STSP-TWPD.
%A further contribution of this work lies in the development of a benchmark generation 
We develop a reproducible benchmark generation procedure for STSP-TWPD instances. This includes the specification of rules for generating time windows, service durations, and pickup-delivery requests, providing a foundation for reproducible and meaningful experimental analysis.

In addition to the classical solution obtained with Gurobi, the study further examines a quantum–hybrid approach for solving the proposed problem.

In recent years, quantum computing has emerged as a novel paradigm for tackling combinatorial optimization problems that are computationally demanding for classical algorithms. Although it has shown potential in this field, handling complex constraint structures remains a major challenge for current quantum optimization approaches. Two main paradigms currently define the landscape of quantum computation: gate-based quantum computing and Quantum Annealing (QA).

Gate-based systems perform universal quantum computation by applying sequences of quantum logic gates to qubits, enabling algorithms such as Shor’s factorization~\citep{shor1994algorithms} and Grover’s search~\citep{grover1996fast}, which theoretically provide exponential or quadratic speed-ups over classical methods. In practice, however, these algorithms require large-scale, fault-tolerant architectures with thousands of error-corrected qubits, resources that remain far beyond current technological capabilities~\citep{preskill2018quantum}. As a result, gate-based methods are mostly limited to proof-of-concept demonstrations and demand specialized expertise in quantum circuit design, state manipulation, and error correction, restricting their use in applied optimization contexts.

QA, on the other hand, represents a more specialized and technologically mature approach tailored for optimization. It exploits quantum tunneling and adiabatic evolution to search for the minimum of an energy landscape encoded as an Ising or Quadratic Unconstrained Binary Optimization (QUBO) model~\citep{glover2018tutorial,glover2020quantum}. This paradigm is particularly appealing because many combinatorial optimization problems can, in principle, be formulated as QUBOs, and modern hardware platforms such as D-Wave’s quantum annealers already allow the execution of optimization instances. QA has therefore been explored in a wide range of combinatorial optimization problems, including scheduling~\citep{venturelli2016job, perez2024solving, ciacco2025Educational}, routing~\citep{holliday2025advanced, osaba2026quantum}, facility location~\citep{ciacco2026facility, malviya2023logistics}, and packing~\citep{de2022hybrid, garcia2022comparative}.

Despite these efforts, practical limitations persist. The associated energy landscapes often exhibit high ruggedness and numerous local minima, which can hinder the convergence of QA and variational algorithms~\citep{abbas2024challenges}. Recent analyses have also emphasized the gap between demonstrating \emph{feasibility} of quantum formulations and achieving any meaningful notion of \emph{quantum advantage}. In particular, Smith-Miles et al.~\citep{smith2025travelling} argue that current quantum optimization pipelines remain highly sensitive to modeling choices, penalty parameter tuning, hardware noise, and embedding constraints, and therefore exhibit no clear scaling benefits over state-of-the-art classical methods. Their conclusions reinforce the need for a cautious and honest interpretation of existing QA results, especially in routing problems where instance sizes remain small.

Although we are aware of the current technical limitations of quantum optimization, in this work we nevertheless explore the use of a Constrained Quadratic Model (CQM)-based approach to solve the problem. To this end, we employ the CQM framework implemented through D-Wave’s \texttt{LeapCQMHybrid} solver. Unlike traditional QUBO formulations, the CQM framework preserves explicit linear and quadratic constraints alongside the quadratic objective function, enabling feasibility to be enforced directly rather than through penalty terms. The hybrid quantum–classical architecture delegates constraint handling to a classical optimization layer while leveraging the quantum annealer to explore the combinatorial search space. This makes the CQM approach a more stable and expressive modeling environment for complex constrained optimization problems, reducing the reliance on manual penalty tuning and preserving the structural integrity of the original formulation.

The main contributions of this work are as follows:
\begin{itemize}
%%\color{red}
\item We introduce a novel %framework 
and previously unaddressed variant of the STSP, the STSP-TWPD, which extends the STSP by incorporating time windows and pickup and delivery constraints;
\item We propose two novel formulations for the STSP-TWPD: %, which simultaneously integrate spatial coverage, scheduling and logistical constraints
    an arc-based model, ABF, and  % extends existing formulations from the literature, while the 
    a node-based model, NBF;
    %is specifically designed to improve performance within CQM-based quantum solvers.
    \item We formally define and analyze two additional Steiner-based routing variants (STSP-TW and STSP-PD), which, to the best of our knowledge, have not been previously addressed in the literature;
    \item We present a reduction method that enhances the models by eliminating unnecessary arcs and minimizing model complexity;
    
\item We introduce a novel procedure for generating STSP-TWPD instances. In particular, we define a method for the creation of time windows, service times and customer demands for STSP-TWPD. %a contribution that, to the best of our knowledge, has not yet been addressed in the current literature.
%%\color{black}
        \item We implement and solve the proposed models using classical optimization tools as well as D-Wave’s hybrid quantum–classical solver, assessing their applicability to instances of practical relevance. %(\texttt{LeapCQMHybrid}). %, demonstrating their practical feasibility on problem instances of realistic size.    
\end{itemize}

The remaining of this paper is structured as follows. In Section~\ref{sec:Related works}, we review the scientific literature related to STSP, VRP-TWPD and quantum approaches for combinatorial optimization problems. 
Section~\ref{sec:problem_statement} provides a detailed description of the problem under study, including its main characteristics and motivation. Section~\ref{sec:mathematical_formulation} presents the formal mathematical formulations of the problem, for both the ABF and NBF variants, along with a discussion on their equivalence, dimensionality, and linearization mechanisms. Two simplified problem variants, namely STSP-TW and STSP-PD, are also derived from the general STSP-TWPD formulation. Section~\ref{sec:preprocessing} introduces the arcs filtering and graph reduction method aimed at improving solution efficiency. Section~\ref{sec:complexity} analyzes the dimensionality and computational complexity of the formulations, both before and after the application of the preprocessing method. 
Section~\ref{sec:experiments} reports the computational experiments, including the instance generation process, solver setup, and numerical analysis performed with both classical and quantum approaches. 
Section~\ref{sec:conclusions} summarizes the concluding remarks and prospects for future studies.

\section{Related works}\label{sec:Related works}
This section presents a brief %structured and comprehensive 
overview of the current state of the art, aiming to position the proposed research within the broader scientific context. To the best of our knowledge, no prior work has directly addressed the STSP-TWPD, neither through classical optimization methods nor via quantum computing approaches such as QA.

%\color{red} 
To bridge this gap, the literature is reviewed along two key dimensions. First, we examine classical methods developed for the STSP and the VRP-TWPD, which form the methodological foundation for our work. Second, we discuss recent routing studies that employ quantum or hybrid quantum–classical optimization approaches, highlighting current progress and remaining limitations.
%\color{black}
%combinatorial structure of the model under study. Second, we survey recent advancements in the application of quantum computing, particularly QA, to combinatorial optimization problems, including routing and scheduling, highlighting works that demonstrate the potential of hybrid quantum-classical techniques.
The analysis provides insights into the evolution of relevant problem formulations and solution strategies, offering the conceptual and technical foundation for the novel contribution presented in this paper. %Additionally, we describe the main scientific contributions of this work.
%%Special attention is given to the only existing work that addresses the STSP using QA, namely \cite{Ciacco2025STSP}, where the authors formulate the problem as a QUBO and apply a novel arc-reduction preprocessing technique. This review concludes with a discussion of quantum-specific techniques and solvers, providing a foundation for the approach proposed in this work.
\subsection{Classical approaches}
\subsubsection{Works on the STSP}
%\paragraph{Historical background.}
The STSP was introduced by Cornuéjols et al.~\cite{cornuejols1985traveling} and Fleischmann~\cite{fleischmann1985cutting}. Cornuéjols et al.~\cite{cornuejols1985traveling} presented the Graphical Traveling Salesman Problem, a generalization of the TSP defined on arbitrary graphs, where nodes and edges can be traversed multiple times. The goal is to find a tour that visits all nodes at least once, allowing for repeated traversals. This formulation is particularly suited for structured environments with constrained connectivity, such as city maps or factory layouts.
Fleischmann~\cite{fleischmann1985cutting} proposed an exact solution method for the STSP based on a cutting plane approach. His model only considers the subset of required nodes and relaxes the requirement of Hamiltonian cycles, allowing for shorter tours. The formulation introduces valid inequalities tailored to the STSP to strengthen the linear relaxation, enabling exact solution of instances on realistic road networks.
%\paragraph{Compact mathematical formulations.}
Letchford et al.~\cite{letchford2013compact} introduced three compact formulations for the STSP, leveraging techniques developed for the classical TSP. These include the single-commodity flow, multi-commodity flow, and time-staged formulations. Computational results indicate that the multi-commodity formulation is especially powerful, capable of solving instances with over 200 nodes using standard branch-and-bound algorithms. Notably, it achieves the same linear programming lower bound as Fleischmann’s exponential model, representing a significant computational advancement. 
Rodríguez-Pereira et al.~\cite{rodriguez2019steiner} proposed a recent compact ILP formulation based on the observation that any optimal solution to the STSP traverses each edge at most twice. Their model employs two sets of binary variables to represent the first and second traversal of each edge, significantly reducing model size while preserving accuracy. This formulation enables exact solution of instances with up to 500 nodes, outperforming prior approaches.
%\paragraph{Heuristic approaches.}
Interian and Ribeiro~\cite{interian2017grasp} developed a metaheuristic based on the Greedy Randomized Adaptive Search Procedure, enhanced with path-relinking and periodic restarts. Their approach incorporates a reduced 2-opt neighborhood, problem-specific greedy criteria, and local search intensification. The use of path-relinking allows structured exploration of elite solution paths, improving overall solution quality. Experimental results confirm its effectiveness on large, sparse STSP instances.
Álvarez-Miranda et al. (2019)~\cite{alvarez2019note} proposed transforming the STSP into a standard TSP by computing the metric closure over the required nodes, where each edge represents the shortest-path distance in the original graph. 

%\paragraph{Multi-agent and dynamic variants.}
%Liu et al.~\cite{liu2021m} formally introduced the \emph{multi-agent Steiner Traveling Salesman Problem with online edge blockages}, motivated by real-world urban delivery scenarios. The authors examine two variants: \emph{MinMax}, which minimizes the maximum cost among $m$ agents, and \emph{MinSum}, which minimizes the total cost. They establish theoretical lower bounds on the competitive ratios for online algorithms, $\lfloor k/m \rfloor + 1$ for MinMax and $k/m + 1$ for MinSum, and propose two algorithms: \emph{ForestTraversal} for MinMax (with competitive ratio $\leq 1.5k + 5$) and \emph{Retrace} for MinSum (with ratio $\leq 2k + 2$). These algorithms rely on depth-first traversals and dynamically reassign tasks when unexpected edge failures occur.
%Zhang et al.~\cite{zhang2022asymptotically} improved upon Liu et al.’s work by presenting the \emph{Traversal} algorithm for the online MinMax STSP. This algorithm achieves an asymptotically tight competitive ratio of $\lfloor k/m \rfloor + 10$ and can be adapted to the MinSum version with a ratio of at most $5m + k - 1$. The main innovation lies in the balanced and dynamic task distribution among agents, which allows for improved resilience and performance under uncertainty due to road blockages.

\subsubsection{Works on the VRP-TWPDP}
%The VRP-TWPDP has been extensively studied, especially in the context of dynamic and real-time logistics, where operational constraints and customer demands evolve continuously.
%\paragraph{Exact and dynamic methods.}
Dumas et al. (1991)~\cite{dumas1991pickup} proposed a foundational exact method for the VRP-TWPDP, incorporating vehicle capacity, pickup and delivery precedence, and time window constraints. Their approach is based on a column generation scheme applied to a set-partitioning formulation, where each column represents a feasible route. The subproblem, formulated as a resource-constrained shortest path, is solved via dynamic programming and provides strong lower bounds.
Chang et al. (2003)~\cite{chang2003real} extended the VRP-TWPDP into a real-time environment, introducing a formulation where customer requests arrive dynamically and must be handled immediately. The authors proposed a heuristic composed of three main components: route construction, route improvement using an anytime algorithm, and a tabu search procedure. Parameters for the tabu search are optimized via Taguchi orthogonal arrays. Their results show that the method is efficient in reducing operational costs under real-time constraints. Both approaches provide solid foundations for practical vehicle routing under temporal and resource constraints.
%\paragraph{Routing under real-time and time-window constraints.}
%Caramia et al. (2002)~\cite{caramia2002routing} addressed a dynamic many-to-many dial-a-ride variant of the VRP-TWPDP for urban taxi fleet management, using the road network of Rome as a testbed. Their heuristic relies on an iterative improvement framework triggered by real-time events such as new requests or vehicle failures. At its core lies the {Single Cab Routing} routine, which computes optimal single-vehicle routes using a dynamic programming algorithm enhanced with A* search.
Fabri and Recht (2006) \cite{fabri2006dynamic} proposed a dynamic extension of the VRP-TWPDP in which each customer request includes two separate time windows, one for pickup and one for delivery. Departing from the stretch-factor approach in~\cite{caramia2002routing}, their model allows explicit time window constraints and vehicle waiting. The solution combines exact single-vehicle routing via dynamic programming and a fast heuristic to allocate requests across vehicles.
As shown in Table~\ref{TabellaStatoArte1}, the classical STSPVRP-PD (i.e., the variant without time-window constraints or paired pickup-and-delivery requirements) has been addressed at very large scales using both exact algorithms and sophisticated heuristics.

The problem considered in this work, however, presents a substantially richer structure. The addition of time windows combined with pickup-and-delivery operations introduces temporal feasibility constraints and precedence relationships that significantly alter the underlying combinatorial landscape.
A related line of research is the VRP-TWPD, which represents the closest classical counterpart. Early exact methods, such as those by Dumas et al.~\cite{dumas1991pickup}, already demonstrated how the interaction between time-window feasibility and pickup–delivery pairing severely restricts tractable instance sizes, with exact algorithms typically limited to a few dozen requests. Subsequent heuristic and metaheuristic studies achieve larger scales, up to a few hundred requests, yet consistently highlight the intrinsic difficulty of jointly managing precedence, time windows, and routing decisions.

{\small
\begin{table}[ht]
\centering
\caption{
Comparison of the main contributions in the literature on the STSP and its variants and on VRP-TWPD and its related extensions. The table includes the following columns: 
\textbf{Reference} lists the original authors; 
\textbf{STSP} indicates whether the work addresses the classical STSP; 
\textbf{TW} specifies whether time windows are considered; 
\textbf{PD} identifies whether pickup-and-delivery constraints are included; 
\textbf{Max size} reports the largest instance solved (in number of nodes); 
\textbf{Approach} summarizes the nature of the method proposed (theoretical, exact, or heuristic).}\label{TabellaStatoArte1}
\renewcommand{\arraystretch}{1.25}
\begin{tabular}{
    L{4.5cm}   % Riferimento
    C{1.2cm}   % STSP
    C{1.2cm}   % TW
    C{1.2cm}   % PD
    C{2.8cm}   % Max size
    C{2.6cm}   % Metodo
}
\hline
\textbf{Reference} &
\textbf{STSP} & \textbf{TW} & \textbf{PD} &
\textbf{Max size} &
\textbf{Approach} \\
\hline

Cornuéjols et al.~\cite{cornuejols1985traveling} & \cmark & \xmark & \xmark & -- & theoretical \\

Fleischmann~\cite{fleischmann1985cutting} & \cmark & \xmark & \xmark & 292 nodes & exact \\

Letchford et al.~\cite{letchford2013compact} & \cmark & \xmark & \xmark & 200 nodes & exact \\

Rodríguez-Pereira et al.~\cite{rodriguez2019steiner} & \cmark & \xmark & \xmark & 500 nodes & exact \\

Interian and Ribeiro~\cite{interian2017grasp} & \cmark & \xmark & \xmark & 3353 nodes & heuristic \\

Álvarez-Miranda et al.~\cite{alvarez2019note} & \cmark & \xmark & \xmark & 300 nodes & exact \\

Dumas et al.~\cite{dumas1991pickup} & \xmark & \cmark & \cmark & 15 nodes & exact \\

Chang et al.~\cite{chang2003real} & \xmark & \cmark & \cmark & 100 nodes & heuristic \\

Fabri and Recht~\cite{fabri2006dynamic} & \xmark & \cmark & \cmark & 1000 nodes & heuristic \\

\hline
\end{tabular}
\end{table}
}

\subsection{Quantum Approaches}\label{sec:quantum}

A broad range of combinatorial optimization problems have been explored through QA. For a comprehensive and structured discussion of existing QA applications, 
we refer the reader to the surveys by Ciacco et al.~\cite{ciacco2025review},  
Osaba et al.~\cite{osaba2022systematic}, Yulianti and Surendro~\cite{yulianti2022annealing} and Muhamediyeva et al.~\cite{muhamediyeva2026optimization}. Ciacco et al.~\cite{ciacco2025review} offer a detailed examination of quantum algorithm design across multiple domains, including healthcare, finance, production planning, and logistics, together with a systematic classification of methodological frameworks. 
Osaba et al.~\cite{osaba2022systematic} present a unified and end-to-end review of 
eighteen years of research (2004–2021) at the intersection between quantum computing and routing problems, analysing 53 contributions and identifying emerging trends and open challenges.
In addition, Yulianti and Surendro~\cite{yulianti2022annealing} provide a systematic review of QA implementations, mapping 229 studies across thirteen application areas and four methodological families. Their work emphasizes how QA research has expanded in parallel with the evolution of commercial quantum annealers, highlighting both performance gains and persistent hardware-driven limitations.
Muhamediyeva et al.~\cite{muhamediyeva2026optimization} examines the practical limitations of both quantum annealing and gate-based frameworks, and provides an accessible discussion of the quantum approximation optimization and quantum alternating operator ansatz paradigms.

Within this wider landscape, the only quantum-based study explicitly addressing the STSP is  presented by Ciacco et al.~\cite{ciacco2025steiner}. In their work, the authors reformulate the STSP as a QUBO model derived from an integer linear programming formulation,  introducing a dedicated preprocessing phase aimed at removing redundant arcs and 
reducing the dimensionality of the problem before its quantum optimisation. The resulting model is evaluated using both the D-Wave LeapBQM simulator and a Quantum Processing Unit (QPU), demonstrating that the preprocessing step substantially enhances computational performance and contributes to obtaining higher-quality solutions.
%Against this broader backdrop, we focus here on reviewing recent and particularly significant QA-based contributions that are most relevant to the problem addressed in this study, placing specific emphasis on routing-related formulations. This targeted selection reflects the current state of the art in QA for routing problems, while avoiding the redundancy of replicating the exhaustive analyses already provided by the aforementioned surveys.

Against this broader backdrop, we focus here on reviewing recent and particularly significant 
QA-based contributions that are most relevant to the problem addressed in this study, placing 
specific emphasis on routing-related formulations. In particular, our analysis concentrates on the most recent developments appearing after 2022, thereby capturing the latest progress in the use of QA for routing problems. This targeted selection reflects the current state of the art while avoiding the redundancy of replicating the exhaustive analyses already provided by the aforementioned surveys.

%%\color{black}
%Venturelli et al.~\cite{venturelli2016job} and Pérez Armas et al.~\cite{perez2024solving} investigate the Job Shop Scheduling Problem (JSSP). In their work, the JSSP is formulated as a QUBO model and solved on D-Wave quantum annealers using different pre-processing strategies to optimize scheduling instances. The results are then compared with state-of-the-art classical approaches to assess the effectiveness of quantum methods. Similarly, Bożejko et al.~\cite{bozejko2024scheduling} propose a hybrid quantum-classical branch-and-bound algorithm for the single-machine total weighted tardy jobs problem, while Orts et al.~\cite{orts2023parallel} present a compact QUBO formulation for the unrelated parallel machine scheduling problem with priorities and switching delays, explicitly tailored to hardware limitations and validated on the D-Wave platform.
%Beyond scheduling, QA has also been explored in logistics and facility location. Ding et al.~\cite{ding2021implementation} apply it to logistic network design, whereas Malviya et al.~\cite{malviya2023logistics} focus on optimizing the location of distribution centers in package delivery. Ciacco et al.~\cite{ciacco2026facility} investigate the Two-Level Facility Location Problem, introducing a preprocessing strategy that reduces network size and improves scalability on quantum devices.
%Additional applications include the one-dimensional bin packing problem (1-BPP), studied by García De Andoin et al.~\cite{de2022hybrid,garcia2022comparative}, and VRPs. 
Tambunan et al.~\cite{tambunan2022quantum} investigate QA approaches to the VRP with Weighted Road Segments, formulating the problem as a QUBO and solving it on a D-Wave QPU. %Holliday et al.~\cite{holliday2025advanced} extend the use of QA to the VRP-TW, 
%employing D-Wave’s hybrid CQM solver to handle time-window constraints efficiently. 
Le et al.~\cite{le2023quantum} address the Orienteering Problem, another routing-related variant, through a QUBO formulation executed on D-Wave systems. 
Sinno et al.~\cite{sinno2023performance} provide an extensive empirical evaluation of commercial QA platforms on the Capacitated VRP (CVRP), conducting over 30 hours of experiments using D-Wave hybrid solvers. Their findings reveal that while QPU time remains extremely small, solution quality degrades with increasing constraint density rather than with problem size alone. In particular, the authors highlighting how modelling choices, and especially the number and structure of constraints in the QUBO or CQM formulation, critically influence performance. 
Further contributions include the work of Mori and Furukawa~\cite{10.3389/fphy.2023.1129594}, who introduce the Adjuster Routing Problem, a disaster-response VRP in which insurance companies must deploy large numbers of adjusters across affected areas. Their QUBO formulation incorporates multiple operational constraints, and numerical experiments on D-Wave demonstrate that both solution quality and computational effort are highly sensitive to parameter selection.
Recent contributions have further extended QA to realistic logistics settings. Osaba et al.~\cite{osaba2024solving} consider advanced delivery settings involving heterogeneous vehicle fleets, delivery priorities, and dual capacity constraints (weight and volume). Their experiments on real-world instances using D-Wave’s hybrid CQM solver confirm the 
practical relevance of quantum–classical methods for operational logistics. 
Mario et al.~\cite{mario2024hybrid} propose two hybrid classical-quantum strategies for real-time route optimisation. Their methods decompose the CVRP into clustering and routing phases, using classical Fuzzy C-Means for capacitated clustering and QA for solving either CVRP or TSP subproblems. Osaba et al.~\cite{osaba2025package} investigate the VRP-TWPD, integrating realistic constraints such as simultaneous pickup–delivery operations, time-window restrictions, and vehicle-specific mobility limitations. Their results, obtained through 
D-Wave’s Hybrid CQM Leap framework on seven classes of instances, reinforce the feasibility and robustness of hybrid QA approaches for complex logistics and routing environments.

Table~\ref{tab:QuantumLiterature} summarizes the key characteristics of the QA-based studies discussed above, including the type of problem addressed, the largest instance size tested, and the quantum approach employed.

%\color{black}
%For a comprehensive overview of existing applications of QA in transportation and logistics, we refer the reader to the surveys by Ciacco et al.~\cite{ciacco2025review} and Osaba et al.~\cite{osaba2022systematic}. These reviews analyze the evolution of hybrid quantum-classical methodologies, identify current hardware limitations, and outline key future research directions, particularly in the context of smart logistics systems.

{\small
\begin{table}[ht]
\centering
\caption{Overview of key QA studies in routing-related combinatorial problems. The table includes the following columns: \textbf{Reference}, which lists the authors of each study; 
\textbf{Problem}, which specifies the combinatorial optimization problem addressed; 
\textbf{Max size}, which reports the largest instance solved; 
\textbf{Quantum Approach}, which summarizes the type of quantum method employed.
}
\label{tab:QuantumLiterature}
\renewcommand{\arraystretch}{1.25}
\begin{tabular}{
    L{4.5cm}
    C{4.0cm}
    C{2.6cm}
    C{3cm}
}
\toprule
\textbf{Reference} & \textbf{Problem} & \textbf{Max size} & \textbf{Quantum Approach} \\
\midrule

\multirow{2}{*}{Ciacco et al.~\cite{ciacco2025steiner}} &
\multirow{2}{*}{STSP} &
9 nodes &
Hybrid BQM D-Wave Leap \\
\cline{3-4}
& & 5 nodes & QPU D-Wave \\
\hline
Tambunan et al.~\cite{tambunan2022quantum} & VRP with Weighted Road Segments & 5 vehicles & QPU D-Wave \\
\hline
Le et al.~\cite{le2023quantum} & Orienteering Problem & 7 cities & QPU D-Wave \\
\hline
Sinno et al.~\cite{sinno2023performance} & CVRP & 80 customers & Hybrid CQM D-Wave Leap \\
\hline
Mori and Furukawa~\cite{10.3389/fphy.2023.1129594} & Adjuster Routing Problem & 20 adjusters & Hybrid CQM D-Wave Leap \\
\hline
Osaba et al.~\cite{osaba2024solving} & VRP  & 29 customers & Hybrid CQM D-Wave Leap \\
\hline
Mario et al.~\cite{mario2024hybrid} & CVRP & 37 customers & QPU D-Wave \\
\bottomrule
Osaba et al.~\cite{osaba2025package} & VRP-TWPD & 24 nodes & Hybrid CQM D-Wave Leap \\

\bottomrule

\end{tabular}
\end{table}
}

\section{Problem definition}\label{sec:problem_statement}
%\subsection{Characteristics of the problem}\label{subsec:characteristics_of_problem}
We introduce the STSP-TWPD, that integrates two important real-world constraints: time windows and pickup and delivery requirements. 
In the STSP-TWPD, a single vehicle starts and ends at a central depot and must serve all required nodes, i.e. customers that may request either pickups or deliveries. %and must serve all required nodes, which include customers that may request either pickups or deliveries.% which include customers and pickup/delivery points. %Each required node has an associated time window, a predefined interval during which service must begin
Each required node is associated with a time window, defined by a lower and upper bound, during which service must begin, reflecting real-world constraints such as store opening hours or scheduled delivery slots. %The pickup and delivery aspect imposes precedence constraints, requiring that goods picked up at certain nodes be delivered to specific destinations, with the vehicle visiting each pickup location before its corresponding delivery location while respecting all time windows.
The pickup and delivery component imposes precedence constraints, whereby certain nodes correspond to pickup requests and others to delivery requests. Importantly, this differs from the standard VRP-PD, where goods are transported between specific origin and destination locations. %Instead, we address a variant tailored to our application context, in which pickups and deliveries are independent and non-simultaneous. 
In our application context, a customer cannot request both a pickup and a delivery service at the same time. Pickup must precede a delivery, and all operations must be carried out within the time window associated with each customer.

The objective of the STSP-TWPD is to determine a minimum route starting and ending at the depot that visits all required nodes within their respective time windows and satisfies all pickup and delivery precedence constraints. The route may traverse optional nodes to build a feasible and efficient path. 

%Pickup and delivery constraints arise from requests involving paired locations: a pickup site where goods or items are loaded onto the vehicle and a delivery site where these goods are unloaded. These pairs introduce precedence constraints, pickup must occur before delivery and vehicle capacity constraints that must be respected throughout the route to maintain feasibility.

The STSP-TWPD is defined on a directed not complete graph $G = (V, A)$, where $V$ is the set of all nodes, including the depot (0) and customer nodes and $V_r \subseteq V$ is the subset of customer nodes (required nodes) that must be visited.

Each customer node $i \in V_r$ is associated with a time window $[a_i, b_i]$, where $a_i$ denotes the earliest time at which service at node $i$ can start and $b_i$ denotes the latest acceptable start time. %This interval represents the earliest ($a_i$) and latest ($b_i$) allowable time at which service can begin at that location. 
Each customer node is also characterized by a service time $s_i$, which represents the time required to complete the pickup or delivery operation at node $i$. Furthermore, each node is associated with a demand value $d_i$, which quantifies the amount of goods handled at that location. A positive value of $d_i$ indicates a pickup, while a negative value indicates a delivery. 
%Each node $i \in V$ is associated with a time window $[a_i, b_i]$, where $a_i$ and $b_i$ represent the earliest and latest allowable start times for service at node $i$, respectively. The vehicle must begin servicing node $i$ within this interval. The service duration at node $i$ is denoted by $s_i$.\\
The travel time associated with arc $k$ is denoted by $l_k$. 
Equivalently, the travel time to traverse the arc from node $i$ to node $j$ is denoted by $l_{ij}$. 
We assume a unit-speed vehicle, so that the travel time is numerically identical to the arc length.

% and the cost of traversing arc $k \in A$ (connecting nodes $i$ and $j$) is given by $c_k$.
The vehicle has a total capacity $Q$ and %each node $i \in V_r$ has a demand $d_i$, representing pickup or delivery quantities. T
the route must ensure that the vehicle's load never exceeds capacity $Q$ at any point.

We denote by $\delta^+(i)$ the set of arcs leaving node $i$ and by $\delta^-(i)$ the set of arcs entering node $i$.

%The goal is to find a minimum-cost route for a single vehicle that starts and ends at the depot node $0 \in V$, visits all nodes in $V_r$ almost once and services each node $i \in V_r$ within its time window $[a_i, b_i]$. 
The objective of the problem is to determine a route that minimizes the total traversal time, defined as the sum of the travel times associated with the selected arcs, while ensuring that all customer nodes in $V_r$ are visited within their respective time windows $[a_i, b_i]$. The solution must also respect vehicle capacity limits throughout the route and properly handle pickup and delivery operations.

The parameters used in the formulation of the STSP are summarized in Table~\ref{tab:notations}.

\begin{table}[H]
\centering
\begin{tabular}{>{\centering\arraybackslash}m{3cm}>{\arraybackslash}m{8cm}}
\toprule
\textbf{Notation}  &   \textbf{Description} \\
\midrule
$V$ & set of all nodes (including depot and customers) \\
$V_r$ & set of customer nodes (required nodes) \\
$A$  & Set of arcs in the graph \\
$\delta^+(i)$ & set of arcs leaving node $i$ \\
$\delta^-(i)$ & set of arcs entering node $i$ \\
\hline
%$p \subseteq V_r$ & set of pickup nodes \\
%$d \subseteq V_r$ & set of delivery nodes \\
$a_i$ & earliest time to begin service at node $i$ \\
$b_i$ & latest time to begin service at node $i$ \\
$s_i$ & service time at node $i$ \\
$d_i$ & demand at node $i$ \\
$l_{ij}$ & travel time between node $i$ and node $j$\\
$l_k$ & travel time of arc $k \in A$ \\
$Q$ & vehicle capacity\\ 
\bottomrule
\end{tabular}
\caption{Notation used in the problem formulation.} \label{tab:notations}
\end{table}

\section{Mathematical Formulation}\label{sec:mathematical_formulation}
%This section presents the two mathematical formulations for the STSP-TWPD. In both cases, the objective is to determine a minimum route for a vehicle that:
%\begin{itemize}
  %  \item starts and ends at at the depot node;
  %  \item visits all required customer nodes within their respective time windows;
 %   \item respects pickup-and-delivery precedence and demand constraints;
 %   \item never exceeds the vehicle's capacity at any point along the route.
%\end{itemize}
The first formulation, referred to as the ABF, is an arc-based model, where decision variables are defined on the arcs of the graph to capture both vehicle movement and the sequence of visits. The second formulation, called the NBF, adopts a node, based approach. Here, variables are associated with nodes. %, which simplifies the representation of several aspects of the problem, particularly the management of time windows, precedence relations, and pickup-and-delivery constraints.
\subsection{Arc-Based Formulation - ABF}\label{subsec:abf}

The ABF models the routing decisions using binary variables associated with arcs and time steps. This formulation closely follows classical STSP representations, while extending them to incorporate time windows and pickup-and-delivery constraints. By indexing arcs over time, ABF enables a precise representation of routing sequences and transitions between nodes.

\subsubsection{Decision Variables.}
The variables used to define the model are:
\begin{description}
   \item[$y_k^t$] $\begin{cases} 
    1, & \text{if arc } k \in A \text{ is traversed at time step } t \\
    0, & \text{otherwise}
    \end{cases}$
    
   \item[$x_i^t$] $\begin{cases} 
    1, & \text{if a pickup or delivery operation is performed at node } i \text{ at time } t \\
    0, & \text{otherwise}
    \end{cases}$
   \item[$\tau_i^t$:] depart time at node $i$ during time step $t$

   \item[$q_i^t$:] capacity of the vehicle after servicing node $i$ at time $t$
\end{description}
The time index $t$ is defined over the interval $1 \leq t \leq |A|$, where $|A|$ is the total number of arcs in the graph. This upper bound ensures that the model can capture any feasible route, including the longest possible path that traverses all arcs. By associating a time step to each potential arc traversal, the model preserves temporal consistency and allows for accurate sequencing of operations throughout the routing process.
\subsubsection{Objective Function.} 
\begin{align}
    \min \quad & \sum_{t=1}^{|A|} \sum_{k \in A} l_k \: y_k^t \label{eq:objective1}
\end{align}
The objective function \eqref{eq:objective1} minimizes the total travel time, calculated as the sum of the travel times $l_k$ of all arcs traversed across all time steps. %, weighted by the routing variables $y_k^t$. 
%The objective function \eqref{eq:objective} minimizes the total travel distance (or cost), computed by summing the lengths $l_k$ of all arcs traversed over all time steps, weighted by the routing variables $y_k^t$. This encourages the selection of shorter and more efficient routes, thus reducing the overall operational cost.

\subsubsection{Routing Constraints.}

\begin{align}
    \sum_{k \in \delta^+(0)} y_k^1 = 1% \sum_{t=1}^{|A|} \sum_{k \in \delta^+(0)} y_{0j}^t=1 
    \label{eq:depot_start}
\end{align}

Constraints~\eqref{eq:depot_start} ensure that the vehicle departs from the depot (node 0) exactly once at the beginning of the planning horizon ($t=1$), thereby establishing the starting point of the route.
\begin{align}
    y_k^1 = 0, \quad \forall k \in A \setminus \delta^+(0) \label{eq:depot_arc_restriction}
\end{align}

Constraints~\eqref{eq:depot_arc_restriction} enforce that, at time $t=1$, only arcs emanating from the depot can be used. All other arcs are restricted at this stage, preventing the vehicle from appearing anywhere other than the depot at the beginning.

\begin{align}
    \sum_{t=1}^{|A|} \sum_{k \in \delta^+(0)} y_k^t = \sum_{t=1}^{|A|} \sum_{k \in \delta^-(0)} y_k^t \label{eq:depot_flow_balance}
\end{align}

Constraints~\eqref{eq:depot_flow_balance} ensure flow conservation at the depot across the entire planning horizon: the number of departures from the depot must equal the number of returns. This guarantees that all vehicle routes are closed and that no vehicle is left outside the depot at the end.

\begin{align}
    \sum_{k \in \delta^-(i)} y_k^t = \sum_{k \in \delta^+(i)} y_k^{t+1}, \quad \forall i \in V \setminus \{0\},\; 1 \leq t < |A|-1 \label{eq:flow_conservation}
\end{align}

Constraints \eqref{eq:flow_conservation} enforce route continuity for every node $i$ except the depot, at consecutive time steps. It balances incoming and outgoing arcs so that if the vehicle arrives at $i$ at time $t$, it must also depart at time $t+1$, avoiding breaks or gaps in the route.
\subsubsection{Single service for customer.}

\begin{align}
   x_{i}^t \leq M \: \sum_{k \in \delta_i^-} y_{k}^t , \quad \forall i \in V,\; 1 \leq t < |A| \label{eq:xy_link}
\end{align}
Constraints~\eqref{eq:xy_link} establish a link between the service variable $x_i^t$ and the routing variable $y_k^t$. It states that a service at node $i$ and time $t$ (i.e., $x_i^t = 1$) is only allowed if the vehicle actually arrives at node $i$ at that time $t$. %via some incoming arc.

\begin{align}
    \sum_{t=1}^{|A|} x_{i}^t = 1, \quad \forall i \in V_r
    \label{eq:unique_visit}
\end{align}

Constraints \eqref{eq:unique_visit} require each customer node $i \in V_r$ to be visited exactly once within the planning horizon, ensuring all demands are served for problem feasibility.

\subsubsection{Time Constraints.}

\begin{align}
    \tau_j^{t+1} &\geq \tau_i^t  + s_i \:x_j^{t+1}+ l_{k}- M (1 - y_k^{t+1}), %\left( \sum_{k \in \delta^+(i)} y_k^{t+1} + \sum_{k \in \delta^-(j)} y_k^{t+1}\right)\right), 
    \notag \\
   & \quad \forall k \in A: k \in \delta_i^+, k \in \delta_j^-, i \in V, j \in V, \; 0 \leq t < |A|-1 \label{eq:time_consistency}
\end{align}

Constraints~\eqref{eq:time_consistency} enforce temporal consistency between consecutive node visits. If the vehicle travels from node $i$ to node $j$ at time $t+1$ via arc $k$, the arrival time at $j$ must respect both the service time $s_i$ at $i$ and the travel time $l_k$. The Big-M term ensures the constraint is inactive when the arc is not used.%The Big-M term relaxes the constraint when the arc $(i,j)$ is not traversed.

\begin{align}
   \tau_{i}^t \leq M \: \sum_{k \in \delta_i^-} y_{k}^t , \quad \forall i \in V,\; 1 \leq t < |A| \label{eq:xy_link22}
\end{align}
Constraint~\eqref{eq:xy_link22} links the time variable $\tau_i^t$ to the routing decision variables $y_k^t$. It ensures that a time value is assigned to node $i$ at time $t$ only if the vehicle arrives at node $i$ via an incoming arc at that time. The use of the Big-M constant prevents the assignment of a non-zero time when no vehicle reaches the node, effectively deactivating the constraint in such cases. %This helps maintain consistency between routing and timing decisions.

\begin{align}
    \tau_{i}^t \geq (a_i+s_i) \; x_{i}^t, \quad \forall i \in V_r,\; 1 \leq t \leq |A| \label{eq:tau_lower} \\
    \tau_{i}^t \leq (b_i+s_i) \; x_{i}^t + M (1 - x_{i}^t), \quad \forall i \in V_r,\; 1 \leq t \leq |A| \label{eq:tau_upper}
\end{align}
Constraints~\eqref{eq:tau_lower} and~\eqref{eq:tau_upper} impose time window constraints. If service at customer $i$ is scheduled at time $t$, the corresponding time variable $\tau_i^t$ must lie within the valid time window $[a_i + s_i, b_i + s_i]$. The Big-M term relaxes the upper bound when the node is not serviced.
\subsubsection{Capacity Constraints.}
\begin{equation}
  q_0^0 = Q
    \label{eq:initial_load}
\end{equation}
Constraint~\eqref{eq:initial_load} sets the initial load of the vehicle at the depot equal to its full capacity $Q$. %This defines the starting condition for load propagation.
We assume that the vehicle starts empty and has its full capacity available for pickups or deliveries.
\begin{equation}
  q_i^0 \leq 0, \quad \forall i \in V\setminus\{0\}
    \label{eq:initial_load_customer}
\end{equation}
Constraints~\eqref{eq:initial_load_customer} ensure that no node other than the depot has any available load capacity at time zero.

\begin{align}
   q_{i}^t \leq Q \: \sum_{k \in \delta_i^-} y_{k}^t, \quad \forall i \in V,\; 1 \leq t < |A| \label{eq:capacity_3}
\end{align}
Constraints~\eqref{eq:capacity_3} couple the load variable $q_i^t$ with the routing decisions. They ensure that a non-zero load can be assigned to node $i$ at time $t$ only if the vehicle actually arrives at that node via some incoming arc. The load is also bounded above by the vehicle's capacity $Q$, and the Big-M logic embedded via $Q$ deactivates the constraint when the node is not visited at that time.

\begin{align}
q_j^{t+1} &\geq q_i^t - d_j\: x_j^{t+1}- Q \left(1 - y_k^{t+1}\right), \notag \\
   &  \quad \forall\:  k \in A: k \in \delta_i^+, k \in \delta_j^-, i \in V, j \in V;\:  0 \leq t < |A|-1
\label{eq:load_propagation}
\end{align}
Constraints~\eqref{eq:load_propagation} govern the dynamic update of the vehicle's load as it moves through the network over time. Specifically, if a transfer from node $i$ to node $j$ takes place via an active arc $k \in A$ (i.e., $k \in \delta_i^+$ and $k \in \delta_j^-$ with $y_k^{t+1} = 1$), then the load at node $j$ at time $t+1$ ($q_j^{t+1}$) must be at least the load at node $i$ at time $t$ ($q_i^t$), plus the quantity $d_j$ associated with a pickup or delivery at node $j$, provided that $x_j^{t+1} = 1$. The term $-Q(1 - y_k^{t+1})$ acts as a big-M relaxation, deactivating the constraint when arc $k$ is not used. 
%This constraint ensures that the vehicle's load evolves sequentially and realistically along the route, reflecting service operations performed at each node.
\begin{align}
    \sum_{t=1}^{|A|} q_{0}^t = Q-\sum_{i \in V_r} d_i
    \label{eq:capacity total}
\end{align}
Constraint~\eqref{eq:capacity total} ensures that the load carried by the vehicle upon returning to the depot reflects the total quantity delivered throughout the planning horizon. Specifically, this condition enforces that the cumulative load on arcs entering the depot equals the initial capacity $Q$ minus the total customer demand $\sum_{i \in V_r} d_i$. In practice, this constraint guarantees that no load is "lost" during the route and that the residual load at the final return to the depot matches the expected leftover capacity, preserving load consistency and avoiding artificial capacity loss.

%\begin{align}
%    0 \leq q_i^t \leq Q, \quad \forall i \in V,\; 0 \leq t \leq |A| \label{eq:capacity_bounds}
%\end{align}

%Constraint \eqref{eq:capacity_bounds} ensures the vehicle load stays within feasible limits, never exceeding capacity $Q$ nor becoming negative, across all nodes and times.

\subsubsection{Variable Domains.}

\begin{align}
    &y_k^t \in \{0, 1\}, \quad \forall k \in A,\; 1 \leq t \leq |A| \label{eq:binary_def} \\
    &x_{i}^t \in \{0, 1\}, \quad \forall i \in V_r,\; 1 \leq t \leq |A| \label{eq:x_binary_def} \\
    &\tau_i^t \geq 0, \quad \forall i \in V,\; 0 \leq t \leq |A| \label{eq:tau_nonneg} \\
    &0 \leq q_i^t \leq Q, \quad \forall i \in V,\; 0 \leq t \leq |A| \label{eq:load_bounds}
\end{align}

These domain constraints \eqref{eq:binary_def}--\eqref{eq:load_bounds} define the feasible values of the decision variables. The binary variables $y_k^t$ and $x_i^t$ encode route choices and service decisions, while $\tau_i^t$ and $q_i^t$ represent continuous time and load states, bounded to preserve problem realism.
%%\color{red}
\subsection{Node-Based Formulation - NBF}\label{subsec:nbf}

The NBF is a novel approach in which routing decisions are defined over the nodes forming the arcs of the network. %NBF is specifically structured to enhance compatibility with quantum and hybrid solvers, while still enforcing all classical routing and capacity constraints.
\subsubsection{Decision Variables.} The variables used to define the model are:
\begin{description}
   \item[$y_{ij}^t$] $\begin{cases} 
    1, & \text{if arc } (i,j) \in A \text{ is traversed at the period } t  \\
    0, & \text{otherwise}
    \end{cases}$
    
   \item[$x_i^t$] $\begin{cases} 
    1, & \text{if a pickup or delivery operation is performed at node } i \text{ at time } t \\
    0, & \text{otherwise}
    \end{cases}$
   \item[$\tau_i^t$:] depart time at node $i$ during time step $t$

   \item[$q_i^t$:] capacity of the vehicle after servicing node $i$ at time $t$
\end{description}
The time index $t$ is retained in the NBF to capture the temporal evolution of routing and service decisions. Although constraints are structured at the node level, time-indexed variables are necessary to model sequencing, time-window feasibility, and load propagation across consecutive steps.
As defined in ABF formulation, the time index $t$ is in the range $1 \leq t \leq |A|$, which guarantees sufficient temporal solution to represent any feasible route, including the longest one that traverses all arcs in the graph.

\subsubsection{Objective Function.}
\begin{equation}
    \min \sum_{(i,j) \in A} \sum_{t=1}^{|A|}  l_{ij} \;y_{ij}^t \label{eq:objective2}
\end{equation}

The objective function (\ref{eq:objective2}) minimizes the total travel time by summing the travel time $l_{ij}$ of all arcs $(i,j)$ that are traversed.

\subsubsection{Routing Constraints}
\begin{equation}
    \sum_{j \in V: (0,j) \in A} y_{0j}^1 = \sum_{t=1}^{|A|} \sum_{j \in V: (0,j) \in A} y_{0j}^t = 1
    \label{eq:depot_departure}
\end{equation}
Constraints~\eqref{eq:depot_departure} ensure that the vehicle departs from the depot exactly once, and that this departure occurs at the beginning of the planning horizon ($t = 1$). % The constraint imposes both temporal and structural consistency, establishing a unique starting point for the route.
\begin{equation}
    \sum_{j \in V: (i,j) \in A} y_{ij}^1 = 0, \quad \forall i \in V \setminus \{0\}
    \label{eq:node_start}
\end{equation}
Constraints~\eqref{eq:node_start} prevent any node other than the depot from initiating a route. At time step $t = 1$, only the depot is allowed to have outgoing arcs. All other nodes must have zero outbound flow, ensuring a valid and centralized route start.
\begin{align}
    \sum_{t=1}^{|A|} \sum_{j \in V: (0,j) \in A} y_{0j}^t =     \sum_{t=1}^{|A|} \sum_{j \in V: (i,0) \in A} y_{i0}^t \label{eq:depot_flow_balance22}
\end{align}
Constraint~\eqref{eq:depot_flow_balance22} ensures flow balance at the depot over the entire planning horizon. It states that the total number of departures from the depot (i.e., arcs of the form $(0,j)$ used at any time $t$) must equal the total number of returns to the depot (i.e., arcs of the form $(i,0)$). This guarantees that vehicle that leaves the depot also returns, preventing route fragments that do not end at the depot and ensuring closed routes.

\begin{equation}
    \sum_{j \in V: (i,j) \in A} y_{ij}^{t+1} = \sum_{j \in V: (j,i) \in A} y_{ji}^t, \quad \forall i \in V \setminus \{0\} ,\; 1 \leq t < |A| 
    \label{eq:flow_conservation1}
\end{equation}

Constraints~\eqref{eq:flow_conservation1} enforce flow conservation at each non-depot node across time. Specifically, the vehicle can only leave a node if it previously arrived there, enforcing temporal consistency and eliminating disconnected paths.

\subsubsection{Single service for customer.}

\begin{align}
   x_{j}^t \leq M \; \sum_{i \in V: (i,j) \in A} y_{ij}^t , \quad \forall j \in V,\; 1 \leq t < |A| \label{eq:linking_x_y}
\end{align}
Constraints~\eqref{eq:linking_x_y} ensure a logical link between the service variable $x_j^t$ and the routing decision variables $y_{ij}^t$. They ensure that service at node $j$ at time $t$ can only occur if the vehicle actually arrives at node $j$ through some incoming arc $(i,j)$ at that time. The use of the Big-M constant allows the constraint to be satisfied when no incoming arc is used (i.e., when the node is not visited), thereby avoiding infeasibility. 
\begin{equation}
    \sum_{t=1}^{|A|} x_i^t = 1, \quad \forall i \in V_r
    \label{eq:single_service}
\end{equation}

Constraints~\eqref{eq:single_service} guarantee that each required node $i \in V_r$ is visited and serviced exactly once during the planning horizon. This requirement is fundamental to the feasibility of the solution, as it ensures full coverage of all customer demands.

\subsubsection{Time constraints.}
The following timing constraints match those in the ABF formulation and guarantee correct scheduling and adherence to time windows for vehicle visits along the route.

\begin{equation}
    \tau_j^{t+1} \geq \tau_i^t + s_j \:x_j^{t+1} + l_{ij}- M(1 - y_{ij}^{t+1}), 
    \quad \forall (i,j) \in A,\; 0 \leq t \leq |A|-1
    \label{eq:time_consistency1}
\end{equation}
Constraints~\eqref{eq:time_consistency1} ensure time feasibility between two consecutive nodes. If arc $(i,j)$ is traversed, the arrival time at node $j$ must be consistent with the service duration at $j$, the travel time and the departure from $i$. The Big-M term disables the constraint when arc $(i,j)$ is not used.

\begin{align}
   \tau_{i}^t \leq M \: \sum_{j \in V: (ji) \in A} y_{ji}^t , \quad \forall i \in V,\; 1 \leq t < |A| \label{eq:xy_link_22}
\end{align}
Constraints~\eqref{eq:xy_link_22} link the arrival time variable $\tau_i^t$ to the routing decisions $y_{ji}^t$, ensuring time consistency with vehicle movements. Specifically, they enforce that a valid arrival time at node $i$ at time $t$ can only be assigned if the vehicle actually arrives at $i$ via some incoming arc $(j,i)$ at that time. The Big-M constant deactivates the constraint when no such arc is used, preventing $\tau_i^t$ from taking a non-zero value unless node $i$ is truly visited.

\begin{align}
    \tau_i^t &\geq (a_i + s_i) \; x_i^t, 
    \quad \forall i \in V_r,\; 1 \leq t \leq |A| \label{eq:time_window_lower} \\
    \tau_i^t &\leq (b_i + s_i) \; x_i^t + M(1 - x_i^t), \quad \forall i \in V_r,\; 1 \leq t \leq |A|
    \label{eq:time_window_upper}
\end{align}

Constraints~\eqref{eq:time_window_lower} and~\eqref{eq:time_window_upper} enforce that, if a required node $i$ is serviced at time $t$, the associated service must begin within its predefined time window $[a_i, b_i]$. The lower bound ensures the service does not start too early, while the upper bound prevents late arrivals. The inclusion of the service time $s_i$ within both bounds aligns the constraint with the full duration of the stop. The Big-M relaxation deactivates the constraint when no service occurs at that time.

%%\color{red}
\subsubsection{Capacity Constraints.}
The following capacity constraints are similar to those in the ABF formulation, ensure proper load management throughout the route.
\begin{equation}
  q_0^0 = Q
  \label{eq:initial_vehicle_load}
\end{equation}
Constraint~\eqref{eq:initial_vehicle_load} sets the initial load of the vehicle at the depot to its full capacity $Q$. This defines the starting inventory available for deliveries at the beginning of the planning horizon.
\begin{equation}
  q_i^0 \leq 0, \quad \forall i \in V\setminus\{0\}
  \label{eq:customer_initial_bound}
\end{equation}
Constraints~\eqref{eq:customer_initial_bound} enforces that only the depot can have a load at time zero. %%impose an upper bound on the initial load value $q_i^0$ for all nodes $i$ excluding the depot. It prevents infeasible load assignments at customer nodes by ensuring that no node (other than the depot) holds a quantity exceeding the vehicle’s total capacity at the initial time step.
\begin{align}
   q_{i}^t \leq Q \; \sum_{j \in V: (j,i) \in A} y_{ji}^t, \quad \forall i \in V,\; 1 \leq t < |A|
   \label{eq:load_routing_link}
\end{align}
Constraints~\eqref{eq:load_routing_link} enforce consistency between the load variables and routing decisions. They allow the load $q_i^t$ at node $i$ and time $t$ to be positive only if the vehicle arrives at that node through an incoming arc. When no vehicle enters node $i$ at time $t$, the constraint ensures that $q_i^t = 0$. The upper bound $Q$ also ensures that the carried load never exceeds the vehicle's capacity.

\begin{align}
q_j^{t+1} &\geq q_i^t - d_j\: x_j^{t+1}- Q \left(1 - y_{ij}^{t+1}\right), \quad \forall \:i \in V, \forall \: j \in V; 0 \leq t < |A|-1
\label{eq:load_propagation2}
\end{align}
Constraints~\eqref{eq:load_propagation2} ensure consistency in the propagation of load across the network over time. If there is a transfer from node $i$ to node $j$ via arc $(i,j)$ at time $t+1$, then the inventory at node $j$ at time $t+1$ ($q_j^{t+1}$) must be at least the inventory at node $i$ at time $t$ ($q_i^t$), plus any quantity $d_j$ delivered or picked up at $j$ (when $x_j^{t+1} = 1$). The term $-Q(1 - y_{ij}^{t+1})$ ensures the constraint is only enforced when the arc is active.
\begin{align}
    \sum_{t=1}^{|A|} q_{0}^t = Q - \sum_{i \in V_r} d_i
    \label{eq:return_capacity_consistency}
\end{align}
Constraint~\eqref{eq:return_capacity_consistency} ensure that the total load associated with returns to the depot is consistent with the vehicle’s initial inventory and the deliveries made. It guarantees that the load remaining on the vehicle upon return equals the unused portion of its capacity, i.e., the initial load $Q$ minus the total customer demand $\sum_{i \in V_r} d_i$. %This avoids artificial loss or gain of load during the route and preserves the integrity of the capacity accounting throughout the planning horizon.

\subsubsection{Variable Domains.}
\begin{align}
    &y_{ij}^t \in \{0,1\}, \quad \forall (i,j) \in A,\; 1 \leq t \leq |A| \label{eq:domain_y2} \\
    &x_i^t \in \{0,1\}, \quad \forall i \in V,\; 1 \leq t \leq |A| \label{eq:domain_x2} \\
    &\tau_i^t \geq 0, \quad \forall i \in V,\;0 \leq t \leq |A| \label{eq:domain_tau2} \\
    &0 \leq q_i^t \leq Q, \quad \forall i \in V,\;0 \leq t \leq |A| \label{eq:domain_q2}
\end{align}

Constraints~\eqref{eq:domain_y2} and \eqref{eq:domain_x2} defines the decision variables binary. Constraints~\eqref{eq:domain_tau2} ensure that times are non-negative for all nodes and time steps. Constraints~\eqref{eq:domain_q2} bound the vehicle load between 0 and the maximum capacity $Q$ at all nodes and time steps.
%\color{red}
\subsection{Considerations on the ABF and NBF Formulations}
This section clarifies the distinctive features of the ABF and NBF formulations, examines their structural dimensionality, and discusses the role of linearization parameters, such as the Big-$M$ and capacity constants, that ensure model consistency and numerical stability.
\subsubsection{Equivalence Between ABF and NBF Formulations}\label{proof}
%\color{black}
We prove that the ABF and the NBF are mathematically equivalent representations of the STSP-TWPD. Despite their structural differences, ABF defines binary arc-time variables $y_k^t $, while the NBF employs node-time variables $y_{ij}^t $, both models encode the same feasible space and lead to the same optimal solutions. We establish this equivalence by constructing bijective mappings between their variable sets, constraints, and objective functions.

\paragraph{Variable Correspondence} we define a one-to-one mapping between the variables used in the two models:
\begin{itemize}
    \item \textbf{Arc variables:} 
    $
    y_{ij}^t \equiv y_k^t, \quad \text{where } k = (i,j)
    $.
    In both models, $ y_{ij}^t $ (or $ y_k^t $) denotes whether arc $ (i,j) $ is traversed at time $ t $. Thus, the arc-time variables are directly shared.
    
    \item \textbf{Service variables:}
    $
    x_i^t \text{ are used identically in both ABF and NBF.}
    $
    
    \item \textbf{Arrival time and load variables:}
    $
    \tau_i^t \text{ and } q_i^t \text{ are defined identically across both formulations.}
    $
\end{itemize}

\paragraph{Objective Function Correspondence}
the objective functions in both ABF~\eqref{eq:objective1} and NBF~\eqref{eq:objective2} are equivalent. Therefore, minimizing the total travel time in~\eqref{eq:objective1} is equivalent to minimizing the total travel time in~\eqref{eq:objective2}.
%as they both aim to minimize the total distance traveled by the vehicle.

\paragraph{Constraints Set Correspondence:}
we verify that, for each category of constraints, the two models impose structurally identical restrictions. In particular:
\begin{itemize}
    \item \textbf{Routing and flow conservation:} both models enforce that each route starts and ends at the depot and that flow is preserved throughout the route (ABF: Eq.~\eqref{eq:depot_start}--\eqref{eq:flow_conservation}, NBF: Eq.~\eqref{eq:depot_departure}--\eqref{eq:flow_conservation1});
    
    \item \textbf{Customer service:} in both models, each request is served exactly once, and service variables are linked to routing decisions (ABF: Eq.~\eqref{eq:xy_link} and \eqref{eq:unique_visit}; NBF: Eq.~\eqref{eq:linking_x_y} and \eqref{eq:single_service});
    
    \item \textbf{Time and scheduling constraints:} the models ensure time consistency across nodes and arcs and impose service within the given time windows, only when service is actually performed (ABF: Eq.~\eqref{eq:time_consistency}--\eqref{eq:tau_upper}; NBF: Eq.~\eqref{eq:time_consistency1}--\eqref{eq:time_window_upper});
    
    \item \textbf{Load and capacity:} both formulations ensure that load propagation is consistent, delivery and pickup amounts are feasible and capacity limits are never exceeded (ABF: Eq.~\eqref{eq:initial_load}--\eqref{eq:capacity total}; NBF: Eq.~\eqref{eq:initial_vehicle_load}--\eqref{eq:return_capacity_consistency});
    
    \item \textbf{Variable domains:} the domain definitions are the same in both models, including binary routing and service variables and bounded continuous time and load variables.
\end{itemize}

%\color{red}
\paragraph{Model Dimensionality Comparison}
Both the ABF and NBF have the same dimensionality when expressed in terms of the actual network size. 
Let $|V|$ be the number of nodes (including the depot), $|A|$ the number of directed arcs of the graph, $r=|V_r|$ the number of required nodes, and let $T$ denote the number of time layers used in the model (in our setting, $T=|A|$ as a safe upper bound).
In the ABF, binary routing variables are defined for each arc–time pair, yielding $\Theta(|A|\cdot T)$ variables of type $y^t_{\cdot}$, plus $\Theta(r\cdot T)$ service variables $x_i^t$. 
Continuous variables for time and load, $\tau_i^t$ and $q_i^t$, contribute $\Theta(|V|\cdot T)$. 
The number of linear constraints is of the same orders, i.e., combinations of $\Theta(|A|\cdot T)$ and $\Theta(|V|\cdot T)$ terms due to routing/flow, timing, and capacity relations.
The NBF employs the same indexing scheme in practice (arc–time for routing activation, node–time for service, time, and load), resulting in identical counts: $\Theta(|A|\cdot T)$ binary routing variables, $\Theta(r\cdot T)$ service binaries, and $\Theta(|V|\cdot T)$ continuous variables, with a comparable number of constraints. 
Hence, the two formulations are equivalent not only in feasibility and optimality but also in dimensionality. 

The numerical results comparing the two formulations are presented in subsection ~\ref{sec:results}, where we assess their performance and verify the correspondence of optimal solutions.

\subsubsection{Linearization Parameters and Numerical Considerations}
%\paragraph{Role and Impact of Big-$M$ and Capacity Parameters}
The constant $M$ is employed in both formulations to model time and capacity relations when the corresponding routing or service variables are not active. A similar reasoning applies to the vehicle capacity constant $Q$, which acts as a natural upper bound and plays a role analogous to that of $M$ in relaxing capacity-related constraints. While $Q$ has a clear physical interpretation, both constants function as upper-bounding parameters that indirectly affect the tightness of the relaxation and the numerical scaling of the model.
However, the inclusion of such parameters inevitably increases the overall model complexity. 
Each constraint containing a Big-$M$ or $Q$ term introduces additional conditional logic that must be explicitly represented within the linear structure of the formulation. 
As a result, the number of constraints grows, and their coefficients span a wider numerical range, which can enlarge the branch-and-bound search tree and increase the solver’s computational burden. 
This effect becomes particularly evident in large-scale instances, where the accumulation of these upper-bounding constraints amplifies both the size of the constraint matrix and the number of potential branching points.
To clarify their practical role within ABF and NBF models, the specific constraints in which these parameters are used are identified below.
In the ABF formulation, $M$ appears in constraints~\eqref{eq:xy_link}, \eqref{eq:xy_link22}, and \eqref{eq:tau_upper}, while the capacity constant $Q$ is used in this role in constraints~\eqref{eq:capacity_3} and~\eqref{eq:load_propagation}. 
Similarly, in the NBF formulation, $M$ appears in constraints~\eqref{eq:linking_x_y}, \eqref{eq:xy_link_22}, and \eqref{eq:time_window_upper}, whereas $Q$ is used in constraints~\eqref{eq:load_routing_link} and~\eqref{eq:load_propagation2}. 
In all these cases, the Big-$M$ and $Q$ parameters allow the corresponding constraint to be automatically relaxed when the related routing or service decision variable is inactive, preserving linearity without introducing nonlinear or logical formulations.Similarly, in the NBF formulation, $M$ appears in constraints~\eqref{eq:linking_x_y}, \eqref{eq:xy_link_22}, and \eqref{eq:time_window_upper}, whereas $Q$ is used in this role in constraints~\eqref{eq:load_routing_link} and~\eqref{eq:load_propagation2}. 
In all these cases, the Big-$M$ and $Q$ parameters allow the corresponding constraint to be automatically relaxed when the related routing or service decision variable is inactive. 
This mechanism is a standard linearization technique in mixed-integer linear programming (MILP), widely adopted to preserve model linearity and to avoid the introduction of logical or conditional expressions that would otherwise result in nonlinear formulations.
Indeed, in the absence of this type of relaxation, several constraints would need to be expressed in a nonlinear form, involving products of binary and continuous variables.

For example, constraint~\eqref{eq:time_consistency} in the ABF model, which is written as:
\begin{equation}
    \tau_j^{t+1} \geq \tau_i^t + s_i x_j^{t+1} + l_k - M(1 - y_k^{t+1}),
\end{equation}
would instead take the nonlinear form:
\begin{equation}
    \tau_j^{t+1} \geq (\tau_i^t + s_i x_j^{t+1} + l_k)\; y_k^{t+1}.
\end{equation}
Similarly, for the load propagation constraint~\eqref{eq:load_propagation}:
\begin{equation}
    q_j^{t+1} \geq q_i^t - d_j x_j^{t+1} - Q(1 - y_k^{t+1})
\end{equation}
avoids the nonlinear form
\begin{equation}
    q_j^{t+1} \geq (q_i^t - d_j x_j^{t+1}) y_k^{t+1},
\end{equation}
which would yield a non-convex feasible region. 

Such nonlinearities would substantially increase computational complexity and reduce compatibility with classical MILP solvers as well as hybrid quantum optimization frameworks. 
Therefore, although the adoption of Big-$M$ and $Q$ parameters introduces known drawbacks in terms of numerical conditioning and model size, their inclusion is justified by the need to maintain a fully linear and computationally tractable formulation. 

\subsection{Problem Variants: STSP-PD and STSP-TW}\label{sec:new_variants}
The purpose of introducing these variants is twofold: to provide new benchmark instances for Steiner-based routing problems under temporal and operational constraints, and to isolate the computational effects of time-window and pickup-and-delivery conditions.
\subsubsection{STSP-PD}
To the best of our knowledge, the STSP-PD has not been previously formulated or analyzed as an independent variant of the STSP. 
This variant is derived directly from the ABF and NBF formulations presented in Subsection~\ref{sec:mathematical_formulation}, with the sole modification of removing the time-window constraints (Eqs.~\eqref{eq:tau_lower}–\eqref{eq:tau_upper} in the ABF and Eqs.~\eqref{eq:time_window_lower}–\eqref{eq:time_window_upper} in the NBF). 
All other constraints remain unchanged from the full STSP-TWPD model.
The STSP-PD variant thus provides a methodological baseline for evaluating how temporal feasibility constraints influence computational effort and model structure. 
\subsubsection{STSP-TW}
Similarly, to the best of our knowledge, the STSP-TW has not been previously defined or investigated as an independent problem in the Steiner routing literature. 
We formally introduce this variant as a delivery-only extension of the STSP framework with time windows. 
While the STSP-TWPD jointly accounts for pickup-and-delivery precedence and temporal feasibility, the STSP-TW focuses exclusively on the temporal dimension, allowing us to isolate and examine the effects of time windows in the absence of pickup operations.
From a modeling perspective, this requires adjustments to both instance generation and capacity initialization. 
In the standard STSP-TWPD setting, 50\% of required nodes correspond to pickup requests (demands uniformly sampled from $\{10, \dots, 20\}$) and 50\% to deliveries (demands sampled from $\{-5, \dots, -1\}$), while all remaining nodes have zero demand. 
In the delivery-only case, all customer demands are instead drawn from $\{-5, \dots, -1\}$, effectively eliminating pickup operations.
The vehicle, which in the STSP-TWPD starts empty with full capacity available ($q_0 = Q$), must now begin fully loaded to perform only deliveries. 
Accordingly, the model initialization is modified as:
\begin{equation}
q_0 = 0,
\end{equation}
and the flow conservation at the depot is updated to preserve consistency between total delivered and residual load:
\begin{equation}
\sum_{t=1}^{|A|} q_0^t = Q + \sum_i d_i.
\end{equation}
These changes ensure that the STSP-TW remains feasible and structurally consistent with the underlying Steiner framework while capturing the pure delivery routing scenario.
%\color{black}

\section{Arcs Filtering and Graph Reduction Method} \label{sec:preprocessing}
The Arcs Filtering and Graph Reduction (AFGR) method is a preprocessing technique applied to directed non-complete graphs, aimed at simplifying the original problem without losing essential information. Specifically, AFGR eliminates arcs that are either irrelevant or excessively costly (in terms of their travel time), while ensuring that the connectivity between required nodes, those that must be visited or serve a purpose in the solution, is fully preserved. By doing so, the graph is effectively reduced in size, which simplifies the underlying optimization problem.
This simplification is fundamental for improving the efficiency of both the problem formulation and its subsequent solution process. By reducing the size of the arc set A, AFGR removes unnecessary complexity and focuses computational effort on the most promising parts of the search space. The resulting smaller and cleaner model leads to faster computations and more scalable algorithms. Although developed in the context of the STSP-TWPD, the AFGR procedure is general and can be readily adapted to other routing and network design problems that involve sparse or partially connected graphs, where preserving essential connectivity while reducing model complexity is crucial.

The method operates in several stages as described below:
\begin{enumerate}
    \item {\bf Elimination of irrelevant arcs:} we remove arcs that do not involve any required nodes. Specifically, only those arcs whose endpoints include at least one node from the required node set $V_r$ or the depot node $0$ are retained.
    %\item {\bf Threshold-Based Length Filtering:} a length threshold $\alpha$ is computed to identify and remove arcs that are more long. This threshold is defined as: $ \alpha = m + 0.5 \; m$, where $m$ is the average length of the arcs retained after the first step. Arcs with a length greater than or equal to $\alpha$ are marked for removal. However, two important exceptions are made:
    %\begin{enumerate}
   %     \item Arcs are retained if they are incident to a relevant node (i.e., either endpoint is in $V_r$ or is the depot node $0$).
    %    \item Arcs are also retained if their removal would result in the disconnection of one of their endpoints from the graph, which is checked through a connectivity validation procedure.
   % \end{enumerate}
    \item \textbf{Shortest-Path-Based Arc Filtering:} for every pair of required nodes (i.e., all nodes in $V_r$ and the depot node $0$), the algorithm computes shortest paths using Dijkstra's algorithm. Any arc that participates in at least one such shortest path is considered essential and retained.
    \item \textbf{Retention of Useful Arcs:} only the arcs involved in shortest paths between required node pairs are preserved. All other arcs are removed, since they are not part of any shortest route between critical nodes.
    \item {\bf Eliminating Isolated Steiner Nodes:} after arc filtering, the algorithm checks for isolated Steiner nodes, nodes not in $V_r$ and not incident to any remaining arc. These nodes are removed from the node set $V$. % All arcs incident to such isolated nodes are removed, as they do not contribute to the solution.
    \item {\bf Update graph structures:} the remaining arcs are renumbered with sequential IDs to maintain a consistent indexing scheme. The graph's internal data structures, including arc sets and the dictionaries for arc lengths, are updated accordingly. %Additionally, the incoming ($D^-$) and outgoing ($D^+$) arc sets for each node are rebuilt to reflect the new graph structure.
\end{enumerate}
%The EFGR method returns the final filtered arc set $A$.
To provide a clear and executable representation, the above steps are detailed in the pseudocode of \ref{afgr}.
\begin{algorithm}[H]
\caption{AFGR}
\label{afgr}
\begin{algorithmic}[1]
\State \textbf{Input:} $G = (V, A)$: directed graph, $V_r$: set of required nodes, $0$: depot node
\State \textbf{Output:} $A'$: filtered arc set
\State \textbf{Step 1: Elimination of irrelevant arcs}
\State $A_1 \gets \{ (i,j) \in A \mid i \in V_r \cup \{0\} \ \text{or} \ j \in V_r \cup \{0\} \}$
\State \textbf{Step 2: Shortest-Path-Based arc filtering}
\For{each $u \in V_r \cup \{0\}$}
    \State $(\text{dist}, \text{prev}) \gets \text{Dijkstra}(G, u)$
    \For{each $v \in V_r \cup \{0\}, v \neq u$}
        \If{$v \in \text{prev}$}
            \State Reconstruct path $u \rightarrow v$ using $\text{prev}$
            \State Mark all arcs in path as useful
        \EndIf
    \EndFor
\EndFor
\State \textbf{Step 3: Retention of useful arcs}
\State $A' \gets$ all arcs that were marked as useful
\State \textbf{Step 4: Eliminating isolated Steiner nodes}
\State $V' \gets$ all nodes incident to arcs in $A'$ plus $V_r \cup \{0\}$
\State Remove any node $v \notin V_r \cup \{0\}$ that has no incident arc in $A'$
\State \textbf{Step 5: Update graph structures}
\State Reindex arcs in $A'$ sequentially
\State Update graph structures
%%\State Rebuild $\delta ^+$ and $\delta^-$ sets
\State \Return $V', A'$
\end{algorithmic}
\end{algorithm}
%\color{red}
A detailed quantitative analysis of the dimensionality reduction achieved by the AFGR procedure is presented in subsection~(\ref{sec:complexity}). In that part of the paper, we evaluate how the application of AFGR affects the structural size of both mathematical formulations (ABF and NBF) by comparing the number of variables and constraints before and after preprocessing.

It is important to note that, although the AFGR procedure preserves full connectivity among required nodes, it does not provide a formal guarantee that the global optimal solution of the original problem is always retained after arc reduction. The method is therefore heuristic in nature, designed to improve computational tractability while maintaining the essential structural properties of the problem. Nevertheless, computational experiments indicate that the reduced graphs consistently yield optimal solutions across all tested instances.

\section{Computational study} \label{sec:experiments} 
This section is divided into four main parts. The first part, \textit{Computational setup and solver configuration}, describes the hardware and software environment, along with the solver details. The second part, \textit{Generation of instances}, explains how the test instances for the experiments were created. The third part, \textit{Model dimensionality and computational complexity}, analyzes the structural growth of the ABF and NBF formulations. Finally, the fourth part, \textit{Numerical results}, presents and discusses the computational outcomes, highlighting the performance of the models and solvers. 
%%\color{red}
%\color{black}
\subsection{Computational setup and solver configuration}
All computational experiments are conducted on a Windows 10 machine equipped with an Intel processor (Family 6, Model 126, Stepping 5), featuring 4 physical cores (8 logical threads) and 15.6 GB of RAM. The implementation is developed in Python 3 using Jupyter Notebook. For exact optimization experiments, we employ Gurobi Optimizer version 11.0.1.

The CQM formulation is implemented using D-Wave’s Ocean SDK, specifically \texttt{dimod} version 0.12.18 and \texttt{dwave-system} version 1.28.0. All problem instances are solved through D-Wave's hybrid solver service for CQMs, identified as \texttt{hybrid\allowbreak\_constrained\allowbreak\_quadratic\allowbreak\_model\allowbreak\_version1p}. We impose a \texttt{time\_limit} of 5 seconds. The CQM framework represents a generalization of the traditional QUBO approach, designed to support optimization problems that include both objective functions and explicit constraints. Unlike QUBO models, where all constraints must be manually embedded into the objective function through penalty terms with user-defined coefficients, the CQM formulation allows equality and inequality constraints to be declared directly within the model. This provides a higher level of abstraction and avoids the empirical and problem-dependent process of penalty weight tuning, which often affects convergence and solution quality.

In practice, the CQM framework automatically converts each constraint into an energy-contributing term within the Hamiltonian by means of an internal penalty calibration mechanism. This automated process determines appropriate penalty scales that balance constraint satisfaction and objective minimization, ensuring that feasible solutions are prioritized without introducing numerical instability. As a result, users can focus on expressing the optimization model (defining variables, constraints, and objectives) without manually managing penalty coefficients.
In our implementation, all decision variables are represented as binary or integer types, and both the objective and constraints are formulated using D-Wave’s \texttt{dimod.ConstrainedQuadraticModel()} interface, as documented in the official D-Wave Ocean SDK manual~\cite{DWaveCQMdocs}. The completed model is automatically compiled and submitted to the hybrid solver, which orchestrates the interplay between classical and quantum resources to enforce constraints and minimize the Hamiltonian energy.

From a methodological perspective, it is important to clarify the rationale for adopting the CQM framework instead of a standard QUBO formulation.  A pure QUBO model represents an unconstrained binary quadratic problem in which all linear and mixed-integer constraints must be embedded into the objective function through quadratic penalty terms.  This transformation introduces additional binary variables to encode continuous quantities (such as time and load) and requires large penalty coefficients to enforce feasibility. 

The resulting energy landscape becomes highly ill-conditioned, and the appropriate penalty magnitudes must be empirically tuned to balance constraint satisfaction and objective optimization. 
Such manual calibration is problem-dependent and can strongly affect numerical stability, relaxation tightness, and convergence behaviour, issues that are further amplified in large-scale or hybrid quantum computing settings. In contrast, the CQM formulation allows constraints to be defined explicitly, maintaining the original linear structure of the MILP while avoiding artificial penalty weights and full binary encoding of continuous variables. This leads to a more stable and interpretable optimization model, in which feasibility is handled natively by the solver through an internal penalty-scaling mechanism. 
%\color{black}

\begin{figure}[!ht]
    \centering
  \includegraphics[width=0.55\linewidth]{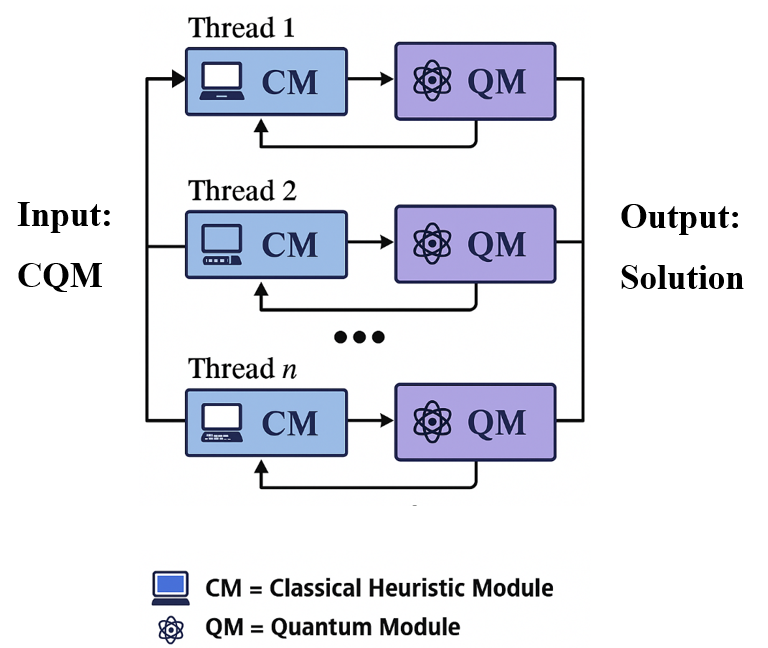}
    \caption{General structure of D-Wave’s hybrid solvers, adapted from \cite{osaba2025d}. 
    Each branch includes a CM and a QM that interact asynchronously to refine candidate solutions.}
    \label{fig:hybrid_architecture}
\end{figure}

We utilize the LeapCQMHybrid solver, a proprietary hybrid optimization tool developed by D-Wave Systems,to solve the CQM models defined for the problems considered in this work. This solver is %part of the broader Hybrid Solver Service, which encompasses a suite of problem-specific algorithms that integrate classical and quantum computing techniques. Hybrid Solver Service is designed to tackle problem instances that exceed the capacity of current quantum processing units (QPUs). 
a cloud-based framework designed to integrate classical and quantum computing resources within a unified optimization pipeline. As described in recent works on hybrid architectures \cite{osaba2025d, bertuzzi2024evaluation, willsch2022benchmarking}, the Hybrid Solver Service (HSS) follows a modular asynchronous workflow that combines the strengths of classical heuristics with quantum annealing-based exploration.
From a structural perspective, each instance submitted to the hybrid solver is decomposed into multiple branches, which are executed in parallel on independent computational threads. Each branch contains two tightly coupled modules:
\begin{itemize}
    \item \textbf{Classical Module (CM):} responsible for large-scale exploration of the solution space. The CM implements advanced metaheuristics, including local search, tabu search, and simulated annealing, to generate feasible candidate solutions and to construct reduced subproblems suitable for quantum processing;
    \item \textbf{Quantum Module (QM):} in charge of performing quantum-guided refinement. The QM executes quantum annealing cycles on the \texttt{D-Wave Advantage2\_system1.4} Quantum Processing Units (QPUs), which features 4596 qubits arranged in a Zephyr topology. Each quantum query corresponds to a partial embedding of the global CQM into the QPU’s architecture. The resulting quantum samples are analyzed and ranked according to their energy levels, then asynchronously sent back to the CM.
\end{itemize}
The CM and QM operate asynchronously, communicating through an adaptive feedback mechanism. The CM monitors convergence of its classical heuristics and invokes the QM whenever search stagnation is detected or diversification is required. This mechanism allows the quantum annealer to explore alternative regions of the energy landscape, providing probabilistic samples that can help the CM escape local minima. The asynchronous workflow also ensures that latency in one branch does not block the global progress of the solver as a major improvement over earlier synchronous hybrid frameworks.
Figure~\ref{fig:hybrid_architecture} could schematically represent this interaction, showing how the CM initiates quantum subproblem queries, receives low-energy configurations from the QM, and updates its search trajectory based on quantum feedback.

The proportion of computation devoted to classical versus quantum phases is dynamically determined by the solver’s internal scheduler and is not user-controllable. According to D-Wave’s technical documentation~\cite{DWave2022performance} the frequency of quantum queries depends on problem size, constraint density, and classical progress metrics. In typical workloads, quantum queries account for a minority of total iterations but play a crucial exploratory role, guiding the search toward low-energy basins that are difficult to reach through purely classical heuristics.
Although the internal configuration of the hybrid solver is proprietary, its design philosophy follows that of D-Wave’s hybrid architecture, where quantum and classical resources cooperate asynchronously. The classical component ensures scalability and constraint management, while the quantum component contributes through stochastic sampling based on quantum tunneling. This mechanism allows the solver to explore low-energy regions of the search space that are typically difficult to reach with classical heuristics alone.
The use of the quantum module is therefore not aimed at achieving quantum speedup in a strict complexity-theoretic sense, but at improving diversification and refinement within the hybrid optimization loop. As evidenced in recent benchmark studies~\cite{osaba2025d, bertuzzi2024evaluation, willsch2022benchmarking}, such hybridization can enhance convergence reliability and solution quality for large combinatorial problems that exceed the capacity of current QPUs.

It is therefore appropriate to regard the \texttt{LeapCQMHybrid} as a hybrid algorithm, where the quantum-guided component acts as an accelerator for diversification and refinement rather than as a full solver. The CM ensures scalability and feasibility management, while the QM provides stochastic sampling capabilities rooted in quantum tunneling effects. Together, these mechanisms form a synergistic loop that leverages both computational paradigms to achieve improved convergence speed and solution quality on combinatorial problems exceeding the current QPU capacity.
%\color{black}
%LeapCQMHybrid employs a multi-threaded architecture where each computational thread independently explores the solution space. Each thread combines a classical heuristic component, responsible for performing the main optimization steps using advanced classical methods, with a quantum-guided component that leverages quantum hardware to explore promising solution regions and refine candidate solutions. This cooperative interplay between classical and quantum elements enhances both global exploration and local optimization phases.

In our experiments, quantum queries are executed on \texttt{D-Wave's Advantage2\_system1.4 QPU}, the most advanced QPU architecture currently available. This system consists of 4596 qubits arranged according to a Zephyr connectivity graph. Solver parameters are kept at their default values, with the only tunable parameter, \texttt{time\_limit}, left unchanged across all instances to ensure consistency and reproducibility.
Due to the proprietary nature of LeapCQMHybrid, detailed information about its internal quantum routines, resource allocation, and the exact number of qubits used per instance is not disclosed. For further technical insights and practical applications, readers are referred to D-Wave’s official documentation and recent literature on hybrid quantum computing \cite{osaba2025d, bertuzzi2024evaluation, willsch2022benchmarking}.

\subsection{Generation of instances}
The generation of test instances plays a critical role in assessing the robustness and generalizability of the proposed formulations. 
This subsection describes the procedure adopted to construct the benchmark instances used in the computational study. 
In particular, we detail how the graph topology is generated and how all problem components, including time windows, service times, pickup–delivery demands, and vehicle capacity, are defined. 
The overall process is designed to produce different, controlled, and fully reproducible instances.
For clarity, we summarize the instance generation procedure in a step-by-step manner:

\begin{enumerate}
    \item Construction of the underlying graph structure;
    \item Selection of required nodes;
    \item Generation of travel times based on Euclidean distances;
    \item Generation of time windows;
    \item Assignment of service times;
    \item Generation of pickup and delivery demands;
    \item Definition of vehicle capacity;
    \item Definition of auxiliary parameters (e.g., Big-M constant).
\end{enumerate}

The design of the instance generator is guided by three main principles. 
First, realism: the generated instances aim to reflect key characteristics of real-world routing problems, including heterogeneous time windows, service times, and mixed pickup–delivery operations. 
Second, controllability: the generation process allows us to systematically vary structural and operational parameters (e.g., graph sparsity and proportion of required nodes) in order to analyze their impact on problem difficulty. 
Third, reproducibility: all random components are generated using a fixed seed, ensuring that the instances can be exactly replicated. 
%While our framework builds upon the procedure introduced by Letchford et al.~\cite{letchford2013compact}, we extend it to incorporate additional variability in both structural and parametric components, enabling a more comprehensive experimental evaluation.

\paragraph{Graph construction}
The node coordinates are generated by adapting the procedure proposed by Letchford et al.~\cite{letchford2013compact}, a widely adopted benchmark method for constructing instances of the STSP. 
Specifically, we consider a set of $n$ vertices $V = \{v_1, v_2, \dots, v_n\}$ uniformly positioned along the circumference of a circle with fixed radius. Each vertex $v_i$ is assigned a Cartesian coordinate $(x_i, y_i)$ computed as: 
\begin{equation}
    x_i = R \cos\left( \frac{2\pi i}{n} \right), \quad 
    y_i = R \sin\left( \frac{2\pi i}{n} \right),
\end{equation}
where $R = 100$ and $i \in \{0, \dots, n-1\}$. 
From this layout, we consider all possible undirected edges between vertices and compute the Euclidean distance between endpoints as the base length of each arc. The candidate edges are sorted in ascending order of their length and incrementally added to the graph.
Following Letchford et al.~\cite{letchford2013compact}, edges are inserted while enforcing geometric constraints to preserve a sparse and planar-like structure. In particular, each edge is accepted only if:
\begin{itemize}
    \item it does not intersect any previously selected edge;
    \item the angle between any two adjacent edges incident to the same vertex is not smaller than a given threshold $\theta$.
\end{itemize}
%\cite{letchford2013compact} impose a angle of $\theta = 60^\circ$ between any two adjacent edges incident to the same vertex. 
In the benchmark proposed by \cite{letchford2013compact}, an angle of $\theta = 60^\circ$ is imposed between any two adjacent edges incident to the same vertex.
We extend the generation scheme by considering an additional angular threshold $\theta = 80^\circ$. This allows us to generate graph topologies with different levels of sparsity. In particular, larger values of $\theta$ restrict the number of feasible edges, resulting in sparser and more constrained network structures.
To ensure connectivity, additional arcs are inserted when necessary. Specifically, for any vertex without incident arcs, at least one arc is added by connecting it to a nearby vertex, while preserving the geometric consistency of the graph. These arcs are assigned a cost equal to their Euclidean distance. Duplicate arcs are removed to avoid redundancy.

The set of required nodes $V_r \subseteq V$ is determined by selecting a fixed proportion of nodes. While the \cite{letchford2013compact} benchmark considers a single ratio $|V_r|/|V| = 0.3$, we broaden this setting by considering $|V_r|/|V| \in {0.2, 0.3, 0.4, 0.5, 0.6}$, thereby enabling a more comprehensive analysis of how the proportion of required nodes affects the problem structure and solution characteristics.

The joint variation of the angular constraint and the required-node ratio defines a structured family of instance classes, summarized in Table~\ref{tab:instance_classes}. These parameters jointly control the structural properties of the graph. In particular, larger angular thresholds produce sparser and more constrained networks, while higher values of $|V_r|/|V|$ increase the number of mandatory visits, leading to denser routing requirements.

The resulting classes are grouped into two families: A1--A4 for $\theta = 60^\circ$ and B1--B5 for $\theta = 80^\circ$. Compared to the baseline instances, this extended design enables a systematic investigation of the interplay between graph sparsity and the density of required nodes.

\begin{table}[h]
\centering
\renewcommand{\arraystretch}{1.3}
\resizebox{0.7\textwidth}{!}{
\begin{tabular}{>{\centering\arraybackslash}p{2cm}ccc}
\toprule
\textbf{Class} & \textbf{Angle constraint} & \textbf{$|V_r|/|V|$} & \textbf{Topology effect} \\
\midrule

A1 & $60^\circ$ & 0.20 & Very sparse \\
A2 (\cite{letchford2013compact}'s instances) & $60^\circ$ & 0.30 & Moderate sparsity \\
A3 & $60^\circ$ & 0.40 & Sparse \\
A4 & $60^\circ$ & 0.50 & Moderate density \\
A5 & $60^\circ$ & 0.60 & Dense \\

\midrule

B1 & $80^\circ$ & 0.20 & Very sparse \\
B2 & $80^\circ$ & 0.30 & Sparse \\
B3 & $80^\circ$ & 0.40 & Sparse \\
B4 & $80^\circ$ & 0.50 & Highly constrained \\
B5 & $80^\circ$ & 0.60 & Very highly constrained \\

\bottomrule
\end{tabular}}
\caption{Instance classes obtained by varying the angular constraint and the proportion of required nodes.}
\label{tab:instance_classes}
\end{table}

To the best of our knowledge, no benchmark generation procedure currently exists for the STSP-TWPD, nor for its individual variants STSP-TW and STSP-PD. To fill this gap, we propose a novel instance generation framework specifically tailored to these problem settings.
The proposed generator combines graph-based structural information with controlled randomization to produce realistic and diverse instances, while ensuring full reproducibility through the use of a fixed random seed. This design allows us to systematically control the temporal and operational characteristics of the problem, enabling a meaningful and comprehensive computational evaluation.

\paragraph{Time windows}
Let $l_{\min}[i]$ denote the shortest path distance from the depot to node $i$, computed via Dijkstra’s algorithm, and let $l_{\max}[i]$ denote a longest-path estimate obtained through a Bellman-Ford-based procedure.
For each node $i \in V_r$, two reference values $l^{(1)}_i$ and $l^{(2)}_i$ are derived by combining $l_{\min}[i]$ and $l_{\max}[i]$ as to produce two distinct reference levels, with $l^{(1)}_i \leq l^{(2)}_i$. 
The set of required nodes $V_r$ is partitioned into two equally sized subsets for the generation of early time windows.
\begin{itemize}
    \item For 50\% of the nodes in $V_r$, the earliest service start time $a_i$ is defined as
    \begin{align*}
        a_i = l^{(1)}_i + \epsilon_i,
    \end{align*}
    where $\epsilon_i$ is a random value uniformly distributed in the interval $(0,\; 0.5 \cdot l^{(1)}_i)$.
    
    \item For the remaining 50\% of nodes, the earliest service start time is generated as
    \begin{align*}
        a_i = l^{(2)}_i + \epsilon_i,
    \end{align*}
    where $\epsilon_i$ is uniformly drawn from $(0,\; 0.5 \cdot l^{(2)}_i)$.
\end{itemize}

The latest service start time $b_i$ is then computed by adding a random value $w_i$ to $a_i$, where $w_i$ is a random offset, uniformly drawn from the interval $(l^{(1)}_i,\; l^{(2)}_i)$, ensuring that $b_i \geq a_i$:
\begin{align*}
    b_i = a_i + w_i.
\end{align*}
For non-required nodes $i \in V \setminus V_r$, we set $a_i = 0$ and $b_i = 10^5$.

\paragraph{Service times}
For each $i \in V_r$, the service time is sampled as $s_i \in \{5, 10\}$. For all $i \in V \setminus V_r$, $s_i = 0$.

\paragraph{Demands}
The set of required nodes is partitioned into two equally sized subsets, corresponding to pickup and delivery operations. For the first subset (pickup nodes), positive demands are assigned by sampling uniformly from the interval $\{10, \dots, 20\}$. For the second subset (delivery nodes), negative demands are generated uniformly from $\{-5, \dots, -1\}$. All non-required nodes are assigned zero demand.
\color{black}
\paragraph{Vehicle capacity}
The capacity is set to match the total absolute demand, with a small random offset:
       \begin{align*}
        Q = \sum_{i \in V} |d_i| + \theta, 
        \end{align*}
        where $\theta $ is random value in $\left(\min_i |d_i|,\; \max_i d_i\right)$.

    This promotes tight feasibility and favors the existence of feasible solutions.

\paragraph{Big-M constant}
The constant $M$ is defined to upper-bound any feasible arrival time:
   \begin{align*}
        M = \max_{i \in V} b_i + \max_{i \in V} s_i + \max_{(i,j) \in A} l_{ij}
\end{align*}
This expression combines three worst-case components:
\begin{itemize}
\item $\displaystyle \max_{i \in V} b_i$: the latest possible time window upper bound across all nodes;
\item $\displaystyle \max_{i \in V} s_i$: the maximum service time at any node;
\item $\displaystyle \max_{(i,j) \in A} l_{ij}$: the longest travel time between any pair of connected nodes.
\end{itemize}
By summing these values, $M$ provides a safe upper bound on the arrival time at any node in any feasible route. 
\color{black}
Table~\ref{tab:network_parameters} summarizes the generation rules and notation.

\begin{table}[ht]
\centering
\caption{ Parameters and generation rules for STSP-TWPD instances.}
\label{tab:network_parameters}
\renewcommand{\arraystretch}{1.3}
\resizebox{0.7\textwidth}{!}{
\begin{tabular}{>{\centering\arraybackslash}m{2.5cm} >{\arraybackslash}m{11.5cm}}
\toprule
\textbf{Notation} & \textbf{Description} \\
\midrule

\multicolumn{2}{c}{\textit{Graph structure}} \\
\midrule
$V$ & Set of nodes \\
$V_r$ & Subset of required nodes, with $|V_r|/|V| \in \{0.2, 0.3, 0.4, 0.5, 0.6\}$ \\
$A$ & Set of arcs derived from a sparse graph \\
$l_{ij}$ & Euclidean distance between nodes $(i,j)$ \\
$l_k$ & Travel time of arc $k \in A$ \\

\midrule
\multicolumn{2}{c}{\textit{Distance measures}} \\
\midrule
$l_{\min}[i]$ & Shortest-path distance from the depot to node $i$ (Dijkstra) \\
$l_{\max}[i]$ & Longest-path estimate from the depot to node $i$ (Bellman-Ford-based) \\

\midrule
\multicolumn{2}{c}{\textit{Time windows}} \\
\midrule
$a_i$ & $\forall i \in V_r$: earliest service time generated as combinations of $l_{\min}[i]$ and $l_{\max}[i]$, with an additional random perturbation \newline
$\forall i \in V \setminus V_r$: $a_i = 0$ \\

$b_i$ & $\forall i \in V_r$: $b_i = a_i + w_i$, where $w_i$ is a random offset \newline
$\forall i \in V \setminus V_r$: $b_i = 10^5$ \\

$s_i$ & $\forall i \in V_r$: $s_i \in \{5, 10\}$ \newline
$\forall i \in V \setminus V_r$: $s_i = 0$ \\

\midrule
\multicolumn{2}{c}{\textit{Demands and capacity}} \\
\midrule
$d_i$ & $\forall i \in V_r$: 50\% pickups $[10,20]$, 50\% deliveries $[-5,-1]$ \newline
$\forall i \in V \setminus V_r$: $d_i = 0$ \\

$Q$ & Vehicle capacity: $Q = \sum_{i \in V} |d_i| + \theta$, with $\theta$ a small random offset \\

\midrule
\multicolumn{2}{c}{\textit{Auxiliary parameter}} \\
\midrule
$M$ & Big-M constant: $M = \max_i b_i + \max_i s_i + \max_{(i,j)\in A} l_{ij}$ \\

\bottomrule
\end{tabular}}
\end{table}
To enhance the replicability of this study, all generated instances and the instance generation code are publicly available at \cite{repository}.

\subsection{Model Dimensionality and Computational Complexity}\label{sec:complexity}

We provide a comprehensive analysis of the dimensional growth and computational complexity of the proposed formulations, ABF and NBF,  for the STSP-TWPD, both in their original form and after the application of the AFGR reduction method. The objective of this analysis is twofold: first, to quantify how the structural size of the models evolves as a function of instance characteristics; second, to rigorously assess the effectiveness of AFGR in controlling such growth across heterogeneous classes of instances.

The analysis is conducted over all instance classes defined in Table~\ref{tab:instance_classes}, and the results are reported in \ref{app:complexitytable} (Tables~\ref{tab:ABF-complexity_A1}--\ref{tab:ABF-complexity_B5}). These tables cover a wide spectrum of structural configurations, ranging from sparse to highly constrained networks. This setting allows us to jointly evaluate the impact of instance size, expressed in terms of the number of nodes $|V|$, while model dimensionality is measured in terms of binary and continuous variables, as well as equality and inequality constraints, thus providing a complete and consistent characterization of each formulation. This comprehensive evaluation framework ensures that the observed trends are robust across heterogeneous structural conditions.

The results in Tables~\ref{tab:ABF-complexity_A1}--\ref{tab:ABF-complexity_B5} clearly indicate that, across all instance classes, the dimensionality of both formulations scales rapidly with the problem size when AFGR is not applied. 
For the ABF formulation, increasing the number of nodes from $|V| = 4$ to $|V| = 20$ leads to a sharp growth in the number of variables (from 145 to 41,185) and constraints (from 258 to 80,853). The largest values are consistently observed in denser configurations such as A1, B1, and B2. This is explained by the higher arc density in these instances, which directly amplifies the number of arc--time combinations and, consequently, the number of routing variables and associated constraints. 
This confirms that arc density plays a critical role in driving model dimensionality, beyond the mere increase in the number of nodes.

A fully analogous behavior is observed for the NBF formulation, confirming that both models share the same structural growth pattern. Although ABF and NBF exhibit the same theoretical order of growth, they differ in their practical dimensionality. In particular, ABF tends to generate a larger number of constraints due to the explicit arc-based indexing, whereas NBF achieves a slightly more compact constraint representation while preserving the same asymptotic complexity. This suggests that, while the two formulations are theoretically equivalent, their structural organization has a non-negligible impact on practical model size. 

The effect of AFGR is both substantial and systematic across all instances. Indeed, the reduction ranges from a minimum of approximately 30\% in the number of variables and 31\% in the number of constraints for instances such as A1 and B1, up to peaks of 71\% (variables) and 73\% (constraints) for instances A5 and B5. 

A more detailed analysis across instance classes reveals a clear and systematic dependency of the reduction effectiveness on the structural properties of the graph. In sparse instances (e.g., A1--A2 and B1--B2), where the ratio $|V_r|/|V|$ is relatively low, the reduction is moderate, typically between 30\% and 44\%. In these cases, a larger portion of the network must be preserved to ensure connectivity among required nodes. This suggests that the reduction potential is inherently limited when the structure of the graph is already close to the minimal connectivity requirements. In intermediate configurations (e.g., A3 and B3), the reduction becomes more pronounced, generally around 55\%. Finally, in dense and highly constrained instances (e.g., A4--A5 and B4--B5), the reduction reaches its highest levels, approaching and in some cases exceeding 70\%. 
It is also important to note that these latter classes correspond to the largest instances ($|V| = 20$), where the reduction becomes even more dramatic, reaching up to approximately 90\% in the number of variables and 91\% in the number of constraints. This provides compelling evidence that AFGR becomes increasingly effective as both instance size and structural density grow.

This behavior is a direct consequence of the AFGR mechanism. By retaining only arcs belonging to shortest paths between required nodes, AFGR systematically removes structurally redundant connections. As the proportion of required nodes increases, the relative number of irrelevant arcs grows significantly, allowing the reduction process to become increasingly selective and effective. This suggests that AFGR implicitly exploits structural redundancy in dense graphs, leading to a more aggressive pruning of the arc set. As a result, the arc set is drastically reduced, leading to a substantial decrease in both variables and constraints.

The detailed growth patterns are further illustrated in \ref{app:complexityfigure} (Figures~\ref{fig:AFGR_A1}--\ref{fig:AFGR_B5}), which provide a visual representation of the evolution of model size across all instance classes. In all cases, the original formulations exhibit a steep and clearly increasing trend in both variables and constraints, with a marked acceleration for $|V| \geq 12$. In contrast, the curves corresponding to the AFGR-enhanced models remain significantly flatter, indicating a much more controlled scaling behavior. This visual evidence reinforces the quantitative results observed in \ref{app:complexitytable}. 
A particularly relevant observation is that the gap between the original and reduced models widens progressively as $|V|$ increases. While the difference is relatively limited for small instances, it becomes substantial for larger ones (e.g., $|V| \geq 15$), where the reduced models require only a fraction of the variables and constraints of the original formulations. This clearly indicates that the benefits of AFGR scale positively with problem size, making it particularly suitable for large-scale instances. This effect is especially evident in Figures~\ref{fig:AFGR_A4}, \ref{fig:AFGR_A5}, \ref{fig:AFGR_B4}, and \ref{fig:AFGR_B5}, corresponding to instances A4, A5, B4, and B5, respectively, where the reduction in both variables and constraints is particularly pronounced. This confirms that AFGR is especially effective in dense and highly constrained scenarios, where structural redundancy is more significant.

\color{black}

\subsection{Numerical results}  \label{sec:results} 
The analysis of the results is organized into two main phases.
%\color{red}
\begin{itemize}
    \item \textbf{Performance analysis with a classical solver:} 
    we assess how the ABF and NBF formulations scale computationally across the three problem variants (STSP-TWPD, STSP-PD, and STSP-TW);
\item \textbf{Performance analysis with a quantum solver:} 
we present and discuss the results obtained for the STSP-TWPD, STSP-PD, and STSP-TW using D-Wave’s hybrid quantum–classical solver. This phase evaluates the capability of the proposed ABF and NBF formulations to operate within a quantum optimization framework, assessing their performance in terms of solution quality, feasibility, and computational efficiency.
\end{itemize} 

In each of the above settings, results are reported for both ABF and NBF models, with and without the AFGR procedure, to evaluate the performance impact of the proposed AFGR method. 

\subsubsection{Performance analysis with a classical solver}
We present a comprehensive computational analysis structured in four main stages. 
First, we examine the performance of the proposed formulations on the complete STSP-TWPD, assessing their scalability and the impact of the AFGR preprocessing method. 
Second and third, we extend the analysis to the two simplified variants, STSP-PD and STSP-TW, respectively. 
Finally, we provide a comparative discussion among the three models (STSP-TWPD, STSP-PD, and STSP-TW), highlighting how their structural and computational characteristics evolve under different constraint configurations.
The computational performance of the three proposed models is evaluated using both the ABF and NBF formulations across all instance classes reported in Table~\ref{tab:instance_classes}.
A time limit of 600 seconds is imposed for all experiments.

\paragraph{STSP-TWPD} \label{subsec:variant-nopickup}
The complete set of results is reported in \ref{app:classic_table_STSP_TWPD} (Tables~\ref{tab:classic_A1}--\ref{tab:classic_B5}). 

A first fundamental observation emerging from the computational analysis is that, for the vast majority of tested instances, both ABF and NBF formulations with and without AFGR, achieve the same objective function value. This result confirms the theoretical equivalence of the two formulations demonstrated in Section~\ref{proof} and, more importantly, shows that the application of the AFGR procedure does not deteriorate solution quality.

A detailed analysis reveals an exception in the most challenging instances. 
In particular, for instance class $A2$, the increasing problem size leads to a progressive deterioration in the performance of the standard formulations. While for smaller instances both ABF and NBF are able to reach optimality, significant differences emerge as $|V|$ increases. 
For larger instances ($|V| \geq 14$), the standard formulations begin to exhibit non-zero optimality gaps, indicating a deterioration in solution quality. 

For example, at $|V| = 14$, the standard NBF formulation reaches an objective value of 1352,80 with a gap of 11\%, whereas the reduced model attains the optimal value of 1350,57 with zero gap.
This discrepancy becomes more pronounced for larger instances. At $|V| = 16$, the standard formulations show gaps up to 27--28\%, with objective values of 1551,41, while the reduced models significantly improve the solution quality, reaching values as low as 1498,74 with zero gap.

A similar trend is observed for $|V| = 17$: the standard formulations exhibit optimality gaps of approximately 23--24\%, with an objective value of 1295,56, whereas the AFGR-based formulations attain optimal solutions (gap equal to 0\%) and a slightly improved objective value of 1294,31. 
An even more significant improvement is observed at $|V| = 18$, where the standard formulations report gaps up to 34\%, while the reduced models converge to the optimal value of 1250,49 with zero gap.
At $|V| = 19$, the standard formulations show gaps up to 35\%, while AFGR reduces both the objective value and the gap (e.g., for ABF formulation, from 1317,96 to 1308,23, with gap reduction from 35\% to 6\%). 

Finally, for $|V| = 20$, both standard formulations (ABF and NBF) attain an objective value of 1380.82. By contrast, the application of AFGR leads to a significant improvement, reducing the objective value to 1358.66 for ABF (23\% reduction) and to 1348.40 for NBF (19\% reduction).

A similar behaviour can be observed for instance class $B2$. 
For instance, at $|V| = 14$, the standard NBF formulation attains an objective value of 1352,80 with a gap of 9\%, whereas the reduced model achieves the optimal value of 1350,57 with zero gap. This behaviour becomes more pronounced as the instance size increases.

At $|V| = 16$, both standard formulations exhibit a gap of 27\%, with an objective value of 1551,41, while the reduced models reach a significantly better solution of 1498,74 with zero gap. 
A similar trend is observed for $|V| = 17$ and $|V| = 18$. Focusing on the ABF formulation, the standard models exhibit optimality gaps of up to 15\% and 34\%, respectively, whereas the AFGR-based models consistently achieve optimal solutions (zero gap) together with improved objective values (e.g., from 1295.56 to 1294.31 for $|V| = 17$, and from 1251.55 to 1250.49 for $|V| = 18$).
The advantage remains evident for the largest instances. At $|V| = 19$, the standard formulations report gaps up to 28\%, whereas the reduced models either achieve optimality or significantly reduce the gap.

Finally, for $|V| = 20$, the standard formulations exhibit optimality gaps as high as 37--40\%. In contrast, the application of AFGR leads to substantially improved solutions, reducing both the optimality gap (down to 18--12\%) and the objective value (from 1376,16 to 1358,66 for ABF, and from 1376,16 to 1348,40 for NBF).

These results clearly demonstrate that the reduced models obtained through AFGR consistently achieve solutions with zero or significantly lower optimality gaps, often improving upon the solutions found by the standard formulations.

From a computational perspective, the impact of AFGR on solution times is particularly significant. The reduction procedure leads to substantial improvements in computational efficiency for both formulations, with time savings that become increasingly pronounced as the problem size grows. More specifically, the average time reduction achieved by AFGR ranges between approximately 52\% and 98\%, depending on the instance class and formulation. For smaller instances (e.g., class $A1$ and $B1$), the average time saving is already considerable, reaching about 61\% for ABF and 66\% for NBF. As the problem size increases, the benefits of AFGR become even more evident. For medium-sized instances (classes $A2$--$B3$), the reduction exceeds 52\% and reaches up to 95\%. Finally, for the largest instances (classes $A4$--$B5$), the computational gain becomes dramatic, with average time savings of up to 98\% for both formulations.

Overall, these results clearly highlight the effectiveness of the AFGR preprocessing method. While preserving optimality across all tested instances, it drastically reduces computational effort, making the solution of large-scale STSP-TWPD instances significantly more tractable. This confirms that preprocessing techniques such as AFGR are not only beneficial but essential when dealing with complex routing problems characterized by time windows and pickup-and-delivery constraints.

\paragraph{STSP-PD}
This part of the paper presents the computational results for the STSP-PD variant, with particular focus on the comparison between standard and AFGR-based formulations. 
We also analyze the impact of removing time-window constraints on model performance and solution quality.
The full set of results is provided in \ref{app:classic_table_STSP_PD} (Tables~\ref{tab:classic_PD_A1}--\ref{tab:classic_PD_B5}). 

A first key observation is that both formulations consistently achieve identical objective function values across all tested instances. This behavior can be attributed to the reduced complexity of the problem setting, as the absence of time window constraints makes the model significantly easier to solve compared to the full STSP-TWPD variant.

From a computational standpoint, the impact of AFGR on solution times is substantial. The preprocessing procedure yields very significant reductions in computational time for both formulations, with improvements that increase as the instance size grows. For the smallest instances (classes $A1$ and $B1$), the average time savings are already notable, reaching approximately 68\% for ABF and 77\% for NBF. As the instance size increases (classes $A2$--$B3$), the computational gains become even more pronounced, with reductions ranging between 85\% and over 90\%. In particular, for class $A3$ and $B3$, the time savings reach up to 93\% for the NBF formulation.
The most remarkable improvements are observed in the largest instances (classes $A4$--$B5$), where the reduction becomes dramatic. In these cases, AFGR allows for average time savings between 93\% and 97\% for both formulations, effectively reducing solution times by more than an order of magnitude. This highlights the crucial role of preprocessing in enhancing scalability for larger and more complex instances.

A comparison of the average objective function values between STSP-TWPD and STSP-PD reveals a clear impact of time window constraints on solution quality. In particular, for several instance classes (e.g., A2 and B2), the STSP-TWPD exhibits significantly higher objective values than STSP-PD, reflecting the additional routing restrictions imposed by time windows.
Moreover, while STSP-PD consistently yields identical objective values across all formulations and configurations, STSP-TWPD shows slight variations between standard and reduced models, with AFGR providing marginal improvements. 

A further comparison between the STSP-TWPD and STSP-PD variants highlights a substantial difference in computational effort. As expected, the presence of time-window constraints in STSP-TWPD leads to significantly higher solution times compared to the STSP-PD model. This gap is particularly evident in medium-sized instances. For example, in class $A2$, the average solution time for the standard formulation reaches approximately 982 seconds for ABF and over 1000 seconds for NBF in STSP-TWPD, whereas the corresponding values for STSP-PD remain below 50 seconds. A similar behavior is observed for class $B2$, where the STSP-TWPD requires more than 1000 seconds for all instance sizes, while the STSP-PD remains around 50 seconds. This clearly indicates that the inclusion of time-window constraints dramatically increases the computational complexity of the problem.
Even for instances classes $A3$ and $B3$, the difference remains significant. STSP-TWPD requires between 118 and 129 seconds in the standard formulation for all instance sizes, while STSP-PD solves the same instances in approximately half the time (around 50--78 seconds). 
For instances classes $A1$, $B1$, $A4$--$B5$, the gap is less extreme but still consistent. In all cases, STSP-TWPD systematically requires more computational time than STSP-PD, confirming the additional burden introduced by temporal constraints. 
The impact of AFGR is remarkable in both variants. For STSP-TWPD, the preprocessing reduces solution times from hundreds or even thousands of seconds to single-digit values in many cases (e.g., from over 1000 seconds to less than 10 seconds in classes $A2$ and $B2$). A similar effect is observed for STSP-PD, although starting from much lower initial times.

Thus, as expected, the STSP-TWPD variant generally requires higher computational times in its standard formulation, due to the additional complexity introduced by time-window constraints. However, despite this increased modeling complexity, the AFGR procedure effectively mitigates the computational burden, yielding performance gains that are fully comparable to those observed in the STSP-PD case.

\paragraph{STSP-TW} 
We report the computational results for the STSP-TW variant, focusing on the impact of time-window constraints on model performance. 
The complete results are presented in \ref{app:classic_table_STSP_TW} (Tables~\ref{tab:classic_TW_A1}--\ref{tab:classic_TW_B5}). 

As in the STSP-TWPD case, the STSP-PD results show that, for almost all instance classes, all tested configurations (ABF and NBF, with and without AFGR) yield identical objective function values with zero optimality gap. 
Nevertheless, for instance classes A2 and B2, a different trend is observed in larger instances. This behavior reflects the increased computational difficulty associated with the corresponding STSP-TWPD variant, where time-window constraints play a critical role. In particular, for class A2 with $|V| = 20$, the standard ABF formulation attains an objective value of 1352.61 with an optimality gap of 38\%, while the standard NBF formulation performs even worse, reaching an objective value of 1360.02 with a gap of 41\%. These results clearly indicate that, without preprocessing, the solver struggles to approach high-quality solutions within the time limit.
The application of AFGR leads to a substantial improvement in both formulations. For ABF, the objective value is reduced to 1349,17, while the optimality gap decreases to 17\%. A similar improvement is observed for NBF, where the objective value decreases to 1352,61 and the gap is reduced to 23\%.
A similar behaviour can be observed for class $B2$ when focusing on the largest instance ($|V| = 20$).
The standard ABF formulation attains an objective value of 1352,61 with a gap of 38\%, while the standard NBF performs worse, reaching 1360,02 with a gap of 41\%. These values indicate that, without preprocessing, the solver struggles to identify high-quality solutions within the given time limits.
The introduction of AFGR leads to a clear improvement in both formulations. For ABF, the objective value is reduced to 1349,17, while the optimality gap decreases from 38\% to 17\%. For NBF, the improvement is even more evident in terms of solution quality, as the objective value decreases to 1349,17, accompanied by a reduction of the gap from 41\% to 18\%.

From a computational perspective, the data clearly highlight the effectiveness of the AFGR preprocessing across all instance classes and for both formulations.
For the instances classes $A1$ and $B1$, the reduction already provides a noticeable improvement, with average time savings of approximately 57--59\%. 
For instances classes $A2$ and $B2$, the standard formulations require several hundreds of seconds (around 660--700 seconds), whereas the reduced models lower the computational time to approximately 180--210 seconds. This corresponds to average savings between 69\% and 73\%, confirming that AFGR effectively mitigates the rapid growth in computational effort.
For classes $A3$ and $B3$, the reduction becomes even more effective, with time savings reaching 85--87\%. In these cases, the solution time drops from around 6--8 seconds to less than 1 second, demonstrating that AFGR is able to drastically simplify the problem even when the initial computational burden is moderate.
The most remarkable improvements are observed for the instances classes $A4$--$B5$. Here, the reduction achieves savings between 93\% and 98\%, effectively reducing solution times from several seconds to fractions of a second. In particular, for classes $A5$ and $B5$, the computational time is reduced by up to 98\%, highlighting the scalability of the preprocessing approach.

These results show a clear and consistent trend: the effectiveness of AFGR increases with the size and complexity of the instance. While the preprocessing already provides benefits for small problems, it becomes essential for larger instances, where it dramatically reduces computational time and ensures tractability.

A further comparison can be conducted by analyzing the objective function values obtained for the STSP-TWPD and STSP-TW variants. The results reveal a clear and consistent pattern across all instance classes.
In particular, for every tested instance, the STSP-TW formulation systematically achieves lower objective function values compared to STSP-TWPD. This behaviour is expected, as the STSP-TWPD model includes additional pickup-and-delivery constraints, which restrict the feasible solution space and force the adoption of longer routes.
For example, in class $A2$, the objective value decreases from approximately 1300 in STSP-TWPD to about 1200 in STSP-TW. A similar reduction is observed in class $B2$, confirming that the removal of pickup-and-delivery constraints allows for significantly more efficient routing.
This trend remains consistent across all instance classes. For $A3$ and $B3$, the objective value decreases from approximately 920,98 (STSP-TWPD) to 882,77 (STSP-TW). For $A4$ and $B4$, the reduction continues, with values decreasing from about 913,35 to 858,89. Finally, for $A5$ and $B5$, the objective value further decreases from 850,07 to approximately 818,57.

These results clearly demonstrate that pickup-and-delivery constraints have a significant impact on the routing cost, consistently increasing the objective function value. In contrast, time-window constraints alone do not impose the same level of structural restriction, allowing the STSP-TW model to achieve more efficient solutions. This highlights the strong influence of pickup-and-delivery requirements on the complexity and cost structure of Steiner routing problems.

A complementary analysis can be performed by comparing the computational times of the STSP-TWPD and STSP-TW variants. The STSP-TWPD model requires significantly higher computational effort than STSP-TW, due to the additional pickup-and-delivery constraints that substantially increase the complexity of the problem. This difference is already visible in the instances classes $A1$ and $B1$, where STSP-TWPD requires approximately 16--23 seconds, while STSP-TW solves the same instances in less than 6 seconds.
The gap becomes more pronounced for $A2$ instances where, the standard formulations for STSP-TWPD require around 980--1000 seconds, whereas STSP-TW reduces this time to approximately 670 seconds. A similar behaviour is observed in class $B2$, where STSP-TWPD exceeds 1000 seconds, while STSP-TW remains around 660 seconds. 
The most striking differences emerge in classes $A3$ and $B3$, where STSP-TWPD requires between 118 and 129 seconds, while STSP-TW solves the same instances in less than 8 seconds. 
In classes $A4$ and $B4$, STSP-TWPD requires approximately 40--52 seconds, while STSP-TW reduces the time to about 3--7 seconds. In $A5$ and $B5$ instances, the computational effort remains consistently lower for STSP-TW.

Overall, these results clearly indicate that pickup-and-delivery constraints have a substantial impact on computational complexity, leading to a significant increase in solution times. In contrast, the STSP-TW model, while still constrained by time windows, remains considerably more tractable. This further confirms that pickup-and-delivery requirements represent the primary driver of computational difficulty in these problem settings.

\paragraph{Route-Based comparison of STSP-TWPD, STSP-PD and STSP-TW} 
\color{black}
The main objective of this set of experiments is to understand the role of the different constraints that define the STSP-TWPD problem. Since we propose STSP-TWPD as the most realistic variant, we compare it against its simplified versions (STSP-PD and STSP-TW) in order to assess how the solution changes when either time windows or pickup-and-delivery constraints are removed. 
We compare the performance of the three models, STSP-TWPD, STSP-PD and STSP-TW, by analyzing the routes generated for different instances. To this end, we consider instances up to $|V| = 10$, focusing on class A2, which is widely used as a benchmark in the literature.
We visualize the optimal routes and analyze their characteristics. This visual inspection allows us to understand how time windows and pickup-delivery constraints influence the route structure.
Moreover, we provide a quantitative comparison by plotting the objective function values for all models across the different instances. This comparison highlights how performance evolves as the number of nodes increases.

The figures \ref{app:classicfigure} (Figures~\ref{fig:route_comparison_V7}--\ref{fig:route_comparison_V10}) show the routes generated by each model for each instance size. Each subfigure corresponds to a specific model and instance size, and all are followed by a joint discussion comparing the structural differences among the models.

Figure~\ref{fig:route_comparison_V7} illustrates the optimal routes generated by the three variants of the problem: STSP-TWPD, STSP-PD, and STSP-TW, for an instance with $V = 7$ nodes. The visual differences in the route structures provide clear insight into the distinct effects of the modeled constraints.
The route generated by the STSP-TWPD model appears highly structured and operationally realistic. The solution exhibits a tightly controlled path that respects both the precedence between pickup and delivery pairs and the feasibility of visiting nodes within their respective time windows. Arcs are shorter, the sequence is more intuitive, and the route avoids unnecessary detours, reflecting a careful optimization process under strict feasibility conditions.
In contrast, the STSP-PD model omits time window constraints. While it continues to respect pickup-before-delivery precedence, the lack of temporal feasibility checks introduces greater spatial flexibility. This is reflected visually by more dispersed routing, with longer arcs and a less compact structure. Although the model adheres to precedence rules, the absence of time constraints can lead to suboptimal arrival times in practical settings.
The STSP-TW model, on the other hand, enforces time windows but disregards pickup-and-delivery dependencies. As a result, the route structure prioritizes visiting nodes within their allowed time intervals, but without enforcing any service order between related pickup and delivery nodes. This leads to the most irregular and unstructured routing among the three models. Nodes may be visited in a sequence that is infeasible from a logistical standpoint, though still valid temporally.

Figure~\ref{fig:route_comparison_V8} presents the optimal routes derived from the STSP-TWPD, STSP-PD, and STSP-TW models for an instance with $V = 8$ nodes. The differences in routing patterns highlight the impact of the specific constraints imposed by each model.
The STSP-TWPD solution, which integrates both time windows and pickup-and-delivery constraints, displays a well-organized and realistic route. The path maintains strict adherence to pickup-before-delivery precedence and ensures all visits occur within designated time windows. The result is a compact, efficient route that minimizes unnecessary movement.
In the STSP-PD model, time windows are excluded, leaving only the precedence relationships between pickup and delivery pairs. This relaxation introduces more spatial freedom, leading to visibly longer and more dispersed arcs. While precedence is preserved, the lack of temporal constraints may reduce practical feasibility in time-sensitive scenarios.
The STSP-TW model enforces time windows but disregards the service order between pickup and delivery nodes. Consequently, while the route complies with timing requirements, it may violate logical delivery flows. This often results in suboptimal or counterintuitive sequences, underscoring the limitations of omitting precedence relationships.

Figure~\ref{fig:route_comparison_V9} shows the optimal routing solutions for the STSP-TWPD, STSP-PD, and STSP-TW models on an instance with $V = 9$ nodes. As the problem size increases, the differences introduced by each model's constraints become even more evident in the routing structures.
The STSP-TWPD model, which simultaneously considers both time windows and pickup-and-delivery constraints, generates the most coherent and operationally sound route. The red arcs form a compact and efficient sequence, carefully obeying both temporal feasibility and precedence relations. The path avoids detours and demonstrates strong spatial and logical organization.
The STSP-PD model relaxes the time window requirements while preserving pickup-before-delivery precedence. The resulting route reflects more flexibility in arc selection, leading to longer connections and a less compact structure. Although logically correct, the solution may be inefficient in time-sensitive operations due to its disregard for temporal constraints.
The STSP-TW model enforces time windows but omits pickup-and-delivery relationships. This can be seen in the irregularity of the node visitation order, although all nodes are served within their respective time windows, there is no guarantee that pickups precede their associated deliveries. This leads to a route that may be temporally valid but infeasible from a logistics perspective.

Figure~\ref{fig:route_comparison_V10} presents the optimal routing solutions produced by the STSP-TWPD, STSP-PD, and STSP-TW models for an instance with $V = 10$ nodes. As problem complexity increases, the distinctions introduced by each set of constraints become even more prominent.
In the STSP-TWPD model, which integrates both time window and pickup-and-delivery constraints, the route is highly structured and operationally consistent. The solution respects the required precedence relationships and ensures that each node is visited within its feasible time window. The resulting route is compact, with efficient use of arcs and minimal detours, reflecting a balanced solution in terms of both spatial and temporal feasibility.
In the STSP-PD model, which omits time windows, the route remains logically sound in terms of pickup-before-delivery ordering but is more spatially dispersed. The lack of time constraints allows for longer arcs and less compactness, which may lead to inefficiencies in real-world applications where timing is critical.
The STSP-TW model, conversely, enforces time windows but ignores pickup-and-delivery precedence. The resulting route fulfills all temporal constraints but may violate service logic, visiting deliveries before their associated pickups. This leads to a less practical route, even if temporally valid, and highlights the limitations of ignoring precedence in time-constrained settings.

\begin{figure}[htbp]
    \centering
    \includegraphics[width=0.7\textwidth]{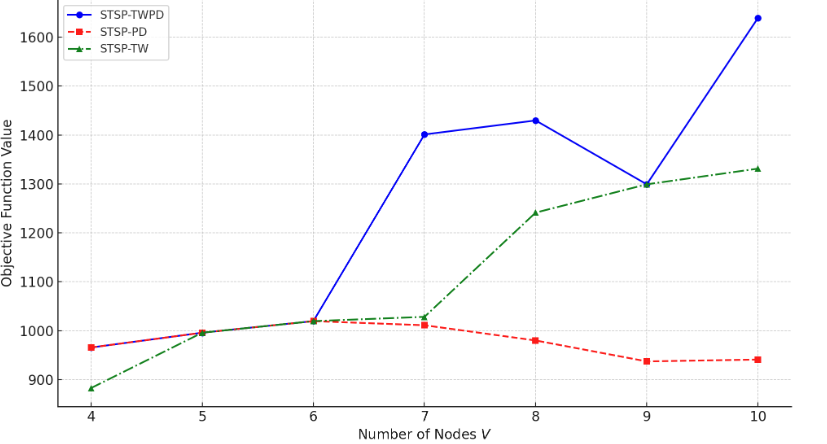}
 \caption{Objective function values for STSP-TWPD, STSP-PD, and STSP-TW models as a function of $V$.}    \label{fig:objective_vs_v}
\end{figure}

Figure~\ref{fig:objective_vs_v} clearly illustrates the evolution of the objective function values as the number of nodes $V$ increases. While all three models show similar performance for small problem sizes ($V \leq 6$), their behavior diverges significantly for larger instances.
The STSP-TWPD model exhibits a sharply increasing trend, particularly for $V > 6$. This reflects the compounded complexity introduced by simultaneously enforcing both time windows and pickup-and-delivery precedence. These combined constraints drastically reduce the solution space, often forcing the solver to take more costly routes in order to satisfy both feasibility requirements.
In contrast, the STSP-PD model consistently yields much lower objective values beyond $V = 6$. Since it does not impose any time window constraints, the model benefits from a more flexible temporal domain. This enables shorter and more direct routes, even while still respecting pickup-before-delivery precedence. The relative flatness of the red curve highlights the reduced sensitivity of STSP-PD to instance size compared to the more constrained models.
The STSP-TW model occupies an intermediate position between the two extremes. It enforces time windows but ignores pickup-and-delivery ordering. As a result, the green curve often lies between the STSP-PD and STSP-TWPD lines. Although it adheres to temporal feasibility, the absence of service order constraints can lead to logically infeasible solutions, such as delivering before picking up, that reduce the objective cost but render the route impractical in real-world logistics settings.
This comparison highlights the trade-off between solution quality (in terms of objective cost) and operational feasibility. While relaxing constraints such as time windows or precedence can yield cheaper solutions, only the STSP-TWPD model guarantees operationally realistic results under strict feasibility requirements.

\subsubsection{Performance Analysis with a quantum solver}%Problem Analysis} \label{sec:full-problem-analysis}
%We present the experimental evaluation of the STSP-TWPD using a quantum solver. %, specifically designed to leverage quantum computing capabilities, aiming to assess both the quality of the solutions obtained and the computational efficiency of the model under different configurations, including the presence or absence of time windows and pickup-and-delivery constraints.
%\color{red}
The objective is to assess the feasibility of expressing and solving the proposed formulations within a constrained quantum optimization framework. The experiments are intended as a proof of concept, providing insight into how the STSP-TWPD, and also STSP-PD and STSP-TW variants, can be represented and tackled through emerging hybrid quantum–classical architectures, specifically D-Wave’s \texttt{LeapCQMHybrid} platform.
%Due to the limited computational capabilities of current quantum hardware, we restrict the quantum experiments to instances from class A2, which are commonly used as benchmark instances in the literature.

The quantum experiments are conducted on the A2 benchmark class. This class was selected as a representative and widely used benchmark set, enabling a consistent evaluation of the proposed formulations and solution approaches within a homogeneous experimental framework.

This part of the work is structured into three main phases. 
First, we analyze the STSP-TWPD problem, under both the ABF and NBF formulations. Second, we consider the STSP-PD variant and, finally, we analyze the STSP-TW variant.

All tables share the same structure. Each row corresponds to a specific problem size, indicated by $V$. The columns are organized into two main groups, comparing the performance with and without the use of AFGR.
For both configurations, the reported metrics include: the average and standard deviation of the objective function value (OF Avg and OF Std), the percentage of runs leading to a feasible or optimal solution over ten repetitions per instance (\% Solved), as well as the average and standard deviation of computation time (Time Avg and Time Std).

%These statistics are based on 10 independent runs per configuration.\\
We have conducted 10 tests for each instance in order to evaluate the performance and robustness of the model. In all experiments presented in this section, the solver time limit is fixed to 10 seconds for every instance and configuration. Furthermore, whenever no feasible solution is obtained in any of the 10 runs, the corresponding entries in the table are marked with a dash ``--'' to indicate the absence of valid results. 

\paragraph{STSP-TWPD} 
The experimental results are reported in \ref{app:quantum_table_STSP_TWPD} (Table~\ref{tab:performance-comparison1}) indicate that the AFGR technique contributes to improving the scalability and numerical stability
of both the ABF and NBF formulations for the STSP-TWPD.

For the ABF formulation, AFGR consistently improves the solution quality and feasibility across all instance sizes.
 On small instances ($V=4$ and $V=5$), both configurations solve 100\% of the problems; however, AFGR provides a slightly better average GAP and more stable solution times. As the problem size increases ($V \geq 6$), the benefit of AFGR becomes even more pronounced: without AFGR, no feasible solutions are found for $V=7$ to $V=10$, while with AFGR, feasible solutions are obtained for $V=7$, $V=8$, and $V=9$, with solving rates up to 20\%. Notably, the GAP with AFGR remains close to or slightly lower than in the baseline formulation when a solution is available (e.g., 46–48\% for $V=6$–$V=9$), whereas the baseline often fails entirely. Additionally, average solution times are comparable or even lower with AFGR for larger instances (e.g., $V=9$: 15,32s vs 20,49s), despite the increased complexity of the feasible repair mechanism. This demonstrates that AFGR enhances both efficiency and effectiveness.

Similarly, the NBF formulation benefits from AFGR across the full range of tested sizes. 
The results for the STSP-TWPD solved using the NBF formulation, both with and without the AFGR method.
While both variants perform similarly for small instances ($V=4$ to $V=6$), AFGR slightly reduces the optimality gap (e.g., 47\% vs 51\% for $V=6$), improves the percentage of solved instances (e.g., 100\% vs 90\% for $V=5$), and maintains acceptable solution times. For larger sizes ($V \geq 7$), the baseline fails to find feasible solutions, whereas AFGR achieves partial solving success (e.g., 20–30\% for $V=7$–$V=9$) with significantly improved GAP values (e.g., 38–48\%). %Despite solving harder instances, AFGR often yields lower average solution times, such as for $V=8$ (15,81s vs 21,37s) and $V=9$ (15,95s vs 22,33s), showing its computational advantage. The reduction in standard deviation of solution times in many cases also confirms a more stable and predictable solving process.

Overall, these results confirm that the integration of AFGR contributes to improving both the feasibility and quality of solutions in both ABF and NBF formulations. AFGR increases the percentage of solved instances, generally reduces the optimality GAP and often leads to comparable or faster solution times. 

\paragraph{STSP-PD}
This part of the work analyzes STSP-PD problem. At this stage, we focus solely on the ABF formulation, rather than testing both ABF and NBF, as their equivalence has already been established in subsection (\ref{proof}). 

The results presented in \ref{app:quantum_table_STSP_PD} (Table~\ref{tab:performance-comparisonTW}) suggest that the AFGR technique provides computational advantages, particularly in terms of model reduction and improved solver tractability.
%The experimental results reported in Table~\ref{tab:performance-comparisonTW} demonstrate the effectiveness of the AFGR technique. 
The introduction of the AFGR module leads to a systematic improvement in the average objective function value, with an average reduction in the optimality gap ranging between 2\% and 10\% across the tested instances. For instance, in the case with $V=6$, the gap decreases from 5\% without AFGR to 2\% with AFGR, highlighting the ability of AFGR to guide the search process toward higher-quality solutions.
In terms of solution stability, the standard deviations of the objective values decrease significantly when AFGR is activated. This indicates greater robustness of the algorithm with respect to the inherent randomness of the heuristic optimization process. Although the addition of the AFGR mechanism introduces a slight increase in the average computational time, the total runtime remains substantially lower than that of exact approaches and is largely justified by the quality gains obtained.
A more detailed analysis confirms that AFGR consistently reduces the average optimality gap across all problem sizes. Specifically, for $V=4$, the gap drops from 4\% without AFGR to 1\% with AFGR. For $V=6$, it decreases from 5\% to 2\%, and for $V=10$, a particularly challenging case, the improvement is even more pronounced, from 14\% without AFGR to 6\% with AFGR. Overall, the AFGR mechanism yields a gap reduction of approximately 40\% to 70\% compared to the baseline method, resulting in solutions that are significantly closer to the known optimum.
Moreover, the percentage of successfully solved instances improves with the inclusion of AFGR, which confirms the greater efficiency of the adaptive repair mechanism in handling infeasible or borderline-feasible configurations. 
Integrating the AFGR module into the model without time windows leads to a substantial improvement in solution quality, enhanced robustness, and overall increased reliability of the heuristic procedure, while keeping the computational effort within acceptable limits.

The comparison be time window seen the models with and without time windows reveals several general trends regarding performance, solution quality, and computational behavior.
The inclusion of time windows constraints significantly increases the problem complexity, as reflected by higher average objective function values and increased computational times, as we expected. On average, the presence of time windows leads to a deterioration of approximately 15--30\% in solution quality, when comparing configurations with and without time windows under the same algorithmic settings. This performance drop is primarily due to the additional temporal feasibility constraints, which restrict the solution space and limit the flexibility of routing decisions.
Despite the added difficulty introduced by time windows, the AFGR mechanism remains effective. Although the average optimality gap tends to be higher in the presence of time windows, as expected due to the stricter feasibility requirements and the reduced number of high-quality feasible solutions, AFGR consistently contributes to narrowing the gap with respect to the optimal benchmark (as computed by Gurobi). Moreover, the use of AFGR leads to a marked increase in the percentage of solved instances, particularly in the most constrained time windows cases.
From a computational perspective, AFGR proves to be essential for recovering feasible solutions in time windows instances, where the baseline approach often either fails to find a solution or produces highly unstable outcomes. Execution times tend to increase further in time windows scenarios, especially in configurations without AFGR, which require more iterations and repair steps to reach feasibility. The AFGR enhanced configurations achieve a better trade-off be time window seen feasibility and runtime, supporting the robustness of the proposed mechanism in handling both temporal and spatial constraints simultaneously.

\paragraph{STSP-TW} This portion focuses on evaluating instances where the problem is limited to delivery tasks only, excluding pickup operations. The variant analyzed here corresponds to the configuration described in paragraph \textit{STSP-TWPD model without pickup operations} in the subsection~\ref{subsec:variant-nopickup}. We focus solely on the ABF formulation, rather than testing both ABF and NBF, since their equivalence has already been established in Section~\ref{proof}. The performance of the model is assessed under this configuration, considering both the presence and absence of the AFGR preprocessing strategy.

The results reported in \ref{app:quantum_table_STSP_TW} (Table~\ref{tab:performance-comparisonPD}) highlight the impact of excluding pickup operations on the behavior of the ABF model, both with and without the AFGR mechanism. For small-size instances ($V=4$), the performance is already strong even without AFGR: the average optimality gap is close to the reference value ($GAP=4\%$) with 100\% of instances solved. Nevertheless, the inclusion of AFGR further improves the results by reducing the average gap to 2\% and ensuring greater stability, as reflected in the lower standard deviation. In this case, the slight increase in average computational time is negligible and fully justified by the gain in robustness.  
When the problem size increases ($V=5$), the model becomes less stable: the gap rises to 28\% without AFGR, with high variability across runs (Std=167,55). The introduction of AFGR reduces the average gap to 23\% and increases solution consistency, although variability is not entirely eliminated. Moreover, the average computational time decreases significantly (from 54,48s to 39,15s). %, suggesting that AFGR preprocessing helps to better guide the search process and avoid unproductive iterations.  
For $V=6$ and $V=7$, the problem becomes more challenging. Without AFGR, the gap reaches 51--60\%, with only 60\% and 30\% of instances solved, respectively. The integration of AFGR does not reduce the gap, which remains high, but it considerably improves the percentage of solved instances (80\% for $V=6$ and 50\% for $V=7$) and leads to more stable outcomes, as shown by the reduction in standard deviation. This indicates that in more complex scenarios AFGR plays a crucial role primarily in ensuring feasibility and reducing variability rather than directly improving the objective function value.  
For larger instances ($V \geq 8$), the difficulty further increases. Without AFGR, the algorithm fails to produce any feasible solution (0\% solved). With AFGR, non-optimal but consistent solutions are obtained: for $V=8$ the average gap is 57\% with 40\% of instances solved, while for $V=9$ the gap is 47\% with 30\% solved. These results confirm the essential role of AFGR as a recovery mechanism for feasibility in highly complex cases.  
Finally, for $V=10$, neither configuration is able to consistently generate valid solutions. The problem becomes intractable within the given time limits, suggesting that additional acceleration or decomposition strategies would be necessary to deal with instances of this scale.  

Overall, the analysis shows that AFGR becomes increasingly beneficial as instance complexity increases. For the smallest cases, it yields only marginal improvements in solution quality, whereas for more challenging configurations it plays a crucial role in obtaining feasible and stable solutions. AFGR also consistently reduces variability, even when the optimality gap remains significant, thereby enhancing the robustness of the solving process. Nevertheless, the limitations of current quantum technology become evident as the problem size increases: beyond $V=8$, even with AFGR, the solutions deviate substantially from the optimum and the percentage of solved instances decreases. This confirms the role of AFGR as an effective support mechanism for maintaining feasibility and stability, while acknowledging that overall scalability remains constrained by the capabilities of present hybrid quantum–classical solvers.

\section{Conclusions and Further Work}\label{sec:conclusions} 

This work introduces the STSP-TWPD, a novel and practically relevant extension of the classical STSP. By integrating time windows, pickup and delivery operations and vehicle capacity constraints within a Steiner graph structure, the STSP-TWPD captures the multifaceted challenges faced in real-world logistics applications such as last-mile delivery, reverse logistics, and time-sensitive distribution. To address the inherent computational complexity of this NP-hard problem, we develop two original mathematical formulations: an ABF that extends traditional STSP models by explicitly incorporating temporal and logistical dimensions, and a NBF, a novel modeling representation for the STSP. 
We also introduce formal mathematical formulations for the STSP-PD and the STSP-TW, which, to the best of our knowledge, have not previously appeared in the literature.
Furthermore, we propose a reduction method, the AFGR, a preprocessing technique that reduces problem dimensionality by eliminating redundant arcs, while preserving feasibility, significantly enhancing scalability and solver performance. %%\color{black}
We generated an extensive set of benchmark instance classes specifically designed to systematically test the proposed models and solution methodologies under varying levels of complexity.
We solved the proposed models using classical solver and additionally assessed the feasibility and compatibility of the STSP-TWPD formulations within a hybrid quantum–classical computing framework. By implementing our models on D-Wave’s \texttt{LeapCQMHybrid} solver, we provide a proof of concept demonstrating that complex routing problems such as the STSP-TWPD can already be represented and solved, at least for small instances, using emerging quantum optimization paradigms.
The experimental results demonstrate that the AFGR method can reduce the model size by more than 70\%, while consistently achieving objective values that are equal to or even better than those obtained without AFGR.

Looking ahead, several avenues for future research emerge from this work. From a modeling perspective, incorporating additional realistic constraints, such as stochastic travel times, multiple vehicles, dynamic request arrivals, or customer priority levels, would extend the applicability of the STSP-TWPD to more complex and operationally relevant logistics scenarios. A particularly meaningful direction is to relax the current assumption that each customer issues either a pickup or a delivery request, with pickups required to precede deliveries. Exploring more general pickup-and-delivery structures, such as simultaneous pickup and delivery, flexible precedence patterns, or multi-commodity flows, could significantly broaden the modeling capabilities of the STSP-TWPD framework.

Another promising research direction concerns the generation of benchmark instances with more diverse underlying graph structures. While the benchmark instances considered in this work allow the systematic variation of key problem parameters, future studies could investigate alternative graph-generation mechanisms involving clustered node distributions, heterogeneous spatial layouts, varying graph densities, and different connectivity patterns. Exploring these structural characteristics may lead to more challenging and realistic benchmark instances, enabling a deeper assessment of the robustness, scalability, and performance of both classical and quantum solution approaches under a wider range of operating conditions.

On the computational side, this work also opens promising opportunities for advancing quantum-assisted optimization. Further investigation of hybrid quantum–classical solvers is warranted, including comparative analyses across quantum platforms, embedding techniques, and solver configurations. Such studies would help clarify how improvements in coherence time, qubit connectivity, and sampling strategies may enhance the scalability and performance of STSP-TWPD formulations in quantum settings.

%Finally, the development of standardized benchmark datasets will support reproducibility and facilitate the evaluation of emerging quantum and hybrid optimization techniques. Overall, this study lays the groundwork for a new generation of optimization models and algorithms capable of addressing the complexity of modern routing systems through the combined power of rigorous modeling, reduction techniques, and hybrid quantum-classical computing.

%\color{black}

\section*{Declaration of competing interest}
The authors declare that they have no known competing financial interests or personal relationships that could have appeared to
influence the work reported in this paper.

\section*{Data availability}
The data used in this study are available in the repository provided in this work. The benchmark datasets employed are included and referenced within the repository.

\section*{Acknowledgments}
Eneko Osaba acknowledges support from the Basque Government through ELKARTEK Program under Grants KK-2025/00074 (Newhegaz project).

\appendix
\section{Model Dimensionality and Computational Complexity -- Tables}
\label{app:complexitytable}

\begin{table}[h]
\centering
\caption{
Detailed comparison of ABF and NBF complexity with and without the AFGR method across A1 istances (as defined in Table~\ref{tab:instance_classes}). For each instance size, we report:
\textbf{Bin}: number of binary decision variables,
   \textbf{Cont}: number of continuous variables,
    \textbf{Tot Vars}: total number of variables (binary + continuous),
   $Eq:$ number of equality constraints,
    $\leq$: number of inequality constraints,
    $\geq$: number of inequality constraints,
    \textbf{Tot C}: total number of constraints.
The columns labeled ``With AFGR'' show the reduced model size after applying the AFGR. The final columns indicate the percentage reduction in total variables and constraints. }\label{tab:ABF-complexity_A1}
% ------------------ ABF ------------------
\resizebox{\columnwidth}{!}{
\begin{tabular}{c|ccc|cccc|ccc|cccc|c|c}
\toprule
\multicolumn{17}{c}{\textbf{ABF}} \\
\toprule
\multirow{2}{*}{$V$} & \multicolumn{7}{c}{\textbf{Without AFGR}} & \multicolumn{7}{c}{\textbf{With AFGR}} & \multirow{2}{*}{\begin{tabular}[c]{@{}c@{}}\% Red.\\ Vars\end{tabular}} & \multirow{2}{*}{\begin{tabular}[c]{@{}c@{}}\% Red.\\ Constr.\end{tabular}} \\
\cmidrule(lr){2-8} \cmidrule(lr){9-15}
 & Bin & Cont & Tot Vars& $Eq$ & $\leq$ & $\geq$ & Tot C & Bin & Cont & Tot Vars & $Eq$ & $\leq$ & $\geq$ & Tot C &  \\
\midrule
   
    4     & 81    & 64    & 145   & 31    & 157   & 70    & 258   & 81    & 64    & 145   & 31    & 157   & 70    & 258   & 0\%   & 0\% \\
    5     & 155   & 110   & 265   & 52    & 294   & 140   & 486   & 155   & 110   & 265   & 52    & 294   & 140   & 486   & 0\%   & 0\% \\
    6     & 286   & 180   & 466   & 84    & 509   & 252   & 845   & 286   & 180   & 466   & 84    & 509   & 252   & 845   & 0\%   & 0\% \\
    7     & 501   & 280   & 781   & 132   & 861   & 456   & 1449  & 501   & 280   & 781   & 132   & 861   & 456   & 1449  & 0\%   & 0\% \\
    8     & 833   & 416   & 1249  & 198   & 1382  & 775   & 2355  & 833   & 416   & 1249  & 198   & 1382  & 775   & 2355  & 0\%   & 0\% \\
    9     & 1321  & 594   & 1915  & 285   & 2120  & 1248  & 3653  & 859   & 468   & 1327  & 223   & 1483  & 800   & 2506  & 31\%  & 31\% \\
    10    & 2010  & 820   & 2830  & 396   & 3129  & 1920  & 5445  & 1354  & 660   & 2014  & 317   & 2249  & 1280  & 3846  & 29\%  & 29\% \\
    11    & 2951  & 1100  & 4051  & 533   & 4420  & 2793  & 7746  & 1387  & 726   & 2113  & 348   & 2346  & 1280  & 3974  & 48\%  & 49\% \\
    12    & 4201  & 1440  & 5641  & 701   & 6147  & 4012  & 10860 & 2092  & 984   & 3076  & 475   & 3411  & 1960  & 5846  & 45\%  & 46\% \\
    13    & 5823  & 1846  & 7669  & 902   & 8342  & 5600  & 14844 & 3051  & 1300  & 4351  & 631   & 4814  & 2891  & 8336  & 43\%  & 44\% \\
    14    & 7886  & 2324  & 10210 & 1139  & 11083 & 7626  & 19848 & 4321  & 1680  & 6001  & 819   & 6621  & 4130  & 11570 & 41\%  & 42\% \\
    15    & 10465 & 2880  & 13345 & 1415  & 14454 & 10165 & 26034 & 5965  & 2130  & 8095  & 1042  & 8904  & 5740  & 15686 & 39\%  & 40\% \\
    16    & 13641 & 3520  & 17161 & 1732  & 18436 & 13189 & 33357 & 6036  & 2272  & 8308  & 1111  & 9115  & 5740  & 15966 & 52\%  & 52\% \\
    17    & 17501 & 4250  & 21751 & 2095  & 23328 & 16988 & 42411 & 8135  & 2822  & 10957 & 1384  & 11988 & 7790  & 21162 & 50\%  & 50\% \\
    18    & 22138 & 5076  & 27214 & 2506  & 29137 & 21560 & 53203 & 10753 & 3456  & 14209 & 1699  & 15502 & 10355 & 27556 & 48\%  & 48\% \\
    19    & 27651 & 6004  & 33655 & 2968  & 35971 & 27004 & 65943 & 13971 & 4180  & 18151 & 2059  & 19747 & 13516 & 35322 & 46\%  & 46\% \\
    20    & 34145 & 7040  & 41185 & 3484  & 43944 & 33425 & 80853 & 17876 & 5000  & 22876 & 2467  & 24819 & 17360 & 44646 & 44\%  & 45\% \\
  
\midrule
\multicolumn{15}{r}{\textbf{Average Reduction:}} & \textbf{30\%} & \textbf{31\%} \\
\bottomrule
\end{tabular}
}

\vspace{0.5cm}

% ------------------ NBF ------------------
\resizebox{\columnwidth}{!}{
\begin{tabular}{c|ccc|cccc|ccc|cccc|c|c}
\toprule
\multicolumn{17}{c}{\textbf{NBF}} \\
\toprule
\multirow{2}{*}{$V$} & \multicolumn{7}{c}{\textbf{Without AFGR}} & \multicolumn{7}{c}{\textbf{With AFGR}} & \multirow{2}{*}{\begin{tabular}[c]{@{}c@{}}\% Red.\\ Vars\end{tabular}} & \multirow{2}{*}{\begin{tabular}[c]{@{}c@{}}\% Red.\\ Constr.\end{tabular}} \\
\cmidrule(lr){2-8} \cmidrule(lr){9-15}
 & Bin & Cont & Tot Vars& $Eq$ & $\leq$ & $\geq$ & Tot C & Bin & Cont & Tot Vars & $Eq$ & $\leq$ & $\geq$ & Tot C &  \\
\midrule
  
    4     & 81    & 64    & 145   & 29    & 157   & 70    & 256   & 81    & 64    & 145   & 29    & 157   & 70    & 256   & 0\%   & 0\% \\
    5     & 155   & 110   & 265   & 49    & 294   & 140   & 483   & 155   & 110   & 265   & 49    & 294   & 140   & 483   & 0\%   & 0\% \\
    6     & 286   & 180   & 466   & 79    & 509   & 252   & 840   & 286   & 180   & 466   & 79    & 509   & 252   & 840   & 0\%   & 0\% \\
    7     & 501   & 280   & 781   & 124   & 861   & 456   & 1441  & 501   & 280   & 781   & 124   & 861   & 456   & 1441  & 0\%   & 0\% \\
    8     & 833   & 416   & 1249  & 186   & 1382  & 775   & 2343  & 833   & 416   & 1249  & 186   & 1382  & 775   & 2343  & 0\%   & 0\% \\
    9     & 1321  & 594   & 1915  & 268   & 2120  & 1248  & 3636  & 859   & 468   & 1327  & 212   & 1483  & 800   & 2495  & 31\%  & 31\% \\
    10    & 2010  & 820   & 2830  & 373   & 3129  & 1920  & 5422  & 1354  & 660   & 2014  & 301   & 2249  & 1280  & 3830  & 29\%  & 29\% \\
    11    & 2951  & 1100  & 4051  & 503   & 4420  & 2793  & 7716  & 1387  & 726   & 2113  & 333   & 2346  & 1280  & 3959  & 48\%  & 49\% \\
    12    & 4201  & 1440  & 5641  & 663   & 6147  & 4012  & 10822 & 2092  & 984   & 3076  & 454   & 3411  & 1960  & 5825  & 45\%  & 46\% \\
    13    & 5823  & 1846  & 7669  & 855   & 8342  & 5600  & 14797 & 3051  & 1300  & 4351  & 603   & 4814  & 2891  & 8308  & 43\%  & 44\% \\
    14    & 7886  & 2324  & 10210 & 1082  & 11083 & 7626  & 19791 & 4321  & 1680  & 6001  & 783   & 6621  & 4130  & 11534 & 41\%  & 42\% \\
    15    & 10465 & 2880  & 13345 & 1347  & 14454 & 10165 & 25966 & 5965  & 2130  & 8095  & 997   & 8904  & 5740  & 15641 & 39\%  & 40\% \\
    16    & 13641 & 3520  & 17161 & 1652  & 18436 & 13189 & 33277 & 6036  & 2272  & 8308  & 1067  & 9115  & 5740  & 15922 & 52\%  & 52\% \\
    17    & 17501 & 4250  & 21751 & 2002  & 23328 & 16988 & 42318 & 8135  & 2822  & 10957 & 1330  & 11988 & 7790  & 21108 & 50\%  & 50\% \\
    18    & 22138 & 5076  & 27214 & 2399  & 29137 & 21560 & 53096 & 10753 & 3456  & 14209 & 1634  & 15502 & 10355 & 27491 & 48\%  & 48\% \\
    19    & 27651 & 6004  & 33655 & 2846  & 35971 & 27004 & 65821 & 13971 & 4180  & 18151 & 1982  & 19747 & 13516 & 35245 & 46\%  & 46\% \\
    20    & 34145 & 7040  & 41185 & 3346  & 43944 & 33425 & 80715 & 17876 & 5000  & 22876 & 2377  & 24819 & 17360 & 44556 & 44\%  & 45\% \\
  \midrule
\multicolumn{15}{r}{\textbf{Average Reduction:}} & \textbf{30\%} & \textbf{31\%} \\
\bottomrule
\end{tabular}}
\end{table}

\begin{table}[h]
\centering
\caption{
Detailed comparison of ABF and NBF complexity with and without the AFGR method across A2 istances (as defined in Table~\ref{tab:instance_classes}). For each instance size, we report:
\textbf{Bin}: number of binary decision variables,
   \textbf{Cont}: number of continuous variables,
    \textbf{Tot Vars}: total number of variables (binary + continuous),
   $Eq:$ number of equality constraints,
    $\leq$: number of inequality constraints,
    $\geq$: number of inequality constraints,
    \textbf{Tot C}: total number of constraints.
The columns labeled ``With AFGR'' show the reduced model size after applying the AFGR. The final columns indicate the percentage reduction in total variables and constraints. }\label{tab:ABF-complexit_A3}
% ------------------ ABF ------------------
\resizebox{\columnwidth}{!}{
\begin{tabular}{c|ccc|cccc|ccc|cccc|c|c}
\toprule
\multicolumn{17}{c}{\textbf{ABF}} \\
\toprule
\multirow{2}{*}{$V$} & \multicolumn{7}{c}{\textbf{Without AFGR}} & \multicolumn{7}{c}{\textbf{With AFGR}} & \multirow{2}{*}{\begin{tabular}[c]{@{}c@{}}\% Red.\\ Vars\end{tabular}} & \multirow{2}{*}{\begin{tabular}[c]{@{}c@{}}\% Red.\\ Constr.\end{tabular}} \\
\cmidrule(lr){2-8} \cmidrule(lr){9-15}
 & Bin & Cont & Tot Vars& $Eq$ & $\leq$ & $\geq$ & Tot C & Bin & Cont & Tot Vars & $Eq$ & $\leq$ & $\geq$ & Tot C &  \\
\midrule
   
    4     & 81    & 64    & 145   & 31    & 157   & 70    & 258   & 81    & 64    & 145   & 31    & 157   & 70    & 258   & 0\%   & 0\% \\
    5     & 155   & 110   & 265   & 51    & 284   & 130   & 465   & 155   & 110   & 265   & 51    & 284   & 130   & 465   & 0\%   & 0\% \\
    6     & 286   & 180   & 466   & 84    & 509   & 252   & 845   & 286   & 180   & 466   & 84    & 509   & 252   & 845   & 0\%   & 0\% \\
    7     & 501   & 280   & 781   & 132   & 861   & 456   & 1449  & 501   & 280   & 781   & 132   & 861   & 456   & 1449  & 0\%   & 0\% \\
    8     & 833   & 416   & 1249  & 197   & 1357  & 750   & 2304  & 521   & 320   & 841   & 150   & 919   & 456   & 1525  & 33\%  & 34\% \\
    9     & 1321  & 594   & 1915  & 284   & 2088  & 1216  & 3588  & 541   & 360   & 901   & 169   & 996   & 475   & 1640  & 53\%  & 54\% \\
    10    & 2010  & 820   & 2830  & 395   & 3089  & 1880  & 5364  & 885   & 520   & 1405  & 247   & 1559  & 800   & 2606  & 50\%  & 51\% \\
    11    & 2951  & 1100  & 4051  & 532   & 4371  & 2744  & 7647  & 911   & 572   & 1483  & 271   & 1635  & 800   & 2706  & 63\%  & 65\% \\
    12    & 4201  & 1440  & 5641  & 700   & 6088  & 3953  & 10741 & 1420  & 792   & 2212  & 379   & 2443  & 1280  & 4102  & 61\%  & 62\% \\
    13    & 5823  & 1846  & 7669  & 901   & 8272  & 5530  & 14703 & 2133  & 1066  & 3199  & 514   & 3532  & 1960  & 6006  & 58\%  & 59\% \\
    14    & 7886  & 2324  & 10210 & 1138  & 11001 & 7544  & 19683 & 3101  & 1400  & 4501  & 679   & 4962  & 2891  & 8532  & 56\%  & 57\% \\
    15    & 10465 & 2880  & 13345 & 1413  & 14264 & 9975  & 25652 & 3151  & 1500  & 4651  & 727   & 5110  & 2891  & 8728  & 65\%  & 66\% \\
    16    & 13641 & 3520  & 17161 & 1731  & 18327 & 13080 & 33138 & 4441  & 1920  & 6361  & 935   & 6977  & 4130  & 12042 & 63\%  & 64\% \\
    17    & 17501 & 4250  & 21751 & 2094  & 23204 & 16864 & 42162 & 6107  & 2414  & 8521  & 1180  & 9326  & 5740  & 16246 & 61\%  & 61\% \\
    18    & 22138 & 5076  & 27214 & 2504  & 28857 & 21280 & 52641 & 8218  & 2988  & 11206 & 1464  & 12153 & 7708  & 21325 & 59\%  & 59\% \\
    19    & 27651 & 6004  & 33655 & 2966  & 35657 & 26690 & 65313 & 10849 & 3648  & 14497 & 1792  & 15693 & 10260 & 27745 & 57\%  & 58\% \\
    20    & 34145 & 7040  & 41185 & 3482  & 43594 & 33075 & 80151 & 14081 & 4400  & 18481 & 2166  & 19966 & 13407 & 35539 & 55\%  & 56\% \\
    \midrule
\multicolumn{15}{r}{\textbf{Average Reduction:}} & \textbf{43\%} & \textbf{44\%} \\
\bottomrule
\end{tabular}
}

\vspace{0.5cm}

% ------------------ NBF ------------------
\resizebox{\columnwidth}{!}{
\begin{tabular}{c|ccc|cccc|ccc|cccc|c|c}
\toprule
\multicolumn{17}{c}{\textbf{NBF}} \\
\toprule
\multirow{2}{*}{$V$} & \multicolumn{7}{c}{\textbf{Without AFGR}} & \multicolumn{7}{c}{\textbf{With AFGR}} & \multirow{2}{*}{\begin{tabular}[c]{@{}c@{}}\% Red.\\ Vars\end{tabular}} & \multirow{2}{*}{\begin{tabular}[c]{@{}c@{}}\% Red.\\ Constr.\end{tabular}} \\
\cmidrule(lr){2-8} \cmidrule(lr){9-15}
 & Bin & Cont & Tot Vars& $Eq$ & $\leq$ & $\geq$ & Tot C & Bin & Cont & Tot Vars & $Eq$ & $\leq$ & $\geq$ & Tot C &  \\
\midrule
 
    4     & 81    & 64    & 145   & 31    & 157   & 70    & 258   & 81    & 64    & 145   & 31    & 157   & 70    & 258   & 0\%   & 0\% \\
    5     & 155   & 110   & 265   & 51    & 284   & 130   & 465   & 155   & 110   & 265   & 51    & 284   & 130   & 465   & 0\%   & 0\% \\
    6     & 286   & 180   & 466   & 84    & 509   & 252   & 845   & 286   & 180   & 466   & 84    & 509   & 252   & 845   & 0\%   & 0\% \\
    7     & 501   & 280   & 781   & 132   & 861   & 456   & 1449  & 501   & 280   & 781   & 132   & 861   & 456   & 1449  & 0\%   & 0\% \\
    8     & 833   & 416   & 1249  & 197   & 1357  & 750   & 2304  & 521   & 320   & 841   & 150   & 919   & 456   & 1525  & 33\%  & 34\% \\
    9     & 1321  & 594   & 1915  & 284   & 2088  & 1216  & 3588  & 541   & 360   & 901   & 169   & 996   & 475   & 1640  & 53\%  & 54\% \\
    10    & 2010  & 820   & 2830  & 395   & 3089  & 1880  & 5364  & 885   & 520   & 1405  & 247   & 1559  & 800   & 2606  & 50\%  & 51\% \\
    11    & 2951  & 1100  & 4051  & 532   & 4371  & 2744  & 7647  & 911   & 572   & 1483  & 271   & 1635  & 800   & 2706  & 63\%  & 65\% \\
    12    & 4201  & 1440  & 5641  & 700   & 6088  & 3953  & 10741 & 1420  & 792   & 2212  & 379   & 2443  & 1280  & 4102  & 61\%  & 62\% \\
    13    & 5823  & 1846  & 7669  & 901   & 8272  & 5530  & 14703 & 2133  & 1066  & 3199  & 514   & 3532  & 1960  & 6006  & 58\%  & 59\% \\
    14    & 7886  & 2324  & 10210 & 1138  & 11001 & 7544  & 19683 & 3101  & 1400  & 4501  & 679   & 4962  & 2891  & 8532  & 56\%  & 57\% \\
    15    & 10465 & 2880  & 13345 & 1413  & 14264 & 9975  & 25652 & 3151  & 1500  & 4651  & 727   & 5110  & 2891  & 8728  & 65\%  & 66\% \\
    16    & 13641 & 3520  & 17161 & 1731  & 18327 & 13080 & 33138 & 4441  & 1920  & 6361  & 935   & 6977  & 4130  & 12042 & 63\%  & 64\% \\
    17    & 17501 & 4250  & 21751 & 2094  & 23204 & 16864 & 42162 & 6107  & 2414  & 8521  & 1180  & 9326  & 5740  & 16246 & 61\%  & 61\% \\
    18    & 22138 & 5076  & 27214 & 2504  & 28857 & 21280 & 52641 & 8218  & 2988  & 11206 & 1464  & 12153 & 7708  & 21325 & 59\%  & 59\% \\
    19    & 27651 & 6004  & 33655 & 2966  & 35657 & 26690 & 65313 & 10849 & 3648  & 14497 & 1792  & 15693 & 10260 & 27745 & 57\%  & 58\% \\
    20    & 34145 & 7040  & 41185 & 3482  & 43594 & 33075 & 80151 & 14081 & 4400  & 18481 & 2166  & 19966 & 13407 & 35539 & 55\%  & 56\% \\
  \midrule
\multicolumn{15}{r}{\textbf{Average Reduction:}} & \textbf{43\%} & \textbf{44\%} \\
\bottomrule
\end{tabular}}
\end{table}

\begin{table}[h]
\centering
\caption{
Detailed comparison of ABF and NBF complexity with and without the AFGR method across A3 istances (as defined in Table~\ref{tab:instance_classes}). For each instance size, we report:
\textbf{Bin}: number of binary decision variables,
   \textbf{Cont}: number of continuous variables,
    \textbf{Tot Vars}: total number of variables (binary + continuous),
   $Eq:$ number of equality constraints,
    $\leq$: number of inequality constraints,
    $\geq$: number of inequality constraints,
    \textbf{Tot C}: total number of constraints.
The columns labeled ``With AFGR'' show the reduced model size after applying the AFGR. The final columns indicate the percentage reduction in total variables and constraints. }\label{tab:ABF-complexity_A5}
% ------------------ ABF ------------------
\resizebox{\columnwidth}{!}{
\begin{tabular}{c|ccc|cccc|ccc|cccc|c|c}
\toprule
\multicolumn{17}{c}{\textbf{ABF}} \\
\toprule
\multirow{2}{*}{$V$} & \multicolumn{7}{c}{\textbf{Without AFGR}} & \multicolumn{7}{c}{\textbf{With AFGR}} & \multirow{2}{*}{\begin{tabular}[c]{@{}c@{}}\% Red.\\ Vars\end{tabular}} & \multirow{2}{*}{\begin{tabular}[c]{@{}c@{}}\% Red.\\ Constr.\end{tabular}} \\
\cmidrule(lr){2-8} \cmidrule(lr){9-15}
 & Bin & Cont & Tot Vars& $Eq$ & $\leq$ & $\geq$ & Tot C & Bin & Cont & Tot Vars & $Eq$ & $\leq$ & $\geq$ & Tot C &  \\
\midrule
   
    4     & 81    & 64    & 145   & 30    & 150   & 63    & 243   & 81    & 64    & 145   & 30    & 150   & 63    & 243   & 0\%   & 0\% \\
    5     & 155   & 110   & 265   & 51    & 284   & 130   & 465   & 155   & 110   & 265   & 51    & 284   & 130   & 465   & 0\%   & 0\% \\
    6     & 286   & 180   & 466   & 83    & 495   & 238   & 816   & 222   & 156   & 378   & 72    & 401   & 180   & 653   & 19\%  & 20\% \\
    7     & 501   & 280   & 781   & 131   & 842   & 437   & 1410  & 301   & 210   & 511   & 97    & 552   & 252   & 901   & 35\%  & 36\% \\
    8     & 833   & 416   & 1249  & 197   & 1357  & 750   & 2304  & 521   & 320   & 841   & 150   & 919   & 456   & 1525  & 33\%  & 34\% \\
    9     & 1321  & 594   & 1915  & 283   & 2056  & 1184  & 3523  & 331   & 270   & 601   & 124   & 652   & 266   & 1042  & 69\%  & 70\% \\
    10    & 2010  & 820   & 2830  & 394   & 3049  & 1840  & 5283  & 561   & 400   & 961   & 187   & 1054  & 475   & 1716  & 66\%  & 68\% \\
    11    & 2951  & 1100  & 4051  & 531   & 4322  & 2695  & 7548  & 581   & 440   & 1021  & 205   & 1112  & 475   & 1792  & 75\%  & 76\% \\
    12    & 4201  & 1440  & 5641  & 699   & 6029  & 3894  & 10622 & 1420  & 792   & 2212  & 378   & 2411  & 1248  & 4037  & 61\%  & 62\% \\
    13    & 5823  & 1846  & 7669  & 900   & 8202  & 5460  & 14562 & 2133  & 1066  & 3199  & 513   & 3492  & 1920  & 5925  & 58\%  & 59\% \\
    14    & 7886  & 2324  & 10210 & 1136  & 10837 & 7380  & 19353 & 2174  & 1148  & 3322  & 552   & 3613  & 1920  & 6085  & 67\%  & 69\% \\
    15    & 10465 & 2880  & 13345 & 1412  & 14169 & 9880  & 25461 & 3151  & 1500  & 4651  & 726   & 5061  & 2842  & 8629  & 65\%  & 66\% \\
    16    & 13641 & 3520  & 17161 & 1729  & 18109 & 12862 & 32700 & 3201  & 1600  & 4801  & 774   & 5209  & 2842  & 8825  & 72\%  & 73\% \\
    17    & 17501 & 4250  & 21751 & 2092  & 22956 & 16616 & 41664 & 4501  & 2040  & 6541  & 992   & 7096  & 4071  & 12159 & 70\%  & 71\% \\
    18    & 22138 & 5076  & 27214 & 2503  & 28717 & 21140 & 52360 & 6178  & 2556  & 8734  & 1248  & 9467  & 5670  & 16385 & 68\%  & 69\% \\
    19    & 27651 & 6004  & 33655 & 2964  & 35343 & 26376 & 64683 & 6249  & 2698  & 8947  & 1317  & 9678  & 5670  & 16665 & 73\%  & 74\% \\
    20    & 34145 & 7040  & 41185 & 3480  & 43244 & 32725 & 79449 & 8384  & 3320  & 11704 & 1626  & 12647 & 7708  & 21981 & 72\%  & 72\% \\
  \midrule
\multicolumn{15}{r}{\textbf{Average Reduction:}} & \textbf{53\%} & \textbf{54\%} \\
\bottomrule
\end{tabular}
}

\vspace{0.5cm}

% ------------------ NBF ------------------
\resizebox{\columnwidth}{!}{
\begin{tabular}{c|ccc|cccc|ccc|cccc|c|c}
\toprule
\multicolumn{17}{c}{\textbf{NBF}} \\
\toprule
\multirow{2}{*}{$V$} & \multicolumn{7}{c}{\textbf{Without AFGR}} & \multicolumn{7}{c}{\textbf{With AFGR}} & \multirow{2}{*}{\begin{tabular}[c]{@{}c@{}}\% Red.\\ Vars\end{tabular}} & \multirow{2}{*}{\begin{tabular}[c]{@{}c@{}}\% Red.\\ Constr.\end{tabular}} \\
\cmidrule(lr){2-8} \cmidrule(lr){9-15}
 & Bin & Cont & Tot Vars& $Eq$ & $\leq$ & $\geq$ & Tot C & Bin & Cont & Tot Vars & $Eq$ & $\leq$ & $\geq$ & Tot C &  \\
\midrule
4     & 81    & 64    & 145   & 28    & 150   & 63    & 241   & 81    & 64    & 145   & 28    & 150   & 63    & 241   & 0\%   & 0\% \\
    5     & 155   & 110   & 265   & 48    & 284   & 130   & 462   & 155   & 110   & 265   & 48    & 284   & 130   & 462   & 0\%   & 0\% \\
    6     & 286   & 180   & 466   & 78    & 495   & 238   & 811   & 222   & 156   & 378   & 68    & 401   & 180   & 649   & 19\%  & 20\% \\
    7     & 501   & 280   & 781   & 123   & 842   & 437   & 1402  & 301   & 210   & 511   & 93    & 552   & 252   & 897   & 35\%  & 36\% \\
    8     & 833   & 416   & 1249  & 185   & 1357  & 750   & 2292  & 521   & 320   & 841   & 143   & 919   & 456   & 1518  & 33\%  & 34\% \\
    9     & 1321  & 594   & 1915  & 266   & 2056  & 1184  & 3506  & 331   & 270   & 601   & 122   & 652   & 266   & 1040  & 69\%  & 70\% \\
    10    & 2010  & 820   & 2830  & 371   & 3049  & 1840  & 5260  & 561   & 400   & 961   & 182   & 1054  & 475   & 1711  & 66\%  & 67\% \\
    11    & 2951  & 1100  & 4051  & 501   & 4322  & 2695  & 7518  & 581   & 440   & 1021  & 201   & 1112  & 475   & 1788  & 75\%  & 76\% \\
    12    & 4201  & 1440  & 5641  & 661   & 6029  & 3894  & 10584 & 1420  & 792   & 2212  & 364   & 2411  & 1248  & 4023  & 61\%  & 62\% \\
    13    & 5823  & 1846  & 7669  & 853   & 8202  & 5460  & 14515 & 2133  & 1066  & 3199  & 493   & 3492  & 1920  & 5905  & 58\%  & 59\% \\
    14    & 7886  & 2324  & 10210 & 1079  & 10837 & 7380  & 19296 & 2174  & 1148  & 3322  & 533   & 3613  & 1920  & 6066  & 67\%  & 69\% \\
    15    & 10465 & 2880  & 13345 & 1344  & 14169 & 9880  & 25393 & 3151  & 1500  & 4651  & 700   & 5061  & 2842  & 8603  & 65\%  & 66\% \\
    16    & 13641 & 3520  & 17161 & 1649  & 18109 & 12862 & 32620 & 3201  & 1600  & 4801  & 749   & 5209  & 2842  & 8800  & 72\%  & 73\% \\
    17    & 17501 & 4250  & 21751 & 1999  & 22956 & 16616 & 41571 & 4501  & 2040  & 6541  & 959   & 7096  & 4071  & 12126 & 70\%  & 71\% \\
    18    & 22138 & 5076  & 27214 & 2396  & 28717 & 21140 & 52253 & 6178  & 2556  & 8734  & 1206  & 9467  & 5670  & 16343 & 68\%  & 69\% \\
    19    & 27651 & 6004  & 33655 & 2842  & 35343 & 26376 & 64561 & 6249  & 2698  & 8947  & 1276  & 9678  & 5670  & 16624 & 73\%  & 74\% \\
    20    & 34145 & 7040  & 41185 & 3342  & 43244 & 32725 & 79311 & 8384  & 3320  & 11704 & 1575  & 12647 & 7708  & 21930 & 72\%  & 72\% \\
   
  \midrule
\multicolumn{15}{r}{\textbf{Average Reduction:}} & \textbf{53\%} & \textbf{54\%} \\
\bottomrule
\end{tabular}}
\end{table}
\begin{table}[h]
\centering
\caption{
Detailed comparison of ABF and NBF complexity with and without the AFGR method across A4 istances (as defined in Table~\ref{tab:instance_classes}). For each instance size, we report:
\textbf{Bin}: number of binary decision variables,
   \textbf{Cont}: number of continuous variables,
    \textbf{Tot Vars}: total number of variables (binary + continuous),
   $Eq:$ number of equality constraints,
    $\leq$: number of inequality constraints,
    $\geq$: number of inequality constraints,
    \textbf{Tot C}: total number of constraints.
The columns labeled ``With AFGR'' show the reduced model size after applying the AFGR. The final columns indicate the percentage reduction in total variables and constraints. }\label{tab:ABF-complexity_B2}
% ------------------ ABF ------------------
\resizebox{\columnwidth}{!}{
\begin{tabular}{c|ccc|cccc|ccc|cccc|c|c}
\toprule
\multicolumn{17}{c}{\textbf{ABF}} \\
\toprule
\multirow{2}{*}{$V$} & \multicolumn{7}{c}{\textbf{Without AFGR}} & \multicolumn{7}{c}{\textbf{With AFGR}} & \multirow{2}{*}{\begin{tabular}[c]{@{}c@{}}\% Red.\\ Vars\end{tabular}} & \multirow{2}{*}{\begin{tabular}[c]{@{}c@{}}\% Red.\\ Constr.\end{tabular}} \\
\cmidrule(lr){2-8} \cmidrule(lr){9-15}
 & Bin & Cont & Tot Vars& $Eq$ & $\leq$ & $\geq$ & Tot C & Bin & Cont & Tot Vars & $Eq$ & $\leq$ & $\geq$ & Tot C &  \\
\midrule
  
    4     & 81    & 64    & 145   & 30    & 136   & 63    & 229   & 81    & 64    & 145   & 30    & 136   & 63    & 229   & 0\%   & 0\% \\
    5     & 155   & 110   & 265   & 50    & 254   & 120   & 424   & 89    & 80    & 169   & 36    & 158   & 63    & 257   & 36\%  & 39\% \\
    6     & 286   & 180   & 466   & 83    & 453   & 238   & 774   & 222   & 156   & 378   & 72    & 365   & 180   & 617   & 19\%  & 20\% \\
    7     & 501   & 280   & 781   & 130   & 766   & 418   & 1314  & 177   & 154   & 331   & 69    & 316   & 130   & 515   & 58\%  & 61\% \\
    8     & 833   & 416   & 1249  & 196   & 1232  & 725   & 2153  & 316   & 240   & 556   & 110   & 539   & 252   & 901   & 55\%  & 58\% \\
    9     & 1321  & 594   & 1915  & 282   & 1896  & 1152  & 3330  & 331   & 270   & 601   & 123   & 582   & 252   & 957   & 69\%  & 71\% \\
    10    & 2010  & 820   & 2830  & 393   & 2809  & 1800  & 5002  & 346   & 300   & 646   & 137   & 625   & 266   & 1028  & 77\%  & 79\% \\
    11    & 2951  & 1100  & 4051  & 530   & 4028  & 2646  & 7204  & 581   & 440   & 1021  & 204   & 998   & 456   & 1658  & 75\%  & 77\% \\
    12    & 4201  & 1440  & 5641  & 698   & 5616  & 3835  & 10149 & 937   & 624   & 1561  & 294   & 1536  & 775   & 2605  & 72\%  & 74\% \\
    13    & 5823  & 1846  & 7669  & 898   & 7642  & 5320  & 13860 & 963   & 676   & 1639  & 318   & 1612  & 775   & 2705  & 79\%  & 80\% \\
    14    & 7886  & 2324  & 10210 & 1135  & 10181 & 7298  & 18614 & 1486  & 924   & 2410  & 440   & 2381  & 1248  & 4069  & 76\%  & 78\% \\
    15    & 10465 & 2880  & 13345 & 1410  & 13314 & 9690  & 24414 & 1519  & 990   & 2509  & 471   & 2478  & 1248  & 4197  & 81\%  & 83\% \\
    16    & 13641 & 3520  & 17161 & 1728  & 17128 & 12753 & 31609 & 2256  & 1312  & 3568  & 630   & 3535  & 1920  & 6085  & 79\%  & 81\% \\
    17    & 17501 & 4250  & 21751 & 2090  & 21716 & 16368 & 40174 & 3025  & 1632  & 4657  & 788   & 4622  & 2585  & 7995  & 79\%  & 80\% \\
    18    & 22138 & 5076  & 27214 & 2501  & 27177 & 20860 & 50538 & 3301  & 1800  & 5101  & 870   & 5064  & 2842  & 8776  & 81\%  & 83\% \\
    19    & 27651 & 6004  & 33655 & 2962  & 33616 & 26062 & 62640 & 3351  & 1900  & 5251  & 918   & 5212  & 2842  & 8972  & 84\%  & 86\% \\
    20    & 34145 & 7040  & 41185 & 3478  & 41144 & 32375 & 76997 & 4681  & 2400  & 7081  & 1166  & 7040  & 4071  & 12277 & 83\%  & 84\% \\
    \midrule
\multicolumn{15}{r}{\textbf{Average Reduction:}} & \textbf{65\%} & \textbf{67\%} \\
\bottomrule
\end{tabular}
}

\vspace{0.5cm}

% ------------------ NBF ------------------
\resizebox{\columnwidth}{!}{
\begin{tabular}{c|ccc|cccc|ccc|cccc|c|c}
\toprule
\multicolumn{17}{c}{\textbf{NBF}} \\
\toprule
\multirow{2}{*}{$V$} & \multicolumn{7}{c}{\textbf{Without AFGR}} & \multicolumn{7}{c}{\textbf{With AFGR}} & \multirow{2}{*}{\begin{tabular}[c]{@{}c@{}}\% Red.\\ Vars\end{tabular}} & \multirow{2}{*}{\begin{tabular}[c]{@{}c@{}}\% Red.\\ Constr.\end{tabular}} \\
\cmidrule(lr){2-8} \cmidrule(lr){9-15}
 & Bin & Cont & Tot Vars& $Eq$ & $\leq$ & $\geq$ & Tot C & Bin & Cont & Tot Vars & $Eq$ & $\leq$ & $\geq$ & Tot C &  \\
\midrule
  4     & 81    & 64    & 145   & 28    & 150   & 63    & 241   & 81    & 64    & 145   & 28    & 150   & 63    & 241   & 0\%   & 0\% \\
    5     & 155   & 110   & 265   & 47    & 274   & 120   & 441   & 89    & 80    & 169   & 35    & 172   & 63    & 270   & 36\%  & 39\% \\
    6     & 286   & 180   & 466   & 78    & 495   & 238   & 811   & 222   & 156   & 378   & 68    & 401   & 180   & 649   & 19\%  & 20\% \\
    7     & 501   & 280   & 781   & 122   & 823   & 418   & 1363  & 177   & 154   & 331   & 68    & 346   & 130   & 544   & 58\%  & 60\% \\
    8     & 833   & 416   & 1249  & 184   & 1332  & 725   & 2241  & 316   & 240   & 556   & 107   & 595   & 252   & 954   & 55\%  & 57\% \\
    9     & 1321  & 594   & 1915  & 265   & 2024  & 1152  & 3441  & 331   & 270   & 601   & 121   & 638   & 252   & 1011  & 69\%  & 71\% \\
    10    & 2010  & 820   & 2830  & 370   & 3009  & 1800  & 5179  & 346   & 300   & 646   & 136   & 695   & 266   & 1097  & 77\%  & 79\% \\
    11    & 2951  & 1100  & 4051  & 500   & 4273  & 2646  & 7419  & 581   & 440   & 1021  & 200   & 1093  & 456   & 1749  & 75\%  & 76\% \\
    12    & 4201  & 1440  & 5641  & 660   & 5970  & 3835  & 10465 & 937   & 624   & 1561  & 286   & 1686  & 775   & 2747  & 72\%  & 74\% \\
    13    & 5823  & 1846  & 7669  & 851   & 8062  & 5320  & 14233 & 963   & 676   & 1639  & 311   & 1762  & 775   & 2848  & 79\%  & 80\% \\
    14    & 7886  & 2324  & 10210 & 1078  & 10755 & 7298  & 19131 & 1486  & 924   & 2410  & 428   & 2605  & 1248  & 4281  & 76\%  & 78\% \\
    15    & 10465 & 2880  & 13345 & 1342  & 13979 & 9690  & 25011 & 1519  & 990   & 2509  & 460   & 2702  & 1248  & 4410  & 81\%  & 82\% \\
    16    & 13641 & 3520  & 17161 & 1648  & 18000 & 12753 & 32401 & 2256  & 1312  & 3568  & 613   & 3855  & 1920  & 6388  & 79\%  & 80\% \\
    17    & 17501 & 4250  & 21751 & 1997  & 22708 & 16368 & 41073 & 3025  & 1632  & 4657  & 765   & 4998  & 2585  & 8348  & 79\%  & 80\% \\
    18    & 22138 & 5076  & 27214 & 2394  & 28437 & 20860 & 51691 & 3301  & 1800  & 5101  & 847   & 5505  & 2842  & 9194  & 81\%  & 82\% \\
    19    & 27651 & 6004  & 33655 & 2840  & 35029 & 26062 & 63931 & 3351  & 1900  & 5251  & 896   & 5653  & 2842  & 9391  & 84\%  & 85\% \\
    20    & 34145 & 7040  & 41185 & 3340  & 42894 & 32375 & 78609 & 4681  & 2400  & 7081  & 1136  & 7630  & 4071  & 12837 & 83\%  & 84\% \\
 
  \midrule
\multicolumn{15}{r}{\textbf{Average Reduction:}} & \textbf{65\%} & \textbf{66\%} \\
\bottomrule
\end{tabular}}
\end{table}
\begin{table}[h]
\centering
\caption{
Detailed comparison of ABF and NBF complexity with and without the AFGR method across A5 istances (as defined in Table~\ref{tab:instance_classes}). For each instance size, we report:
\textbf{Bin}: number of binary decision variables,
   \textbf{Cont}: number of continuous variables,
    \textbf{Tot Vars}: total number of variables (binary + continuous),
   $Eq:$ number of equality constraints,
    $\leq$: number of inequality constraints,
    $\geq$: number of inequality constraints,
    \textbf{Tot C}: total number of constraints.
The columns labeled ``With AFGR'' show the reduced model size after applying the AFGR. The final columns indicate the percentage reduction in total variables and constraints. }\label{tab:ABF-complexity_B4}
% ------------------ ABF ------------------
\resizebox{\columnwidth}{!}{
\begin{tabular}{c|ccc|cccc|ccc|cccc|c|c}
\toprule
\multicolumn{17}{c}{\textbf{ABF}} \\
\toprule
\multirow{2}{*}{$V$} & \multicolumn{7}{c}{\textbf{Without AFGR}} & \multicolumn{7}{c}{\textbf{With AFGR}} & \multirow{2}{*}{\begin{tabular}[c]{@{}c@{}}\% Red.\\ Vars\end{tabular}} & \multirow{2}{*}{\begin{tabular}[c]{@{}c@{}}\% Red.\\ Constr.\end{tabular}} \\
\cmidrule(lr){2-8} \cmidrule(lr){9-15}
 & Bin & Cont & Tot Vars& $Eq$ & $\leq$ & $\geq$ & Tot C & Bin & Cont & Tot Vars & $Eq$ & $\leq$ & $\geq$ & Tot C &  \\
\midrule
   
    4     & 81    & 64    & 145   & 30    & 136   & 63    & 229   & 81    & 64    & 145   & 30    & 136   & 63    & 229   & 0\%   & 0\% \\
    5     & 155   & 110   & 265   & 50    & 254   & 120   & 424   & 89    & 80    & 169   & 36    & 158   & 63    & 257   & 36\%  & 39\% \\
    6     & 286   & 180   & 466   & 82    & 453   & 224   & 759   & 118   & 108   & 226   & 48    & 213   & 80    & 341   & 52\%  & 55\% \\
    7     & 501   & 280   & 781   & 130   & 766   & 418   & 1314  & 177   & 154   & 331   & 69    & 316   & 130   & 515   & 58\%  & 61\% \\
    8     & 833   & 416   & 1249  & 195   & 1232  & 700   & 2127  & 188   & 176   & 364   & 78    & 347   & 130   & 555   & 71\%  & 74\% \\
    9     & 1321  & 594   & 1915  & 282   & 1896  & 1152  & 3330  & 331   & 270   & 601   & 123   & 582   & 252   & 957   & 69\%  & 71\% \\
    10    & 2010  & 820   & 2830  & 392   & 2809  & 1760  & 4961  & 346   & 300   & 646   & 136   & 625   & 252   & 1013  & 77\%  & 80\% \\
    11    & 2951  & 1100  & 4051  & 529   & 4028  & 2597  & 7154  & 361   & 330   & 691   & 149   & 668   & 252   & 1069  & 83\%  & 85\% \\
    12    & 4201  & 1440  & 5641  & 697   & 5616  & 3776  & 10089 & 601   & 480   & 1081  & 222   & 1056  & 456   & 1734  & 81\%  & 83\% \\
    13    & 5823  & 1846  & 7669  & 897   & 7642  & 5250  & 13789 & 727   & 572   & 1299  & 266   & 1272  & 546   & 2084  & 83\%  & 85\% \\
    14    & 7886  & 2324  & 10210 & 1134  & 10181 & 7216  & 18531 & 989   & 728   & 1717  & 342   & 1688  & 775   & 2805  & 83\%  & 85\% \\
    15    & 10465 & 2880  & 13345 & 1409  & 13314 & 9595  & 24318 & 1015  & 780   & 1795  & 366   & 1764  & 775   & 2905  & 87\%  & 88\% \\
    16    & 13641 & 3520  & 17161 & 1726  & 17128 & 12535 & 31389 & 1396  & 992   & 2388  & 470   & 2355  & 1080  & 3905  & 86\%  & 88\% \\
    17    & 17501 & 4250  & 21751 & 2089  & 21716 & 16244 & 40049 & 2107  & 1326  & 3433  & 635   & 3398  & 1710  & 5743  & 84\%  & 86\% \\
    18    & 22138 & 5076  & 27214 & 2499  & 27177 & 20580 & 50256 & 2146  & 1404  & 3550  & 672   & 3513  & 1710  & 5895  & 87\%  & 88\% \\
    19    & 27651 & 6004  & 33655 & 2961  & 33616 & 25905 & 62482 & 2379  & 1558  & 3937  & 747   & 3898  & 1920  & 6565  & 88\%  & 89\% \\
    20    & 34145 & 7040  & 41185 & 3476  & 41144 & 32025 & 76645 & 2420  & 1640  & 4060  & 786   & 4019  & 1920  & 6725  & 90\%  & 91\% \\
    
\midrule
\multicolumn{15}{r}{\textbf{Average Reduction:}} & \textbf{71\%} & \textbf{73\%} \\
\bottomrule
\end{tabular}
}

\vspace{0.5cm}

% ------------------ NBF ------------------
\resizebox{\columnwidth}{!}{
\begin{tabular}{c|ccc|cccc|ccc|cccc|c|c}
\toprule
\multicolumn{17}{c}{\textbf{NBF}} \\
\toprule
\multirow{2}{*}{$V$} & \multicolumn{7}{c}{\textbf{Without AFGR}} & \multicolumn{7}{c}{\textbf{With AFGR}} & \multirow{2}{*}{\begin{tabular}[c]{@{}c@{}}\% Red.\\ Vars\end{tabular}} & \multirow{2}{*}{\begin{tabular}[c]{@{}c@{}}\% Red.\\ Constr.\end{tabular}} \\
\cmidrule(lr){2-8} \cmidrule(lr){9-15}
 & Bin & Cont & Tot Vars& $Eq$ & $\leq$ & $\geq$ & Tot C & Bin & Cont & Tot Vars & $Eq$ & $\leq$ & $\geq$ & Tot C &  \\
\midrule
  
    4     & 81    & 64    & 145   & 28    & 150   & 63    & 241   & 81    & 64    & 145   & 28    & 150   & 63    & 241   & 0\%   & 0\% \\
    5     & 155   & 110   & 265   & 47    & 274   & 120   & 441   & 89    & 80    & 169   & 35    & 172   & 63    & 270   & 36\%  & 39\% \\
    6     & 286   & 180   & 466   & 77    & 481   & 224   & 782   & 118   & 108   & 226   & 47    & 229   & 80    & 356   & 52\%  & 54\% \\
    7     & 501   & 280   & 781   & 122   & 823   & 418   & 1363  & 177   & 154   & 331   & 68    & 346   & 130   & 544   & 58\%  & 60\% \\
    8     & 833   & 416   & 1249  & 183   & 1307  & 700   & 2190  & 188   & 176   & 364   & 78    & 377   & 130   & 585   & 71\%  & 73\% \\
    9     & 1321  & 594   & 1915  & 265   & 2024  & 1152  & 3441  & 331   & 270   & 601   & 121   & 638   & 252   & 1011  & 69\%  & 71\% \\
    10    & 2010  & 820   & 2830  & 369   & 2969  & 1760  & 5098  & 346   & 300   & 646   & 135   & 681   & 252   & 1068  & 77\%  & 79\% \\
    11    & 2951  & 1100  & 4051  & 499   & 4224  & 2597  & 7320  & 361   & 330   & 691   & 149   & 724   & 252   & 1125  & 83\%  & 85\% \\
    12    & 4201  & 1440  & 5641  & 659   & 5911  & 3776  & 10346 & 601   & 480   & 1081  & 219   & 1151  & 456   & 1826  & 81\%  & 82\% \\
    13    & 5823  & 1846  & 7669  & 850   & 7992  & 5250  & 14092 & 727   & 572   & 1299  & 262   & 1377  & 546   & 2185  & 83\%  & 84\% \\
    14    & 7886  & 2324  & 10210 & 1077  & 10673 & 7216  & 18966 & 989   & 728   & 1717  & 336   & 1838  & 775   & 2949  & 83\%  & 84\% \\
    15    & 10465 & 2880  & 13345 & 1341  & 13884 & 9595  & 24820 & 1015  & 780   & 1795  & 361   & 1914  & 775   & 3050  & 87\%  & 88\% \\
    16    & 13641 & 3520  & 17161 & 1646  & 17782 & 12535 & 31963 & 1396  & 992   & 2388  & 461   & 2535  & 1080  & 4076  & 86\%  & 87\% \\
    17    & 17501 & 4250  & 21751 & 1996  & 22584 & 16244 & 40824 & 2107  & 1326  & 3433  & 620   & 3664  & 1710  & 5994  & 84\%  & 85\% \\
    18    & 22138 & 5076  & 27214 & 2392  & 28157 & 20580 & 51129 & 2146  & 1404  & 3550  & 658   & 3779  & 1710  & 6147  & 87\%  & 88\% \\
    19    & 27651 & 6004  & 33655 & 2839  & 34872 & 25905 & 63616 & 2379  & 1558  & 3937  & 733   & 4218  & 1920  & 6871  & 88\%  & 89\% \\
    20    & 34145 & 7040  & 41185 & 3338  & 42544 & 32025 & 77907 & 2420  & 1640  & 4060  & 773   & 4339  & 1920  & 7032  & 90\%  & 91\% \\
 
  \midrule
\multicolumn{15}{r}{\textbf{Average Reduction:}} & \textbf{71\%} & \textbf{73\%} \\
\bottomrule
\end{tabular}}
\end{table}

\begin{table}[h]
\centering
\caption{
Detailed comparison of ABF and NBF complexity with and without the AFGR method across B1 istances (as defined in Table~\ref{tab:instance_classes}). For each instance size, we report:
\textbf{Bin}: number of binary decision variables,
   \textbf{Cont}: number of continuous variables,
    \textbf{Tot Vars}: total number of variables (binary + continuous),
   $Eq:$ number of equality constraints,
    $\leq$: number of inequality constraints,
    $\geq$: number of inequality constraints,
    \textbf{Tot C}: total number of constraints.
The columns labeled ``With AFGR'' show the reduced model size after applying the AFGR. The final columns indicate the percentage reduction in total variables and constraints. }\label{tab:ABF-complexity_A2}
% ------------------ ABF ------------------
\resizebox{\columnwidth}{!}{
\begin{tabular}{c|ccc|cccc|ccc|cccc|c|c}
\toprule
\multicolumn{17}{c}{\textbf{ABF}} \\
\toprule
\multirow{2}{*}{$V$} & \multicolumn{7}{c}{\textbf{Without AFGR}} & \multicolumn{7}{c}{\textbf{With AFGR}} & \multirow{2}{*}{\begin{tabular}[c]{@{}c@{}}\% Red.\\ Vars\end{tabular}} & \multirow{2}{*}{\begin{tabular}[c]{@{}c@{}}\% Red.\\ Constr.\end{tabular}} \\
\cmidrule(lr){2-8} \cmidrule(lr){9-15}
 & Bin & Cont & Tot Vars& $Eq$ & $\leq$ & $\geq$ & Tot C & Bin & Cont & Tot Vars & $Eq$ & $\leq$ & $\geq$ & Tot C &  \\
\midrule
   4     & 81    & 64    & 145   & 31    & 157   & 70    & 258   & 81    & 64    & 145   & 31    & 157   & 70    & 258   & 0\%   & 0\% \\
    5     & 155   & 110   & 265   & 52    & 294   & 140   & 486   & 155   & 110   & 265   & 52    & 294   & 140   & 486   & 0\%   & 0\% \\
    6     & 286   & 180   & 466   & 84    & 509   & 252   & 845   & 286   & 180   & 466   & 84    & 509   & 252   & 845   & 0\%   & 0\% \\
    7     & 501   & 280   & 781   & 132   & 861   & 456   & 1449  & 501   & 280   & 781   & 132   & 861   & 456   & 1449  & 0\%   & 0\% \\
    8     & 833   & 416   & 1249  & 198   & 1382  & 775   & 2355  & 833   & 416   & 1249  & 198   & 1382  & 775   & 2355  & 0\%   & 0\% \\
    9     & 1321  & 594   & 1915  & 285   & 2120  & 1248  & 3653  & 859   & 468   & 1327  & 223   & 1483  & 800   & 2506  & 31\%  & 31\% \\
    10    & 2010  & 820   & 2830  & 396   & 3129  & 1920  & 5445  & 1354  & 660   & 2014  & 317   & 2249  & 1280  & 3846  & 29\%  & 29\% \\
    11    & 2951  & 1100  & 4051  & 533   & 4420  & 2793  & 7746  & 1387  & 726   & 2113  & 348   & 2346  & 1280  & 3974  & 48\%  & 49\% \\
    12    & 4201  & 1440  & 5641  & 701   & 6147  & 4012  & 10860 & 2092  & 984   & 3076  & 475   & 3411  & 1960  & 5846  & 45\%  & 46\% \\
    13    & 5823  & 1846  & 7669  & 902   & 8342  & 5600  & 14844 & 3051  & 1300  & 4351  & 631   & 4814  & 2891  & 8336  & 43\%  & 44\% \\
    14    & 7886  & 2324  & 10210 & 1139  & 11083 & 7626  & 19848 & 4321  & 1680  & 6001  & 819   & 6621  & 4130  & 11570 & 41\%  & 42\% \\
    15    & 10465 & 2880  & 13345 & 1415  & 14454 & 10165 & 26034 & 5965  & 2130  & 8095  & 1042  & 8904  & 5740  & 15686 & 39\%  & 40\% \\
    16    & 13641 & 3520  & 17161 & 1732  & 18436 & 13189 & 33357 & 6036  & 2272  & 8308  & 1111  & 9115  & 5740  & 15966 & 52\%  & 52\% \\
    17    & 17501 & 4250  & 21751 & 2095  & 23328 & 16988 & 42411 & 8135  & 2822  & 10957 & 1384  & 11988 & 7790  & 21162 & 50\%  & 50\% \\
    18    & 22138 & 5076  & 27214 & 2506  & 29137 & 21560 & 53203 & 10753 & 3456  & 14209 & 1699  & 15502 & 10355 & 27556 & 48\%  & 48\% \\
    19    & 27651 & 6004  & 33655 & 2968  & 35971 & 27004 & 65943 & 13971 & 4180  & 18151 & 2059  & 19747 & 13516 & 35322 & 46\%  & 46\% \\
    20    & 34145 & 7040  & 41185 & 3484  & 43944 & 33425 & 80853 & 17876 & 5000  & 22876 & 2467  & 24819 & 17360 & 44646 & 44\%  & 45\% \\
 \midrule
\multicolumn{15}{r}{\textbf{Average Reduction:}} & \textbf{30\%} & \textbf{31\%} \\
\bottomrule
\end{tabular}
}

\vspace{0.5cm}

% ------------------ NBF ------------------
\resizebox{\columnwidth}{!}{
\begin{tabular}{c|ccc|cccc|ccc|cccc|c|c}
\toprule
\multicolumn{17}{c}{\textbf{NBF}} \\
\toprule
\multirow{2}{*}{$V$} & \multicolumn{7}{c}{\textbf{Without AFGR}} & \multicolumn{7}{c}{\textbf{With AFGR}} & \multirow{2}{*}{\begin{tabular}[c]{@{}c@{}}\% Red.\\ Vars\end{tabular}} & \multirow{2}{*}{\begin{tabular}[c]{@{}c@{}}\% Red.\\ Constr.\end{tabular}} \\
\cmidrule(lr){2-8} \cmidrule(lr){9-15}
 & Bin & Cont & Tot Vars& $Eq$ & $\leq$ & $\geq$ & Tot C & Bin & Cont & Tot Vars & $Eq$ & $\leq$ & $\geq$ & Tot C &  \\
\midrule
  
    4     & 81    & 64    & 145   & 31    & 157   & 70    & 258   & 81    & 64    & 145   & 31    & 157   & 70    & 258   & 0\%   & 0\% \\
    5     & 155   & 110   & 265   & 52    & 294   & 140   & 486   & 155   & 110   & 265   & 52    & 294   & 140   & 486   & 0\%   & 0\% \\
    6     & 286   & 180   & 466   & 84    & 509   & 252   & 845   & 286   & 180   & 466   & 84    & 509   & 252   & 845   & 0\%   & 0\% \\
    7     & 501   & 280   & 781   & 132   & 861   & 456   & 1449  & 501   & 280   & 781   & 132   & 861   & 456   & 1449  & 0\%   & 0\% \\
    8     & 833   & 416   & 1249  & 198   & 1382  & 775   & 2355  & 833   & 416   & 1249  & 198   & 1382  & 775   & 2355  & 0\%   & 0\% \\
    9     & 1321  & 594   & 1915  & 285   & 2120  & 1248  & 3653  & 859   & 468   & 1327  & 223   & 1483  & 800   & 2506  & 31\%  & 31\% \\
    10    & 2010  & 820   & 2830  & 396   & 3129  & 1920  & 5445  & 1354  & 660   & 2014  & 317   & 2249  & 1280  & 3846  & 29\%  & 29\% \\
    11    & 2951  & 1100  & 4051  & 533   & 4420  & 2793  & 7746  & 1387  & 726   & 2113  & 348   & 2346  & 1280  & 3974  & 48\%  & 49\% \\
    12    & 4201  & 1440  & 5641  & 701   & 6147  & 4012  & 10860 & 2092  & 984   & 3076  & 475   & 3411  & 1960  & 5846  & 45\%  & 46\% \\
    13    & 5823  & 1846  & 7669  & 902   & 8342  & 5600  & 14844 & 3051  & 1300  & 4351  & 631   & 4814  & 2891  & 8336  & 43\%  & 44\% \\
    14    & 7886  & 2324  & 10210 & 1139  & 11083 & 7626  & 19848 & 4321  & 1680  & 6001  & 819   & 6621  & 4130  & 11570 & 41\%  & 42\% \\
    15    & 10465 & 2880  & 13345 & 1415  & 14454 & 10165 & 26034 & 5965  & 2130  & 8095  & 1042  & 8904  & 5740  & 15686 & 39\%  & 40\% \\
    16    & 13641 & 3520  & 17161 & 1732  & 18436 & 13189 & 33357 & 6036  & 2272  & 8308  & 1111  & 9115  & 5740  & 15966 & 52\%  & 52\% \\
    17    & 17501 & 4250  & 21751 & 2095  & 23328 & 16988 & 42411 & 8135  & 2822  & 10957 & 1384  & 11988 & 7790  & 21162 & 50\%  & 50\% \\
    18    & 22138 & 5076  & 27214 & 2506  & 29137 & 21560 & 53203 & 10753 & 3456  & 14209 & 1699  & 15502 & 10355 & 27556 & 48\%  & 48\% \\
    19    & 27651 & 6004  & 33655 & 2968  & 35971 & 27004 & 65943 & 13971 & 4180  & 18151 & 2059  & 19747 & 13516 & 35322 & 46\%  & 46\% \\
    20    & 34145 & 7040  & 41185 & 3484  & 43944 & 33425 & 80853 & 17876 & 5000  & 22876 & 2467  & 24819 & 17360 & 44646 & 44\%  & 45\% \\
  \midrule
\multicolumn{15}{r}{\textbf{Average Reduction:}} & \textbf{30\%} & \textbf{31\%} \\
\bottomrule
\end{tabular}}
\end{table}

\begin{table}[h]
\centering
\caption{
Detailed comparison of ABF and NBF complexity with and without the AFGR method across B2 istances (as defined in Table~\ref{tab:instance_classes}). For each instance size, we report:
\textbf{Bin}: number of binary decision variables,
   \textbf{Cont}: number of continuous variables,
    \textbf{Tot Vars}: total number of variables (binary + continuous),
   $Eq:$ number of equality constraints,
    $\leq$: number of inequality constraints,
    $\geq$: number of inequality constraints,
    \textbf{Tot C}: total number of constraints.
The columns labeled ``With AFGR'' show the reduced model size after applying the AFGR. The final columns indicate the percentage reduction in total variables and constraints. }\label{tab:ABF-complexity_A4}
% ------------------ ABF ------------------
\resizebox{\columnwidth}{!}{
\begin{tabular}{c|ccc|cccc|ccc|cccc|c|c}
\toprule
\multicolumn{17}{c}{\textbf{ABF}} \\
\toprule
\multirow{2}{*}{$V$} & \multicolumn{7}{c}{\textbf{Without AFGR}} & \multicolumn{7}{c}{\textbf{With AFGR}} & \multirow{2}{*}{\begin{tabular}[c]{@{}c@{}}\% Red.\\ Vars\end{tabular}} & \multirow{2}{*}{\begin{tabular}[c]{@{}c@{}}\% Red.\\ Constr.\end{tabular}} \\
\cmidrule(lr){2-8} \cmidrule(lr){9-15}
 & Bin & Cont & Tot Vars& $Eq$ & $\leq$ & $\geq$ & Tot C & Bin & Cont & Tot Vars & $Eq$ & $\leq$ & $\geq$ & Tot C &  \\
\midrule
   
    4     & 81    & 64    & 145   & 31    & 157   & 70    & 258   & 81    & 64    & 145   & 31    & 157   & 70    & 258   & 0\%   & 0\% \\
    5     & 155   & 110   & 265   & 51    & 284   & 130   & 465   & 155   & 110   & 265   & 51    & 284   & 130   & 465   & 0\%   & 0\% \\
    6     & 286   & 180   & 466   & 84    & 509   & 252   & 845   & 286   & 180   & 466   & 84    & 509   & 252   & 845   & 0\%   & 0\% \\
    7     & 501   & 280   & 781   & 132   & 861   & 456   & 1449  & 501   & 280   & 781   & 132   & 861   & 456   & 1449  & 0\%   & 0\% \\
    8     & 833   & 416   & 1249  & 197   & 1357  & 750   & 2304  & 521   & 320   & 841   & 150   & 919   & 456   & 1525  & 33\%  & 34\% \\
    9     & 1321  & 594   & 1915  & 284   & 2088  & 1216  & 3588  & 541   & 360   & 901   & 169   & 996   & 475   & 1640  & 53\%  & 54\% \\
    10    & 2010  & 820   & 2830  & 395   & 3089  & 1880  & 5364  & 885   & 520   & 1405  & 247   & 1559  & 800   & 2606  & 50\%  & 51\% \\
    11    & 2951  & 1100  & 4051  & 532   & 4371  & 2744  & 7647  & 911   & 572   & 1483  & 271   & 1635  & 800   & 2706  & 63\%  & 65\% \\
    12    & 4201  & 1440  & 5641  & 700   & 6088  & 3953  & 10741 & 1420  & 792   & 2212  & 379   & 2443  & 1280  & 4102  & 61\%  & 62\% \\
    13    & 5823  & 1846  & 7669  & 901   & 8272  & 5530  & 14703 & 2133  & 1066  & 3199  & 514   & 3532  & 1960  & 6006  & 58\%  & 59\% \\
    14    & 7886  & 2324  & 10210 & 1138  & 11001 & 7544  & 19683 & 3101  & 1400  & 4501  & 679   & 4962  & 2891  & 8532  & 56\%  & 57\% \\
    15    & 10465 & 2880  & 13345 & 1413  & 14264 & 9975  & 25652 & 3151  & 1500  & 4651  & 727   & 5110  & 2891  & 8728  & 65\%  & 66\% \\
    16    & 13641 & 3520  & 17161 & 1731  & 18327 & 13080 & 33138 & 4441  & 1920  & 6361  & 935   & 6977  & 4130  & 12042 & 63\%  & 64\% \\
    17    & 17501 & 4250  & 21751 & 2094  & 23204 & 16864 & 42162 & 6107  & 2414  & 8521  & 1180  & 9326  & 5740  & 16246 & 61\%  & 61\% \\
    18    & 22138 & 5076  & 27214 & 2504  & 28857 & 21280 & 52641 & 8218  & 2988  & 11206 & 1464  & 12153 & 7708  & 21325 & 59\%  & 59\% \\
    19    & 27651 & 6004  & 33655 & 2966  & 35657 & 26690 & 65313 & 10849 & 3648  & 14497 & 1792  & 15693 & 10260 & 27745 & 57\%  & 58\% \\
    20    & 34145 & 7040  & 41185 & 3482  & 43594 & 33075 & 80151 & 14081 & 4400  & 18481 & 2166  & 19966 & 13407 & 35539 & 55\%  & 56\% \\
  
\midrule
\multicolumn{15}{r}{\textbf{Average Reduction:}} & \textbf{43\%} & \textbf{44\%} \\
\bottomrule
\end{tabular}
}

\vspace{0.5cm}

% ------------------ NBF ------------------
\resizebox{\columnwidth}{!}{
\begin{tabular}{c|ccc|cccc|ccc|cccc|c|c}
\toprule
\multicolumn{17}{c}{\textbf{NBF}} \\
\toprule
\multirow{2}{*}{$V$} & \multicolumn{7}{c}{\textbf{Without AFGR}} & \multicolumn{7}{c}{\textbf{With AFGR}} & \multirow{2}{*}{\begin{tabular}[c]{@{}c@{}}\% Red.\\ Vars\end{tabular}} & \multirow{2}{*}{\begin{tabular}[c]{@{}c@{}}\% Red.\\ Constr.\end{tabular}} \\
\cmidrule(lr){2-8} \cmidrule(lr){9-15}
 & Bin & Cont & Tot Vars& $Eq$ & $\leq$ & $\geq$ & Tot C & Bin & Cont & Tot Vars & $Eq$ & $\leq$ & $\geq$ & Tot C &  \\
\midrule
  
    4     & 81    & 64    & 145   & 29    & 157   & 70    & 256   & 81    & 64    & 145   & 29    & 157   & 70    & 256   & 0\%   & 0\% \\
    5     & 155   & 110   & 265   & 48    & 284   & 130   & 462   & 155   & 110   & 265   & 48    & 284   & 130   & 462   & 0\%   & 0\% \\
    6     & 286   & 180   & 466   & 79    & 509   & 252   & 840   & 286   & 180   & 466   & 79    & 509   & 252   & 840   & 0\%   & 0\% \\
    7     & 501   & 280   & 781   & 124   & 861   & 456   & 1441  & 501   & 280   & 781   & 124   & 861   & 456   & 1441  & 0\%   & 0\% \\
    8     & 833   & 416   & 1249  & 185   & 1357  & 750   & 2292  & 521   & 320   & 841   & 143   & 919   & 456   & 1518  & 33\%  & 34\% \\
    9     & 1321  & 594   & 1915  & 267   & 2088  & 1216  & 3571  & 541   & 360   & 901   & 163   & 996   & 475   & 1634  & 53\%  & 54\% \\
    10    & 2010  & 820   & 2830  & 372   & 3089  & 1880  & 5341  & 885   & 520   & 1405  & 237   & 1559  & 800   & 2596  & 50\%  & 51\% \\
    11    & 2951  & 1100  & 4051  & 502   & 4371  & 2744  & 7617  & 911   & 572   & 1483  & 262   & 1635  & 800   & 2697  & 63\%  & 65\% \\
    12    & 4201  & 1440  & 5641  & 662   & 6088  & 3953  & 10703 & 1420  & 792   & 2212  & 365   & 2443  & 1280  & 4088  & 61\%  & 62\% \\
    13    & 5823  & 1846  & 7669  & 854   & 8272  & 5530  & 14656 & 2133  & 1066  & 3199  & 494   & 3532  & 1960  & 5986  & 58\%  & 59\% \\
    14    & 7886  & 2324  & 10210 & 1081  & 11001 & 7544  & 19626 & 3101  & 1400  & 4501  & 652   & 4962  & 2891  & 8505  & 56\%  & 57\% \\
    15    & 10465 & 2880  & 13345 & 1345  & 14264 & 9975  & 25584 & 3151  & 1500  & 4651  & 701   & 5110  & 2891  & 8702  & 65\%  & 66\% \\
    16    & 13641 & 3520  & 17161 & 1651  & 18327 & 13080 & 33058 & 4441  & 1920  & 6361  & 901   & 6977  & 4130  & 12008 & 63\%  & 64\% \\
    17    & 17501 & 4250  & 21751 & 2001  & 23204 & 16864 & 42069 & 6107  & 2414  & 8521  & 1137  & 9326  & 5740  & 16203 & 61\%  & 61\% \\
    18    & 22138 & 5076  & 27214 & 2397  & 28857 & 21280 & 52534 & 8218  & 2988  & 11206 & 1411  & 12153 & 7708  & 21272 & 59\%  & 60\% \\
    19    & 27651 & 6004  & 33655 & 2844  & 35657 & 26690 & 65191 & 10849 & 3648  & 14497 & 1728  & 15693 & 10260 & 27681 & 57\%  & 58\% \\
    20    & 34145 & 7040  & 41185 & 3344  & 43594 & 33075 & 80013 & 14081 & 4400  & 18481 & 2090  & 19966 & 13407 & 35463 & 55\%  & 56\% \\
  \midrule
\multicolumn{15}{r}{\textbf{Average Reduction:}} & \textbf{43\%} & \textbf{44\%} \\
\bottomrule
\end{tabular}}
\end{table}

\begin{table}[h]
\centering
\caption{
Detailed comparison of ABF and NBF complexity with and without the AFGR method across B3 istances (as defined in Table~\ref{tab:instance_classes}). For each instance size, we report:
\textbf{Bin}: number of binary decision variables,
   \textbf{Cont}: number of continuous variables,
    \textbf{Tot Vars}: total number of variables (binary + continuous),
   $Eq:$ number of equality constraints,
    $\leq$: number of inequality constraints,
    $\geq$: number of inequality constraints,
    \textbf{Tot C}: total number of constraints.
The columns labeled ``With AFGR'' show the reduced model size after applying the AFGR. The final columns indicate the percentage reduction in total variables and constraints. }\label{tab:ABF-complexity_B1}
% ------------------ ABF ------------------
\resizebox{\columnwidth}{!}{
\begin{tabular}{c|ccc|cccc|ccc|cccc|c|c}
\toprule
\multicolumn{17}{c}{\textbf{ABF}} \\
\toprule
\multirow{2}{*}{$V$} & \multicolumn{7}{c}{\textbf{Without AFGR}} & \multicolumn{7}{c}{\textbf{With AFGR}} & \multirow{2}{*}{\begin{tabular}[c]{@{}c@{}}\% Red.\\ Vars\end{tabular}} & \multirow{2}{*}{\begin{tabular}[c]{@{}c@{}}\% Red.\\ Constr.\end{tabular}} \\
\cmidrule(lr){2-8} \cmidrule(lr){9-15}
 & Bin & Cont & Tot Vars& $Eq$ & $\leq$ & $\geq$ & Tot C & Bin & Cont & Tot Vars & $Eq$ & $\leq$ & $\geq$ & Tot C &  \\
\midrule
  4     & 81    & 64    & 145   & 30    & 150   & 63    & 243   & 81    & 64    & 145   & 30    & 150   & 63    & 243   & 0\%   & 0\% \\
    5     & 155   & 110   & 265   & 51    & 284   & 130   & 465   & 155   & 110   & 265   & 51    & 284   & 130   & 465   & 0\%   & 0\% \\
    6     & 286   & 180   & 466   & 83    & 495   & 238   & 816   & 222   & 156   & 378   & 72    & 401   & 180   & 653   & 19\%  & 20\% \\
    7     & 501   & 280   & 781   & 131   & 842   & 437   & 1410  & 301   & 210   & 511   & 97    & 552   & 252   & 901   & 35\%  & 36\% \\
    8     & 833   & 416   & 1249  & 197   & 1357  & 750   & 2304  & 521   & 320   & 841   & 150   & 919   & 456   & 1525  & 33\%  & 34\% \\
    9     & 1321  & 594   & 1915  & 283   & 2056  & 1184  & 3523  & 331   & 270   & 601   & 124   & 652   & 266   & 1042  & 69\%  & 70\% \\
    10    & 2010  & 820   & 2830  & 394   & 3049  & 1840  & 5283  & 561   & 400   & 961   & 187   & 1054  & 475   & 1716  & 66\%  & 68\% \\
    11    & 2951  & 1100  & 4051  & 531   & 4322  & 2695  & 7548  & 581   & 440   & 1021  & 205   & 1112  & 475   & 1792  & 75\%  & 76\% \\
    12    & 4201  & 1440  & 5641  & 699   & 6029  & 3894  & 10622 & 1420  & 792   & 2212  & 378   & 2411  & 1248  & 4037  & 61\%  & 62\% \\
    13    & 5823  & 1846  & 7669  & 900   & 8202  & 5460  & 14562 & 2133  & 1066  & 3199  & 513   & 3492  & 1920  & 5925  & 58\%  & 59\% \\
    14    & 7886  & 2324  & 10210 & 1136  & 10837 & 7380  & 19353 & 2174  & 1148  & 3322  & 552   & 3613  & 1920  & 6085  & 67\%  & 69\% \\
    15    & 10465 & 2880  & 13345 & 1412  & 14169 & 9880  & 25461 & 3151  & 1500  & 4651  & 726   & 5061  & 2842  & 8629  & 65\%  & 66\% \\
    16    & 13641 & 3520  & 17161 & 1729  & 18109 & 12862 & 32700 & 3201  & 1600  & 4801  & 774   & 5209  & 2842  & 8825  & 72\%  & 73\% \\
    17    & 17501 & 4250  & 21751 & 2092  & 22956 & 16616 & 41664 & 4501  & 2040  & 6541  & 992   & 7096  & 4071  & 12159 & 70\%  & 71\% \\
    18    & 22138 & 5076  & 27214 & 2503  & 28717 & 21140 & 52360 & 6178  & 2556  & 8734  & 1248  & 9467  & 5670  & 16385 & 68\%  & 69\% \\
    19    & 27651 & 6004  & 33655 & 2964  & 35343 & 26376 & 64683 & 6249  & 2698  & 8947  & 1317  & 9678  & 5670  & 16665 & 73\%  & 74\% \\
    20    & 34145 & 7040  & 41185 & 3480  & 43244 & 32725 & 79449 & 8384  & 3320  & 11704 & 1626  & 12647 & 7708  & 21981 & 72\%  & 72\% \\
    \midrule
\multicolumn{15}{r}{\textbf{Average Reduction:}} & \textbf{53\%} & \textbf{54\%} \\
\bottomrule
\end{tabular}
}

\vspace{0.5cm}

% ------------------ NBF ------------------
\resizebox{\columnwidth}{!}{
\begin{tabular}{c|ccc|cccc|ccc|cccc|c|c}
\toprule
\multicolumn{17}{c}{\textbf{NBF}} \\
\toprule
\multirow{2}{*}{$V$} & \multicolumn{7}{c}{\textbf{Without AFGR}} & \multicolumn{7}{c}{\textbf{With AFGR}} & \multirow{2}{*}{\begin{tabular}[c]{@{}c@{}}\% Red.\\ Vars\end{tabular}} & \multirow{2}{*}{\begin{tabular}[c]{@{}c@{}}\% Red.\\ Constr.\end{tabular}} \\
\cmidrule(lr){2-8} \cmidrule(lr){9-15}
 & Bin & Cont & Tot Vars& $Eq$ & $\leq$ & $\geq$ & Tot C & Bin & Cont & Tot Vars & $Eq$ & $\leq$ & $\geq$ & Tot C &  \\
\midrule
  4     & 81    & 64    & 145   & 28    & 150   & 63    & 241   & 81    & 64    & 145   & 28    & 150   & 63    & 241   & 0\%   & 0\% \\
    5     & 155   & 110   & 265   & 48    & 284   & 130   & 462   & 155   & 110   & 265   & 48    & 284   & 130   & 462   & 0\%   & 0\% \\
    6     & 286   & 180   & 466   & 78    & 495   & 238   & 811   & 222   & 156   & 378   & 68    & 401   & 180   & 649   & 19\%  & 20\% \\
    7     & 501   & 280   & 781   & 123   & 842   & 437   & 1402  & 301   & 210   & 511   & 93    & 552   & 252   & 897   & 35\%  & 36\% \\
    8     & 833   & 416   & 1249  & 185   & 1357  & 750   & 2292  & 521   & 320   & 841   & 143   & 919   & 456   & 1518  & 33\%  & 34\% \\
    9     & 1321  & 594   & 1915  & 266   & 2056  & 1184  & 3506  & 331   & 270   & 601   & 122   & 652   & 266   & 1040  & 69\%  & 70\% \\
    10    & 2010  & 820   & 2830  & 371   & 3049  & 1840  & 5260  & 561   & 400   & 961   & 182   & 1054  & 475   & 1711  & 66\%  & 67\% \\
    11    & 2951  & 1100  & 4051  & 501   & 4322  & 2695  & 7518  & 581   & 440   & 1021  & 201   & 1112  & 475   & 1788  & 75\%  & 76\% \\
    12    & 4201  & 1440  & 5641  & 661   & 6029  & 3894  & 10584 & 1420  & 792   & 2212  & 364   & 2411  & 1248  & 4023  & 61\%  & 62\% \\
    13    & 5823  & 1846  & 7669  & 853   & 8202  & 5460  & 14515 & 2133  & 1066  & 3199  & 493   & 3492  & 1920  & 5905  & 58\%  & 59\% \\
    14    & 7886  & 2324  & 10210 & 1079  & 10837 & 7380  & 19296 & 2174  & 1148  & 3322  & 533   & 3613  & 1920  & 6066  & 67\%  & 69\% \\
    15    & 10465 & 2880  & 13345 & 1344  & 14169 & 9880  & 25393 & 3151  & 1500  & 4651  & 700   & 5061  & 2842  & 8603  & 65\%  & 66\% \\
    16    & 13641 & 3520  & 17161 & 1649  & 18109 & 12862 & 32620 & 3201  & 1600  & 4801  & 749   & 5209  & 2842  & 8800  & 72\%  & 73\% \\
    17    & 17501 & 4250  & 21751 & 1999  & 22956 & 16616 & 41571 & 4501  & 2040  & 6541  & 959   & 7096  & 4071  & 12126 & 70\%  & 71\% \\
    18    & 22138 & 5076  & 27214 & 2396  & 28717 & 21140 & 52253 & 6178  & 2556  & 8734  & 1206  & 9467  & 5670  & 16343 & 68\%  & 69\% \\
    19    & 27651 & 6004  & 33655 & 2842  & 35343 & 26376 & 64561 & 6249  & 2698  & 8947  & 1276  & 9678  & 5670  & 16624 & 73\%  & 74\% \\
    20    & 34145 & 7040  & 41185 & 3342  & 43244 & 32725 & 79311 & 8384  & 3320  & 11704 & 1575  & 12647 & 7708  & 21930 & 72\%  & 72\% \\
  \midrule
\multicolumn{15}{r}{\textbf{Average Reduction:}} & \textbf{53\%} & \textbf{54\%} \\
\bottomrule
\end{tabular}}
\end{table}

\begin{table}[h]
\centering
\caption{
Detailed comparison of ABF and NBF complexity with and without the AFGR method across B4 istances (as defined in Table~\ref{tab:instance_classes}). For each instance size, we report:
\textbf{Bin}: number of binary decision variables,
   \textbf{Cont}: number of continuous variables,
    \textbf{Tot Vars}: total number of variables (binary + continuous),
   $Eq:$ number of equality constraints,
    $\leq$: number of inequality constraints,
    $\geq$: number of inequality constraints,
    \textbf{Tot C}: total number of constraints.
The columns labeled ``With AFGR'' show the reduced model size after applying the AFGR. The final columns indicate the percentage reduction in total variables and constraints. }\label{tab:ABF-complexity_B3}
% ------------------ ABF ------------------
\resizebox{\columnwidth}{!}{
\begin{tabular}{c|ccc|cccc|ccc|cccc|c|c}
\toprule
\multicolumn{17}{c}{\textbf{ABF}} \\
\toprule
\multirow{2}{*}{$V$} & \multicolumn{7}{c}{\textbf{Without AFGR}} & \multicolumn{7}{c}{\textbf{With AFGR}} & \multirow{2}{*}{\begin{tabular}[c]{@{}c@{}}\% Red.\\ Vars\end{tabular}} & \multirow{2}{*}{\begin{tabular}[c]{@{}c@{}}\% Red.\\ Constr.\end{tabular}} \\
\cmidrule(lr){2-8} \cmidrule(lr){9-15}
 & Bin & Cont & Tot Vars& $Eq$ & $\leq$ & $\geq$ & Tot C & Bin & Cont & Tot Vars & $Eq$ & $\leq$ & $\geq$ & Tot C &  \\
\midrule
  
    4     & 81    & 64    & 145   & 30    & 150   & 63    & 243   & 81    & 64    & 145   & 30    & 150   & 63    & 243   & 0\%   & 0\% \\
    5     & 155   & 110   & 265   & 50    & 274   & 120   & 444   & 89    & 80    & 169   & 36    & 172   & 63    & 271   & 36\%  & 39\% \\
    6     & 286   & 180   & 466   & 83    & 495   & 238   & 816   & 222   & 156   & 378   & 72    & 401   & 180   & 653   & 19\%  & 20\% \\
    7     & 501   & 280   & 781   & 130   & 823   & 418   & 1371  & 177   & 154   & 331   & 69    & 346   & 130   & 545   & 58\%  & 60\% \\
    8     & 833   & 416   & 1249  & 196   & 1332  & 725   & 2253  & 316   & 240   & 556   & 110   & 595   & 252   & 957   & 55\%  & 58\% \\
    9     & 1321  & 594   & 1915  & 282   & 2024  & 1152  & 3458  & 331   & 270   & 601   & 123   & 638   & 252   & 1013  & 69\%  & 71\% \\
    10    & 2010  & 820   & 2830  & 393   & 3009  & 1800  & 5202  & 346   & 300   & 646   & 137   & 695   & 266   & 1098  & 77\%  & 79\% \\
    11    & 2951  & 1100  & 4051  & 530   & 4273  & 2646  & 7449  & 581   & 440   & 1021  & 204   & 1093  & 456   & 1753  & 75\%  & 76\% \\
    12    & 4201  & 1440  & 5641  & 698   & 5970  & 3835  & 10503 & 937   & 624   & 1561  & 294   & 1686  & 775   & 2755  & 72\%  & 74\% \\
    13    & 5823  & 1846  & 7669  & 898   & 8062  & 5320  & 14280 & 963   & 676   & 1639  & 318   & 1762  & 775   & 2855  & 79\%  & 80\% \\
    14    & 7886  & 2324  & 10210 & 1135  & 10755 & 7298  & 19188 & 1486  & 924   & 2410  & 440   & 2605  & 1248  & 4293  & 76\%  & 78\% \\
    15    & 10465 & 2880  & 13345 & 1410  & 13979 & 9690  & 25079 & 1519  & 990   & 2509  & 471   & 2702  & 1248  & 4421  & 81\%  & 82\% \\
    16    & 13641 & 3520  & 17161 & 1728  & 18000 & 12753 & 32481 & 2256  & 1312  & 3568  & 630   & 3855  & 1920  & 6405  & 79\%  & 80\% \\
    17    & 17501 & 4250  & 21751 & 2090  & 22708 & 16368 & 41166 & 3025  & 1632  & 4657  & 788   & 4998  & 2585  & 8371  & 79\%  & 80\% \\
    18    & 22138 & 5076  & 27214 & 2501  & 28437 & 20860 & 51798 & 3301  & 1800  & 5101  & 870   & 5505  & 2842  & 9217  & 81\%  & 82\% \\
    19    & 27651 & 6004  & 33655 & 2962  & 35029 & 26062 & 64053 & 3351  & 1900  & 5251  & 918   & 5653  & 2842  & 9413  & 84\%  & 85\% \\
    20    & 34145 & 7040  & 41185 & 3478  & 42894 & 32375 & 78747 & 4681  & 2400  & 7081  & 1166  & 7630  & 4071  & 12867 & 83\%  & 84\% \\
    
\midrule
\multicolumn{15}{r}{\textbf{Average Reduction:}} & \textbf{65\%} & \textbf{66\%} \\
\bottomrule
\end{tabular}
}

\vspace{0.5cm}

% ------------------ NBF ------------------
\resizebox{\columnwidth}{!}{
\begin{tabular}{c|ccc|cccc|ccc|cccc|c|c}
\toprule
\multicolumn{17}{c}{\textbf{NBF}} \\
\toprule
\multirow{2}{*}{$V$} & \multicolumn{7}{c}{\textbf{Without AFGR}} & \multicolumn{7}{c}{\textbf{With AFGR}} & \multirow{2}{*}{\begin{tabular}[c]{@{}c@{}}\% Red.\\ Vars\end{tabular}} & \multirow{2}{*}{\begin{tabular}[c]{@{}c@{}}\% Red.\\ Constr.\end{tabular}} \\
\cmidrule(lr){2-8} \cmidrule(lr){9-15}
 & Bin & Cont & Tot Vars& $Eq$ & $\leq$ & $\geq$ & Tot C & Bin & Cont & Tot Vars & $Eq$ & $\leq$ & $\geq$ & Tot C &  \\
\midrule
  
    4     & 81    & 64    & 145   & 28    & 150   & 63    & 241   & 81    & 64    & 145   & 28    & 150   & 63    & 241   & 0\%   & 0\% \\
    5     & 155   & 110   & 265   & 47    & 274   & 120   & 441   & 89    & 80    & 169   & 35    & 172   & 63    & 270   & 36\%  & 39\% \\
    6     & 286   & 180   & 466   & 78    & 495   & 238   & 811   & 222   & 156   & 378   & 68    & 401   & 180   & 649   & 19\%  & 20\% \\
    7     & 501   & 280   & 781   & 122   & 823   & 418   & 1363  & 177   & 154   & 331   & 68    & 346   & 130   & 544   & 58\%  & 60\% \\
    8     & 833   & 416   & 1249  & 184   & 1332  & 725   & 2241  & 316   & 240   & 556   & 107   & 595   & 252   & 954   & 55\%  & 57\% \\
    9     & 1321  & 594   & 1915  & 265   & 2024  & 1152  & 3441  & 331   & 270   & 601   & 121   & 638   & 252   & 1011  & 69\%  & 71\% \\
    10    & 2010  & 820   & 2830  & 370   & 3009  & 1800  & 5179  & 346   & 300   & 646   & 136   & 695   & 266   & 1097  & 77\%  & 79\% \\
    11    & 2951  & 1100  & 4051  & 500   & 4273  & 2646  & 7419  & 581   & 440   & 1021  & 200   & 1093  & 456   & 1749  & 75\%  & 76\% \\
    12    & 4201  & 1440  & 5641  & 660   & 5970  & 3835  & 10465 & 937   & 624   & 1561  & 286   & 1686  & 775   & 2747  & 72\%  & 74\% \\
    13    & 5823  & 1846  & 7669  & 851   & 8062  & 5320  & 14233 & 963   & 676   & 1639  & 311   & 1762  & 775   & 2848  & 79\%  & 80\% \\
    14    & 7886  & 2324  & 10210 & 1078  & 10755 & 7298  & 19131 & 1486  & 924   & 2410  & 428   & 2605  & 1248  & 4281  & 76\%  & 78\% \\
    15    & 10465 & 2880  & 13345 & 1342  & 13979 & 9690  & 25011 & 1519  & 990   & 2509  & 460   & 2702  & 1248  & 4410  & 81\%  & 82\% \\
    16    & 13641 & 3520  & 17161 & 1648  & 18000 & 12753 & 32401 & 2256  & 1312  & 3568  & 613   & 3855  & 1920  & 6388  & 79\%  & 80\% \\
    17    & 17501 & 4250  & 21751 & 1997  & 22708 & 16368 & 41073 & 3025  & 1632  & 4657  & 765   & 4998  & 2585  & 8348  & 79\%  & 80\% \\
    18    & 22138 & 5076  & 27214 & 2394  & 28437 & 20860 & 51691 & 3301  & 1800  & 5101  & 847   & 5505  & 2842  & 9194  & 81\%  & 82\% \\
    19    & 27651 & 6004  & 33655 & 2840  & 35029 & 26062 & 63931 & 3351  & 1900  & 5251  & 896   & 5653  & 2842  & 9391  & 84\%  & 85\% \\
    20    & 34145 & 7040  & 41185 & 3340  & 42894 & 32375 & 78609 & 4681  & 2400  & 7081  & 1136  & 7630  & 4071  & 12837 & 83\%  & 84\% \\
  \midrule
\multicolumn{15}{r}{\textbf{Average Reduction:}} & \textbf{65\%} & \textbf{66\%} \\
\bottomrule
\end{tabular}}
\end{table}

\begin{table}[h]
\centering
\caption{
Detailed comparison of ABF and NBF complexity with and without the AFGR method across B5 istances (as defined in Table~\ref{tab:instance_classes}). For each instance size, we report:
\textbf{Bin}: number of binary decision variables,
   \textbf{Cont}: number of continuous variables,
    \textbf{Tot Vars}: total number of variables (binary + continuous),
   $Eq:$ number of equality constraints,
    $\leq$: number of inequality constraints,
    $\geq$: number of inequality constraints,
    \textbf{Tot C}: total number of constraints.
The columns labeled ``With AFGR'' show the reduced model size after applying the AFGR. The final columns indicate the percentage reduction in total variables and constraints. }\label{tab:ABF-complexity_B5}
% ------------------ ABF ------------------
\resizebox{\columnwidth}{!}{
\begin{tabular}{c|ccc|cccc|ccc|cccc|c|c}
\toprule
\multicolumn{17}{c}{\textbf{ABF}} \\
\toprule
\multirow{2}{*}{$V$} & \multicolumn{7}{c}{\textbf{Without AFGR}} & \multicolumn{7}{c}{\textbf{With AFGR}} & \multirow{2}{*}{\begin{tabular}[c]{@{}c@{}}\% Red.\\ Vars\end{tabular}} & \multirow{2}{*}{\begin{tabular}[c]{@{}c@{}}\% Red.\\ Constr.\end{tabular}} \\
\cmidrule(lr){2-8} \cmidrule(lr){9-15}
 & Bin & Cont & Tot Vars& $Eq$ & $\leq$ & $\geq$ & Tot C & Bin & Cont & Tot Vars & $Eq$ & $\leq$ & $\geq$ & Tot C &  \\
\midrule
   
    4     & 81    & 64    & 145   & 30    & 150   & 63    & 243   & 81    & 64    & 145   & 30    & 150   & 63    & 243   & 0\%   & 0\% \\
    5     & 155   & 110   & 265   & 50    & 274   & 120   & 444   & 89    & 80    & 169   & 36    & 172   & 63    & 271   & 36\%  & 39\% \\
    6     & 286   & 180   & 466   & 82    & 481   & 224   & 787   & 118   & 108   & 226   & 48    & 229   & 80    & 357   & 52\%  & 55\% \\
    7     & 501   & 280   & 781   & 130   & 823   & 418   & 1371  & 177   & 154   & 331   & 69    & 346   & 130   & 545   & 58\%  & 60\% \\
    8     & 833   & 416   & 1249  & 195   & 1307  & 700   & 2202  & 188   & 176   & 364   & 78    & 377   & 130   & 585   & 71\%  & 73\% \\
    9     & 1321  & 594   & 1915  & 282   & 2024  & 1152  & 3458  & 331   & 270   & 601   & 123   & 638   & 252   & 1013  & 69\%  & 71\% \\
    10    & 2010  & 820   & 2830  & 392   & 2969  & 1760  & 5121  & 346   & 300   & 646   & 136   & 681   & 252   & 1069  & 77\%  & 79\% \\
    11    & 2951  & 1100  & 4051  & 529   & 4224  & 2597  & 7350  & 361   & 330   & 691   & 149   & 724   & 252   & 1125  & 83\%  & 85\% \\
    12    & 4201  & 1440  & 5641  & 697   & 5911  & 3776  & 10384 & 601   & 480   & 1081  & 222   & 1151  & 456   & 1829  & 81\%  & 82\% \\
    13    & 5823  & 1846  & 7669  & 897   & 7992  & 5250  & 14139 & 727   & 572   & 1299  & 266   & 1377  & 546   & 2189  & 83\%  & 85\% \\
    14    & 7886  & 2324  & 10210 & 1134  & 10673 & 7216  & 19023 & 989   & 728   & 1717  & 342   & 1838  & 775   & 2955  & 83\%  & 84\% \\
    15    & 10465 & 2880  & 13345 & 1409  & 13884 & 9595  & 24888 & 1015  & 780   & 1795  & 366   & 1914  & 775   & 3055  & 87\%  & 88\% \\
    16    & 13641 & 3520  & 17161 & 1726  & 17782 & 12535 & 32043 & 1396  & 992   & 2388  & 470   & 2535  & 1080  & 4085  & 86\%  & 87\% \\
    17    & 17501 & 4250  & 21751 & 2089  & 22584 & 16244 & 40917 & 2107  & 1326  & 3433  & 635   & 3664  & 1710  & 6009  & 84\%  & 85\% \\
    18    & 22138 & 5076  & 27214 & 2499  & 28157 & 20580 & 51236 & 2146  & 1404  & 3550  & 672   & 3779  & 1710  & 6161  & 87\%  & 88\% \\
    19    & 27651 & 6004  & 33655 & 2961  & 34872 & 25905 & 63738 & 2379  & 1558  & 3937  & 747   & 4218  & 1920  & 6885  & 88\%  & 89\% \\
    20    & 34145 & 7040  & 41185 & 3476  & 42544 & 32025 & 78045 & 2420  & 1640  & 4060  & 786   & 4339  & 1920  & 7045  & 90\%  & 91\% \\
 
\midrule
\multicolumn{15}{r}{\textbf{Average Reduction:}} & \textbf{71\%} & \textbf{73\%} \\
\bottomrule
\end{tabular}
}

\vspace{0.5cm}

% ------------------ NBF ------------------
\resizebox{\columnwidth}{!}{
\begin{tabular}{c|ccc|cccc|ccc|cccc|c|c}
\toprule
\multicolumn{17}{c}{\textbf{NBF}} \\
\toprule
\multirow{2}{*}{$V$} & \multicolumn{7}{c}{\textbf{Without AFGR}} & \multicolumn{7}{c}{\textbf{With AFGR}} & \multirow{2}{*}{\begin{tabular}[c]{@{}c@{}}\% Red.\\ Vars\end{tabular}} & \multirow{2}{*}{\begin{tabular}[c]{@{}c@{}}\% Red.\\ Constr.\end{tabular}} \\
\cmidrule(lr){2-8} \cmidrule(lr){9-15}
 & Bin & Cont & Tot Vars& $Eq$ & $\leq$ & $\geq$ & Tot C & Bin & Cont & Tot Vars & $Eq$ & $\leq$ & $\geq$ & Tot C &  \\
\midrule
  
    4     & 81    & 64    & 145   & 28    & 150   & 63    & 241   & 81    & 64    & 145   & 28    & 150   & 63    & 241   & 0\%   & 0\% \\
    5     & 155   & 110   & 265   & 47    & 274   & 120   & 441   & 89    & 80    & 169   & 35    & 172   & 63    & 270   & 36\%  & 39\% \\
    6     & 286   & 180   & 466   & 77    & 481   & 224   & 782   & 118   & 108   & 226   & 47    & 229   & 80    & 356   & 52\%  & 54\% \\
    7     & 501   & 280   & 781   & 122   & 823   & 418   & 1363  & 177   & 154   & 331   & 68    & 346   & 130   & 544   & 58\%  & 60\% \\
    8     & 833   & 416   & 1249  & 183   & 1307  & 700   & 2190  & 188   & 176   & 364   & 78    & 377   & 130   & 585   & 71\%  & 73\% \\
    9     & 1321  & 594   & 1915  & 265   & 2024  & 1152  & 3441  & 331   & 270   & 601   & 121   & 638   & 252   & 1011  & 69\%  & 71\% \\
    10    & 2010  & 820   & 2830  & 369   & 2969  & 1760  & 5098  & 346   & 300   & 646   & 135   & 681   & 252   & 1068  & 77\%  & 79\% \\
    11    & 2951  & 1100  & 4051  & 499   & 4224  & 2597  & 7320  & 361   & 330   & 691   & 149   & 724   & 252   & 1125  & 83\%  & 85\% \\
    12    & 4201  & 1440  & 5641  & 659   & 5911  & 3776  & 10346 & 601   & 480   & 1081  & 219   & 1151  & 456   & 1826  & 81\%  & 82\% \\
    13    & 5823  & 1846  & 7669  & 850   & 7992  & 5250  & 14092 & 727   & 572   & 1299  & 262   & 1377  & 546   & 2185  & 83\%  & 84\% \\
    14    & 7886  & 2324  & 10210 & 1077  & 10673 & 7216  & 18966 & 989   & 728   & 1717  & 336   & 1838  & 775   & 2949  & 83\%  & 84\% \\
    15    & 10465 & 2880  & 13345 & 1341  & 13884 & 9595  & 24820 & 1015  & 780   & 1795  & 361   & 1914  & 775   & 3050  & 87\%  & 88\% \\
    16    & 13641 & 3520  & 17161 & 1646  & 17782 & 12535 & 31963 & 1396  & 992   & 2388  & 461   & 2535  & 1080  & 4076  & 86\%  & 87\% \\
    17    & 17501 & 4250  & 21751 & 1996  & 22584 & 16244 & 40824 & 2107  & 1326  & 3433  & 620   & 3664  & 1710  & 5994  & 84\%  & 85\% \\
    18    & 22138 & 5076  & 27214 & 2392  & 28157 & 20580 & 51129 & 2146  & 1404  & 3550  & 658   & 3779  & 1710  & 6147  & 87\%  & 88\% \\
    19    & 27651 & 6004  & 33655 & 2839  & 34872 & 25905 & 63616 & 2379  & 1558  & 3937  & 733   & 4218  & 1920  & 6871  & 88\%  & 89\% \\
    20    & 34145 & 7040  & 41185 & 3338  & 42544 & 32025 & 77907 & 2420  & 1640  & 4060  & 773   & 4339  & 1920  & 7032  & 90\%  & 91\% \\
      \midrule
\multicolumn{15}{r}{\textbf{Average Reduction:}} & \textbf{71\%} & \textbf{73\%} \\
\bottomrule
\end{tabular}}
\end{table}

\clearpage
\section{Model Dimensionality and Computational Complexity -- Figures}
\label{app:complexityfigure}

Comparison of the total number of variables and constraints in the STSP-TWPD with and without AFGR as a function of $|V|$. Orange lines represent the model without AFGR, while blue lines represent the model with AFGR. Since the ABF and NBF formulations exhibit nearly identical growth patterns in both variables and constraints, only the ABF formulation is reported for clarity.

\begin{figure}[htbp]
\centering
\includegraphics[width=0.8\textwidth]{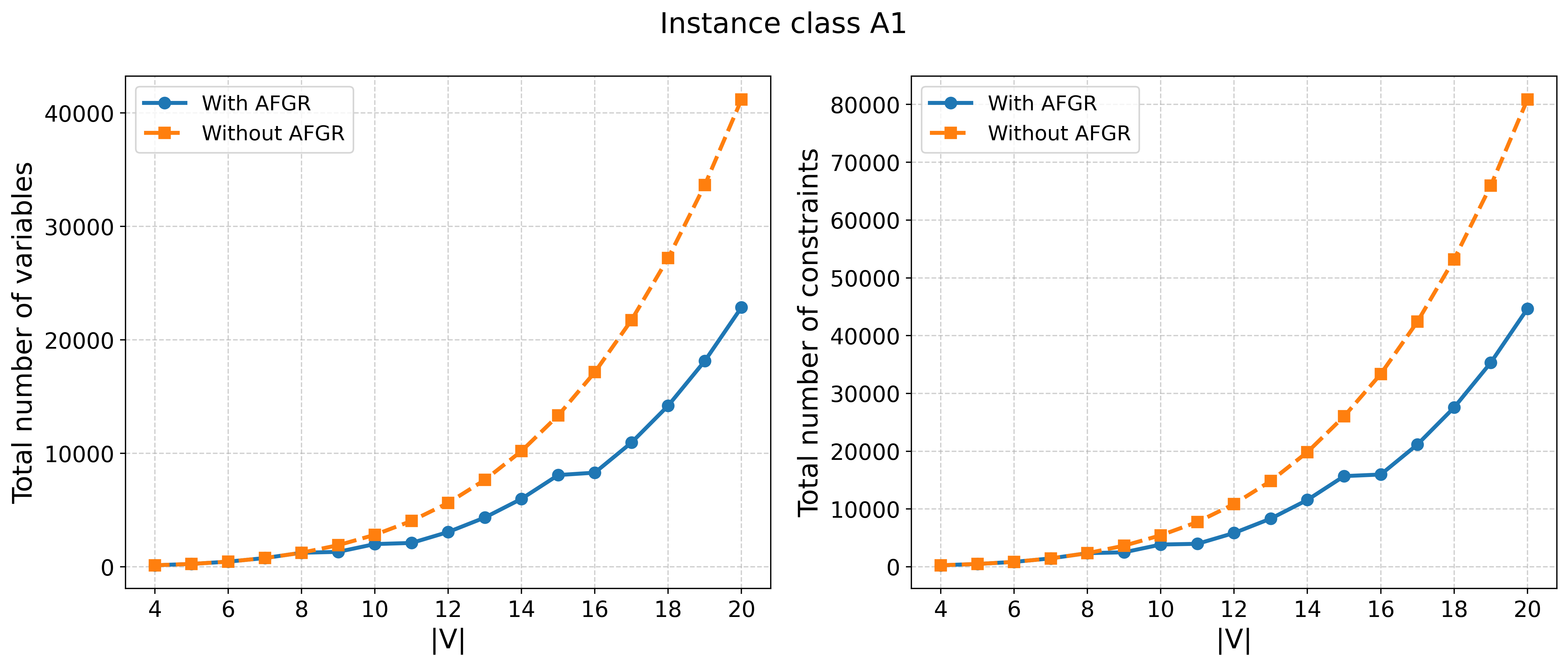}
\caption{Instance class A1: comparison of the total number of variables and constraints in the STSP-TWPD (ABF formulations) with and without AFGR as a function of $|V|$.}
\label{fig:AFGR_A1}
\end{figure}

\begin{figure}[htbp]
\centering
\includegraphics[width=0.8\textwidth]{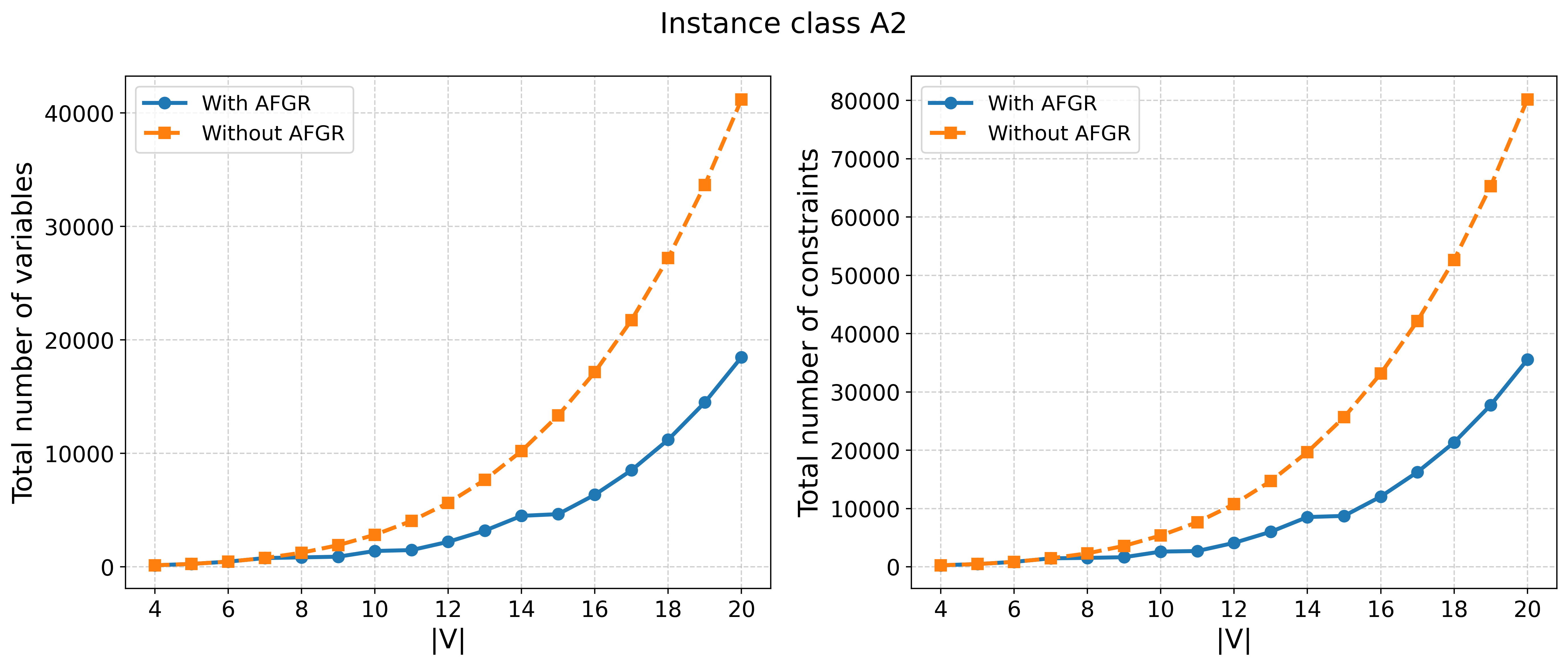}
\caption{Instance class A2: comparison of the total number of variables and constraints in the STSP-TWPD (ABF formulations) with and without AFGR as a function of $|V|$.}
\label{fig:AFGR_A2}
\end{figure}

\begin{figure}[htbp]
\centering
\includegraphics[width=0.8\textwidth]{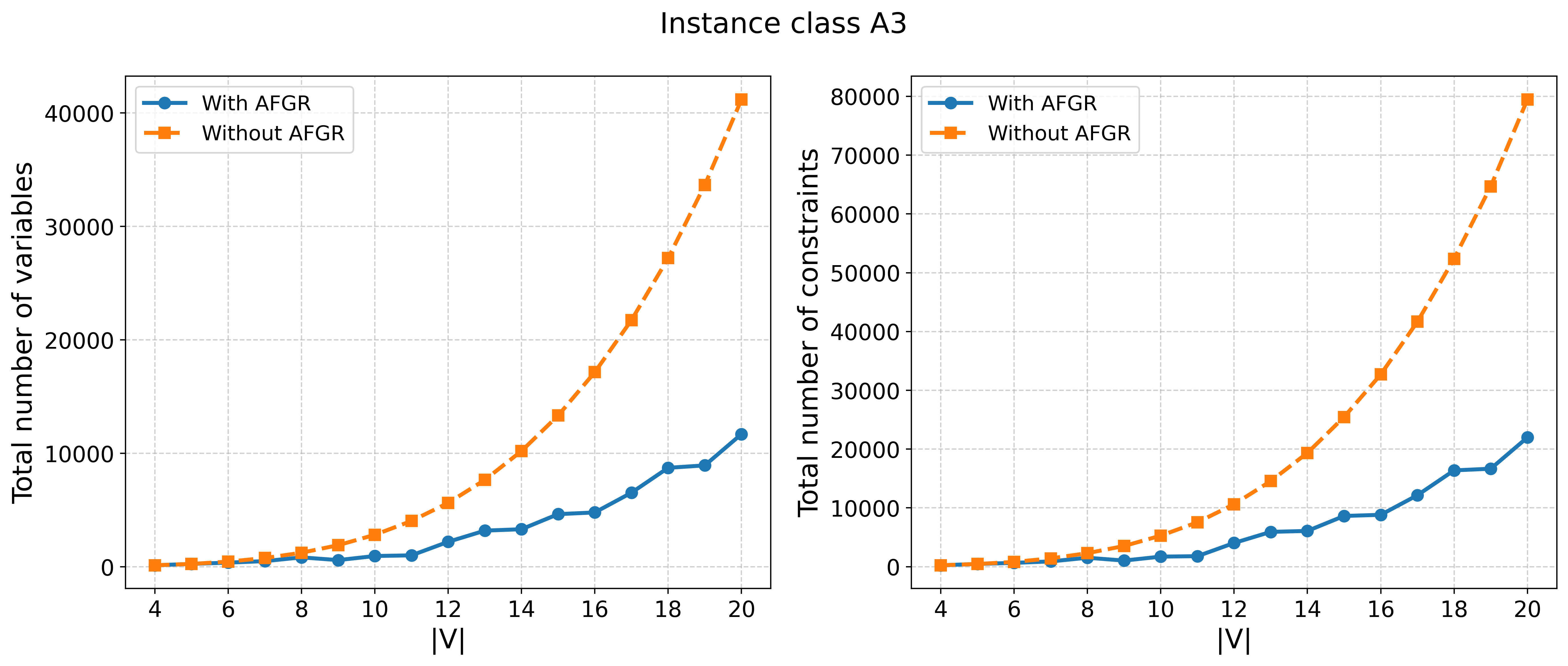}
\caption{Instance class A3: comparison of the total number of variables and constraints in the STSP-TWPD (ABF formulations) with and without AFGR as a function of $|V|$.}
\label{fig:AFGR_A3}
\end{figure}

\begin{figure}[htbp]
\centering
\includegraphics[width=0.8\textwidth]{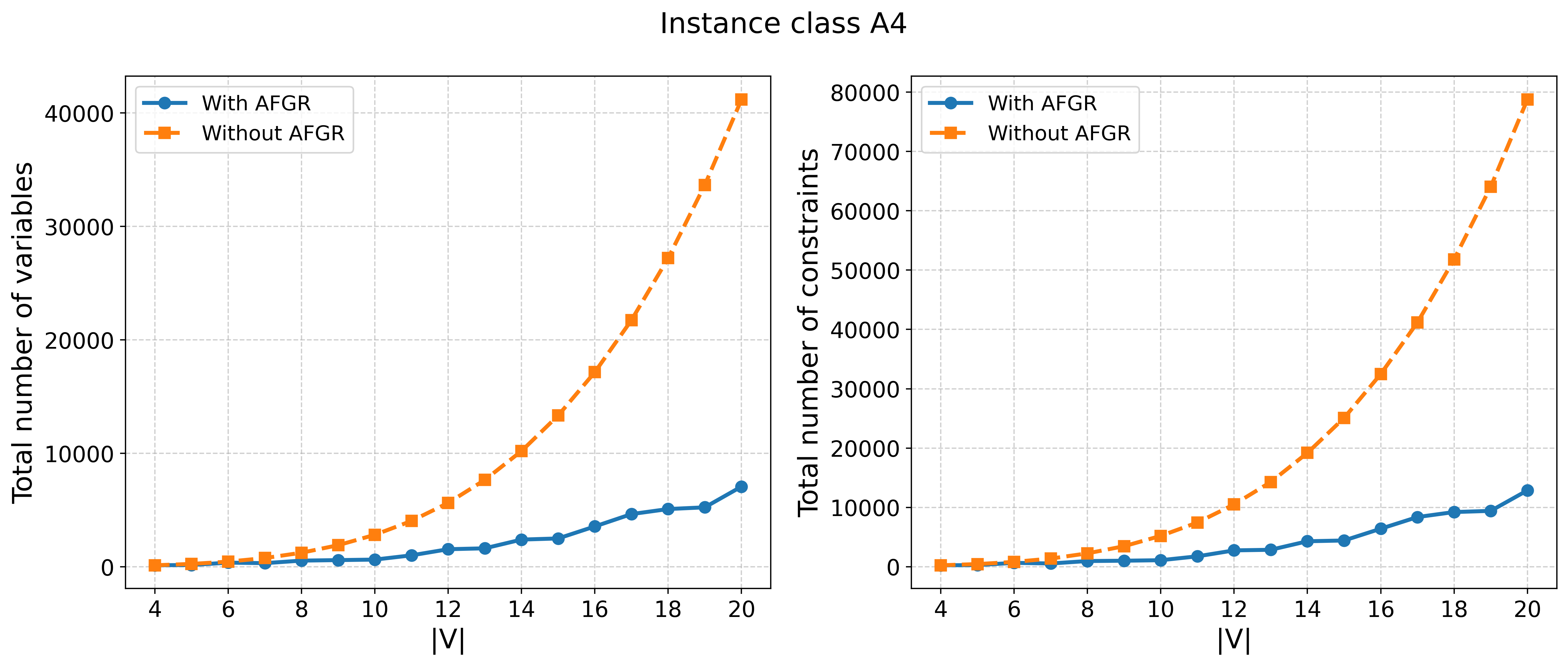}
\caption{Instance class A4: comparison of the total number of variables and constraints in the STSP-TWPD (ABF formulations) with and without AFGR as a function of $|V|$..}
\label{fig:AFGR_A4}
\end{figure}

\begin{figure}[htbp]
\centering
\includegraphics[width=0.8\textwidth]{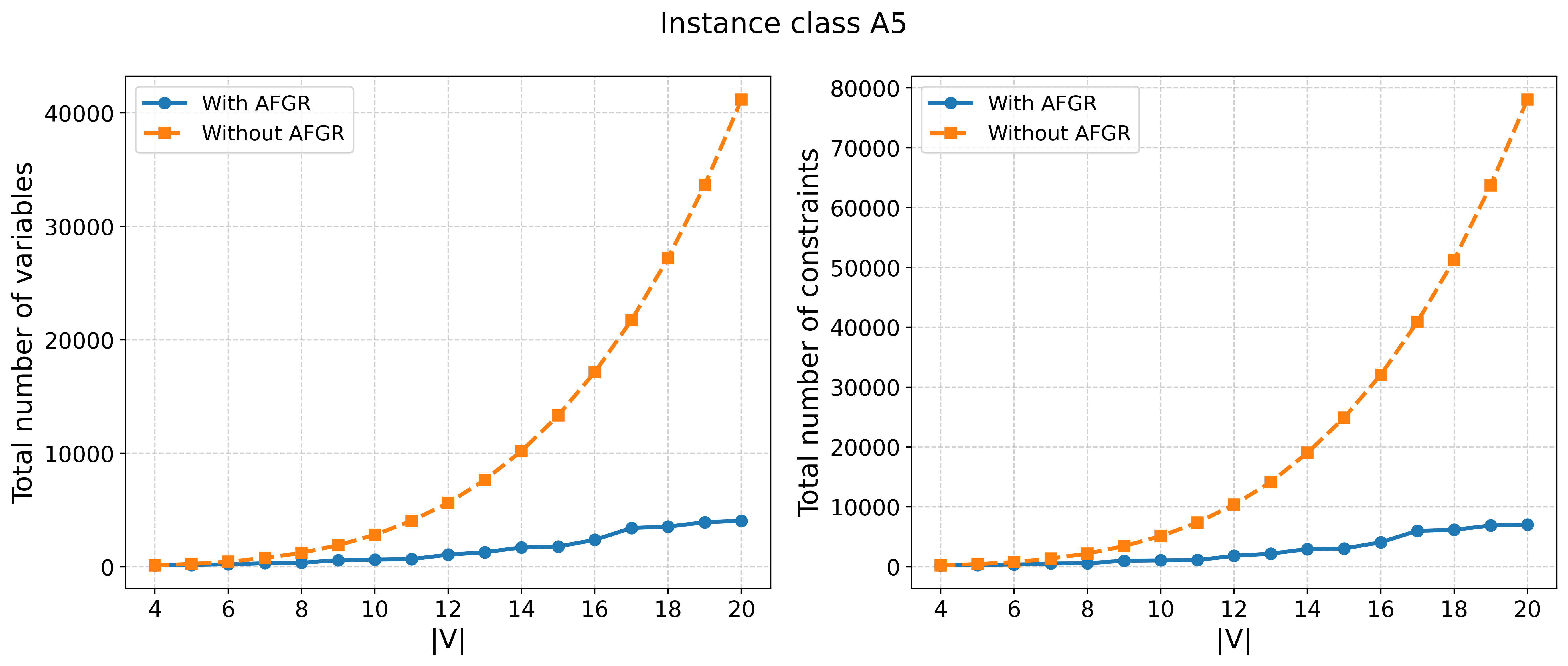}
\caption{Instance class A5: comparison of the total number of variables and constraints in the STSP-TWPD (ABF formulations) with and without AFGR as a function of $|V|$..}
\label{fig:AFGR_A5}
\end{figure}

\begin{figure}[htbp]
\centering
\includegraphics[width=0.8\textwidth]{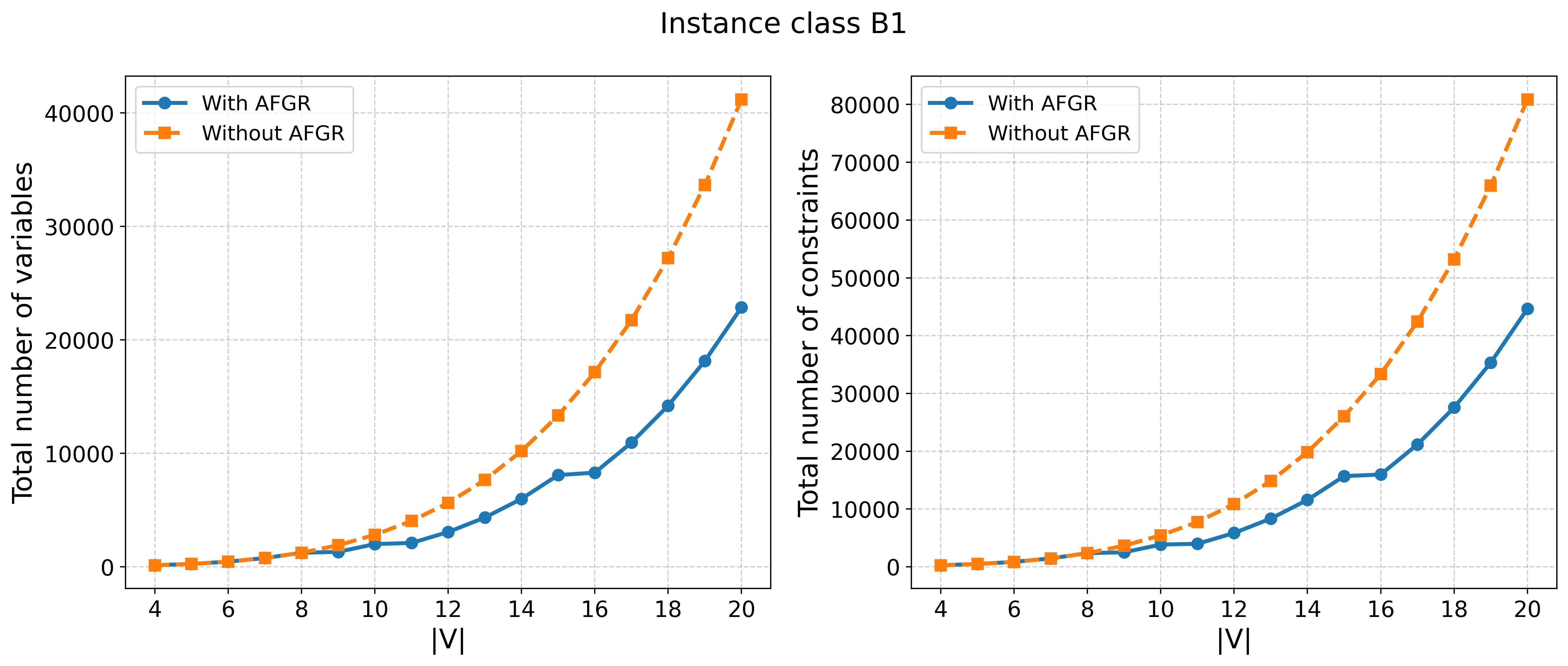}
\caption{Instance class B1: comparison of the total number of variables and constraints in the STSP-TWPD (ABF formulations) with and without AFGR as a function of $|V|$..}
\label{fig:AFGR_B1}
\end{figure}

\begin{figure}[htbp]
\centering
\includegraphics[width=0.8\textwidth]{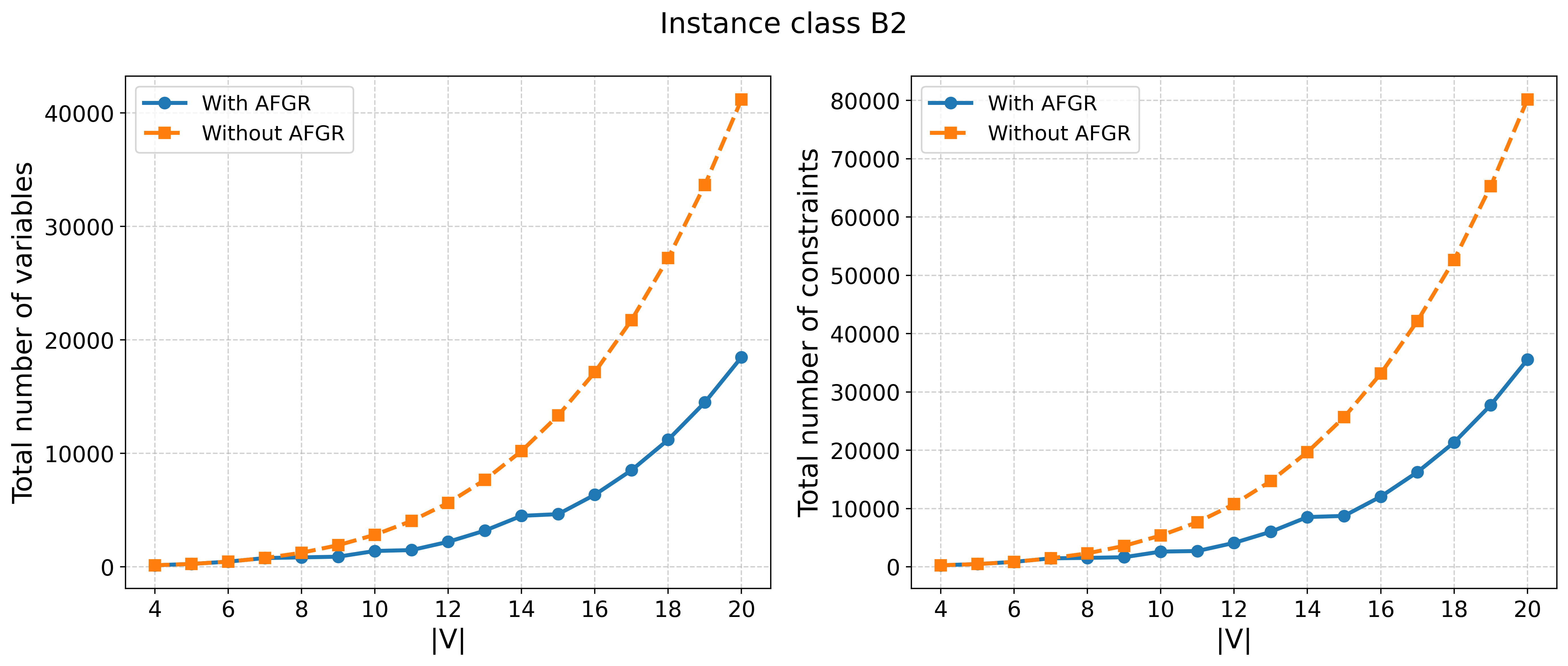}
\caption{Instance class B2: comparison of the total number of variables and constraints in the STSP-TWPD (ABF formulations) with and without AFGR as a function of $|V|$..}
\label{fig:AFGR_B2}
\end{figure}

\begin{figure}[htbp]
\centering
\includegraphics[width=0.8\textwidth]{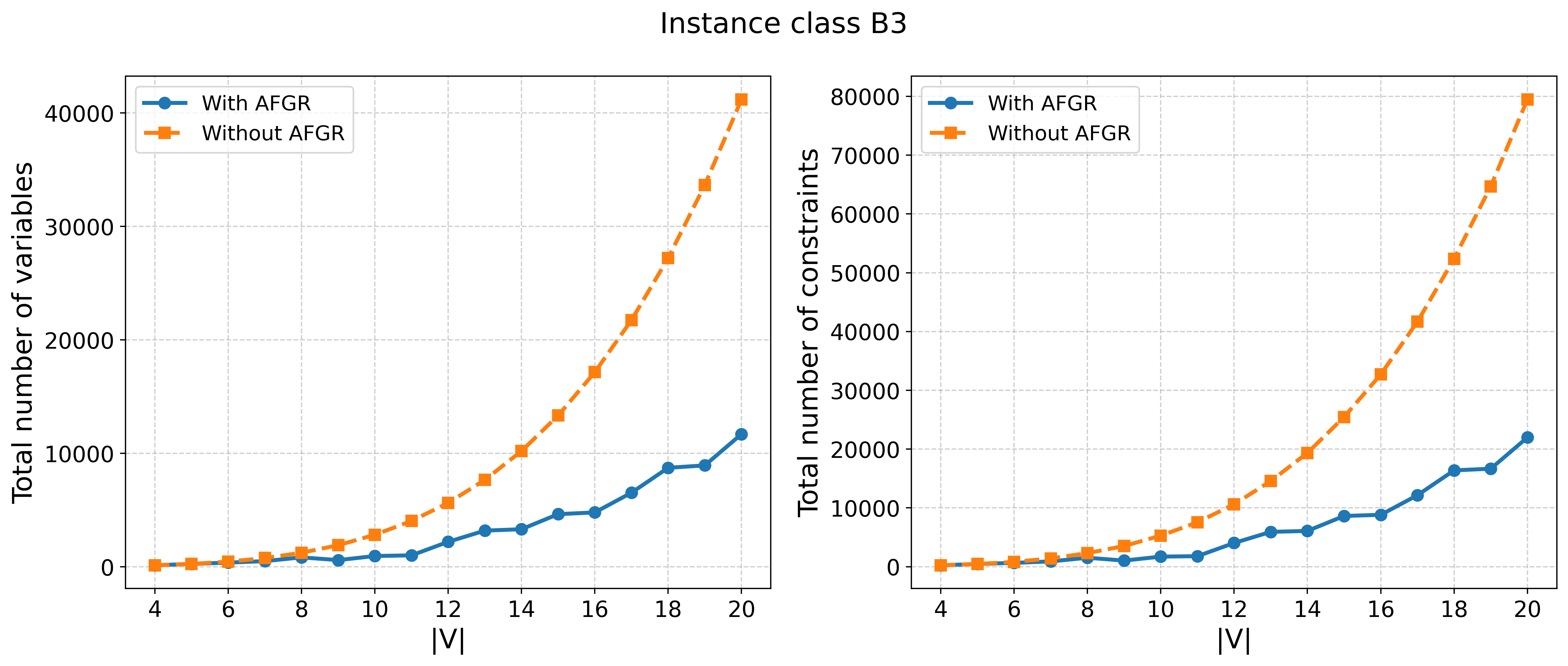}
\caption{Instance class B3: comparison of the total number of variables and constraints in the STSP-TWPD (ABF formulations) with and without AFGR as a function of $|V|$..}
\label{fig:AFGR_B3}
\end{figure}

\begin{figure}[htbp]
\centering
\includegraphics[width=0.8\textwidth]{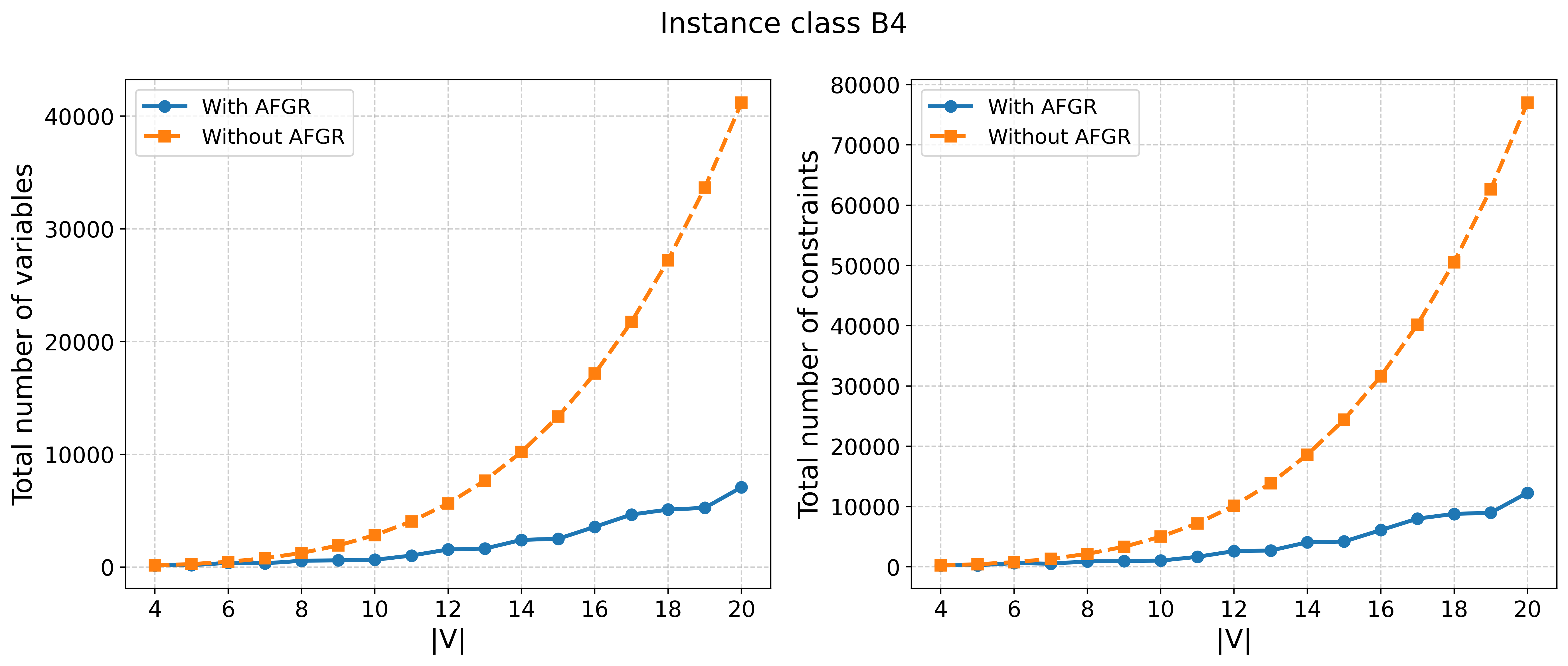}
\caption{Instance class B4: comparison of the total number of variables and constraints in the STSP-TWPD (ABF formulations) with and without AFGR as a function of $|V|$..}
\label{fig:AFGR_B4}
\end{figure}

\begin{figure}[htbp]
\centering
\includegraphics[width=0.8\textwidth]{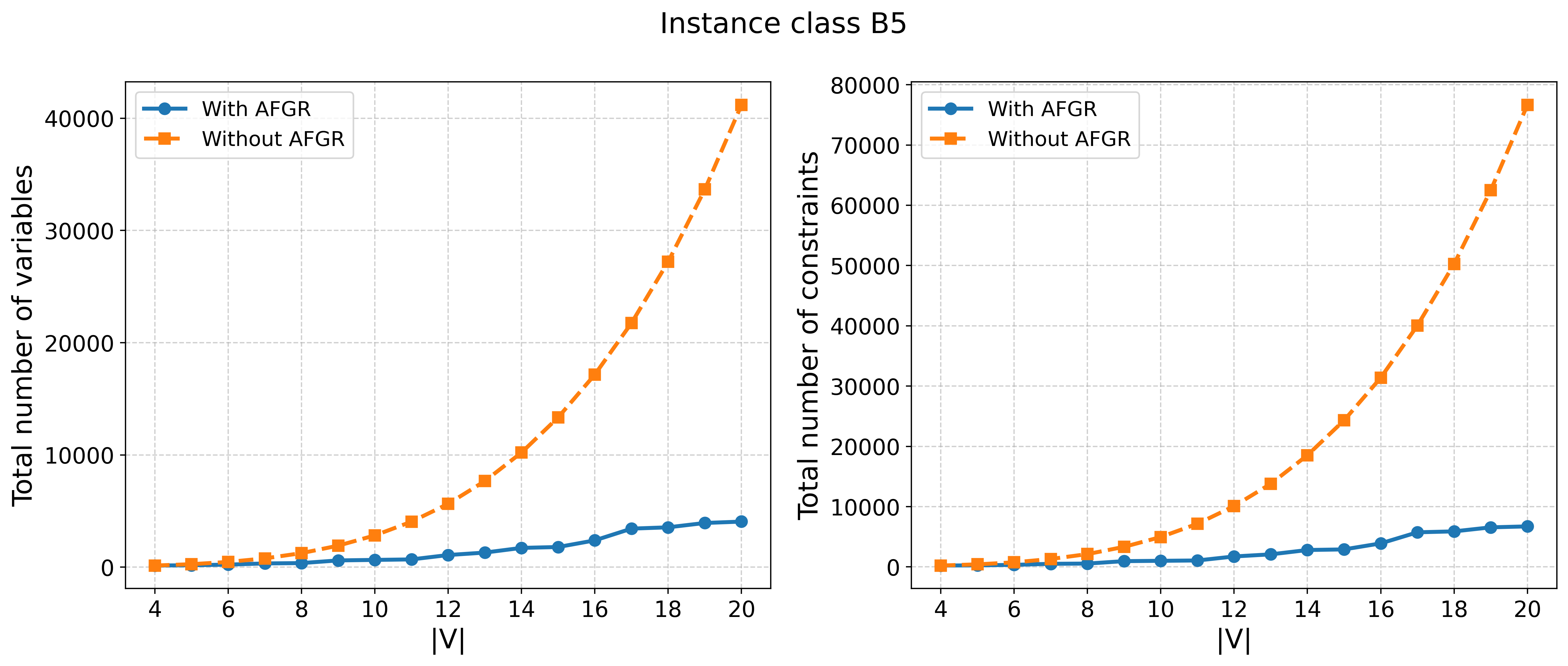}
\caption{Instance class B5: comparison of the total number of variables and constraints in the STSP-TWPD (ABF formulations) with and without AFGR as a function of $|V|$..}
\label{fig:AFGR_B5}
\end{figure}
\clearpage
\section{Computational Results with the Classical Solver -- Tables}
\label{app:classic_table}
\subsection{STSP-TWPD}\label{app:classic_table_STSP_TWPD}
\begin{table}[H]
\centering
\caption{ABF and NBF formulations of the STSP-TWPD model, with and without the AFGR module, for instance class A1. Each row corresponds to an instance with $V$ customers. For each formulation (ABF and NBF), results are reported both in the original setting and after applying AFGR. The reported metrics are as follows:
\textbf{OF}: average objective function value across ten runs,
\textbf{Time}: average solution time in seconds,
\textbf{GAP opt}: average optimality gap, defined as the relative difference between the best feasible solution and the best lower bound at solver termination (MIP gap).
}
\label{tab:classic_A1}
\resizebox{0.9\textwidth}{!}{
\begin{tabular}{c|
ccc|ccc|
ccc|ccc}
\toprule
\multirow{3}{*}{$V$} 
& \multicolumn{6}{c|}{\textbf{ABF}} 
& \multicolumn{6}{c}{\textbf{NBF}} \\
\cmidrule(lr){2-7} \cmidrule(lr){8-13}
& \multicolumn{3}{c|}{\textbf{Without AFGR}} 
& \multicolumn{3}{c|}{\textbf{With AFGR}} 
& \multicolumn{3}{c|}{\textbf{Without AFGR}} 
& \multicolumn{3}{c}{\textbf{With AFGR}} \\
\cmidrule(lr){2-4} \cmidrule(lr){5-7} \cmidrule(lr){8-10} \cmidrule(lr){11-13}
& OF & Time & GAP opt 
& OF & Time & GAP opt 
& OF & Time & GAP opt 
& OF & Time & GAP opt \\
\midrule

    4     & 965,69 & 0,00  & 0\%   & 965,69 & 0,02  & 0\%   & 965,69 & 0,00  & 0\%   & 965,69 & 0,02  & 0\% \\
    5     & 995,96 & 0,02  & 0\%   & 995,96 & 0,02  & 0\%   & 995,96 & 0,05  & 0\%   & 995,96 & 0,05  & 0\% \\
    6     & 1019,62 & 0,02  & 0\%   & 1019,62 & 0,05  & 0\%   & 1019,62 & 0,05  & 0\%   & 1019,62 & 0,03  & 0\% \\
    7     & 1011,19 & 0,06  & 0\%   & 1011,19 & 0,03  & 0\%   & 1011,19 & 0,06  & 0\%   & 1011,19 & 0,09  & 0\% \\
    8     & 991,72 & 0,14  & 0\%   & 991,72 & 0,09  & 0\%   & 991,72 & 0,22  & 0\%   & 991,72 & 0,28  & 0\% \\
    9     & 969,30 & 1,11  & 0\%   & 969,30 & 0,97  & 0\%   & 969,30 & 2,08  & 0\%   & 969,30 & 1,14  & 0\% \\
    10    & 929,54 & 0,63  & 0\%   & 929,54 & 0,19  & 0\%   & 929,54 & 0,34  & 0\%   & 929,54 & 0,61  & 0\% \\
    11    & 908,42 & 0,45  & 0\%   & 908,42 & 0,17  & 0\%   & 908,42 & 0,64  & 0\%   & 908,42 & 0,48  & 0\% \\
    12    & 893,43 & 0,73  & 0\%   & 893,43 & 0,25  & 0\%   & 893,43 & 1,19  & 0\%   & 893,43 & 0,73  & 0\% \\
    13    & 887,31 & 23,69 & 0\%   & 887,31 & 2,42  & 0\%   & 887,31 & 16,34 & 0\%   & 887,31 & 7,86  & 0\% \\
    14    & 865,76 & 3,55  & 0\%   & 865,76 & 1,39  & 0\%   & 865,76 & 2,02  & 0\%   & 865,76 & 0,92  & 0\% \\
    15    & 853,40 & 5,58  & 0\%   & 853,40 & 2,08  & 0\%   & 853,40 & 12,70 & 0\%   & 853,40 & 4,47  & 0\% \\
    16    & 840,52 & 13,34 & 0\%   & 840,52 & 1,28  & 0\%   & 840,52 & 20,81 & 0\%   & 840,52 & 1,72  & 0\% \\
    17    & 832,78 & 20,22 & 0\%   & 832,78 & 2,53  & 0\%   & 832,78 & 36,83 & 0\%   & 832,78 & 9,77  & 0\% \\
    18    & 822,99 & 39,28 & 0\%   & 822,99 & 9,59  & 0\%   & 822,99 & 66,86 & 0\%   & 822,99 & 12,56 & 0\% \\
    19    & 814,92 & 117,88 & 0\%   & 814,92 & 60,56 & 0\%   & 814,92 & 146,56 & 0\%   & 814,92 & 59,80 & 0\% \\
    20    & 805,77 & 48,78 & 0\%   & 805,77 & 26,27 & 0\%   & 805,77 & 81,05 & 0\%   & 805,77 & 29,72 & 0\% \\
      
\bottomrule
\end{tabular}}
\end{table}

\begin{table}[H]
\centering
\caption{ABF and NBF formulations of the STSP-TWPD model, with and without the AFGR module, for instance class A2. Each row corresponds to an instance with $V$ customers. For each formulation (ABF and NBF), results are reported both in the original setting and after applying AFGR. The reported metrics are as follows:
\textbf{OF}: average objective function value across ten runs,
\textbf{Time}: average solution time in seconds,
\textbf{GAP opt}: average optimality gap, defined as the relative difference between the best feasible solution and the best lower bound at solver termination (MIP gap).
}
\label{tab:classic_A2}
\resizebox{0.9\textwidth}{!}{
\begin{tabular}{c|
ccc|ccc|
ccc|ccc}
\toprule
\multirow{3}{*}{$V$} 
& \multicolumn{6}{c|}{\textbf{ABF}} 
& \multicolumn{6}{c}{\textbf{NBF}} \\
\cmidrule(lr){2-7} \cmidrule(lr){8-13}
& \multicolumn{3}{c|}{\textbf{Without AFGR}} 
& \multicolumn{3}{c|}{\textbf{With AFGR}} 
& \multicolumn{3}{c|}{\textbf{Without AFGR}} 
& \multicolumn{3}{c}{\textbf{With AFGR}} \\
\cmidrule(lr){2-4} \cmidrule(lr){5-7} \cmidrule(lr){8-10} \cmidrule(lr){11-13}
& OF & Time & GAP opt 
& OF & Time & GAP opt 
& OF & Time & GAP opt 
& OF & Time & GAP opt \\
\midrule

    4     & 965,69 & 0,00  & 0\%   & 965,69 & 0,02  & 0\%   & 965,69 & 0,02  & 0\%   & 965,69 & 0,00  & 0\% \\
    5     & 995,96 & 0,02  & 0\%   & 995,96 & 0,05  & 0\%   & 995,96 & 0,02  & 0\%   & 995,96 & 0,02  & 0\% \\
    6     & 1019,62 & 0,05  & 0\%   & 1019,62 & 0,06  & 0\%   & 1019,62 & 0,05  & 0\%   & 1019,62 & 0,05  & 0\% \\
    7     & 1401,16 & 6,50  & 0\%   & 1401,16 & 5,56  & 0\%   & 1401,16 & 4,70  & 0\%   & 1401,16 & 4,22  & 0\% \\
    8     & 1429,73 & 7,08  & 0\%   & 1429,73 & 3,30  & 0\%   & 1429,73 & 6,72  & 0\%   & 1429,73 & 3,58  & 0\% \\
    9     & 1299,21 & 10,50 & 0\%   & 1299,21 & 6,30  & 0\%   & 1299,21 & 7,50  & 0\%   & 1299,21 & 1,88  & 0\% \\
    10    & 1620,41 & 222,86 & 0\%   & 1620,41 & 22,58 & 0\%   & 1620,41 & 232,95 & 0\%   & 1620,41 & 16,66 & 0\% \\
    11    & 1559,06 & 1854,38 & 0\%   & 1559,06 & 23,63 & 0\%   & 1559,06 & 1218,11 & 0\%   & 1559,06 & 19,80 & 0\% \\
    12    & 1324,50 & 115,56 & 0\%   & 1324,50 & 16,80 & 0\%   & 1324,50 & 78,33 & 0\%   & 1324,50 & 11,14 & 0\% \\
    13    & 1062,23 & 75,45 & 0\%   & 1062,23 & 16,91 & 0\%   & 1062,23 & 33,70 & 0\%   & 1062,23 & 13,25 & 0\% \\
    14    & 1350,57 & 1586,38 & 0\%   & 1350,57 & 160,02 & 0\%   & 1352,80 & 2311,27 & 11\%  & 1350,57 & 93,27 & 0\% \\
    15    & 1333,68 & 2218,53 & 6\%   & 1333,68 & 133,86 & 0\%   & 1333,68 & 2320,31 & 1\%   & 1333,68 & 108,22 & 0\% \\
    16    & 1551,41 & 2271,97 & 27\%  & 1500,24 & 2228,23 & 7\%   & 1551,41 & 2271,23 & 28\%  & 1498,74 & 2137,42 & 0\% \\
    17    & 1295,56 & 2232,92 & 24\%  & 1294,31 & 431,80 & 0\%   & 1295,56 & 2266,00 & 23\%  & 1294,31 & 727,11 & 0\% \\
    18    & 2258,75 & 2051,06 & 34\%  & 1250,49 & 621,22 & 0\%   & 1327,89 & 2258,75 & 28\%  & 1250,49 & 532,75 & 0\% \\
    19    & 1317,96 & 2088,05 & 35\%  & 1308,23 & 2308,69 & 12\%  & 1311,80 & 2239,31 & 28\%  & 1308,23 & 2244,75 & 6\% \\
    20    & 1380,82 & 1949,44 & 41\%  & 1358,66 & 2075,39 & 23\%  & 1380,82 & 1825,20 & 41\%  & 1348,40 & 2220,09 & 19\% \\
    
\bottomrule
\end{tabular}}
\end{table}

\begin{table}[H]
\centering
\caption{ABF and NBF formulations of the STSP-TWPD model, with and without the AFGR module, for instance class A3. Each row corresponds to an instance with $V$ customers. For each formulation (ABF and NBF), results are reported both in the original setting and after applying AFGR. The reported metrics are as follows:
\textbf{OF}: average objective function value across ten runs,
\textbf{Time}: average solution time in seconds,
\textbf{GAP opt}: average optimality gap, defined as the relative difference between the best feasible solution and the best lower bound at solver termination (MIP gap).
}
\label{tab:classic_A3}
\resizebox{0.9\textwidth}{!}{
\begin{tabular}{c|
ccc|ccc|
ccc|ccc}
\toprule
\multirow{3}{*}{$V$} 
& \multicolumn{6}{c|}{\textbf{ABF}} 
& \multicolumn{6}{c}{\textbf{NBF}} \\
\cmidrule(lr){2-7} \cmidrule(lr){8-13}
& \multicolumn{3}{c|}{\textbf{Without AFGR}} 
& \multicolumn{3}{c|}{\textbf{With AFGR}} 
& \multicolumn{3}{c|}{\textbf{Without AFGR}} 
& \multicolumn{3}{c}{\textbf{With AFGR}} \\
\cmidrule(lr){2-4} \cmidrule(lr){5-7} \cmidrule(lr){8-10} \cmidrule(lr){11-13}
& OF & Time & GAP opt 
& OF & Time & GAP opt 
& OF & Time & GAP opt 
& OF & Time & GAP opt \\
\midrule
4     & 965,69 & 0,02  & 0\%   & 965,69 & 0,02  & 0\%   & 965,69 & 0,00  & 0\%   & 965,69 & 0,00  & 0\% \\
    5     & 995,96 & 0,02  & 0\%   & 995,96 & 0,05  & 0\%   & 995,96 & 0,02  & 0\%   & 995,96 & 0,02  & 0\% \\
    6     & 1019,62 & 0,08  & 0\%   & 1019,62 & 0,02  & 0\%   & 1019,62 & 0,05  & 0\%   & 1019,62 & 0,02  & 0\% \\
    7     & 994,00 & 0,09  & 0\%   & 994,00 & 0,08  & 0\%   & 994,00 & 0,09  & 0\%   & 994,00 & 0,03  & 0\% \\
    8     & 980,07 & 0,16  & 0\%   & 980,07 & 0,17  & 0\%   & 980,07 & 0,09  & 0\%   & 980,07 & 0,06  & 0\% \\
    9     & 900,90 & 0,77  & 0\%   & 900,90 & 0,17  & 0\%   & 900,90 & 1,11  & 0\%   & 900,90 & 0,16  & 0\% \\
    10    & 890,10 & 0,78  & 0\%   & 890,10 & 0,08  & 0\%   & 890,10 & 0,59  & 0\%   & 890,10 & 0,19  & 0\% \\
    11    & 869,52 & 1,13  & 0\%   & 869,52 & 0,14  & 0\%   & 869,52 & 1,02  & 0\%   & 869,52 & 0,13  & 0\% \\
    12    & 889,90 & 0,97  & 0\%   & 889,90 & 0,22  & 0\%   & 889,90 & 0,78  & 0\%   & 889,90 & 0,31  & 0\% \\
    13    & 876,34 & 2,00  & 0\%   & 876,34 & 0,55  & 0\%   & 876,34 & 2,19  & 0\%   & 876,34 & 0,63  & 0\% \\
    14    & 846,34 & 2,52  & 0\%   & 846,34 & 0,64  & 0\%   & 846,34 & 2,89  & 0\%   & 846,34 & 0,56  & 0\% \\
    15    & 1012,46 & 185,84 & 0\%   & 1012,46 & 20,64 & 0\%   & 1012,46 & 71,13 & 0\%   & 1012,46 & 16,22 & 0\% \\
    16    & 980,44 & 148,08 & 0\%   & 980,44 & 11,88 & 0\%   & 980,44 & 293,06 & 0\%   & 980,44 & 18,86 & 0\% \\
    17    & 823,02 & 64,42 & 0\%   & 823,02 & 7,91  & 0\%   & 823,02 & 51,00 & 0\%   & 823,02 & 3,48  & 0\% \\
    18    & 819,86 & 64,98 & 0\%   & 819,86 & 14,28 & 0\%   & 819,86 & 69,59 & 0\%   & 819,86 & 10,16 & 0\% \\
    19    & 810,45 & 111,09 & 0\%   & 810,45 & 11,11 & 0\%   & 810,45 & 154,53 & 0\%   & 810,45 & 10,81 & 0\% \\
    20    & 982,02 & 1497,00 & 0\%   & 982,02 & 59,83 & 0\%   & 982,02 & 1549,38 & 0\%   & 982,02 & 58,63 & 0\% \\

\bottomrule
\end{tabular}}
\end{table}

\begin{table}[H]
\centering
\caption{ABF and NBF formulations of the STSP-TWPD model, with and without the AFGR module, for instance class A4. Each row corresponds to an instance with $V$ customers. For each formulation (ABF and NBF), results are reported both in the original setting and after applying AFGR. The reported metrics are as follows:
\textbf{OF}: average objective function value across ten runs,
\textbf{Time}: average solution time in seconds,
\textbf{GAP opt}: average optimality gap, defined as the relative difference between the best feasible solution and the best lower bound at solver termination (MIP gap).
}
\label{tab:classic_A4}
\resizebox{0.9\textwidth}{!}{
\begin{tabular}{c|
ccc|ccc|
ccc|ccc}
\toprule
\multirow{3}{*}{$V$} 
& \multicolumn{6}{c|}{\textbf{ABF}} 
& \multicolumn{6}{c}{\textbf{NBF}} \\
\cmidrule(lr){2-7} \cmidrule(lr){8-13}
& \multicolumn{3}{c|}{\textbf{Without AFGR}} 
& \multicolumn{3}{c|}{\textbf{With AFGR}} 
& \multicolumn{3}{c|}{\textbf{Without AFGR}} 
& \multicolumn{3}{c}{\textbf{With AFGR}} \\
\cmidrule(lr){2-4} \cmidrule(lr){5-7} \cmidrule(lr){8-10} \cmidrule(lr){11-13}
& OF & Time & GAP opt 
& OF & Time & GAP opt 
& OF & Time & GAP opt 
& OF & Time & GAP opt \\
\midrule

    4     & 965,69 & 0,00  & 0\%   & 965,69 & 0,02  & 0\%   & 965,69 & 0,00  & 0\%   & 965,69 & 0,02  & 0\% \\
    5     & 995,96 & 0,06  & 0\%   & 995,96 & 0,00  & 0\%   & 995,96 & 0,06  & 0\%   & 995,96 & 0,00  & 0\% \\
    6     & 1019,62 & 0,05  & 0\%   & 1019,62 & 0,03  & 0\%   & 1019,62 & 0,05  & 0\%   & 1019,62 & 0,02  & 0\% \\
    7     & 945,85 & 0,03  & 0\%   & 945,85 & 0,02  & 0\%   & 945,85 & 0,03  & 0\%   & 945,85 & 0,03  & 0\% \\
    8     & 946,89 & 0,16  & 0\%   & 946,89 & 0,03  & 0\%   & 946,89 & 0,16  & 0\%   & 946,89 & 0,08  & 0\% \\
    9     & 832,49 & 0,41  & 0\%   & 832,49 & 0,08  & 0\%   & 832,49 & 0,39  & 0\%   & 832,49 & 0,05  & 0\% \\
    10    & 1334,34 & 16,14 & 0\%   & 1334,34 & 2,17  & 0\%   & 1334,34 & 28,14 & 0\%   & 1334,34 & 1,22  & 0\% \\
    11    & 869,52 & 0,48  & 0\%   & 869,52 & 0,06  & 0\%   & 869,52 & 0,67  & 0\%   & 869,52 & 0,06  & 0\% \\
    12    & 869,92 & 0,81  & 0\%   & 869,92 & 0,25  & 0\%   & 869,92 & 0,83  & 0\%   & 869,92 & 0,22  & 0\% \\
    13    & 829,59 & 1,39  & 0\%   & 829,59 & 0,28  & 0\%   & 829,59 & 1,77  & 0\%   & 829,59 & 0,23  & 0\% \\
    14    & 825,66 & 1,66  & 0\%   & 825,66 & 0,41  & 0\%   & 825,66 & 1,75  & 0\%   & 825,66 & 0,23  & 0\% \\
    15    & 818,69 & 4,08  & 0\%   & 818,69 & 0,17  & 0\%   & 818,69 & 8,77  & 0\%   & 818,69 & 0,33  & 0\% \\
    16    & 813,22 & 9,16  & 0\%   & 813,22 & 0,56  & 0\%   & 813,22 & 11,44 & 0\%   & 813,22 & 0,58  & 0\% \\
    17    & 793,81 & 12,72 & 0\%   & 793,81 & 0,83  & 0\%   & 793,81 & 19,03 & 0\%   & 793,81 & 0,52  & 0\% \\
    18    & 935,20 & 294,25 & 0\%   & 935,20 & 14,31 & 0\%   & 935,20 & 247,72 & 0\%   & 935,20 & 11,06 & 0\% \\
    19    & 932,32 & 418,98 & 0\%   & 932,32 & 39,27 & 0\%   & 932,32 & 310,23 & 0\%   & 932,32 & 17,69 & 0\% \\
    20    & 798,13 & 124,59 & 0\%   & 798,13 & 6,13  & 0\%   & 798,13 & 62,06 & 0\%   & 798,13 & 8,41  & 0\% \\

\bottomrule
\end{tabular}}
\end{table}

\begin{table}[H]
\centering
\caption{ABF and NBF formulations of the STSP-TWPD model, with and without the AFGR module, for instance class A5. Each row corresponds to an instance with $V$ customers. For each formulation (ABF and NBF), results are reported both in the original setting and after applying AFGR. The reported metrics are as follows:
\textbf{OF}: average objective function value across ten runs,
\textbf{Time}: average solution time in seconds,
\textbf{GAP opt}: average optimality gap, defined as the relative difference between the best feasible solution and the best lower bound at solver termination (MIP gap).
}
\label{tab:classic_A5}
\resizebox{0.9\textwidth}{!}{
\begin{tabular}{c|
ccc|ccc|
ccc|ccc}
\toprule
\multirow{3}{*}{$V$} 
& \multicolumn{6}{c|}{\textbf{ABF}} 
& \multicolumn{6}{c}{\textbf{NBF}} \\
\cmidrule(lr){2-7} \cmidrule(lr){8-13}
& \multicolumn{3}{c|}{\textbf{Without AFGR}} 
& \multicolumn{3}{c|}{\textbf{With AFGR}} 
& \multicolumn{3}{c|}{\textbf{Without AFGR}} 
& \multicolumn{3}{c}{\textbf{With AFGR}} \\
\cmidrule(lr){2-4} \cmidrule(lr){5-7} \cmidrule(lr){8-10} \cmidrule(lr){11-13}
& OF & Time & GAP opt 
& OF & Time & GAP opt 
& OF & Time & GAP opt 
& OF & Time & GAP opt \\
\midrule

    4     & 965,69 & 0,02  & 0\%   & 965,69 & 0,02  & 0\%   & 965,69 & 0,00  & 0\%   & 965,69 & 0,02  & 0\% \\
    5     & 995,96 & 0,06  & 0\%   & 995,96 & 0,00  & 0\%   & 995,96 & 0,06  & 0\%   & 995,96 & 0,02  & 0\% \\
    6     & 1019,62 & 0,11  & 0\%   & 1019,62 & 0,02  & 0\%   & 1019,62 & 0,06  & 0\%   & 1019,62 & 0,02  & 0\% \\
    7     & 945,85 & 0,03  & 0\%   & 945,85 & 0,02  & 0\%   & 945,85 & 0,05  & 0\%   & 945,85 & 0,02  & 0\% \\
    8     & 664,05 & 0,06  & 0\%   & 664,05 & 0,03  & 0\%   & 664,05 & 0,06  & 0\%   & 664,05 & 0,03  & 0\% \\
    9     & 832,49 & 0,38  & 0\%   & 832,49 & 0,08  & 0\%   & 832,49 & 0,33  & 0\%   & 832,49 & 0,03  & 0\% \\
    10    & 905,93 & 1,08  & 0\%   & 905,93 & 0,17  & 0\%   & 905,93 & 1,53  & 0\%   & 905,93 & 0,13  & 0\% \\
    11    & 1072,45 & 14,89 & 0\%   & 1072,45 & 0,42  & 0\%   & 1072,45 & 14,41 & 0\%   & 1072,45 & 0,39  & 0\% \\
    12    & 843,12 & 0,66  & 0\%   & 843,12 & 0,06  & 0\%   & 843,12 & 1,06  & 0\%   & 843,12 & 0,11  & 0\% \\
    13    & 643,70 & 1,36  & 0\%   & 643,70 & 0,13  & 0\%   & 643,70 & 0,50  & 0\%   & 643,70 & 0,06  & 0\% \\
    14    & 795,95 & 2,13  & 0\%   & 795,95 & 0,22  & 0\%   & 795,95 & 2,14  & 0\%   & 795,95 & 0,22  & 0\% \\
    15    & 794,11 & 2,69  & 0\%   & 794,11 & 0,23  & 0\%   & 794,11 & 5,92  & 0\%   & 794,11 & 0,28  & 0\% \\
    16    & 808,78 & 7,64  & 0\%   & 808,78 & 0,30  & 0\%   & 808,78 & 9,91  & 0\%   & 808,78 & 0,27  & 0\% \\
    17    & 790,10 & 16,13 & 0\%   & 790,10 & 0,30  & 0\%   & 790,10 & 12,88 & 0\%   & 790,10 & 0,56  & 0\% \\
    18    & 785,74 & 24,06 & 0\%   & 785,74 & 0,25  & 0\%   & 785,74 & 24,38 & 0\%   & 785,74 & 0,27  & 0\% \\
    19    & 797,18 & 25,16 & 0\%   & 797,18 & 0,59  & 0\%   & 797,18 & 32,52 & 0\%   & 797,18 & 0,28  & 0\% \\
    20    & 790,54 & 24,97 & 0\%   & 790,54 & 0,53  & 0\%   & 790,54 & 29,11 & 0\%   & 790,54 & 0,61  & 0\% \\
   
\bottomrule
\end{tabular}}
\end{table}

\begin{table}[H]
\centering
\caption{ABF and NBF formulations of the STSP-TWPD model, with and without the AFGR module, for instance class B1. Each row corresponds to an instance with $V$ customers. For each formulation (ABF and NBF), results are reported both in the original setting and after applying AFGR. The reported metrics are as follows:
\textbf{OF}: average objective function value across ten runs,
\textbf{Time}: average solution time in seconds,
\textbf{GAP opt}: average optimality gap, defined as the relative difference between the best feasible solution and the best lower bound at solver termination (MIP gap).
}
\label{tab:classic_B1}
\resizebox{0.9\textwidth}{!}{
\begin{tabular}{c|
ccc|ccc|
ccc|ccc}
\toprule
\multirow{3}{*}{$V$} 
& \multicolumn{6}{c|}{\textbf{ABF}} 
& \multicolumn{6}{c}{\textbf{NBF}} \\
\cmidrule(lr){2-7} \cmidrule(lr){8-13}
& \multicolumn{3}{c|}{\textbf{Without AFGR}} 
& \multicolumn{3}{c|}{\textbf{With AFGR}} 
& \multicolumn{3}{c|}{\textbf{Without AFGR}} 
& \multicolumn{3}{c}{\textbf{With AFGR}} \\
\cmidrule(lr){2-4} \cmidrule(lr){5-7} \cmidrule(lr){8-10} \cmidrule(lr){11-13}
& OF & Time & GAP opt 
& OF & Time & GAP opt 
& OF & Time & GAP opt 
& OF & Time & GAP opt \\
\midrule

    4     & 965,69 & 0,02  & 0\%   & 965,69 & 0,00  & 0\%   & 965,69 & 0,05  & 0\%   & 965,69 & 0,02  & 0\% \\
    5     & 995,96 & 0,02  & 0\%   & 995,96 & 0,02  & 0\%   & 995,96 & 0,02  & 0\%   & 995,96 & 0,02  & 0\% \\
    6     & 1019,62 & 0,05  & 0\%   & 1019,62 & 0,06  & 0\%   & 1019,62 & 0,03  & 0\%   & 1019,62 & 0,08  & 0\% \\
    7     & 1011,19 & 0,11  & 0\%   & 1011,19 & 0,11  & 0\%   & 1011,19 & 0,11  & 0\%   & 1011,19 & 0,08  & 0\% \\
    8     & 991,72 & 0,23  & 0\%   & 991,72 & 0,17  & 0\%   & 991,72 & 0,41  & 0\%   & 991,72 & 0,61  & 0\% \\
    9     & 969,30 & 2,17  & 0\%   & 969,30 & 1,69  & 0\%   & 969,30 & 2,61  & 0\%   & 969,30 & 1,19  & 0\% \\
    10    & 929,54 & 1,17  & 0\%   & 929,54 & 0,38  & 0\%   & 929,54 & 0,39  & 0\%   & 929,54 & 0,58  & 0\% \\
    11    & 908,42 & 0,67  & 0\%   & 908,42 & 0,30  & 0\%   & 908,42 & 0,70  & 0\%   & 908,42 & 0,53  & 0\% \\
    12    & 893,43 & 1,50  & 0\%   & 893,43 & 0,64  & 0\%   & 893,43 & 0,95  & 0\%   & 893,43 & 0,88  & 0\% \\
    13    & 887,31 & 49,69 & 0\%   & 887,31 & 4,05  & 0\%   & 887,31 & 17,94 & 0\%   & 887,31 & 8,89  & 0\% \\
    14    & 865,76 & 6,03  & 0\%   & 865,76 & 2,13  & 0\%   & 865,76 & 2,20  & 0\%   & 865,76 & 0,86  & 0\% \\
    15    & 853,40 & 9,08  & 0\%   & 853,40 & 3,36  & 0\%   & 853,40 & 12,63 & 0\%   & 853,40 & 4,05  & 0\% \\
    16    & 840,52 & 19,05 & 0\%   & 840,52 & 1,64  & 0\%   & 840,52 & 21,02 & 0\%   & 840,52 & 2,66  & 0\% \\
    17    & 832,78 & 26,38 & 0\%   & 832,78 & 2,77  & 0\%   & 832,78 & 35,92 & 0\%   & 832,78 & 6,61  & 0\% \\
    18    & 822,99 & 44,47 & 0\%   & 822,99 & 9,56  & 0\%   & 822,99 & 68,55 & 0\%   & 822,99 & 10,48 & 0\% \\
    19    & 814,92 & 136,05 & 0\%   & 814,92 & 61,53 & 0\%   & 814,92 & 152,94 & 0\%   & 814,92 & 59,44 & 0\% \\
    20    & 805,77 & 80,27 & 0\%   & 805,77 & 32,63 & 0\%   & 805,77 & 76,66 & 0\%   & 805,77 & 28,95 & 0\% \\
   
\bottomrule
\end{tabular}}
\end{table}

\begin{table}[H]
\centering
\caption{ABF and NBF formulations of the STSP-TWPD model, with and without the AFGR module, for instance class B2. Each row corresponds to an instance with $V$ customers. For each formulation (ABF and NBF), results are reported both in the original setting and after applying AFGR. The reported metrics are as follows:
\textbf{OF}: average objective function value across ten runs,
\textbf{Time}: average solution time in seconds,
\textbf{GAP opt}: average optimality gap, defined as the relative difference between the best feasible solution and the best lower bound at solver termination (MIP gap).
}
\label{tab:classic_B2}
\resizebox{0.9\textwidth}{!}{
\begin{tabular}{c|
ccc|ccc|
ccc|ccc}
\toprule
\multirow{3}{*}{$V$} 
& \multicolumn{6}{c|}{\textbf{ABF}} 
& \multicolumn{6}{c}{\textbf{NBF}} \\
\cmidrule(lr){2-7} \cmidrule(lr){8-13}
& \multicolumn{3}{c|}{\textbf{Without AFGR}} 
& \multicolumn{3}{c|}{\textbf{With AFGR}} 
& \multicolumn{3}{c|}{\textbf{Without AFGR}} 
& \multicolumn{3}{c}{\textbf{With AFGR}} \\
\cmidrule(lr){2-4} \cmidrule(lr){5-7} \cmidrule(lr){8-10} \cmidrule(lr){11-13}
& OF & Time & GAP opt 
& OF & Time & GAP opt 
& OF & Time & GAP opt 
& OF & Time & GAP opt \\
\midrule
4     & 965,69 & 0,00  & 0\%   & 965,69 & 0,02  & 0\%   & 965,69 & 0,02  & 0\%   & 965,69 & 0,00  & 0\% \\
    5     & 995,96 & 0,02  & 0\%   & 995,96 & 0,06  & 0\%   & 995,96 & 0,05  & 0\%   & 995,96 & 0,02  & 0\% \\
    6     & 1019,62 & 0,09  & 0\%   & 1019,62 & 0,08  & 0\%   & 1019,62 & 0,02  & 0\%   & 1019,62 & 0,02  & 0\% \\
    7     & 1401,16 & 4,66  & 0\%   & 1401,16 & 4,44  & 0\%   & 1401,16 & 3,48  & 0\%   & 1401,16 & 3,36  & 0\% \\
    8     & 1429,73 & 5,67  & 0\%   & 1429,73 & 2,73  & 0\%   & 1429,73 & 6,02  & 0\%   & 1429,73 & 3,19  & 0\% \\
    9     & 1299,21 & 8,28  & 0\%   & 1299,21 & 4,30  & 0\%   & 1299,21 & 6,44  & 0\%   & 1299,21 & 1,67  & 0\% \\
    10    & 1620,41 & 211,44 & 0\%   & 1620,41 & 36,92 & 0\%   & 1620,41 & 183,75 & 0\%   & 1620,41 & 14,14 & 0\% \\
    11    & 1559,06 & 2212,17 & 6\%   & 1559,06 & 33,58 & 0\%   & 1559,06 & 1025,38 & 0\%   & 1559,06 & 18,67 & 0\% \\
    12    & 1324,50 & 145,38 & 0\%   & 1324,50 & 23,19 & 0\%   & 1324,50 & 70,28 & 0\%   & 1324,50 & 10,13 & 0\% \\
    13    & 1062,23 & 110,19 & 0\%   & 1062,23 & 22,70 & 0\%   & 1062,23 & 31,25 & 0\%   & 1062,23 & 12,08 & 0\% \\
    14    & 1350,57 & 2091,31 & 0\%   & 1350,57 & 213,72 & 0\%   & 1352,80 & 2318,47 & 9\%   & 1350,57 & 86,20 & 0\% \\
    15    & 1333,68 & 2201,80 & 9\%   & 1333,68 & 130,38 & 0\%   & 1333,68 & 2150,22 & 0\%   & 1333,68 & 101,23 & 0\% \\
    16    & 1551,41 & 2218,80 & 27\%  & 1498,74 & 2163,17 & 0\%   & 1551,41 & 2281,69 & 27\%  & 1498,74 & 2002,67 & 0\% \\
    17    & 1295,56 & 2289,63 & 15\%  & 1294,31 & 388,28 & 0\%   & 1295,56 & 2253,78 & 21\%  & 1294,31 & 609,56 & 0\% \\
    18    & 1251,55 & 2034,39 & 34\%  & 1250,49 & 495,66 & 0\%   & 1251,55 & 2274,48 & 22\%  & 1250,49 & 437,89 & 0\% \\
    19    & 1311,80 & 2081,78 & 28\%  & 1308,23 & 2316,02 & 5\%   & 1311,80 & 2198,03 & 27\%  & 1308,23 & 1849,89 & 0\% \\
    20    & 1376,16 & 2081,14 & 37\%  & 1358,66 & 2274,70 & 18\%  & 1376,16 & 1939,08 & 40\%  & 1348,40 & 2318,77 & 12\% \\
  
\bottomrule
\end{tabular}}
\end{table}

\begin{table}[H]
\centering
\caption{ABF and NBF formulations of the STSP-TWPD model, with and without the AFGR module, for instance class B3. Each row corresponds to an instance with $V$ customers. For each formulation (ABF and NBF), results are reported both in the original setting and after applying AFGR. The reported metrics are as follows:
\textbf{OF}: average objective function value across ten runs,
\textbf{Time}: average solution time in seconds,
\textbf{GAP opt}: average optimality gap, defined as the relative difference between the best feasible solution and the best lower bound at solver termination (MIP gap).
}
\label{tab:classic_B3}
\resizebox{0.9\textwidth}{!}{
\begin{tabular}{c|
ccc|ccc|
ccc|ccc}
\toprule
\multirow{3}{*}{$V$} 
& \multicolumn{6}{c|}{\textbf{ABF}} 
& \multicolumn{6}{c}{\textbf{NBF}} \\
\cmidrule(lr){2-7} \cmidrule(lr){8-13}
& \multicolumn{3}{c|}{\textbf{Without AFGR}} 
& \multicolumn{3}{c|}{\textbf{With AFGR}} 
& \multicolumn{3}{c|}{\textbf{Without AFGR}} 
& \multicolumn{3}{c}{\textbf{With AFGR}} \\
\cmidrule(lr){2-4} \cmidrule(lr){5-7} \cmidrule(lr){8-10} \cmidrule(lr){11-13}
& OF & Time & GAP opt 
& OF & Time & GAP opt 
& OF & Time & GAP opt 
& OF & Time & GAP opt \\
\midrule

    4     & 965,69 & 0,02  & 0\%   & 965,69 & 0,00  & 0\%   & 965,69 & 0,00  & 0\%   & 965,69 & 0,00  & 0\% \\
    5     & 995,96 & 0,00  & 0\%   & 995,96 & 0,03  & 0\%   & 995,96 & 0,02  & 0\%   & 995,96 & 0,02  & 0\% \\
    6     & 1019,62 & 0,02  & 0\%   & 1019,62 & 0,02  & 0\%   & 1019,62 & 0,05  & 0\%   & 1019,62 & 0,03  & 0\% \\
    7     & 994,00 & 0,06  & 0\%   & 994,00 & 0,02  & 0\%   & 994,00 & 0,06  & 0\%   & 994,00 & 0,02  & 0\% \\
    8     & 980,07 & 0,16  & 0\%   & 980,07 & 0,17  & 0\%   & 980,07 & 0,16  & 0\%   & 980,07 & 0,06  & 0\% \\
    9     & 900,90 & 0,73  & 0\%   & 900,90 & 0,22  & 0\%   & 900,90 & 1,09  & 0\%   & 900,90 & 0,16  & 0\% \\
    10    & 890,10 & 0,83  & 0\%   & 890,10 & 0,08  & 0\%   & 890,10 & 0,56  & 0\%   & 890,10 & 0,14  & 0\% \\
    11    & 869,52 & 1,08  & 0\%   & 869,52 & 0,14  & 0\%   & 869,52 & 0,97  & 0\%   & 869,52 & 0,17  & 0\% \\
    12    & 889,90 & 0,92  & 0\%   & 889,90 & 0,28  & 0\%   & 889,90 & 0,80  & 0\%   & 889,90 & 0,28  & 0\% \\
    13    & 876,34 & 2,11  & 0\%   & 876,34 & 0,56  & 0\%   & 876,34 & 2,17  & 0\%   & 876,34 & 0,64  & 0\% \\
    14    & 846,34 & 2,48  & 0\%   & 846,34 & 0,67  & 0\%   & 846,34 & 2,89  & 0\%   & 846,34 & 0,53  & 0\% \\
    15    & 1012,46 & 180,02 & 0\%   & 1012,46 & 20,31 & 0\%   & 1012,46 & 70,45 & 0\%   & 1012,46 & 16,38 & 0\% \\
    16    & 980,44 & 145,81 & 0\%   & 980,44 & 11,44 & 0\%   & 980,44 & 287,19 & 0\%   & 980,44 & 18,73 & 0\% \\
    17    & 823,02 & 63,16 & 0\%   & 823,02 & 7,63  & 0\%   & 823,02 & 50,72 & 0\%   & 823,02 & 3,63  & 0\% \\
    18    & 819,86 & 63,44 & 0\%   & 819,86 & 13,22 & 0\%   & 819,86 & 69,70 & 0\%   & 819,86 & 10,20 & 0\% \\
    19    & 810,45 & 108,41 & 0\%   & 810,45 & 10,61 & 0\%   & 810,45 & 157,63 & 0\%   & 810,45 & 11,00 & 0\% \\
    20    & 982,02 & 1443,25 & 0\%   & 982,02 & 59,28 & 0\%   & 982,02 & 1552,67 & 0\%   & 982,02 & 58,17 & 0\% \\
    
\bottomrule
\end{tabular}}
\end{table}

\begin{table}[H]
\centering
\caption{ABF and NBF formulations of the STSP-TWPD model, with and without the AFGR module, for instance class B4. Each row corresponds to an instance with $V$ customers. For each formulation (ABF and NBF), results are reported both in the original setting and after applying AFGR. The reported metrics are as follows:
\textbf{OF}: average objective function value across ten runs,
\textbf{Time}: average solution time in seconds,
\textbf{GAP opt}: average optimality gap, defined as the relative difference between the best feasible solution and the best lower bound at solver termination (MIP gap).
}
\label{tab:classic_B4}
\resizebox{0.9\textwidth}{!}{
\begin{tabular}{c|
ccc|ccc|
ccc|ccc}
\toprule
\multirow{3}{*}{$V$} 
& \multicolumn{6}{c|}{\textbf{ABF}} 
& \multicolumn{6}{c}{\textbf{NBF}} \\
\cmidrule(lr){2-7} \cmidrule(lr){8-13}
& \multicolumn{3}{c|}{\textbf{Without AFGR}} 
& \multicolumn{3}{c|}{\textbf{With AFGR}} 
& \multicolumn{3}{c|}{\textbf{Without AFGR}} 
& \multicolumn{3}{c}{\textbf{With AFGR}} \\
\cmidrule(lr){2-4} \cmidrule(lr){5-7} \cmidrule(lr){8-10} \cmidrule(lr){11-13}
& OF & Time & GAP opt 
& OF & Time & GAP opt 
& OF & Time & GAP opt 
& OF & Time & GAP opt \\
\midrule
4     & 965,69 & 0,02  & 0\%   & 965,69 & 0,00  & 0\%   & 965,69 & 0,00  & 0\%   & 965,69 & 0,02  & 0\% \\
    5     & 995,96 & 0,02  & 0\%   & 995,96 & 0,02  & 0\%   & 995,96 & 0,06  & 0\%   & 995,96 & 0,00  & 0\% \\
    6     & 1019,62 & 0,02  & 0\%   & 1019,62 & 0,02  & 0\%   & 1019,62 & 0,06  & 0\%   & 1019,62 & 0,08  & 0\% \\
    7     & 945,85 & 0,06  & 0\%   & 945,85 & 0,06  & 0\%   & 945,85 & 0,09  & 0\%   & 945,85 & 0,02  & 0\% \\
    8     & 946,89 & 0,11  & 0\%   & 946,89 & 0,08  & 0\%   & 946,89 & 0,20  & 0\%   & 946,89 & 0,03  & 0\% \\
    9     & 832,49 & 0,39  & 0\%   & 832,49 & 0,03  & 0\%   & 832,49 & 0,33  & 0\%   & 832,49 & 0,03  & 0\% \\
    10    & 1334,34 & 16,41 & 0\%   & 1334,34 & 2,16  & 0\%   & 1334,34 & 27,89 & 0\%   & 1334,34 & 1,25  & 0\% \\
    11    & 869,52 & 0,47  & 0\%   & 869,52 & 0,05  & 0\%   & 869,52 & 0,73  & 0\%   & 869,52 & 0,05  & 0\% \\
    12    & 869,92 & 0,92  & 0\%   & 869,92 & 0,22  & 0\%   & 869,92 & 0,92  & 0\%   & 869,92 & 0,27  & 0\% \\
    13    & 829,59 & 1,44  & 0\%   & 829,59 & 0,28  & 0\%   & 829,59 & 1,75  & 0\%   & 829,59 & 0,25  & 0\% \\
    14    & 825,66 & 1,64  & 0\%   & 825,66 & 0,42  & 0\%   & 825,66 & 1,81  & 0\%   & 825,66 & 0,20  & 0\% \\
    15    & 818,69 & 4,25  & 0\%   & 818,69 & 0,22  & 0\%   & 818,69 & 8,73  & 0\%   & 818,69 & 0,39  & 0\% \\
    16    & 813,22 & 8,95  & 0\%   & 813,22 & 0,59  & 0\%   & 813,22 & 11,56 & 0\%   & 813,22 & 0,55  & 0\% \\
    17    & 793,81 & 13,03 & 0\%   & 793,81 & 0,80  & 0\%   & 793,81 & 18,88 & 0\%   & 793,81 & 0,56  & 0\% \\
    18    & 935,20 & 300,45 & 0\%   & 935,20 & 14,02 & 0\%   & 935,20 & 250,77 & 0\%   & 935,20 & 11,64 & 0\% \\
    19    & 932,32 & 423,19 & 0\%   & 932,32 & 39,13 & 0\%   & 932,32 & 311,88 & 0\%   & 932,32 & 14,66 & 0\% \\
    20    & 798,13 & 124,50 & 0\%   & 798,13 & 6,06  & 0\%   & 798,13 & 61,58 & 0\%   & 798,13 & 8,58  & 0\% \\
    
\bottomrule
\end{tabular}}
\end{table}

\begin{table}[H]
\centering
\caption{ABF and NBF formulations of the STSP-TWPD model, with and without the AFGR module, for instance class B5. Each row corresponds to an instance with $V$ customers. For each formulation (ABF and NBF), results are reported both in the original setting and after applying AFGR. The reported metrics are as follows:
\textbf{OF}: average objective function value across ten runs,
\textbf{Time}: average solution time in seconds,
\textbf{GAP opt}: average optimality gap, defined as the relative difference between the best feasible solution and the best lower bound at solver termination (MIP gap).
}
\label{tab:classic_B5}
\resizebox{0.9\textwidth}{!}{
\begin{tabular}{c|
ccc|ccc|
ccc|ccc}
\toprule
\multirow{3}{*}{$V$} 
& \multicolumn{6}{c|}{\textbf{ABF}} 
& \multicolumn{6}{c}{\textbf{NBF}} \\
\cmidrule(lr){2-7} \cmidrule(lr){8-13}
& \multicolumn{3}{c|}{\textbf{Without AFGR}} 
& \multicolumn{3}{c|}{\textbf{With AFGR}} 
& \multicolumn{3}{c|}{\textbf{Without AFGR}} 
& \multicolumn{3}{c}{\textbf{With AFGR}} \\
\cmidrule(lr){2-4} \cmidrule(lr){5-7} \cmidrule(lr){8-10} \cmidrule(lr){11-13}
& OF & Time & GAP opt 
& OF & Time & GAP opt 
& OF & Time & GAP opt 
& OF & Time & GAP opt \\
\midrule
   4     & 965,69 & 0,00  & 0\%   & 965,69 & 0,00  & 0\%   & 965,69 & 0,02  & 0\%   & 965,69 & 0,02  & 0\% \\
    5     & 995,96 & 0,02  & 0\%   & 995,96 & 0,02  & 0\%   & 995,96 & 0,05  & 0\%   & 995,96 & 0,00  & 0\% \\
    6     & 1019,62 & 0,08  & 0\%   & 1019,62 & 0,02  & 0\%   & 1019,62 & 0,03  & 0\%   & 1019,62 & 0,02  & 0\% \\
    7     & 945,85 & 0,08  & 0\%   & 945,85 & 0,02  & 0\%   & 945,85 & 0,06  & 0\%   & 945,85 & 0,00  & 0\% \\
    8     & 664,05 & 0,11  & 0\%   & 664,05 & 0,03  & 0\%   & 664,05 & 0,13  & 0\%   & 664,05 & 0,02  & 0\% \\
    9     & 832,49 & 0,44  & 0\%   & 832,49 & 0,08  & 0\%   & 832,49 & 0,39  & 0\%   & 832,49 & 0,08  & 0\% \\
    10    & 905,93 & 1,09  & 0\%   & 905,93 & 0,16  & 0\%   & 905,93 & 1,47  & 0\%   & 905,93 & 0,17  & 0\% \\
    11    & 1072,45 & 15,92 & 0\%   & 1072,45 & 0,41  & 0\%   & 1072,45 & 14,55 & 0\%   & 1072,45 & 0,42  & 0\% \\
    12    & 843,12 & 0,64  & 0\%   & 843,12 & 0,08  & 0\%   & 843,12 & 1,09  & 0\%   & 843,12 & 0,11  & 0\% \\
    13    & 643,70 & 1,33  & 0\%   & 643,70 & 0,08  & 0\%   & 643,70 & 0,53  & 0\%   & 643,70 & 0,06  & 0\% \\
    14    & 795,95 & 2,14  & 0\%   & 795,95 & 0,17  & 0\%   & 795,95 & 2,14  & 0\%   & 795,95 & 0,22  & 0\% \\
    15    & 794,11 & 2,69  & 0\%   & 794,11 & 0,28  & 0\%   & 794,11 & 6,11  & 0\%   & 794,11 & 0,22  & 0\% \\
    16    & 808,78 & 7,73  & 0\%   & 808,78 & 0,30  & 0\%   & 808,78 & 9,91  & 0\%   & 808,78 & 0,27  & 0\% \\
    17    & 790,10 & 16,16 & 0\%   & 790,10 & 0,25  & 0\%   & 790,10 & 13,59 & 0\%   & 790,10 & 0,47  & 0\% \\
    18    & 785,74 & 24,06 & 0\%   & 785,74 & 0,25  & 0\%   & 785,74 & 24,13 & 0\%   & 785,74 & 0,25  & 0\% \\
    19    & 797,18 & 26,16 & 0\%   & 797,18 & 0,64  & 0\%   & 797,18 & 32,77 & 0\%   & 797,18 & 0,30  & 0\% \\
    20    & 790,54 & 24,89 & 0\%   & 790,54 & 0,44  & 0\%   & 790,54 & 29,75 & 0\%   & 790,54 & 0,55  & 0\% \\
   
\bottomrule
\end{tabular}}
\end{table}
\clearpage
\subsection{STSP-PD}\label{app:classic_table_STSP_PD}
\begin{table}[H]
\centering
\caption{ABF and NBF formulations of the STSP-PD model, with and without the AFGR module, for instance class A1. Each row corresponds to an instance with $V$ customers. For each formulation (ABF and NBF), results are reported both in the original setting and after applying AFGR. The reported metrics are as follows:
\textbf{OF}: average objective function value across ten runs,
\textbf{Time}: average solution time in seconds,
\textbf{GAP opt}: average optimality gap, defined as the relative difference between the best feasible solution and the best lower bound at solver termination (MIP gap).
}
\label{tab:classic_PD_A1}
\resizebox{0.9\textwidth}{!}{
\begin{tabular}{c|
ccc|ccc|
ccc|ccc}
\toprule
\multirow{3}{*}{$V$} 
& \multicolumn{6}{c|}{\textbf{ABF}} 
& \multicolumn{6}{c}{\textbf{NBF}} \\
\cmidrule(lr){2-7} \cmidrule(lr){8-13}
& \multicolumn{3}{c|}{\textbf{Without AFGR}} 
& \multicolumn{3}{c|}{\textbf{With AFGR}} 
& \multicolumn{3}{c|}{\textbf{Without AFGR}} 
& \multicolumn{3}{c}{\textbf{With AFGR}} \\
\cmidrule(lr){2-4} \cmidrule(lr){5-7} \cmidrule(lr){8-10} \cmidrule(lr){11-13}
& OF & Time & GAP opt 
& OF & Time & GAP opt 
& OF & Time & GAP opt 
& OF & Time & GAP opt \\
\midrule
4     & 965,69 & 0,02  & 0\%     & 965,69 & 0,00  & 0\%     & 965,69 & 0,02  & 0\%     & 965,69 & 0,02  & 0\% \\
    5     & 995,96 & 0,02  & 0\%     & 995,96 & 0,00  & 0\%     & 995,96 & 0,08  & 0\%     & 995,96 & 0,02  & 0\% \\
    6     & 1019,62 & 0,09  & 0\%     & 1019,62 & 0,08  & 0\%     & 1019,62 & 0,11  & 0\%     & 1019,62 & 0,11  & 0\% \\
    7     & 1011,19 & 0,09  & 0\%     & 1011,19 & 0,11  & 0\%     & 1011,19 & 0,06  & 0\%     & 1011,19 & 0,03  & 0\% \\
    8     & 991,72 & 0,20  & 0\%     & 991,72 & 0,20  & 0\%     & 991,72 & 0,22  & 0\%     & 991,72 & 0,20  & 0\% \\
    9     & 969,30 & 1,27  & 0\%     & 969,30 & 0,55  & 0\%     & 969,30 & 2,06  & 0\%     & 969,30 & 0,70  & 0\% \\
    10    & 929,54 & 0,41  & 0\%     & 929,54 & 0,34  & 0\%     & 929,54 & 0,38  & 0\%     & 929,54 & 0,22  & 0\% \\
    11    & 908,42 & 0,66  & 0\%     & 908,42 & 0,34  & 0\%     & 908,42 & 0,56  & 0\%     & 908,42 & 0,31  & 0\% \\
    12    & 893,43 & 0,94  & 0\%     & 893,43 & 0,36  & 0\%     & 893,43 & 0,92  & 0\%     & 893,43 & 0,34  & 0\% \\
    13    & 887,31 & 23,81 & 0\%     & 887,31 & 3,09  & 0\%     & 887,31 & 5,86  & 0\%     & 887,31 & 2,55  & 0\% \\
    14    & 865,76 & 1,53  & 0\%     & 865,76 & 0,70  & 0\%     & 865,76 & 4,03  & 0\%     & 865,76 & 0,56  & 0\% \\
    15    & 853,40 & 3,95  & 0\%     & 853,40 & 0,94  & 0\%     & 853,40 & 7,08  & 0\%     & 853,40 & 0,84  & 0\% \\
    16    & 840,52 & 6,44  & 0\%     & 840,52 & 1,13  & 0\%     & 840,52 & 8,05  & 0\%     & 840,52 & 0,78  & 0\% \\
    17    & 832,78 & 8,27  & 0\%     & 832,78 & 1,20  & 0\%     & 832,78 & 9,14  & 0\%     & 832,78 & 3,97  & 0\% \\
    18    & 822,99 & 13,91 & 0\%     & 822,99 & 2,33  & 0\%     & 822,99 & 24,92 & 0\%     & 822,99 & 6,69  & 0\% \\
    19    & 814,92 & 75,19 & 0\%     & 814,92 & 33,36 & 0\%     & 814,92 & 102,80 & 0\%     & 814,92 & 23,38 & 0\% \\
    20    & 805,77 & 27,89 & 0\%     & 805,77 & 7,98  & 0\%     & 805,77 & 42,20 & 0\%     & 805,77 & 7,27  & 0\% \\
  
\bottomrule
\end{tabular}}
\end{table}

\begin{table}[H]
\centering
\caption{ABF and NBF formulations of the STSP-PD model, with and without the AFGR module, for instance class A2. Each row corresponds to an instance with $V$ customers. For each formulation (ABF and NBF), results are reported both in the original setting and after applying AFGR. The reported metrics are as follows:
\textbf{OF}: average objective function value across ten runs,
\textbf{Time}: average solution time in seconds,
\textbf{GAP opt}: average optimality gap, defined as the relative difference between the best feasible solution and the best lower bound at solver termination (MIP gap).
}
\label{tab:classic_PD_A2}
\resizebox{0.9\textwidth}{!}{
\begin{tabular}{c|
ccc|ccc|
ccc|ccc}
\toprule
\multirow{3}{*}{$V$} 
& \multicolumn{6}{c|}{\textbf{ABF}} 
& \multicolumn{6}{c}{\textbf{NBF}} \\
\cmidrule(lr){2-7} \cmidrule(lr){8-13}
& \multicolumn{3}{c|}{\textbf{Without AFGR}} 
& \multicolumn{3}{c|}{\textbf{With AFGR}} 
& \multicolumn{3}{c|}{\textbf{Without AFGR}} 
& \multicolumn{3}{c}{\textbf{With AFGR}} \\
\cmidrule(lr){2-4} \cmidrule(lr){5-7} \cmidrule(lr){8-10} \cmidrule(lr){11-13}
& OF & Time & GAP opt 
& OF & Time & GAP opt 
& OF & Time & GAP opt 
& OF & Time & GAP opt \\
\midrule
4     & 965,69 & 0,02  & 0\%     & 965,69 & 0,00  & 0\%     & 965,69 & 0,00  & 0\%     & 965,69 & 0,02  & 0\% \\
    5     & 995,96 & 0,06  & 0\%     & 995,96 & 0,03  & 0\%     & 995,96 & 0,02  & 0\%     & 995,96 & 0,02  & 0\% \\
    6     & 1019,62 & 0,08  & 0\%     & 1019,62 & 0,08  & 0\%     & 1019,62 & 0,08  & 0\%     & 1019,62 & 0,09  & 0\% \\
    7     & 1011,19 & 0,09  & 0\%     & 1011,19 & 0,13  & 0\%     & 1011,19 & 0,05  & 0\%     & 1011,19 & 0,06  & 0\% \\
    8     & 980,07 & 0,20  & 0\%     & 980,07 & 0,13  & 0\%     & 980,07 & 0,20  & 0\%     & 980,07 & 0,09  & 0\% \\
    9     & 937,29 & 0,14  & 0\%     & 937,29 & 0,13  & 0\%     & 937,29 & 0,19  & 0\%     & 937,29 & 0,13  & 0\% \\
    10    & 941,05 & 1,27  & 0\%     & 941,05 & 0,45  & 0\%     & 941,05 & 1,20  & 0\%     & 941,05 & 0,78  & 0\% \\
    11    & 917,18 & 3,72  & 0\%     & 917,18 & 0,47  & 0\%     & 917,18 & 4,72  & 0\%     & 917,18 & 0,47  & 0\% \\
    12    & 889,90 & 0,92  & 0\%     & 889,90 & 0,33  & 0\%     & 889,90 & 1,39  & 0\%     & 889,90 & 0,30  & 0\% \\
    13    & 876,34 & 1,13  & 0\%     & 876,34 & 0,22  & 0\%     & 876,34 & 1,75  & 0\%     & 876,34 & 0,53  & 0\% \\
    14    & 870,11 & 36,44 & 0\%     & 870,11 & 2,47  & 0\%     & 870,11 & 18,22 & 0\%     & 870,11 & 2,77  & 0\% \\
    15    & 855,14 & 110,23 & 0\%     & 855,14 & 3,97  & 0\%     & 855,14 & 59,14 & 0\%     & 855,14 & 1,95  & 0\% \\
    16    & 839,02 & 8,98  & 0\%     & 839,02 & 0,52  & 0\%     & 839,02 & 10,69 & 0\%     & 839,02 & 0,56  & 0\% \\
    17    & 831,52 & 11,38 & 0\%     & 831,52 & 1,30  & 0\%     & 831,52 & 10,17 & 0\%     & 831,52 & 0,95  & 0\% \\
    18    & 1021,94 & 593,16 & 0\%     & 1021,94 & 49,23 & 0\%     & 1021,94 & 270,91 & 0\%     & 1021,94 & 64,19 & 0\% \\
    19    & 812,25 & 49,33 & 0\%     & 812,25 & 23,38 & 0\%     & 812,25 & 137,78 & 0\%     & 812,25 & 7,86  & 0\% \\
    20    & 805,00 & 30,30 & 0\%     & 805,00 & 2,66  & 0\%     & 805,00 & 54,25 & 0\%     & 805,00 & 4,45  & 0\% \\
\bottomrule
\end{tabular}}
\end{table}

\begin{table}[H]
\centering
\caption{ABF and NBF formulations of the STSP-PD model, with and without the AFGR module, for instance class A3. Each row corresponds to an instance with $V$ customers. For each formulation (ABF and NBF), results are reported both in the original setting and after applying AFGR. The reported metrics are as follows:
\textbf{OF}: average objective function value across ten runs,
\textbf{Time}: average solution time in seconds,
\textbf{GAP opt}: average optimality gap, defined as the relative difference between the best feasible solution and the best lower bound at solver termination (MIP gap).
}
\label{tab:classic_PD_A3}
\resizebox{0.9\textwidth}{!}{
\begin{tabular}{c|
ccc|ccc|
ccc|ccc}
\toprule
\multirow{3}{*}{$V$} 
& \multicolumn{6}{c|}{\textbf{ABF}} 
& \multicolumn{6}{c}{\textbf{NBF}} \\
\cmidrule(lr){2-7} \cmidrule(lr){8-13}
& \multicolumn{3}{c|}{\textbf{Without AFGR}} 
& \multicolumn{3}{c|}{\textbf{With AFGR}} 
& \multicolumn{3}{c|}{\textbf{Without AFGR}} 
& \multicolumn{3}{c}{\textbf{With AFGR}} \\
\cmidrule(lr){2-4} \cmidrule(lr){5-7} \cmidrule(lr){8-10} \cmidrule(lr){11-13}
& OF & Time & GAP opt 
& OF & Time & GAP opt 
& OF & Time & GAP opt 
& OF & Time & GAP opt \\
\midrule
    4     & 965,69 & 0,02  & 0\%     & 965,69 & 0,00  & 0\%     & 965,69 & 0,02  & 0\%     & 965,69 & 0,05  & 0\% \\
    5     & 995,96 & 0,02  & 0\%     & 995,96 & 0,00  & 0\%     & 995,96 & 0,02  & 0\%     & 995,96 & 0,02  & 0\% \\
    6     & 1019,62 & 0,05  & 0\%     & 1019,62 & 0,02  & 0\%     & 1019,62 & 0,08  & 0\%     & 1019,62 & 0,02  & 0\% \\
    7     & 994,00 & 0,03  & 0\%     & 994,00 & 0,06  & 0\%     & 994,00 & 0,06  & 0\%     & 994,00 & 0,08  & 0\% \\
    8     & 980,07 & 0,20  & 0\%     & 980,07 & 0,09  & 0\%     & 980,07 & 0,20  & 0\%     & 980,07 & 0,17  & 0\% \\
    9     & 900,90 & 0,91  & 0\%     & 900,90 & 0,20  & 0\%     & 900,90 & 1,11  & 0\%     & 900,90 & 0,17  & 0\% \\
    10    & 890,10 & 0,56  & 0\%     & 890,10 & 0,13  & 0\%     & 890,10 & 0,84  & 0\%     & 890,10 & 0,17  & 0\% \\
    11    & 869,52 & 0,89  & 0\%     & 869,52 & 0,13  & 0\%     & 869,52 & 0,63  & 0\%     & 869,52 & 0,13  & 0\% \\
    12    & 889,90 & 0,69  & 0\%     & 889,90 & 0,39  & 0\%     & 889,90 & 1,08  & 0\%     & 889,90 & 0,33  & 0\% \\
    13    & 876,34 & 1,39  & 0\%     & 876,34 & 0,38  & 0\%     & 876,34 & 1,91  & 0\%     & 876,34 & 0,45  & 0\% \\
    14    & 846,34 & 1,45  & 0\%     & 846,34 & 0,45  & 0\%     & 846,34 & 2,77  & 0\%     & 846,34 & 0,50  & 0\% \\
    15    & 1012,46 & 51,91 & 0\%     & 1012,46 & 16,00 & 0\%     & 1012,46 & 54,97 & 0\%     & 1012,46 & 9,28  & 0\% \\
    16    & 980,44 & 246,59 & 0\%     & 980,44 & 18,69 & 0\%     & 980,44 & 156,16 & 0\%     & 980,44 & 21,98 & 0\% \\
    17    & 823,02 & 26,44 & 0\%     & 823,02 & 5,16  & 0\%     & 823,02 & 51,27 & 0\%     & 823,02 & 2,63  & 0\% \\
    18    & 819,86 & 34,69 & 0\%     & 819,86 & 5,55  & 0\%     & 819,86 & 92,66 & 0\%     & 819,86 & 3,97  & 0\% \\
    19    & 810,45 & 50,83 & 0\%     & 810,45 & 4,66  & 0\%     & 810,45 & 135,42 & 0\%     & 810,45 & 4,47  & 0\% \\
    20    & 982,02 & 474,73 & 0\%     & 982,02 & 47,00 & 0\%     & 982,02 & 829,30 & 0\%     & 982,02 & 53,81 & 0\% \\
   
\bottomrule
\end{tabular}}
\end{table}

\begin{table}[H]
\centering
\caption{ABF and NBF formulations of the STSP-PD model, with and without the AFGR module, for instance class A4. Each row corresponds to an instance with $V$ customers. For each formulation (ABF and NBF), results are reported both in the original setting and after applying AFGR. The reported metrics are as follows:
\textbf{OF}: average objective function value across ten runs,
\textbf{Time}: average solution time in seconds,
\textbf{GAP opt}: average optimality gap, defined as the relative difference between the best feasible solution and the best lower bound at solver termination (MIP gap).
}
\label{tab:classic_PD_A4}
\resizebox{0.9\textwidth}{!}{
\begin{tabular}{c|
ccc|ccc|
ccc|ccc}
\toprule
\multirow{3}{*}{$V$} 
& \multicolumn{6}{c|}{\textbf{ABF}} 
& \multicolumn{6}{c}{\textbf{NBF}} \\
\cmidrule(lr){2-7} \cmidrule(lr){8-13}
& \multicolumn{3}{c|}{\textbf{Without AFGR}} 
& \multicolumn{3}{c|}{\textbf{With AFGR}} 
& \multicolumn{3}{c|}{\textbf{Without AFGR}} 
& \multicolumn{3}{c}{\textbf{With AFGR}} \\
\cmidrule(lr){2-4} \cmidrule(lr){5-7} \cmidrule(lr){8-10} \cmidrule(lr){11-13}
& OF & Time & GAP opt 
& OF & Time & GAP opt 
& OF & Time & GAP opt 
& OF & Time & GAP opt \\
\midrule
    4     & 965,69 & 0,02  & 0\%     & 965,69 & 0,02  & 0\%     & 965,69 & 0,02  & 0\%     & 965,69 & 0,00  & 0\% \\
    5     & 995,96 & 0,08  & 0\%     & 995,96 & 0,00  & 0\%     & 995,96 & 0,08  & 0\%     & 995,96 & 0,00  & 0\% \\
    6     & 1019,62 & 0,08  & 0\%     & 1019,62 & 0,02  & 0\%     & 1019,62 & 0,06  & 0\%     & 1019,62 & 0,02  & 0\% \\
    7     & 945,85 & 0,06  & 0\%     & 945,85 & 0,02  & 0\%     & 945,85 & 0,13  & 0\%     & 945,85 & 0,05  & 0\% \\
    8     & 946,89 & 0,14  & 0\%     & 946,89 & 0,05  & 0\%     & 946,89 & 0,09  & 0\%     & 946,89 & 0,11  & 0\% \\
    9     & 832,49 & 0,33  & 0\%     & 832,49 & 0,05  & 0\%     & 832,49 & 0,31  & 0\%     & 832,49 & 0,05  & 0\% \\
    10    & 905,93 & 5,73  & 0\%     & 905,93 & 0,17  & 0\%     & 905,93 & 2,44  & 0\%     & 905,93 & 0,23  & 0\% \\
    11    & 869,52 & 0,59  & 0\%     & 869,52 & 0,09  & 0\%     & 869,52 & 0,50  & 0\%     & 869,52 & 0,14  & 0\% \\
    12    & 869,92 & 0,75  & 0\%     & 869,92 & 0,14  & 0\%     & 869,92 & 1,41  & 0\%     & 869,92 & 0,16  & 0\% \\
    13    & 829,59 & 0,89  & 0\%     & 829,59 & 0,16  & 0\%     & 829,59 & 2,02  & 0\%     & 829,59 & 0,23  & 0\% \\
    14    & 825,66 & 1,41  & 0\%     & 825,66 & 0,34  & 0\%     & 825,66 & 1,83  & 0\%     & 825,66 & 0,33  & 0\% \\
    15    & 818,69 & 4,48  & 0\%     & 818,69 & 0,31  & 0\%     & 818,69 & 5,92  & 0\%     & 818,69 & 0,33  & 0\% \\
    16    & 813,22 & 6,03  & 0\%     & 813,22 & 0,25  & 0\%     & 813,22 & 9,16  & 0\%     & 813,22 & 0,50  & 0\% \\
    17    & 793,81 & 9,38  & 0\%     & 793,81 & 0,36  & 0\%     & 793,81 & 9,81  & 0\%     & 793,81 & 0,48  & 0\% \\
    18    & 935,20 & 124,52 & 0\%     & 935,20 & 10,08 & 0\%     & 935,20 & 95,06 & 0\%     & 935,20 & 15,17 & 0\% \\
    19    & 932,32 & 261,16 & 0\%     & 932,32 & 12,73 & 0\%     & 932,32 & 209,89 & 0\%     & 932,32 & 9,84  & 0\% \\
    20    & 798,13 & 59,59 & 0\%     & 798,13 & 3,94  & 0\%     & 798,13 & 145,52 & 0\%     & 798,13 & 4,14  & 0\% \\

\bottomrule
\end{tabular}}
\end{table}

\begin{table}[H]
\centering
\caption{ABF and NBF formulations of the STSP-PD model, with and without the AFGR module, for instance class A5. Each row corresponds to an instance with $V$ customers. For each formulation (ABF and NBF), results are reported both in the original setting and after applying AFGR. The reported metrics are as follows:
\textbf{OF}: average objective function value across ten runs,
\textbf{Time}: average solution time in seconds,
\textbf{GAP opt}: average optimality gap, defined as the relative difference between the best feasible solution and the best lower bound at solver termination (MIP gap).
}
\label{tab:classic_PD_A5}
\resizebox{0.9\textwidth}{!}{
\begin{tabular}{c|
ccc|ccc|
ccc|ccc}
\toprule
\multirow{3}{*}{$V$} 
& \multicolumn{6}{c|}{\textbf{ABF}} 
& \multicolumn{6}{c}{\textbf{NBF}} \\
\cmidrule(lr){2-7} \cmidrule(lr){8-13}
& \multicolumn{3}{c|}{\textbf{Without AFGR}} 
& \multicolumn{3}{c|}{\textbf{With AFGR}} 
& \multicolumn{3}{c|}{\textbf{Without AFGR}} 
& \multicolumn{3}{c}{\textbf{With AFGR}} \\
\cmidrule(lr){2-4} \cmidrule(lr){5-7} \cmidrule(lr){8-10} \cmidrule(lr){11-13}
& OF & Time & GAP opt 
& OF & Time & GAP opt 
& OF & Time & GAP opt 
& OF & Time & GAP opt \\
\midrule
    4     & 965,69 & 0,00  & 0\%     & 965,69 & 0,02  & 0\%     & 965,69 & 0,02  & 0\%     & 965,69 & 0,00  & 0\% \\
    5     & 995,96 & 0,03  & 0\%     & 995,96 & 0,00  & 0\%     & 995,96 & 0,06  & 0\%     & 995,96 & 0,00  & 0\% \\
    6     & 1019,62 & 0,05  & 0\%     & 1019,62 & 0,02  & 0\%     & 1019,62 & 0,08  & 0\%     & 1019,62 & 0,02  & 0\% \\
    7     & 945,85 & 0,03  & 0\%     & 945,85 & 0,02  & 0\%     & 945,85 & 0,16  & 0\%     & 945,85 & 0,06  & 0\% \\
    8     & 664,05 & 0,05  & 0\%     & 664,05 & 0,05  & 0\%     & 664,05 & 0,09  & 0\%     & 664,05 & 0,02  & 0\% \\
    9     & 832,49 & 0,39  & 0\%     & 832,49 & 0,05  & 0\%     & 832,49 & 0,28  & 0\%     & 832,49 & 0,05  & 0\% \\
    10    & 905,93 & 2,45  & 0\%     & 905,93 & 0,23  & 0\%     & 905,93 & 0,86  & 0\%     & 905,93 & 0,19  & 0\% \\
    11    & 1072,45 & 10,59 & 0\%     & 1072,45 & 0,47  & 0\%     & 1072,45 & 14,47 & 0\%     & 1072,45 & 0,34  & 0\% \\
    12    & 843,12 & 0,89  & 0\%     & 843,12 & 0,09  & 0\%     & 843,12 & 1,05  & 0\%     & 843,12 & 0,13  & 0\% \\
    13    & 643,70 & 0,80  & 0\%     & 643,70 & 0,06  & 0\%     & 643,70 & 0,52  & 0\%     & 643,70 & 0,08  & 0\% \\
    14    & 795,95 & 2,20  & 0\%     & 795,95 & 0,17  & 0\%     & 795,95 & 1,70  & 0\%     & 795,95 & 0,22  & 0\% \\
    15    & 794,11 & 3,64  & 0\%     & 794,11 & 0,16  & 0\%     & 794,11 & 5,19  & 0\%     & 794,11 & 0,13  & 0\% \\
    16    & 808,78 & 6,14  & 0\%     & 808,78 & 0,16  & 0\%     & 808,78 & 7,33  & 0\%     & 808,78 & 0,25  & 0\% \\
    17    & 790,10 & 11,50 & 0\%     & 790,10 & 0,16  & 0\%     & 790,10 & 11,98 & 0\%     & 790,10 & 0,42  & 0\% \\
    18    & 785,74 & 14,42 & 0\%     & 785,74 & 0,28  & 0\%     & 785,74 & 15,81 & 0\%     & 785,74 & 0,20  & 0\% \\
    19    & 797,18 & 21,45 & 0\%     & 797,18 & 0,33  & 0\%     & 797,18 & 18,98 & 0\%     & 797,18 & 0,34  & 0\% \\
    20    & 790,54 & 20,14 & 0\%     & 790,54 & 0,50  & 0\%     & 790,54 & 30,48 & 0\%     & 790,54 & 0,50  & 0\% \\

\bottomrule
\end{tabular}}
\end{table}

\begin{table}[H]
\centering
\caption{ABF and NBF formulations of the STSP-PD model, with and without the AFGR module, for instance class B1. Each row corresponds to an instance with $V$ customers. For each formulation (ABF and NBF), results are reported both in the original setting and after applying AFGR. The reported metrics are as follows:
\textbf{OF}: average objective function value across ten runs,
\textbf{Time}: average solution time in seconds,
\textbf{GAP opt}: average optimality gap, defined as the relative difference between the best feasible solution and the best lower bound at solver termination (MIP gap).
}
\label{tab:classic_PD_B1}
\resizebox{0.9\textwidth}{!}{
\begin{tabular}{c|
ccc|ccc|
ccc|ccc}
\toprule
\multirow{3}{*}{$V$} 
& \multicolumn{6}{c|}{\textbf{ABF}} 
& \multicolumn{6}{c}{\textbf{NBF}} \\
\cmidrule(lr){2-7} \cmidrule(lr){8-13}
& \multicolumn{3}{c|}{\textbf{Without AFGR}} 
& \multicolumn{3}{c|}{\textbf{With AFGR}} 
& \multicolumn{3}{c|}{\textbf{Without AFGR}} 
& \multicolumn{3}{c}{\textbf{With AFGR}} \\
\cmidrule(lr){2-4} \cmidrule(lr){5-7} \cmidrule(lr){8-10} \cmidrule(lr){11-13}
& OF & Time & GAP opt 
& OF & Time & GAP opt 
& OF & Time & GAP opt 
& OF & Time & GAP opt \\
\midrule
    4     & 965,69 & 0,02  & 0\%     & 965,69 & 0,02  & 0\%     & 965,69 & 0,02  & 0\%     & 965,69 & 0,02  & 0\% \\
    5     & 995,96 & 0,02  & 0\%     & 995,96 & 0,02  & 0\%     & 995,96 & 0,02  & 0\%     & 995,96 & 0,03  & 0\% \\
    6     & 1019,62 & 0,08  & 0\%     & 1019,62 & 0,11  & 0\%     & 1019,62 & 0,08  & 0\%     & 1019,62 & 0,16  & 0\% \\
    7     & 1011,19 & 0,13  & 0\%     & 1011,19 & 0,16  & 0\%     & 1011,19 & 0,03  & 0\%     & 1011,19 & 0,05  & 0\% \\
    8     & 991,72 & 0,20  & 0\%     & 991,72 & 0,20  & 0\%     & 991,72 & 0,22  & 0\%     & 991,72 & 0,22  & 0\% \\
    9     & 969,30 & 1,41  & 0\%     & 969,30 & 0,53  & 0\%     & 969,30 & 1,98  & 0\%     & 969,30 & 0,72  & 0\% \\
    10    & 929,54 & 0,41  & 0\%     & 929,54 & 0,41  & 0\%     & 929,54 & 0,34  & 0\%     & 929,54 & 0,22  & 0\% \\
    11    & 908,42 & 0,58  & 0\%     & 908,42 & 0,34  & 0\%     & 908,42 & 0,56  & 0\%     & 908,42 & 0,31  & 0\% \\
    12    & 893,43 & 0,92  & 0\%     & 893,43 & 0,38  & 0\%     & 893,43 & 0,95  & 0\%     & 893,43 & 0,39  & 0\% \\
    13    & 887,31 & 23,78 & 0\%     & 887,31 & 3,13  & 0\%     & 887,31 & 6,09  & 0\%     & 887,31 & 2,61  & 0\% \\
    14    & 865,76 & 1,70  & 0\%     & 865,76 & 0,58  & 0\%     & 865,76 & 4,05  & 0\%     & 865,76 & 0,53  & 0\% \\
    15    & 853,40 & 4,02  & 0\%     & 853,40 & 0,89  & 0\%     & 853,40 & 7,17  & 0\%     & 853,40 & 0,83  & 0\% \\
    16    & 840,52 & 6,86  & 0\%     & 840,52 & 1,22  & 0\%     & 840,52 & 7,98  & 0\%     & 840,52 & 0,80  & 0\% \\
    17    & 832,78 & 8,31  & 0\%     & 832,78 & 1,20  & 0\%     & 832,78 & 9,06  & 0\%     & 832,78 & 3,89  & 0\% \\
    18    & 822,99 & 14,25 & 0\%     & 822,99 & 2,38  & 0\%     & 822,99 & 23,83 & 0\%     & 822,99 & 5,72  & 0\% \\
    19    & 814,92 & 75,55 & 0\%     & 814,92 & 33,53 & 0\%     & 814,92 & 100,88 & 0\%     & 814,92 & 22,80 & 0\% \\
    20    & 805,77 & 28,00 & 0\%     & 805,77 & 8,06  & 0\%     & 805,77 & 41,36 & 0\%     & 805,77 & 7,28  & 0\% \\

\bottomrule
\end{tabular}}
\end{table}

\begin{table}[H]
\centering
\caption{ABF and NBF formulations of the STSP-PD model, with and without the AFGR module, for instance class B2. Each row corresponds to an instance with $V$ customers. For each formulation (ABF and NBF), results are reported both in the original setting and after applying AFGR. The reported metrics are as follows:
\textbf{OF}: average objective function value across ten runs,
\textbf{Time}: average solution time in seconds,
\textbf{GAP opt}: average optimality gap, defined as the relative difference between the best feasible solution and the best lower bound at solver termination (MIP gap).
}
\label{tab:classic_PD_B2}
\resizebox{0.9\textwidth}{!}{
\begin{tabular}{c|
ccc|ccc|
ccc|ccc}
\toprule
\multirow{3}{*}{$V$} 
& \multicolumn{6}{c|}{\textbf{ABF}} 
& \multicolumn{6}{c}{\textbf{NBF}} \\
\cmidrule(lr){2-7} \cmidrule(lr){8-13}
& \multicolumn{3}{c|}{\textbf{Without AFGR}} 
& \multicolumn{3}{c|}{\textbf{With AFGR}} 
& \multicolumn{3}{c|}{\textbf{Without AFGR}} 
& \multicolumn{3}{c}{\textbf{With AFGR}} \\
\cmidrule(lr){2-4} \cmidrule(lr){5-7} \cmidrule(lr){8-10} \cmidrule(lr){11-13}
& OF & Time & GAP opt 
& OF & Time & GAP opt 
& OF & Time & GAP opt 
& OF & Time & GAP opt \\
\midrule
    4     & 965,69 & 0,00  & 0\%     & 965,69 & 0,02  & 0\%     & 965,69 & 0,00  & 0\%     & 965,69 & 0,02  & 0\% \\
    5     & 995,96 & 0,02  & 0\%     & 995,96 & 0,02  & 0\%     & 995,96 & 0,00  & 0\%     & 995,96 & 0,00  & 0\% \\
    6     & 1019,62 & 0,08  & 0\%     & 1019,62 & 0,11  & 0\%     & 1019,62 & 0,09  & 0\%     & 1019,62 & 0,11  & 0\% \\
    7     & 1011,19 & 0,13  & 0\%     & 1011,19 & 0,13  & 0\%     & 1011,19 & 0,06  & 0\%     & 1011,19 & 0,06  & 0\% \\
    8     & 980,07 & 0,20  & 0\%     & 980,07 & 0,11  & 0\%     & 980,07 & 0,22  & 0\%     & 980,07 & 0,13  & 0\% \\
    9     & 937,29 & 0,17  & 0\%     & 937,29 & 0,06  & 0\%     & 937,29 & 0,20  & 0\%     & 937,29 & 0,17  & 0\% \\
    10    & 941,05 & 1,33  & 0\%     & 941,05 & 0,47  & 0\%     & 941,05 & 1,19  & 0\%     & 941,05 & 0,78  & 0\% \\
    11    & 917,18 & 3,61  & 0\%     & 917,18 & 0,47  & 0\%     & 917,18 & 4,88  & 0\%     & 917,18 & 0,45  & 0\% \\
    12    & 889,90 & 0,94  & 0\%     & 889,90 & 0,28  & 0\%     & 889,90 & 1,36  & 0\%     & 889,90 & 0,33  & 0\% \\
    13    & 876,34 & 1,11  & 0\%     & 876,34 & 0,20  & 0\%     & 876,34 & 1,75  & 0\%     & 876,34 & 0,50  & 0\% \\
    14    & 870,11 & 36,38 & 0\%     & 870,11 & 2,63  & 0\%     & 870,11 & 18,66 & 0\%     & 870,11 & 2,84  & 0\% \\
    15    & 855,14 & 111,27 & 0\%     & 855,14 & 4,13  & 0\%     & 855,14 & 59,25 & 0\%     & 855,14 & 1,97  & 0\% \\
    16    & 839,02 & 9,22  & 0\%     & 839,02 & 0,53  & 0\%     & 839,02 & 10,95 & 0\%     & 839,02 & 0,56  & 0\% \\
    17    & 831,52 & 11,45 & 0\%     & 831,52 & 1,34  & 0\%     & 831,52 & 10,06 & 0\%     & 831,52 & 1,03  & 0\% \\
    18    & 1021,94 & 601,95 & 0\%     & 1021,94 & 49,17 & 0\%     & 1021,94 & 268,64 & 0\%     & 1021,94 & 61,61 & 0\% \\
    19    & 812,25 & 49,52 & 0\%     & 812,25 & 23,67 & 0\%     & 812,25 & 136,86 & 0\%     & 812,25 & 7,58  & 0\% \\
    20    & 805,00 & 32,09 & 0\%     & 805,00 & 2,80  & 0\%     & 805,00 & 51,84 & 0\%     & 805,00 & 4,39  & 0\% \\
\bottomrule
\end{tabular}}
\end{table}

\begin{table}[H]
\centering
\caption{ABF and NBF formulations of the STSP-PD model, with and without the AFGR module, for instance class B3. Each row corresponds to an instance with $V$ customers. For each formulation (ABF and NBF), results are reported both in the original setting and after applying AFGR. The reported metrics are as follows:
\textbf{OF}: average objective function value across ten runs,
\textbf{Time}: average solution time in seconds,
\textbf{GAP opt}: average optimality gap, defined as the relative difference between the best feasible solution and the best lower bound at solver termination (MIP gap).
}
\label{tab:classic_PD_B3}
\resizebox{0.9\textwidth}{!}{
\begin{tabular}{c|
ccc|ccc|
ccc|ccc}
\toprule
\multirow{3}{*}{$V$} 
& \multicolumn{6}{c|}{\textbf{ABF}} 
& \multicolumn{6}{c}{\textbf{NBF}} \\
\cmidrule(lr){2-7} \cmidrule(lr){8-13}
& \multicolumn{3}{c|}{\textbf{Without AFGR}} 
& \multicolumn{3}{c|}{\textbf{With AFGR}} 
& \multicolumn{3}{c|}{\textbf{Without AFGR}} 
& \multicolumn{3}{c}{\textbf{With AFGR}} \\
\cmidrule(lr){2-4} \cmidrule(lr){5-7} \cmidrule(lr){8-10} \cmidrule(lr){11-13}
& OF & Time & GAP opt 
& OF & Time & GAP opt 
& OF & Time & GAP opt 
& OF & Time & GAP opt \\
\midrule
4     & 965,69 & 0,03  & 0\%     & 965,69 & 0,02  & 0\%     & 965,69 & 0,00  & 0\%     & 965,69 & 0,02  & 0\% \\
    5     & 995,96 & 0,02  & 0\%     & 995,96 & 0,02  & 0\%     & 995,96 & 0,02  & 0\%     & 995,96 & 0,02  & 0\% \\
    6     & 1019,62 & 0,05  & 0\%     & 1019,62 & 0,02  & 0\%     & 1019,62 & 0,08  & 0\%     & 1019,62 & 0,02  & 0\% \\
    7     & 994,00 & 0,09  & 0\%     & 994,00 & 0,05  & 0\%     & 994,00 & 0,06  & 0\%     & 994,00 & 0,08  & 0\% \\
    8     & 980,07 & 0,20  & 0\%     & 980,07 & 0,13  & 0\%     & 980,07 & 0,20  & 0\%     & 980,07 & 0,13  & 0\% \\
    9     & 900,90 & 0,88  & 0\%     & 900,90 & 0,22  & 0\%     & 900,90 & 1,06  & 0\%     & 900,90 & 0,22  & 0\% \\
    10    & 890,10 & 0,52  & 0\%     & 890,10 & 0,13  & 0\%     & 890,10 & 0,66  & 0\%     & 890,10 & 0,17  & 0\% \\
    11    & 869,52 & 0,89  & 0\%     & 869,52 & 0,13  & 0\%     & 869,52 & 0,66  & 0\%     & 869,52 & 0,13  & 0\% \\
    12    & 889,90 & 0,72  & 0\%     & 889,90 & 0,38  & 0\%     & 889,90 & 1,02  & 0\%     & 889,90 & 0,33  & 0\% \\
    13    & 876,34 & 1,44  & 0\%     & 876,34 & 0,41  & 0\%     & 876,34 & 1,97  & 0\%     & 876,34 & 0,52  & 0\% \\
    14    & 846,34 & 1,42  & 0\%     & 846,34 & 0,50  & 0\%     & 846,34 & 2,47  & 0\%     & 846,34 & 0,52  & 0\% \\
    15    & 1012,46 & 51,81 & 0\%     & 1012,46 & 16,05 & 0\%     & 1012,46 & 55,86 & 0\%     & 1012,46 & 9,41  & 0\% \\
    16    & 980,44 & 248,56 & 0\%     & 980,44 & 18,63 & 0\%     & 980,44 & 155,61 & 0\%     & 980,44 & 21,66 & 0\% \\
    17    & 823,02 & 26,63 & 0\%     & 823,02 & 5,64  & 0\%     & 823,02 & 51,16 & 0\%     & 823,02 & 2,55  & 0\% \\
    18    & 819,86 & 35,09 & 0\%     & 819,86 & 5,59  & 0\%     & 819,86 & 90,22 & 0\%     & 819,86 & 3,84  & 0\% \\
    19    & 810,45 & 50,45 & 0\%     & 810,45 & 4,61  & 0\%     & 810,45 & 131,66 & 0\%     & 810,45 & 4,33  & 0\% \\
    20    & 982,02 & 474,84 & 0\%     & 982,02 & 47,69 & 0\%     & 982,02 & 811,03 & 0\%     & 982,02 & 53,52 & 0\% \\

\bottomrule
\end{tabular}}
\end{table}

\begin{table}[H]
\centering
\caption{ABF and NBF formulations of the STSP-PD model, with and without the AFGR module, for instance class B4. Each row corresponds to an instance with $V$ customers. For each formulation (ABF and NBF), results are reported both in the original setting and after applying AFGR. The reported metrics are as follows:
\textbf{OF}: average objective function value across ten runs,
\textbf{Time}: average solution time in seconds,
\textbf{GAP opt}: average optimality gap, defined as the relative difference between the best feasible solution and the best lower bound at solver termination (MIP gap).
}
\label{tab:classic_PD_B4}
\resizebox{0.9\textwidth}{!}{
\begin{tabular}{c|
ccc|ccc|
ccc|ccc}
\toprule
\multirow{3}{*}{$V$} 
& \multicolumn{6}{c|}{\textbf{ABF}} 
& \multicolumn{6}{c}{\textbf{NBF}} \\
\cmidrule(lr){2-7} \cmidrule(lr){8-13}
& \multicolumn{3}{c|}{\textbf{Without AFGR}} 
& \multicolumn{3}{c|}{\textbf{With AFGR}} 
& \multicolumn{3}{c|}{\textbf{Without AFGR}} 
& \multicolumn{3}{c}{\textbf{With AFGR}} \\
\cmidrule(lr){2-4} \cmidrule(lr){5-7} \cmidrule(lr){8-10} \cmidrule(lr){11-13}
& OF & Time & GAP opt 
& OF & Time & GAP opt 
& OF & Time & GAP opt 
& OF & Time & GAP opt \\
\midrule

    4     & 965,69 & 0,02  & 0\%     & 965,69 & 0,00  & 0\%     & 965,69 & 0,00  & 0\%     & 965,69 & 0,00  & 0\% \\
    5     & 995,96 & 0,08  & 0\%     & 995,96 & 0,02  & 0\%     & 995,96 & 0,08  & 0\%     & 995,96 & 0,02  & 0\% \\
    6     & 1019,62 & 0,08  & 0\%     & 1019,62 & 0,03  & 0\%     & 1019,62 & 0,08  & 0\%     & 1019,62 & 0,06  & 0\% \\
    7     & 945,85 & 0,09  & 0\%     & 945,85 & 0,08  & 0\%     & 945,85 & 0,16  & 0\%     & 945,85 & 0,02  & 0\% \\
    8     & 946,89 & 0,20  & 0\%     & 946,89 & 0,03  & 0\%     & 946,89 & 0,14  & 0\%     & 946,89 & 0,08  & 0\% \\
    9     & 832,49 & 0,38  & 0\%     & 832,49 & 0,08  & 0\%     & 832,49 & 0,38  & 0\%     & 832,49 & 0,09  & 0\% \\
    10    & 905,93 & 5,81  & 0\%     & 905,93 & 0,22  & 0\%     & 905,93 & 2,48  & 0\%     & 905,93 & 0,22  & 0\% \\
    11    & 869,52 & 0,58  & 0\%     & 869,52 & 0,13  & 0\%     & 869,52 & 0,44  & 0\%     & 869,52 & 0,13  & 0\% \\
    12    & 869,92 & 0,81  & 0\%     & 869,92 & 0,14  & 0\%     & 869,92 & 1,33  & 0\%     & 869,92 & 0,16  & 0\% \\
    13    & 829,59 & 0,91  & 0\%     & 829,59 & 0,14  & 0\%     & 829,59 & 1,94  & 0\%     & 829,59 & 0,22  & 0\% \\
    14    & 825,66 & 1,42  & 0\%     & 825,66 & 0,38  & 0\%     & 825,66 & 1,73  & 0\%     & 825,66 & 0,27  & 0\% \\
    15    & 818,69 & 4,36  & 0\%     & 818,69 & 0,31  & 0\%     & 818,69 & 5,69  & 0\%     & 818,69 & 0,27  & 0\% \\
    16    & 813,22 & 6,13  & 0\%     & 813,22 & 0,31  & 0\%     & 813,22 & 9,06  & 0\%     & 813,22 & 0,47  & 0\% \\
    17    & 793,81 & 9,20  & 0\%     & 793,81 & 0,36  & 0\%     & 793,81 & 9,52  & 0\%     & 793,81 & 0,47  & 0\% \\
    18    & 935,20 & 120,11 & 0\%     & 935,20 & 9,95  & 0\%     & 935,20 & 89,70 & 0\%     & 935,20 & 14,78 & 0\% \\
    19    & 932,32 & 259,38 & 0\%     & 932,32 & 12,61 & 0\%     & 932,32 & 213,48 & 0\%     & 932,32 & 10,25 & 0\% \\
    20    & 798,13 & 59,16 & 0\%     & 798,13 & 3,91  & 0\%     & 798,13 & 151,86 & 0\%     & 798,13 & 3,97  & 0\% \\

\bottomrule
\end{tabular}}
\end{table}

\begin{table}[H]
\centering
\caption{ABF and NBF formulations of the STSP-PD model, with and without the AFGR module, for instance class B5. Each row corresponds to an instance with $V$ customers. For each formulation (ABF and NBF), results are reported both in the original setting and after applying AFGR. The reported metrics are as follows:
\textbf{OF}: average objective function value across ten runs,
\textbf{Time}: average solution time in seconds,
\textbf{GAP opt}: average optimality gap, defined as the relative difference between the best feasible solution and the best lower bound at solver termination (MIP gap).
}
\label{tab:classic_PD_B5}
\resizebox{0.9\textwidth}{!}{
\begin{tabular}{c|
ccc|ccc|
ccc|ccc}
\toprule
\multirow{3}{*}{$V$} 
& \multicolumn{6}{c|}{\textbf{ABF}} 
& \multicolumn{6}{c}{\textbf{NBF}} \\
\cmidrule(lr){2-7} \cmidrule(lr){8-13}
& \multicolumn{3}{c|}{\textbf{Without AFGR}} 
& \multicolumn{3}{c|}{\textbf{With AFGR}} 
& \multicolumn{3}{c|}{\textbf{Without AFGR}} 
& \multicolumn{3}{c}{\textbf{With AFGR}} \\
\cmidrule(lr){2-4} \cmidrule(lr){5-7} \cmidrule(lr){8-10} \cmidrule(lr){11-13}
& OF & Time & GAP opt 
& OF & Time & GAP opt 
& OF & Time & GAP opt 
& OF & Time & GAP opt \\
\midrule

    4     & 965,69 & 0,00  & 0\%     & 965,69 & 0,00  & 0\%     & 965,69 & 0,00  & 0\%     & 965,69 & 0,00  & 0\% \\
    5     & 995,96 & 0,06  & 0\%     & 995,96 & 0,00  & 0\%     & 995,96 & 0,03  & 0\%     & 995,96 & 0,00  & 0\% \\
    6     & 1019,62 & 0,08  & 0\%     & 1019,62 & 0,02  & 0\%     & 1019,62 & 0,09  & 0\%     & 1019,62 & 0,00  & 0\% \\
    7     & 945,85 & 0,03  & 0\%     & 945,85 & 0,02  & 0\%     & 945,85 & 0,13  & 0\%     & 945,85 & 0,06  & 0\% \\
    8     & 664,05 & 0,06  & 0\%     & 664,05 & 0,02  & 0\%     & 664,05 & 0,11  & 0\%     & 664,05 & 0,02  & 0\% \\
    9     & 832,49 & 0,39  & 0\%     & 832,49 & 0,09  & 0\%     & 832,49 & 0,39  & 0\%     & 832,49 & 0,08  & 0\% \\
    10    & 905,93 & 2,58  & 0\%     & 905,93 & 0,22  & 0\%     & 905,93 & 0,81  & 0\%     & 905,93 & 0,20  & 0\% \\
    11    & 1072,45 & 10,53 & 0\%     & 1072,45 & 0,45  & 0\%     & 1072,45 & 14,47 & 0\%     & 1072,45 & 0,34  & 0\% \\
    12    & 843,12 & 0,86  & 0\%     & 843,12 & 0,13  & 0\%     & 843,12 & 1,03  & 0\%     & 843,12 & 0,13  & 0\% \\
    13    & 643,70 & 0,92  & 0\%     & 643,70 & 0,05  & 0\%     & 643,70 & 0,58  & 0\%     & 643,70 & 0,11  & 0\% \\
    14    & 795,95 & 2,27  & 0\%     & 795,95 & 0,14  & 0\%     & 795,95 & 1,70  & 0\%     & 795,95 & 0,22  & 0\% \\
    15    & 794,11 & 3,64  & 0\%     & 794,11 & 0,16  & 0\%     & 794,11 & 5,22  & 0\%     & 794,11 & 0,16  & 0\% \\
    16    & 808,78 & 6,27  & 0\%     & 808,78 & 0,16  & 0\%     & 808,78 & 7,41  & 0\%     & 808,78 & 0,23  & 0\% \\
    17    & 790,10 & 11,42 & 0\%     & 790,10 & 0,16  & 0\%     & 790,10 & 12,03 & 0\%     & 790,10 & 0,47  & 0\% \\
    18    & 785,74 & 15,36 & 0\%     & 785,74 & 0,27  & 0\%     & 785,74 & 16,08 & 0\%     & 785,74 & 0,14  & 0\% \\
    19    & 797,18 & 21,44 & 0\%     & 797,18 & 0,31  & 0\%     & 797,18 & 19,23 & 0\%     & 797,18 & 0,39  & 0\% \\
    20    & 790,54 & 20,61 & 0\%     & 790,54 & 0,44  & 0\%     & 790,54 & 32,95 & 0\%     & 790,54 & 0,47  & 0\% \\
  
\bottomrule
\end{tabular}}
\end{table}
\clearpage
\subsection{STSP-TW}\label{app:classic_table_STSP_TW}
\begin{table}[H]
\centering
\caption{ABF and NBF formulations of the STSP-TW model, with and without the AFGR module, for instance class A1. Each row corresponds to an instance with $V$ customers. For each formulation (ABF and NBF), results are reported both in the original setting and after applying AFGR. The reported metrics are as follows:
\textbf{OF}: average objective function value across ten runs,
\textbf{Time}: average solution time in seconds,
\textbf{GAP opt}: average optimality gap, defined as the relative difference between the best feasible solution and the best lower bound at solver termination (MIP gap).
}
\label{tab:classic_TW_A1}
\resizebox{0.9\textwidth}{!}{
\begin{tabular}{c|
ccc|ccc|
ccc|ccc}
\toprule
\multirow{3}{*}{$V$} 
& \multicolumn{6}{c|}{\textbf{ABF}} 
& \multicolumn{6}{c}{\textbf{NBF}} \\
\cmidrule(lr){2-7} \cmidrule(lr){8-13}
& \multicolumn{3}{c|}{\textbf{Without AFGR}} 
& \multicolumn{3}{c|}{\textbf{With AFGR}} 
& \multicolumn{3}{c|}{\textbf{Without AFGR}} 
& \multicolumn{3}{c}{\textbf{With AFGR}} \\
\cmidrule(lr){2-4} \cmidrule(lr){5-7} \cmidrule(lr){8-10} \cmidrule(lr){11-13}
& OF & Time & GAP opt 
& OF & Time & GAP opt 
& OF & Time & GAP opt 
& OF & Time & GAP opt \\
\midrule
4     & 882,84 & 0,02  & 0\%   & 882,84 & 0,00  & 0\%   & 882,84 & 0,00  & 0\%   & 882,84 & 0,02  & 0\% \\
    5     & 995,96 & 0,02  & 0\%   & 995,96 & 0,02  & 0\%   & 995,96 & 0,02  & 0\%   & 995,96 & 0,02  & 0\% \\
    6     & 1019,62 & 0,03  & 0\%   & 1019,62 & 0,02  & 0\%   & 1019,62 & 0,02  & 0\%   & 1019,62 & 0,05  & 0\% \\
    7     & 1011,19 & 0,09  & 0\%   & 1011,19 & 0,11  & 0\%   & 1011,19 & 0,11  & 0\%   & 1011,19 & 0,11  & 0\% \\
    8     & 991,72 & 0,11  & 0\%   & 991,72 & 0,08  & 0\%   & 991,72 & 0,22  & 0\%   & 991,72 & 0,27  & 0\% \\
    9     & 945,55 & 0,19  & 0\%   & 945,55 & 0,19  & 0\%   & 945,55 & 0,20  & 0\%   & 945,55 & 0,20  & 0\% \\
    10    & 929,54 & 0,61  & 0\%   & 929,54 & 0,14  & 0\%   & 929,54 & 0,56  & 0\%   & 929,54 & 0,31  & 0\% \\
    11    & 908,42 & 0,52  & 0\%   & 908,42 & 0,33  & 0\%   & 908,42 & 0,56  & 0\%   & 908,42 & 0,28  & 0\% \\
    12    & 893,43 & 0,64  & 0\%   & 893,43 & 0,22  & 0\%   & 893,43 & 0,39  & 0\%   & 893,43 & 0,50  & 0\% \\
    13    & 879,12 & 0,47  & 0\%   & 879,12 & 0,69  & 0\%   & 879,12 & 0,72  & 0\%   & 879,12 & 0,78  & 0\% \\
    14    & 865,76 & 2,38  & 0\%   & 865,76 & 0,44  & 0\%   & 865,76 & 2,06  & 0\%   & 865,76 & 1,28  & 0\% \\
    15    & 853,40 & 0,98  & 0\%   & 853,40 & 0,45  & 0\%   & 853,40 & 2,66  & 0\%   & 853,40 & 1,33  & 0\% \\
    16    & 840,52 & 4,20  & 0\%   & 840,52 & 0,42  & 0\%   & 840,52 & 1,75  & 0\%   & 840,52 & 0,50  & 0\% \\
    17    & 832,78 & 3,13  & 0\%   & 832,78 & 1,31  & 0\%   & 832,78 & 8,23  & 0\%   & 832,78 & 0,78  & 0\% \\
    18    & 822,99 & 11,08 & 0\%   & 822,99 & 2,88  & 0\%   & 822,99 & 14,78 & 0\%   & 822,99 & 4,05  & 0\% \\
    19    & 814,02 & 10,98 & 0\%   & 814,02 & 4,17  & 0\%   & 814,02 & 16,00 & 0\%   & 814,02 & 7,27  & 0\% \\
    20    & 805,77 & 12,36 & 0\%   & 805,77 & 7,91  & 0\%   & 805,77 & 29,05 & 0\%   & 805,77 & 15,55 & 0\% \\
  
\bottomrule
\end{tabular}}
\end{table}

\begin{table}[H]
\centering
\caption{ABF and NBF formulations of the STSP-TW model, with and without the AFGR module, for instance class A2. Each row corresponds to an instance with $V$ customers. For each formulation (ABF and NBF), results are reported both in the original setting and after applying AFGR. The reported metrics are as follows:
\textbf{OF}: average objective function value across ten runs,
\textbf{Time}: average solution time in seconds,
\textbf{GAP opt}: average optimality gap, defined as the relative difference between the best feasible solution and the best lower bound at solver termination (MIP gap).
}
\label{tab:classic_TW_A2}
\resizebox{0.9\textwidth}{!}{
\begin{tabular}{c|
ccc|ccc|
ccc|ccc}
\toprule
\multirow{3}{*}{$V$} 
& \multicolumn{6}{c|}{\textbf{ABF}} 
& \multicolumn{6}{c}{\textbf{NBF}} \\
\cmidrule(lr){2-7} \cmidrule(lr){8-13}
& \multicolumn{3}{c|}{\textbf{Without AFGR}} 
& \multicolumn{3}{c|}{\textbf{With AFGR}} 
& \multicolumn{3}{c|}{\textbf{Without AFGR}} 
& \multicolumn{3}{c}{\textbf{With AFGR}} \\
\cmidrule(lr){2-4} \cmidrule(lr){5-7} \cmidrule(lr){8-10} \cmidrule(lr){11-13}
& OF & Time & GAP opt 
& OF & Time & GAP opt 
& OF & Time & GAP opt 
& OF & Time & GAP opt \\
\midrule

    4     & 882,84 & 0,02  & 0\%   & 882,84 & 0,00  & 0\%   & 882,84 & 0,02  & 0\%   & 882,84 & 0,00  & 0\% \\
    5     & 995,96 & 0,00  & 0\%   & 995,96 & 0,00  & 0\%   & 995,96 & 0,02  & 0\%   & 995,96 & 0,05  & 0\% \\
    6     & 1019,62 & 0,02  & 0\%   & 1019,62 & 0,03  & 0\%   & 1019,62 & 0,08  & 0\%   & 1019,62 & 0,09  & 0\% \\
    7     & 1028,38 & 0,17  & 0\%   & 1028,38 & 0,19  & 0\%   & 1028,38 & 0,31  & 0\%   & 1028,38 & 0,30  & 0\% \\
    8     & 1241,38 & 0,20  & 0\%   & 1241,38 & 0,22  & 0\%   & 1241,38 & 0,20  & 0\%   & 1241,38 & 0,31  & 0\% \\
    9     & 1299,21 & 2,50  & 0\%   & 1299,21 & 0,88  & 0\%   & 1299,21 & 4,22  & 0\%   & 1299,21 & 1,44  & 0\% \\
    10    & 1331,26 & 3,97  & 0\%   & 1331,26 & 3,14  & 0\%   & 1331,26 & 8,61  & 0\%   & 1331,26 & 4,48  & 0\% \\
    11    & 1299,78 & 14,05 & 0\%   & 1299,78 & 2,72  & 0\%   & 1299,78 & 20,03 & 0\%   & 1299,78 & 3,64  & 0\% \\
    12    & 1289,90 & 27,27 & 0\%   & 1289,90 & 5,80  & 0\%   & 1289,90 & 26,66 & 0\%   & 1289,90 & 4,66  & 0\% \\
    13    & 884,53 & 7,83  & 0\%   & 884,53 & 1,33  & 0\%   & 884,53 & 3,17  & 0\%   & 884,53 & 1,72  & 0\% \\
    14    & 1350,57 & 337,92 & 0\%   & 1350,57 & 67,31 & 0\%   & 1350,57 & 618,56 & 0\%   & 1350,57 & 71,44 & 0\% \\
    15    & 1333,68 & 1295,70 & 0\%   & 1333,68 & 89,06 & 0\%   & 1333,68 & 1393,06 & 0\%   & 1333,68 & 134,75 & 0\% \\
    16    & 1327,57 & 2279,06 & 0\%   & 1327,57 & 128,45 & 0\%   & 1327,57 & 2013,23 & 0\%   & 1327,57 & 127,31 & 0\% \\
    17    & 1260,20 & 2125,03 & 0\%   & 1260,20 & 59,14 & 0\%   & 1260,20 & 1322,69 & 0\%   & 1260,20 & 230,38 & 0\% \\
    18    & 1238,43 & 2017,52 & 0\%   & 1238,43 & 179,22 & 0\%   & 1239,49 & 2216,86 & 22\%  & 1238,43 & 192,14 & 0\% \\
    19    & 1262,45 & 1379,95 & 0\%   & 1262,45 & 278,53 & 0\%   & 1262,45 & 2154,38 & 2\%   & 1262,45 & 574,34 & 0\% \\
    20    & 1352,61 & 1962,98 & 38\%  & 1349,17 & 2269,34 & 17\%  & 1360,02 & 2071,03 & 41\%  & 1352,61 & 2196,95 & 23\% \\
    
\bottomrule
\end{tabular}}
\end{table}

\begin{table}[H]
\centering
\caption{ABF and NBF formulations of the STSP-TW model, with and without the AFGR module, for instance class A3. Each row corresponds to an instance with $V$ customers. For each formulation (ABF and NBF), results are reported both in the original setting and after applying AFGR. The reported metrics are as follows:
\textbf{OF}: average objective function value across ten runs,
\textbf{Time}: average solution time in seconds,
\textbf{GAP opt}: average optimality gap, defined as the relative difference between the best feasible solution and the best lower bound at solver termination (MIP gap).
}
\label{tab:classic_TW_A3}
\resizebox{0.9\textwidth}{!}{
\begin{tabular}{c|
ccc|ccc|
ccc|ccc}
\toprule
\multirow{3}{*}{$V$} 
& \multicolumn{6}{c|}{\textbf{ABF}} 
& \multicolumn{6}{c}{\textbf{NBF}} \\
\cmidrule(lr){2-7} \cmidrule(lr){8-13}
& \multicolumn{3}{c|}{\textbf{Without AFGR}} 
& \multicolumn{3}{c|}{\textbf{With AFGR}} 
& \multicolumn{3}{c|}{\textbf{Without AFGR}} 
& \multicolumn{3}{c}{\textbf{With AFGR}} \\
\cmidrule(lr){2-4} \cmidrule(lr){5-7} \cmidrule(lr){8-10} \cmidrule(lr){11-13}
& OF & Time & GAP opt 
& OF & Time & GAP opt 
& OF & Time & GAP opt 
& OF & Time & GAP opt \\
\midrule
4     & 882,84 & 0,00  & 0\%   & 882,84 & 0,02  & 0\%   & 882,84 & 0,00  & 0\%   & 882,84 & 0,02  & 0\% \\
    5     & 995,96 & 0,00  & 0\%   & 995,96 & 0,02  & 0\%   & 995,96 & 0,06  & 0\%   & 995,96 & 0,05  & 0\% \\
    6     & 1019,62 & 0,02  & 0\%   & 1019,62 & 0,02  & 0\%   & 1019,62 & 0,17  & 0\%   & 1019,62 & 0,03  & 0\% \\
    7     & 994,00 & 0,03  & 0\%   & 994,00 & 0,06  & 0\%   & 994,00 & 0,13  & 0\%   & 994,00 & 0,05  & 0\% \\
    8     & 980,07 & 0,06  & 0\%   & 980,07 & 0,05  & 0\%   & 980,07 & 0,23  & 0\%   & 980,07 & 0,11  & 0\% \\
    9     & 832,49 & 0,11  & 0\%   & 832,49 & 0,02  & 0\%   & 832,49 & 0,45  & 0\%   & 832,49 & 0,05  & 0\% \\
    10    & 890,10 & 0,42  & 0\%   & 890,10 & 0,05  & 0\%   & 890,10 & 1,02  & 0\%   & 890,10 & 0,30  & 0\% \\
    11    & 869,52 & 0,47  & 0\%   & 869,52 & 0,05  & 0\%   & 869,52 & 1,53  & 0\%   & 869,52 & 0,23  & 0\% \\
    12    & 889,90 & 0,94  & 0\%   & 889,90 & 0,16  & 0\%   & 889,90 & 1,72  & 0\%   & 889,90 & 0,39  & 0\% \\
    13    & 876,34 & 1,22  & 0\%   & 876,34 & 0,41  & 0\%   & 876,34 & 2,05  & 0\%   & 876,34 & 0,78  & 0\% \\
    14    & 846,34 & 1,91  & 0\%   & 846,34 & 0,38  & 0\%   & 846,34 & 2,86  & 0\%   & 846,34 & 0,70  & 0\% \\
    15    & 849,76 & 3,03  & 0\%   & 849,76 & 0,38  & 0\%   & 849,76 & 3,47  & 0\%   & 849,76 & 0,84  & 0\% \\
    16    & 827,37 & 4,33  & 0\%   & 827,37 & 0,23  & 0\%   & 827,37 & 4,27  & 0\%   & 827,37 & 0,38  & 0\% \\
    17    & 823,02 & 22,84 & 0\%   & 823,02 & 3,89  & 0\%   & 823,02 & 28,66 & 0\%   & 823,02 & 4,39  & 0\% \\
    18    & 819,86 & 25,48 & 0\%   & 819,86 & 6,27  & 0\%   & 819,86 & 24,16 & 0\%   & 819,86 & 4,19  & 0\% \\
    19    & 809,55 & 12,42 & 0\%   & 809,55 & 1,48  & 0\%   & 809,55 & 28,08 & 0\%   & 809,55 & 1,94  & 0\% \\
    20    & 800,42 & 30,80 & 0\%   & 800,42 & 2,03  & 0\%   & 800,42 & 34,11 & 0\%   & 800,42 & 2,77  & 0\% \\
    
\bottomrule
\end{tabular}}
\end{table}

\begin{table}[H]
\centering
\caption{ABF and NBF formulations of the STSP-TW model, with and without the AFGR module, for instance class A4. Each row corresponds to an instance with $V$ customers. For each formulation (ABF and NBF), results are reported both in the original setting and after applying AFGR. The reported metrics are as follows:
\textbf{OF}: average objective function value across ten runs,
\textbf{Time}: average solution time in seconds,
\textbf{GAP opt}: average optimality gap, defined as the relative difference between the best feasible solution and the best lower bound at solver termination (MIP gap).
}
\label{tab:classic_TW_A4}
\resizebox{0.9\textwidth}{!}{
\begin{tabular}{c|
ccc|ccc|
ccc|ccc}
\toprule
\multirow{3}{*}{$V$} 
& \multicolumn{6}{c|}{\textbf{ABF}} 
& \multicolumn{6}{c}{\textbf{NBF}} \\
\cmidrule(lr){2-7} \cmidrule(lr){8-13}
& \multicolumn{3}{c|}{\textbf{Without AFGR}} 
& \multicolumn{3}{c|}{\textbf{With AFGR}} 
& \multicolumn{3}{c|}{\textbf{Without AFGR}} 
& \multicolumn{3}{c}{\textbf{With AFGR}} \\
\cmidrule(lr){2-4} \cmidrule(lr){5-7} \cmidrule(lr){8-10} \cmidrule(lr){11-13}
& OF & Time & GAP opt 
& OF & Time & GAP opt 
& OF & Time & GAP opt 
& OF & Time & GAP opt \\
\midrule
    4     & 882,84 & 0,02  & 0\%   & 882,84 & 0,02  & 0\%   & 882,84 & 0,00  & 0\%   & 882,84 & 0,00  & 0\% \\
    5     & 878,40 & 0,02  & 0\%   & 878,40 & 0,00  & 0\%   & 878,40 & 0,00  & 0\%   & 878,40 & 0,02  & 0\% \\
    6     & 1019,62 & 0,02  & 0\%   & 1019,62 & 0,05  & 0\%   & 1019,62 & 0,08  & 0\%   & 1019,62 & 0,02  & 0\% \\
    7     & 945,85 & 0,09  & 0\%   & 945,85 & 0,06  & 0\%   & 945,85 & 0,09  & 0\%   & 945,85 & 0,00  & 0\% \\
    8     & 946,89 & 0,06  & 0\%   & 946,89 & 0,08  & 0\%   & 946,89 & 0,08  & 0\%   & 946,89 & 0,02  & 0\% \\
    9     & 832,49 & 0,22  & 0\%   & 832,49 & 0,02  & 0\%   & 832,49 & 0,31  & 0\%   & 832,49 & 0,05  & 0\% \\
    10    & 872,54 & 0,22  & 0\%   & 872,54 & 0,02  & 0\%   & 872,54 & 0,36  & 0\%   & 872,54 & 0,02  & 0\% \\
    11    & 869,52 & 0,59  & 0\%   & 869,52 & 0,09  & 0\%   & 869,52 & 0,59  & 0\%   & 869,52 & 0,06  & 0\% \\
    12    & 869,92 & 0,73  & 0\%   & 869,92 & 0,11  & 0\%   & 869,92 & 1,17  & 0\%   & 869,92 & 0,19  & 0\% \\
    13    & 829,59 & 0,95  & 0\%   & 829,59 & 0,16  & 0\%   & 829,59 & 1,61  & 0\%   & 829,59 & 0,28  & 0\% \\
    14    & 825,66 & 1,06  & 0\%   & 825,66 & 0,23  & 0\%   & 825,66 & 2,92  & 0\%   & 825,66 & 0,38  & 0\% \\
    15    & 818,69 & 3,20  & 0\%   & 818,69 & 0,30  & 0\%   & 818,69 & 4,66  & 0\%   & 818,69 & 0,36  & 0\% \\
    16    & 813,22 & 4,50  & 0\%   & 813,22 & 0,44  & 0\%   & 813,22 & 6,94  & 0\%   & 813,22 & 0,42  & 0\% \\
    17    & 793,81 & 7,84  & 0\%   & 793,81 & 0,48  & 0\%   & 793,81 & 14,39 & 0\%   & 793,81 & 0,55  & 0\% \\
    18    & 800,47 & 7,73  & 0\%   & 800,47 & 0,34  & 0\%   & 800,47 & 20,20 & 0\%   & 800,47 & 0,81  & 0\% \\
    19    & 804,21 & 10,42 & 0\%   & 804,21 & 0,61  & 0\%   & 804,21 & 28,22 & 0\%   & 804,21 & 0,50  & 0\% \\
    20    & 797,36 & 28,03 & 0\%   & 797,36 & 1,41  & 0\%   & 797,36 & 39,59 & 0\%   & 797,36 & 1,20  & 0\% \\
    
\bottomrule
\end{tabular}}
\end{table}

\begin{table}[H]
\centering
\caption{ABF and NBF formulations of the STSP-TW model, with and without the AFGR module, for instance class A5. Each row corresponds to an instance with $V$ customers. For each formulation (ABF and NBF), results are reported both in the original setting and after applying AFGR. The reported metrics are as follows:
\textbf{OF}: average objective function value across ten runs,
\textbf{Time}: average solution time in seconds,
\textbf{GAP opt}: average optimality gap, defined as the relative difference between the best feasible solution and the best lower bound at solver termination (MIP gap).
}
\label{tab:classic_TW_A5}
\resizebox{0.9\textwidth}{!}{
\begin{tabular}{c|
ccc|ccc|
ccc|ccc}
\toprule
\multirow{3}{*}{$V$} 
& \multicolumn{6}{c|}{\textbf{ABF}} 
& \multicolumn{6}{c}{\textbf{NBF}} \\
\cmidrule(lr){2-7} \cmidrule(lr){8-13}
& \multicolumn{3}{c|}{\textbf{Without AFGR}} 
& \multicolumn{3}{c|}{\textbf{With AFGR}} 
& \multicolumn{3}{c|}{\textbf{Without AFGR}} 
& \multicolumn{3}{c}{\textbf{With AFGR}} \\
\cmidrule(lr){2-4} \cmidrule(lr){5-7} \cmidrule(lr){8-10} \cmidrule(lr){11-13}
& OF & Time & GAP opt 
& OF & Time & GAP opt 
& OF & Time & GAP opt 
& OF & Time & GAP opt \\
\midrule

    4     & 882,84 & 0,02  & 0\%   & 882,84 & 0,02  & 0\%   & 882,84 & 0,02  & 0\%   & 882,84 & 0,00  & 0\% \\
    5     & 878,40 & 0,02  & 0\%   & 878,40 & 0,00  & 0\%   & 878,40 & 0,02  & 0\%   & 878,40 & 0,00  & 0\% \\
    6     & 946,41 & 0,02  & 0\%   & 946,41 & 0,00  & 0\%   & 946,41 & 0,03  & 0\%   & 946,41 & 0,02  & 0\% \\
    7     & 945,85 & 0,05  & 0\%   & 945,85 & 0,02  & 0\%   & 945,85 & 0,09  & 0\%   & 945,85 & 0,02  & 0\% \\
    8     & 664,05 & 0,05  & 0\%   & 664,05 & 0,02  & 0\%   & 664,05 & 0,05  & 0\%   & 664,05 & 0,02  & 0\% \\
    9     & 832,49 & 0,22  & 0\%   & 832,49 & 0,02  & 0\%   & 832,49 & 0,38  & 0\%   & 832,49 & 0,02  & 0\% \\
    10    & 872,54 & 0,16  & 0\%   & 872,54 & 0,03  & 0\%   & 872,54 & 0,36  & 0\%   & 872,54 & 0,05  & 0\% \\
    11    & 843,95 & 0,30  & 0\%   & 843,95 & 0,06  & 0\%   & 843,95 & 0,78  & 0\%   & 843,95 & 0,02  & 0\% \\
    12    & 843,12 & 0,73  & 0\%   & 843,12 & 0,09  & 0\%   & 843,12 & 0,98  & 0\%   & 843,12 & 0,17  & 0\% \\
    13    & 643,70 & 0,55  & 0\%   & 643,70 & 0,05  & 0\%   & 643,70 & 0,56  & 0\%   & 643,70 & 0,11  & 0\% \\
    14    & 795,95 & 1,39  & 0\%   & 795,95 & 0,11  & 0\%   & 795,95 & 2,41  & 0\%   & 795,95 & 0,28  & 0\% \\
    15    & 794,11 & 2,22  & 0\%   & 794,11 & 0,14  & 0\%   & 794,11 & 2,36  & 0\%   & 794,11 & 0,22  & 0\% \\
    16    & 808,78 & 3,30  & 0\%   & 808,78 & 0,22  & 0\%   & 808,78 & 5,05  & 0\%   & 808,78 & 0,33  & 0\% \\
    17    & 790,10 & 6,34  & 0\%   & 790,10 & 0,30  & 0\%   & 790,10 & 11,70 & 0\%   & 790,10 & 0,33  & 0\% \\
    18    & 785,74 & 9,45  & 0\%   & 785,74 & 0,17  & 0\%   & 785,74 & 17,83 & 0\%   & 785,74 & 0,23  & 0\% \\
    19    & 797,18 & 13,52 & 0\%   & 797,18 & 0,34  & 0\%   & 797,18 & 35,19 & 0\%   & 797,18 & 0,25  & 0\% \\
    20    & 790,54 & 22,61 & 0\%   & 790,54 & 0,42  & 0\%   & 790,54 & 38,23 & 0\%   & 790,54 & 0,64  & 0\% \\
    
\bottomrule
\end{tabular}}
\end{table}

\begin{table}[H]
\centering
\caption{ABF and NBF formulations of the STSP-TW model, with and without the AFGR module, for instance class B1. Each row corresponds to an instance with $V$ customers. For each formulation (ABF and NBF), results are reported both in the original setting and after applying AFGR. The reported metrics are as follows:
\textbf{OF}: average objective function value across ten runs,
\textbf{Time}: average solution time in seconds,
\textbf{GAP opt}: average optimality gap, defined as the relative difference between the best feasible solution and the best lower bound at solver termination (MIP gap).
}
\label{tab:classic_TW_B1}
\resizebox{0.9\textwidth}{!}{
\begin{tabular}{c|
ccc|ccc|
ccc|ccc}
\toprule
\multirow{3}{*}{$V$} 
& \multicolumn{6}{c|}{\textbf{ABF}} 
& \multicolumn{6}{c}{\textbf{NBF}} \\
\cmidrule(lr){2-7} \cmidrule(lr){8-13}
& \multicolumn{3}{c|}{\textbf{Without AFGR}} 
& \multicolumn{3}{c|}{\textbf{With AFGR}} 
& \multicolumn{3}{c|}{\textbf{Without AFGR}} 
& \multicolumn{3}{c}{\textbf{With AFGR}} \\
\cmidrule(lr){2-4} \cmidrule(lr){5-7} \cmidrule(lr){8-10} \cmidrule(lr){11-13}
& OF & Time & GAP opt 
& OF & Time & GAP opt 
& OF & Time & GAP opt 
& OF & Time & GAP opt \\
\midrule
4     & 882,84 & 0,02  & 0\%   & 882,84 & 0,02  & 0\%   & 882,84 & 0,00  & 0\%   & 882,84 & 0,02  & 0\% \\
    5     & 995,96 & 0,02  & 0\%   & 995,96 & 0,00  & 0\%   & 995,96 & 0,02  & 0\%   & 995,96 & 0,00  & 0\% \\
    6     & 1019,62 & 0,03  & 0\%   & 1019,62 & 0,02  & 0\%   & 1019,62 & 0,05  & 0\%   & 1019,62 & 0,05  & 0\% \\
    7     & 1011,19 & 0,09  & 0\%   & 1011,19 & 0,11  & 0\%   & 1011,19 & 0,11  & 0\%   & 1011,19 & 0,13  & 0\% \\
    8     & 991,72 & 0,08  & 0\%   & 991,72 & 0,13  & 0\%   & 991,72 & 0,22  & 0\%   & 991,72 & 0,22  & 0\% \\
    9     & 945,55 & 0,17  & 0\%   & 945,55 & 0,14  & 0\%   & 945,55 & 0,20  & 0\%   & 945,55 & 0,23  & 0\% \\
    10    & 929,54 & 0,61  & 0\%   & 929,54 & 0,14  & 0\%   & 929,54 & 0,47  & 0\%   & 929,54 & 0,30  & 0\% \\
    11    & 908,42 & 0,50  & 0\%   & 908,42 & 0,38  & 0\%   & 908,42 & 0,61  & 0\%   & 908,42 & 0,34  & 0\% \\
    12    & 893,43 & 0,73  & 0\%   & 893,43 & 0,22  & 0\%   & 893,43 & 0,41  & 0\%   & 893,43 & 0,52  & 0\% \\
    13    & 879,12 & 0,47  & 0\%   & 879,12 & 0,67  & 0\%   & 879,12 & 0,55  & 0\%   & 879,12 & 0,89  & 0\% \\
    14    & 865,76 & 2,30  & 0\%   & 865,76 & 0,42  & 0\%   & 865,76 & 3,19  & 0\%   & 865,76 & 1,69  & 0\% \\
    15    & 853,40 & 0,97  & 0\%   & 853,40 & 0,45  & 0\%   & 853,40 & 2,78  & 0\%   & 853,40 & 1,58  & 0\% \\
    16    & 840,52 & 4,23  & 0\%   & 840,52 & 0,44  & 0\%   & 840,52 & 1,66  & 0\%   & 840,52 & 0,59  & 0\% \\
    17    & 832,78 & 3,14  & 0\%   & 832,78 & 1,34  & 0\%   & 832,78 & 13,70 & 0\%   & 832,78 & 0,94  & 0\% \\
    18    & 822,99 & 10,44 & 0\%   & 822,99 & 2,94  & 0\%   & 822,99 & 20,42 & 0\%   & 822,99 & 5,31  & 0\% \\
    19    & 814,02 & 10,70 & 0\%   & 814,02 & 3,95  & 0\%   & 814,02 & 16,73 & 0\%   & 814,02 & 10,48 & 0\% \\
    20    & 805,77 & 12,38 & 0\%   & 805,77 & 7,78  & 0\%   & 805,77 & 35,53 & 0\%   & 805,77 & 18,09 & 0\% \\
   
\bottomrule
\end{tabular}}
\end{table}

\begin{table}[H]
\centering
\caption{ABF and NBF formulations of the STSP-TW model, with and without the AFGR module, for instance class B2. Each row corresponds to an instance with $V$ customers. For each formulation (ABF and NBF), results are reported both in the original setting and after applying AFGR. The reported metrics are as follows:
\textbf{OF}: average objective function value across ten runs,
\textbf{Time}: average solution time in seconds,
\textbf{GAP opt}: average optimality gap, defined as the relative difference between the best feasible solution and the best lower bound at solver termination (MIP gap).
}
\label{tab:classic_TW_B2}
\resizebox{0.9\textwidth}{!}{
\begin{tabular}{c|
ccc|ccc|
ccc|ccc}
\toprule
\multirow{3}{*}{$V$} 
& \multicolumn{6}{c|}{\textbf{ABF}} 
& \multicolumn{6}{c}{\textbf{NBF}} \\
\cmidrule(lr){2-7} \cmidrule(lr){8-13}
& \multicolumn{3}{c|}{\textbf{Without AFGR}} 
& \multicolumn{3}{c|}{\textbf{With AFGR}} 
& \multicolumn{3}{c|}{\textbf{Without AFGR}} 
& \multicolumn{3}{c}{\textbf{With AFGR}} \\
\cmidrule(lr){2-4} \cmidrule(lr){5-7} \cmidrule(lr){8-10} \cmidrule(lr){11-13}
& OF & Time & GAP opt 
& OF & Time & GAP opt 
& OF & Time & GAP opt 
& OF & Time & GAP opt \\
\midrule
4     & 882,84 & 0,00  & 0\%   & 882,84 & 0,00  & 0\%   & 882,84 & 0,02  & 0\%   & 882,84 & 0,02  & 0\% \\
    5     & 995,96 & 0,00  & 0\%   & 995,96 & 0,00  & 0\%   & 995,96 & 0,00  & 0\%   & 995,96 & 0,02  & 0\% \\
    6     & 1019,62 & 0,03  & 0\%   & 1019,62 & 0,02  & 0\%   & 1019,62 & 0,03  & 0\%   & 1019,62 & 0,05  & 0\% \\
    7     & 1028,38 & 0,17  & 0\%   & 1028,38 & 0,22  & 0\%   & 1028,38 & 0,25  & 0\%   & 1028,38 & 0,31  & 0\% \\
    8     & 1241,38 & 0,16  & 0\%   & 1241,38 & 0,13  & 0\%   & 1241,38 & 0,14  & 0\%   & 1241,38 & 0,33  & 0\% \\
    9     & 1299,21 & 2,48  & 0\%   & 1299,21 & 0,84  & 0\%   & 1299,21 & 4,23  & 0\%   & 1299,21 & 1,44  & 0\% \\
    10    & 1331,26 & 3,58  & 0\%   & 1331,26 & 3,02  & 0\%   & 1331,26 & 8,05  & 0\%   & 1331,26 & 3,41  & 0\% \\
    11    & 1299,78 & 13,39 & 0\%   & 1299,78 & 2,66  & 0\%   & 1299,78 & 17,03 & 0\%   & 1299,78 & 3,23  & 0\% \\
    12    & 1289,90 & 25,94 & 0\%   & 1289,90 & 5,48  & 0\%   & 1289,90 & 31,05 & 0\%   & 1289,90 & 5,06  & 0\% \\
    13    & 884,53 & 7,20  & 0\%   & 884,53 & 1,20  & 0\%   & 884,53 & 3,91  & 0\%   & 884,53 & 1,31  & 0\% \\
    14    & 1350,57 & 320,69 & 0\%   & 1350,57 & 63,19 & 0\%   & 1350,57 & 673,03 & 0\%   & 1350,57 & 79,48 & 0\% \\
    15    & 1333,68 & 1230,53 & 0\%   & 1333,68 & 86,14 & 0\%   & 1333,68 & 1509,47 & 0\%   & 1333,68 & 120,52 & 0\% \\
    16    & 1327,57 & 2189,39 & 0\%   & 1327,57 & 121,69 & 0\%   & 1327,57 & 1611,80 & 0\%   & 1327,57 & 111,91 & 0\% \\
    17    & 1260,20 & 2096,61 & 0\%   & 1260,20 & 59,94 & 0\%   & 1260,20 & 1173,44 & 0\%   & 1260,20 & 183,13 & 0\% \\
    18    & 1238,43 & 2011,86 & 0\%   & 1238,43 & 184,30 & 0\%   & 1239,49 & 2231,22 & 22\%  & 1238,43 & 151,00 & 0\% \\
    19    & 1262,45 & 1378,20 & 0\%   & 1262,45 & 315,50 & 0\%   & 1262,45 & 1878,53 & 0\%   & 1262,45 & 574,80 & 0\% \\
    20    & 1352,61 & 1984,50 & 38\%  & 1349,17 & 2267,52 & 17\%  & 1360,02 & 2081,14 & 41\%  & 1349,17 & 2280,95 & 18\% \\
    
\bottomrule
\end{tabular}}
\end{table}

\begin{table}[H]
\centering
\caption{ABF and NBF formulations of the STSP-TW model, with and without the AFGR module, for instance class B3. Each row corresponds to an instance with $V$ customers. For each formulation (ABF and NBF), results are reported both in the original setting and after applying AFGR. The reported metrics are as follows:
\textbf{OF}: average objective function value across ten runs,
\textbf{Time}: average solution time in seconds,
\textbf{GAP opt}: average optimality gap, defined as the relative difference between the best feasible solution and the best lower bound at solver termination (MIP gap).
}
\label{tab:classic_TW_B3}
\resizebox{0.9\textwidth}{!}{
\begin{tabular}{c|
ccc|ccc|
ccc|ccc}
\toprule
\multirow{3}{*}{$V$} 
& \multicolumn{6}{c|}{\textbf{ABF}} 
& \multicolumn{6}{c}{\textbf{NBF}} \\
\cmidrule(lr){2-7} \cmidrule(lr){8-13}
& \multicolumn{3}{c|}{\textbf{Without AFGR}} 
& \multicolumn{3}{c|}{\textbf{With AFGR}} 
& \multicolumn{3}{c|}{\textbf{Without AFGR}} 
& \multicolumn{3}{c}{\textbf{With AFGR}} \\
\cmidrule(lr){2-4} \cmidrule(lr){5-7} \cmidrule(lr){8-10} \cmidrule(lr){11-13}
& OF & Time & GAP opt 
& OF & Time & GAP opt 
& OF & Time & GAP opt 
& OF & Time & GAP opt \\
\midrule
4     & 882,84 & 0,02  & 0\%   & 882,84 & 0,02  & 0\%   & 882,84 & 0,02  & 0\%   & 882,84 & 0,00  & 0\% \\
    5     & 995,96 & 0,02  & 0\%   & 995,96 & 0,02  & 0\%   & 995,96 & 0,02  & 0\%   & 995,96 & 0,02  & 0\% \\
    6     & 1019,62 & 0,02  & 0\%   & 1019,62 & 0,02  & 0\%   & 1019,62 & 0,08  & 0\%   & 1019,62 & 0,08  & 0\% \\
    7     & 994,00 & 0,03  & 0\%   & 994,00 & 0,05  & 0\%   & 994,00 & 0,03  & 0\%   & 994,00 & 0,05  & 0\% \\
    8     & 980,07 & 0,11  & 0\%   & 980,07 & 0,03  & 0\%   & 980,07 & 0,08  & 0\%   & 980,07 & 0,06  & 0\% \\
    9     & 832,49 & 0,16  & 0\%   & 832,49 & 0,02  & 0\%   & 832,49 & 0,17  & 0\%   & 832,49 & 0,03  & 0\% \\
    10    & 890,10 & 0,44  & 0\%   & 890,10 & 0,11  & 0\%   & 890,10 & 0,53  & 0\%   & 890,10 & 0,16  & 0\% \\
    11    & 869,52 & 0,52  & 0\%   & 869,52 & 0,11  & 0\%   & 869,52 & 0,83  & 0\%   & 869,52 & 0,16  & 0\% \\
    12    & 889,90 & 1,06  & 0\%   & 889,90 & 0,16  & 0\%   & 889,90 & 1,03  & 0\%   & 889,90 & 0,31  & 0\% \\
    13    & 876,34 & 1,17  & 0\%   & 876,34 & 0,42  & 0\%   & 876,34 & 1,30  & 0\%   & 876,34 & 0,42  & 0\% \\
    14    & 846,34 & 1,91  & 0\%   & 846,34 & 0,36  & 0\%   & 846,34 & 2,03  & 0\%   & 846,34 & 0,50  & 0\% \\
    15    & 849,76 & 2,97  & 0\%   & 849,76 & 0,36  & 0\%   & 849,76 & 2,77  & 0\%   & 849,76 & 0,64  & 0\% \\
    16    & 827,37 & 4,16  & 0\%   & 827,37 & 0,28  & 0\%   & 827,37 & 2,80  & 0\%   & 827,37 & 0,38  & 0\% \\
    17    & 823,02 & 22,66 & 0\%   & 823,02 & 4,00  & 0\%   & 823,02 & 26,61 & 0\%   & 823,02 & 4,34  & 0\% \\
    18    & 819,86 & 27,81 & 0\%   & 819,86 & 6,28  & 0\%   & 819,86 & 22,84 & 0\%   & 819,86 & 4,08  & 0\% \\
    19    & 809,55 & 12,73 & 0\%   & 809,55 & 1,45  & 0\%   & 809,55 & 27,72 & 0\%   & 809,55 & 1,67  & 0\% \\
    20    & 800,42 & 31,13 & 0\%   & 800,42 & 2,08  & 0\%   & 800,42 & 35,36 & 0\%   & 800,42 & 2,83  & 0\% \\
    
\bottomrule
\end{tabular}}
\end{table}

\begin{table}[H]
\centering
\caption{ABF and NBF formulations of the STSP-TW model, with and without the AFGR module, for instance class B4. Each row corresponds to an instance with $V$ customers. For each formulation (ABF and NBF), results are reported both in the original setting and after applying AFGR. The reported metrics are as follows:
\textbf{OF}: average objective function value across ten runs,
\textbf{Time}: average solution time in seconds,
\textbf{GAP opt}: average optimality gap, defined as the relative difference between the best feasible solution and the best lower bound at solver termination (MIP gap).
}
\label{tab:classic_TW_B4}
\resizebox{0.9\textwidth}{!}{
\begin{tabular}{c|
ccc|ccc|
ccc|ccc}
\toprule
\multirow{3}{*}{$V$} 
& \multicolumn{6}{c|}{\textbf{ABF}} 
& \multicolumn{6}{c}{\textbf{NBF}} \\
\cmidrule(lr){2-7} \cmidrule(lr){8-13}
& \multicolumn{3}{c|}{\textbf{Without AFGR}} 
& \multicolumn{3}{c|}{\textbf{With AFGR}} 
& \multicolumn{3}{c|}{\textbf{Without AFGR}} 
& \multicolumn{3}{c}{\textbf{With AFGR}} \\
\cmidrule(lr){2-4} \cmidrule(lr){5-7} \cmidrule(lr){8-10} \cmidrule(lr){11-13}
& OF & Time & GAP opt 
& OF & Time & GAP opt 
& OF & Time & GAP opt 
& OF & Time & GAP opt \\
\midrule
4     & 882,84 & 0,00  & 0\%   & 882,84 & 0,00  & 0\%   & 882,84 & 0,02  & 0\%   & 882,84 & 0,02  & 0\% \\
    5     & 878,40 & 0,05  & 0\%   & 878,40 & 0,02  & 0\%   & 878,40 & 0,02  & 0\%   & 878,40 & 0,00  & 0\% \\
    6     & 1019,62 & 0,03  & 0\%   & 1019,62 & 0,00  & 0\%   & 1019,62 & 0,03  & 0\%   & 1019,62 & 0,02  & 0\% \\
    7     & 945,85 & 0,09  & 0\%   & 945,85 & 0,05  & 0\%   & 945,85 & 0,03  & 0\%   & 945,85 & 0,02  & 0\% \\
    8     & 946,89 & 0,05  & 0\%   & 946,89 & 0,03  & 0\%   & 946,89 & 0,08  & 0\%   & 946,89 & 0,06  & 0\% \\
    9     & 832,49 & 0,25  & 0\%   & 832,49 & 0,02  & 0\%   & 832,49 & 0,33  & 0\%   & 832,49 & 0,05  & 0\% \\
    10    & 872,54 & 0,22  & 0\%   & 872,54 & 0,02  & 0\%   & 872,54 & 0,25  & 0\%   & 872,54 & 0,02  & 0\% \\
    11    & 869,52 & 0,64  & 0\%   & 869,52 & 0,09  & 0\%   & 869,52 & 0,42  & 0\%   & 869,52 & 0,05  & 0\% \\
    12    & 869,92 & 0,75  & 0\%   & 869,92 & 0,08  & 0\%   & 869,92 & 1,25  & 0\%   & 869,92 & 0,25  & 0\% \\
    13    & 829,59 & 0,95  & 0\%   & 829,59 & 0,20  & 0\%   & 829,59 & 1,36  & 0\%   & 829,59 & 0,28  & 0\% \\
    14    & 825,66 & 1,05  & 0\%   & 825,66 & 0,20  & 0\%   & 825,66 & 2,42  & 0\%   & 825,66 & 0,28  & 0\% \\
    15    & 818,69 & 3,25  & 0\%   & 818,69 & 0,31  & 0\%   & 818,69 & 3,86  & 0\%   & 818,69 & 0,28  & 0\% \\
    16    & 813,22 & 4,58  & 0\%   & 813,22 & 0,39  & 0\%   & 813,22 & 5,47  & 0\%   & 813,22 & 0,42  & 0\% \\
    17    & 793,81 & 7,53  & 0\%   & 793,81 & 0,44  & 0\%   & 793,81 & 13,27 & 0\%   & 793,81 & 0,52  & 0\% \\
    18    & 800,47 & 7,61  & 0\%   & 800,47 & 0,39  & 0\%   & 800,47 & 16,13 & 0\%   & 800,47 & 0,66  & 0\% \\
    19    & 804,21 & 10,31 & 0\%   & 804,21 & 0,55  & 0\%   & 804,21 & 31,45 & 0\%   & 804,21 & 0,67  & 0\% \\
    20    & 797,36 & 27,97 & 0\%   & 797,36 & 1,28  & 0\%   & 797,36 & 43,75 & 0\%   & 797,36 & 1,30  & 0\% \\
    
\bottomrule
\end{tabular}}
\end{table}

\begin{table}[H]
\centering
\caption{ABF and NBF formulations of the STSP-TW model, with and without the AFGR module, for instance class B5. Each row corresponds to an instance with $V$ customers. For each formulation (ABF and NBF), results are reported both in the original setting and after applying AFGR. The reported metrics are as follows:
\textbf{OF}: average objective function value across ten runs,
\textbf{Time}: average solution time in seconds,
\textbf{GAP opt}: average optimality gap, defined as the relative difference between the best feasible solution and the best lower bound at solver termination (MIP gap).
}
\label{tab:classic_TW_B5}
\resizebox{0.9\textwidth}{!}{
\begin{tabular}{c|
ccc|ccc|
ccc|ccc}
\toprule
\multirow{3}{*}{$V$} 
& \multicolumn{6}{c|}{\textbf{ABF}} 
& \multicolumn{6}{c}{\textbf{NBF}} \\
\cmidrule(lr){2-7} \cmidrule(lr){8-13}
& \multicolumn{3}{c|}{\textbf{Without AFGR}} 
& \multicolumn{3}{c|}{\textbf{With AFGR}} 
& \multicolumn{3}{c|}{\textbf{Without AFGR}} 
& \multicolumn{3}{c}{\textbf{With AFGR}} \\
\cmidrule(lr){2-4} \cmidrule(lr){5-7} \cmidrule(lr){8-10} \cmidrule(lr){11-13}
& OF & Time & GAP opt 
& OF & Time & GAP opt 
& OF & Time & GAP opt 
& OF & Time & GAP opt \\
\midrule
4     & 882,84 & 0,00  & 0\%   & 882,84 & 0,00  & 0\%   & 882,84 & 0,00  & 0\%   & 882,84 & 0,02  & 0\% \\
    5     & 878,40 & 0,00  & 0\%   & 878,40 & 0,00  & 0\%   & 878,40 & 0,02  & 0\%   & 878,40 & 0,00  & 0\% \\
    6     & 946,41 & 0,02  & 0\%   & 946,41 & 0,02  & 0\%   & 946,41 & 0,08  & 0\%   & 946,41 & 0,00  & 0\% \\
    7     & 945,85 & 0,11  & 0\%   & 945,85 & 0,02  & 0\%   & 945,85 & 0,09  & 0\%   & 945,85 & 0,02  & 0\% \\
    8     & 664,05 & 0,03  & 0\%   & 664,05 & 0,02  & 0\%   & 664,05 & 0,09  & 0\%   & 664,05 & 0,02  & 0\% \\
    9     & 832,49 & 0,22  & 0\%   & 832,49 & 0,02  & 0\%   & 832,49 & 0,34  & 0\%   & 832,49 & 0,03  & 0\% \\
    10    & 872,54 & 0,20  & 0\%   & 872,54 & 0,02  & 0\%   & 872,54 & 0,36  & 0\%   & 872,54 & 0,03  & 0\% \\
    11    & 843,95 & 0,33  & 0\%   & 843,95 & 0,02  & 0\%   & 843,95 & 0,80  & 0\%   & 843,95 & 0,05  & 0\% \\
    12    & 843,12 & 0,70  & 0\%   & 843,12 & 0,09  & 0\%   & 843,12 & 1,13  & 0\%   & 843,12 & 0,17  & 0\% \\
    13    & 643,70 & 0,53  & 0\%   & 643,70 & 0,09  & 0\%   & 643,70 & 0,70  & 0\%   & 643,70 & 0,09  & 0\% \\
    14    & 795,95 & 1,36  & 0\%   & 795,95 & 0,08  & 0\%   & 795,95 & 3,02  & 0\%   & 795,95 & 0,25  & 0\% \\
    15    & 794,11 & 2,25  & 0\%   & 794,11 & 0,08  & 0\%   & 794,11 & 2,38  & 0\%   & 794,11 & 0,22  & 0\% \\
    16    & 808,78 & 3,23  & 0\%   & 808,78 & 0,20  & 0\%   & 808,78 & 5,31  & 0\%   & 808,78 & 0,36  & 0\% \\
    17    & 790,10 & 6,27  & 0\%   & 790,10 & 0,28  & 0\%   & 790,10 & 11,83 & 0\%   & 790,10 & 0,33  & 0\% \\
    18    & 785,74 & 9,34  & 0\%   & 785,74 & 0,17  & 0\%   & 785,74 & 17,75 & 0\%   & 785,74 & 0,25  & 0\% \\
    19    & 797,18 & 13,42 & 0\%   & 797,18 & 0,41  & 0\%   & 797,18 & 35,55 & 0\%   & 797,18 & 0,28  & 0\% \\
    20    & 790,54 & 22,88 & 0\%   & 790,54 & 0,45  & 0\%   & 790,54 & 38,56 & 0\%   & 790,54 & 0,61  & 0\% \\

\bottomrule
\end{tabular}}
\end{table}
\clearpage
\section{Computational Results with the Classical Solver -- Figures}
\label{app:classicfigure}
\begin{figure}[htbp]
    \centering
    \includegraphics[width=0.95\textwidth]{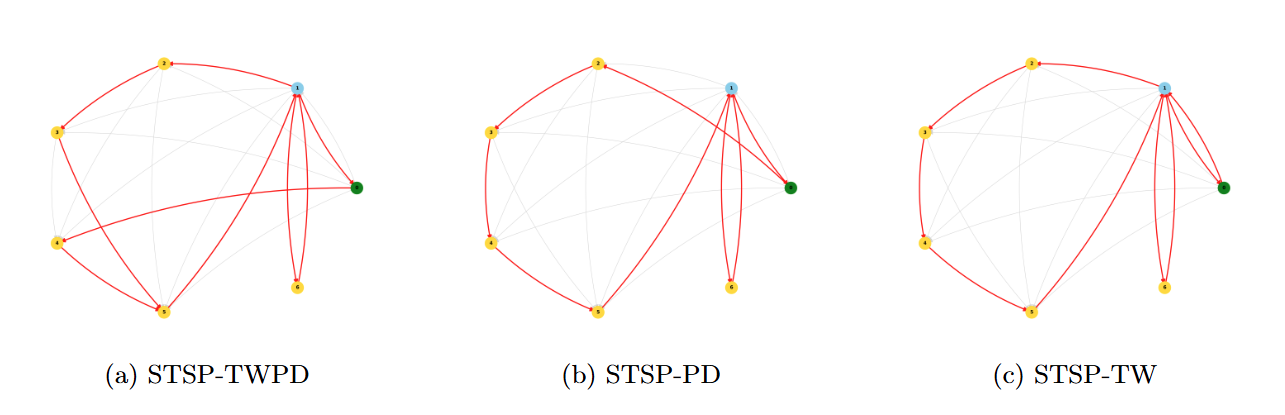}
    \caption{Comparison of optimal routes generated by the STSP-TWPD, STSP-PD and STSP-TW models for $V = 7$.}
    \label{fig:route_comparison_V7}
\end{figure}
\begin{figure}[htbp]
    \centering
    \includegraphics[width=0.95\textwidth]{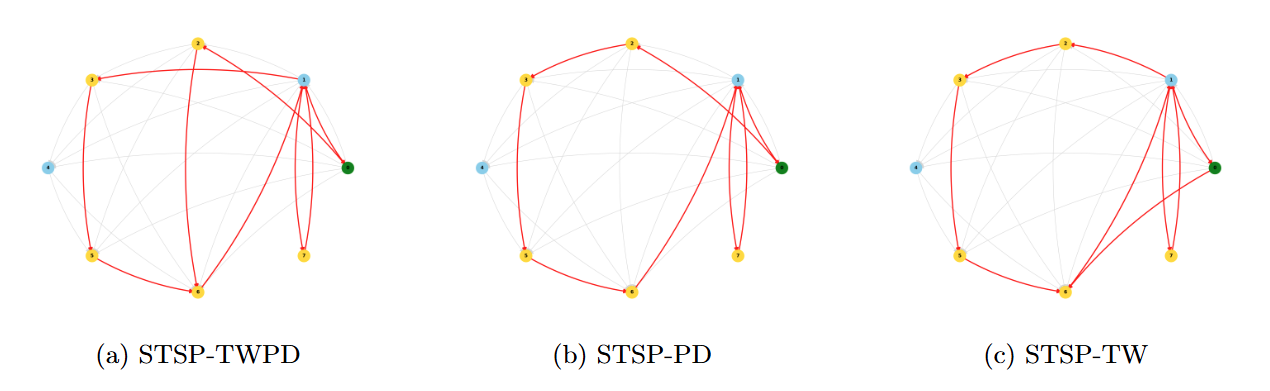}
    \caption{Comparison of optimal routes generated by the STSP-TWPD, STSP-PD, and STSP-TW models for $V = 8$.}
    \label{fig:route_comparison_V8}
\end{figure}
\begin{figure}[htbp]
    \centering
    \includegraphics[width=0.95\textwidth]{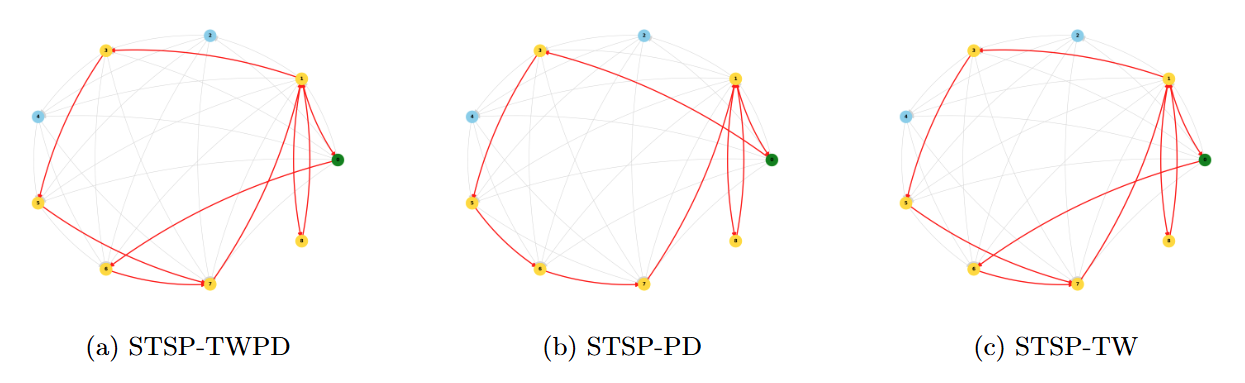}
    \caption{Comparison of optimal routes generated by the STSP-TWPD, STSP-PD, and STSP-TW models for $V = 9$.}
    \label{fig:route_comparison_V9}
\end{figure}
\begin{figure}[htbp]
    \centering
    \includegraphics[width=0.95\textwidth]{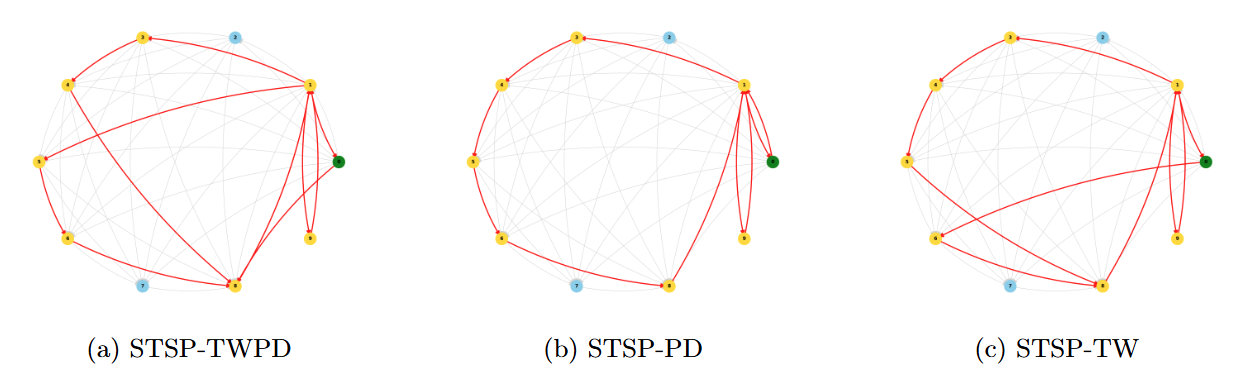}
    \caption{Comparison of optimal routes generated by the STSP-TWPD, STSP-PD, and STSP-TW models for $V = 10$.}
    \label{fig:route_comparison_V10}
\end{figure}
\clearpage
\section{Computational Results with the Quantum Solver -- Tables}
\label{app:quantum_table}
\subsection{STSP-TWPD}\label{app:quantum_table_STSP_TWPD}
\begin{table}[h]
\centering
\caption{
Results for the STSP-TWPD using the ABF and NBF formulation, comparing performance with and without the AFGR method. Each row corresponds to a different instance size, defined by the number of vehicles $V$ and includes:
\textbf{OF Avg}: average OF value across ten runs,
\textbf{OF Std}: standard deviation of the OF value,
\textbf{GAP}: average optimality gap,
\textbf{\% Solved}: percentage of instances solved successfully within the time limit,
\textbf{Time Avg}: average solution time in seconds,
\textbf{Time Std}: standard deviation of the solution time.
A value of ``--'' indicates that no valid solution was obtained in any of the ten runs.}\label{tab:performance-comparison1}

% ------------------ ABF ------------------
\resizebox{\textwidth}{!}{
\begin{tabular}{c|cccc|cc|cccc|cc}
\toprule
\multicolumn{13}{c}{\textbf{ABF}} \\
\toprule
\multirow{2}{*}{$V$} & \multicolumn{6}{c|}{\textbf{Without AFGR}} & \multicolumn{6}{c}{\textbf{With AFGR}} \\
\cmidrule(lr){2-7} \cmidrule(lr){8-13}
 & OF Avg & OF Std & GAP & \% Solved & Time Avg & Time Std 
   & OF Avg & OF Std & GAP & \% Solved & Time Avg & Time Std \\
\midrule
4  & 965,69 & 0,00 & 0\% & 100\% & 11,78 & 0,28  & 965,69 & 0,00 & 0\% & 100\% & 11,99 & 1,10 \\
5  & 1387,73 & 124,50 & 28\% & 100\% & 12,33 & 0,26  & 1333,44 & 183,33 & 25\% & 100\% & 12,46 & 0,59 \\
6  & 1917,22 & 130,94 & 47\% & 30\%  & 13,55 & 0,60  & 1879,42 & 51,44 & 46\% & 60\%  & 13,39 & 0,43 \\
7  & -     & -     & -     & 0\%   & 15,12 & 0,32  & 2715,55 & 0,00 & 48\% & 10\%  & 15,73 & 1,77 \\
8  & -     & -     & -     & 0\%   & 17,48 & 0,93  & 2673,82 & 111,65 & 47\% & 20\%  & 14,87 & 0,25 \\
9  & -     & -     & -     & 0\%   & 20,49 & 0,58  & 2349,56 & 0,00 & 45\% & 10\%  & 15,32 & 0,76 \\
10 & -     & -     & -     & 0\%   & 24,90 & 0,60  & -     & -     & -     & 0\%   & 17,43 & 0,52 \\   
\midrule
\end{tabular}
}

\vspace{0.5cm}

% ------------------ NBF ------------------
\resizebox{\textwidth}{!}{
\begin{tabular}{c|cccc|cc|cccc|cc}
\toprule
\multicolumn{13}{c}{\textbf{NBF}} \\
\toprule
\multirow{2}{*}{$V$} & \multicolumn{6}{c|}{\textbf{Without AFGR}} & \multicolumn{6}{c}{\textbf{With AFGR}} \\
\cmidrule(lr){2-7} \cmidrule(lr){8-13}
 & OF Avg & OF Std & GAP & \% Solved & Time Avg & Time Std 
   & OF Avg & OF Std & GAP & \% Solved & Time Avg & Time Std \\
\midrule
4  & 965,69 & 0,00 & 0\%   & 100\% & 12,05 & 0,45  & 965,69 & 0,00 & 0\%   & 100\% & 11,24 & 2,52 \\
5  & 1390,17 & 248,32 & 28\%  & 90\%  & 13,41 & 1,36  & 1371,49 & 218,9 & 27\%  & 100\% & 16,43 & 11,89 \\
6  & 2079,42 & 183,23 & 51\%  & 60\%  & 15,00 & 2,62  & 1928,23 & 161,03 & 47\%  & 60\%  & 16,07 & 7,41 \\
7  & -    & -    & -      & 0\%   & 19,66 & 11,79 & 2273,17 & 251,44 & 38\%  & 20\%  & 16,77 & 3,91 \\
8  & -     & -     &  -     & 0\%   & 21,37 & 8,46  & 2616,4 & 0,00      & 45\%  & 10\%  & 15,81 & 1,06 \\
9  & -     & -   &  -     & 0\%   & 22,33 & 4,85  & 2508,5 & 119,37 & 48\%  & 30\%  & 15,95 & 1,38 \\
10 & -     & -     &  -     & 0\%   & 26,48 & 6,23  &     -  &   -    &    -   & 0\%   & 19,53 & 3,81 \\
\midrule
\end{tabular}}
\end{table}

\subsection{STSP-PD}\label{app:quantum_table_STSP_PD}
\begin{table}[H]
\centering
\caption{Results of the STSP-PD problem, with and without AFGR. Each row corresponds to a different instance size, defined by the number of vehicles $V$, and includes:
\textbf{OF Avg}: average OF value across ten runs,
\textbf{OF Std}: standard deviation of the OF value,
\textbf{GAP}: average optimality gap,
\textbf{\% Solved}: percentage of instances solved successfully within the time limit,
\textbf{Time Avg}: average solution time in seconds,
\textbf{Time Std}: standard deviation of the solution time.
A value of ``--'' indicates that no valid solution was obtained in any of the ten runs.}
\label{tab:performance-comparisonTW}

\resizebox{\textwidth}{!}{
\begin{tabular}{c|cccc|cc|cccc|cc}
\toprule
\multirow{2}{*}{$V$} & \multicolumn{6}{c|}{\textbf{Without AFGR}} & \multicolumn{6}{c}{\textbf{With AFGR}} \\
\cmidrule(lr){2-7} \cmidrule(lr){8-13}
 & OF Avg & OF Std & GAP & \% Solved & Time Avg & Time Std 
   & OF Avg & OF Std & GAP & \% Solved & Time Avg & Time Std \\
\midrule
4  & 965,69 & 0,00 & 0\%   & 100\% & 13,67 & 5,77  & 965,69 & 0,00  & 0\%   & 100\% & 12,31 & 1,18 \\
5  & 1198,58 & 180,58 & 17\%  & 100\% & 12,83 & 1,36  & 1100,7 & 141,43 & 10\%  & 100\% & 12,65 & 1,51 \\
6  & 1715,50 & 290,64 & 41\%  & 100\% & 14,21 & 1,65  & 1844,78 & 79,30 & 45\%  & 100\% & 15,72 & 4,95 \\
7  & 2523,90 & 166,58 & 60\%  & 100\% & 17,62 & 8,59  & 2517,81 & 248,58 & 60\%  & 100\% & 16,26 & 4,70 \\
8  & 3185,47 & 60,52 & 69\%  & 30\%  & 18,13 & 3,68  & 2614,67 & 165,32 & 63\%  & 100\% & 15,10 & 1,31 \\
9  & -     & -     & -     & 0\%   & 26,21 & 6,63 & 2419,53 & 292,51 & 61\%  & 80\%  & 16,95 & 5,78 \\
10 & -     & -     & -     & 0\%   & 24,91 & 3,35  & 3230,04 & 288,76 & 71\%  & 30\%  & 17,21 & 1,02 \\
11 & -     & -     & -     & 0\%   & 24,91 & 6,41  & -     & -     & -     & 0\%  & 18,35 & 4,23 \\
\bottomrule
\end{tabular}}
\end{table}

\subsection{STSP-TW}\label{app:quantum_table_STSP_TW}
\begin{table}[H]
\centering
\caption{Performance of the STSP-TW problem, evaluated with and without the application of AFGR. The table reports different instance sizes, defined by the number of vehicles $V$, and includes:
\textbf{OF Avg}: average OF value across ten runs,
\textbf{OF Std}: standard deviation of the OF value,
\textbf{GAP}: average optimality gap,
\textbf{\% Solved}: percentage of instances solved successfully within the time limit,
\textbf{Time Avg}: average solution time in seconds,
\textbf{Time Std}: standard deviation of the solution time.
A value of ``--'' indicates that no valid solution was obtained in any of the ten runs.}
\label{tab:performance-comparisonPD}

\resizebox{\textwidth}{!}{
\begin{tabular}{c|cccc|cc|cccc|cc}
\toprule
\multirow{2}{*}{$V$} & \multicolumn{6}{c|}{\textbf{Without AFGR}} & \multicolumn{6}{c}{\textbf{With AFGR}} \\
\cmidrule(lr){2-7} \cmidrule(lr){8-13}
 & OF Avg & OF Std & GAP & \% Solved & Time Avg & Time Std 
   & OF Avg & OF Std & GAP & \% Solved & Time Avg & Time Std \\
\midrule

4  & 915,98 & 42,78 & 4\%   & 100\% & 31,25 & 21,14 & 899,41 & 34,93 & 2\%   & 100\% & 32,23 & 14,82 \\
5  & 1379,81 & 167,55 & 28\%  & 100\% & 54,48 & 43,42 & 1286,42 & 195,08 & 23\%  & 100\% & 39,15 & 20,65 \\
6  & 2077,03 & 130,95 & 51\%  & 60\%  & 30,48 & 20,75 & 2070,27 & 146,15 & 51\%  & 80\%  & 38,53 & 43,52 \\
7  & 2590,66 & 249,51 & 60\%  & 30\%  & 48,51 & 39,63 & 2572,72 & 206,69 & 60\%  & 50\%  & 45,62 & 33,47 \\
8  & -     & -     & -     & 0\%   & 35,69 & 12,13 & 2890,74 & 520,8 & 57\% & 40\%  & 25,97 & 6,13 \\
9  & -     & -     & -     & 0\%   & 21,42 & 1,67  & 2471,66 & 71,11 & 47\% & 30\%  & 19,99 & 1,86 \\
10 & -     & -     & -     & 0\%   & 36,38 & 6,08  & -     & -     & -     & 0\%   & 24,71 & 3,51 \\

\bottomrule
\end{tabular}}
\end{table}

\clearpage

\bibliographystyle{elsarticle-num-names} 
  \bibliography{bibliography}

\end{document}